\documentclass[12pt,a4paper]{article}
\pdfoutput=1
\usepackage{graphicx,epsfig}
\usepackage{dcolumn} 
\usepackage{slashed,color,amsmath,amssymb}
\usepackage{jheppub}
\usepackage{enumitem}
\usepackage{slashbox}

\usepackage{caption, subcaption}
\usepackage{multirow}
\usepackage{bigcenter}
\usepackage{verbatim}
\newcommand{\be}{\begin{equation}}
\newcommand{\ee}{\end{equation}}
\newcommand{\bea}{\begin{eqnarray}}
\newcommand{\eea}{\end{eqnarray}}
\newcommand{\bit}{\begin{itemize}}
\newcommand{\eit}{\end{itemize}}

\def\gsim{\lower0.5ex\hbox{$\:\buildrel >\over\sim\:$}}
\def\lsim{\lower0.5ex\hbox{$\:\buildrel <\over\sim\:$}}

\hypersetup{
   colorlinks=true,       
   linkcolor=blue,        
   citecolor=red,         
   filecolor=magenta,     
}

\usepackage{bm}

\preprint{}

\title{Long-Lived Light Mediators from Higgs boson Decay at HL-LHC, FCC-hh and a Proposal of Dedicated LLP Detectors for FCC-hh}

\author{Biplob Bhattacherjee$^1$, Shigeki Matsumoto$^2$, Rhitaja Sengupta$^1$}

\affiliation{\vspace*{0.1in}$^1$ Centre for High Energy Physics, Indian Institute of Science, Bengaluru 560012, India}
\affiliation{\vspace*{0.1in}$^2$ Kavli IPMU (WPI), UTIAS, University of Tokyo, Kashiwa, Chiba, 277-8583, Japan}

\emailAdd{biplob@iisc.ac.in}
\emailAdd{shigeki.matsumoto@ipmu.jp}
\emailAdd{rhitaja@iisc.ac.in}

\abstract{
We study the pair production of the long-lived mediator particles from the decay of the SM Higgs boson and their subsequent decay into standard model particles. We compute the projected sensitivity, both model-independently and with a minimal model, of using the muon spectrometer of the CMS detector at the HL-LHC experiment for ggF, VBF, and Vh production modes of the Higgs boson and various decay modes of the mediator particle, along with dedicated detectors for LLP searches like CODEX-b and MATHUSLA. Subsequently, we study the improvement with the FCC-hh detector at the 100\,TeV collider experiment for such long-lived mediators, again focusing on the muon spectrometer. We propose dedicated LLP detector designs for the 100\,TeV collider experiment, DELIGHT (\textbf{De}tector for \textbf{l}ong-l\textbf{i}ved particles at hi\textbf{gh} energy of 100\,\textbf{T}eV), and study their sensitivities.
}

\begin{document}

\maketitle


\section{Introduction}
\label{sec:intro}

The standard model (SM) of particle physics has stood the test of time since experiments like the large hadron collider (LHC) experiment have not found any significant deviations from its predictions as of yet. Still, we have concrete reasons to believe that we need to go beyond the SM (BSM) to address many important open questions, of which one of the most pressing questions is the nature of dark matter (DM). The problem of a large amount of missing mass in our universe, {\it a.k.a.} DM, has baffled scientists for more than eight decades now, and particle physicists, both theorists and experimentalists, are working hand in hand to unravel this mystery. A vast variety of DM candidates arising from a multitude of new physics models, like supersymmetry, have been proposed by the theorists\,\cite{Jungman:1995df}. Along with this a large number of experimental collaborations are making enormous efforts to search for these candidates in direct, indirect as well as collider experiments\,\cite{Bertone:2004pz}, and even the negative results are successful in ruling out regions of parameter space of many attractive theories.

Among all the dark matter candidates, the weakly interacting massive particles (WIMPs) which were in thermal equilibrium with SM particles in the early universe are highly motivated since they can explain the present observed relic abundance of DM by the freeze-out mechanism\,\cite{Lee:1977ua}. WIMPs can have masses ranging from $\mathcal{O}(1)$\,MeV\,\cite{Giovanetti:2021izc} to $\mathcal{O}(100)$\,TeV\,\cite{Griest:1989wd}, however collider searches from the LHC experiment and the direct detection experiments have started excluding WIMP masses near the electroweak scale\,\cite{Arcadi:2017kky}. This has brought the lighter (less than a few tens of GeV\,\cite{Battaglieri:2017aum}) and heavier (greater than a TeV\,\cite{Cirelli:2005uq}) WIMPs into the limelight. With the increasing centre of mass energy of colliders, as well as the increasing amount of data that is being collected in direct and collider experiments, we might be able to gain some sensitivity for the heavier WIMPs\,\cite{FCC:2018vvp, DARWIN:2016hyl}. On the other hand, the fact that lighter WIMPs have evaded our experiments till now motivates their search further, and it incites us to search for the so-called light mediator particle, which is predicted in many scenarios for light WIMPs and connecting dark sector particles (i.e.,  dark matter) with those of the SM. For instance, focusing on light fermionic WIMPs, the minimal renormalizable model consistent with Lorentz symmetry requires a bosonic mediator particle connecting the dark matter with the SM\cite{Patt:2006fw,Kim:2006af,Pospelov:2007mp,Pospelov:2008jd,Kanemura:2010sh,Kanemura:2011nm,Pospelov:2011yp,Baek:2011aa,Djouadi:2011aa,Lopez-Honorez:2012tov,Esch:2013rta,Ghorbani:2014qpa,Dutra:2015vca,Freitas:2015hsa,Ghorbani:2016edw,Kamada:2018zxi,Beniwal:2018hyi,Kamada:2018kmi,Matsumoto:2018acr}, where the latter has to be as light as the WIMP to explain the dark matter relic abundance. Moreover, from the viewpoint of the small-scale problems of structure formation in our universe\,\cite{Bullock:2017xww}, a light fermionic WIMP with a light bosonic mediator enjoys particular attention, since it can potentially solve the problem by generating a large and velocity-dependent scattering cross section between the light WIMPs\,\cite{Tulin:2017ara}.

Among various possibilities of the mediator particle, in this paper, we focus on the scalar mediator particle, which has a mixing with the SM Higgs boson. The LHC experiment has searched for such mediators and has pushed the allowed values of a mixing between the SM Higgs boson and the scalar mediator to be small\footnote{For instance, in the minimal model presented in section\,\ref{ssec:decay-med}, the mixing angle squared is constrained to be less than 10$^{-6}$--10$^{-7}$ when the mediator particle is lighter than a few GeV, which is obtained by e.g. the measurement of rare B meson decays at B-factories.\,\cite{Beacham:2019nyx}. Moreover, the mixing angle squared is universally required to be less than $10^{-1}$ without respect to the mediator mass, as a result of precision measurements of the SM Higgs boson at the LHC experiment\,\cite{ATLAS-CONF-2020-027,CMS-PAS-HIG-19-005}.}. Among all the production modes of such a mediator, the one from the decay of an on-shell SM Higgs boson will be the most important for very small values of mixing, because this production does not depend on the mixing. Therefore, we restrict ourselves to this production mode in this paper. The mediator can then decay into SM particles through its mixing with the Higgs boson, and going to smaller values of mixing will make the mediator particle long-lived, provided it is lighter than the WIMP\footnote{Interaction between the mediator particle and the WIMP is in general not suppressed by the mixing, unlike those between the mediator and SM particles. So, if the mediator particle is enough heavier than the WIMP, it could decay into WIMPs and will not have a longer lifetime.}. The branching of the SM Higgs boson decay into the mediator particles is constrained to be below $\sim$ 10\,\%\,\cite{Arcadi:2021mag}, which is projected to come down to $1.9\%$ at HL-LHC\,\cite{deBlas:2019rxi}\footnote{There also exists direct search limits for the decay of SM Higgs boson into a pair of mediator particles which promptly decay into SM leptons. For recent references, please refer Refs\,\cite{ATLAS:2021ldb,cmscollaboration2021search}.}.

LHC detectors were designed to be optimal for new physics particles which decay promptly. However, as we have just witnessed, many BSM theories can predict long-lived particles (LLPs), such as supersymmetry, Higgs-portal, gauge-portal, dark matter, and heavy neutrino theories. In order to ensure that we do not miss out on any signal of new physics, it is essential to make our analyses inclusive to the lifetime frontier, and this has recently gained a lot of attention\,\cite{Alimena:2019zri}. Experimental collaborations, like ATLAS, CMS and LHCb, are making steady efforts in developing search analyses for LLPs covering a wide variety of signatures\,\cite{ATLAS:2015xit,ATLAS:2018rjc,ATLAS:2018niw,ATLAS:2018tup,ATLAS:2019fwx,ATLAS:2019tkk,ATLAS:2019jcm,ATL-PHYS-PUB-2019-002,ATLAS:2020xyo,ATLAS-CONF-2021-032,ATLAS:2021jig,CMS:2014hka,CMS:2017kku,CMS:2018bvr,CMS-PAS-FTR-18-002,CMS:2019zxa,CMS:2020atg,CMS-PAS-EXO-19-021,CMS:2021juv,CMS:2021kdm,CMS:2021yhb,LHCb:2016buh,LHCb:2016inz,LHCb:2016awg,LHCb:2017xxn,LHCb:2019vmc,LHCb:2020akw}. Along with this, there is a host of proposals for dedicated LLP detectors, such as FASER\,\cite{Feng:2017uoz,Boiarska:2019vid}, MATHUSLA\,\cite{Chou:2016lxi,Curtin:2018mvb,Alpigiani:2020iam}, CODEX-b\,\cite{Gligorov:2017nwh,Aielli:2019ivi}, ANUBIS\,\cite{Bauer:2019vqk} detectors, and Forward Multi-particle Spectrometer\,\cite{Albrow:2020duu}, which contribute significantly in probing the lifetime frontier. A lot of phenomenological studies also explore the possibilities of LLP searches spanning a wide variety of models and signatures, such as those in Refs.\,\cite{Ilten:2015hya,Gago:2015vma,Banerjee:2017hmw,KumarBarman:2018hla,Bhattacherjee:2019fpt,CidVidal:2019urm,Banerjee:2019ktv,Jones-Perez:2019plk,Gershtein:2020mwi,Fuchs:2020cmm,Chakraborty:2020cpa,Cheung:2020ndx,Liu:2020vur,Evans:2020aqs,Linthorne:2021oiz,Alimena:2021mdu}.

Present upper limits on the branching of SM Higgs boson to a long-lived scalar mediator particle is around 4\% for a mediator mass of 45\,GeV at proper decay length of 1-2\,cm using the ATLAS inner detector at 13\,TeV and 139\,fb$^{-1}$, when it decays into $b$ quarks\,\cite{ATLAS:2021jig}, and around 1\% for 40\,GeV mediator at proper decay length of 1\,m decaying into hadronic jets in the ATLAS muon spectrometer with 33\,fb$^{-1}$ of data\,\cite{ATLAS:2019jcm}. The CMS detector constrains the SM Higgs boson branching to long-lived particles to be below 6\% for 40\,GeV mediator with proper decay lengths between 1–10\,cm which decays into SM quarks within the CMS tracker using the associated production of the Higgs boson with the $Z$ boson\,\cite{CMS:2021yhb} and 117\,fb$^{-1}$ of data. This constraint goes down to 0.3\% for the same mediator mass around proper decay lengths of 1\,m when displaced activities are searched for in the CMS muon spectrometer with an integrated luminosity of 137\,fb$^{-1}$\,\cite{CMS:2021juv}. When the mediator particle decays into electrons or muons, CMS has ruled out the branching fractions greater than 0.03\% for the SM Higgs boson decaying into two long-lived scalar particles of mass 30\,GeV having proper decay lengths ranging from 0.1-12\,cm with 113-118\,fb$^{-1}$\,\cite{CMS:2021kdm}.

Our endeavour is to study extensively the long-lived scalar mediator particle produced from the SM Higgs boson decay, which decays into various SM particles. We perform our analyses for the high luminosity run of the LHC (HL-LHC) experiment, where the increased luminosity comes with a high amount of pile-up (PU) vertices per bunch crossing. Minimising the effect of PU motivates us to look for decays of the long-lived mediator particle in the muon spectrometer of the CMS detector. For all cases, we consider the three dominant production modes of the SM Higgs boson; the gluon-gluon fusion (ggF), vector boson fusion (VBF) and associated production with a vector boson (Vh) processes. We first perform a model-independent analysis assuming 100\,\% branching to each decay mode of the mediator particle into SM particles, which we later combine those using the branching ratios predicted in the minimal model. Due to the presence of multiple production modes of the Higgs boson, and various decay modes of a scalar mediator particle which mixes with the Higgs boson, a study of individual production and decay mode is required to understand how each of them affect the combined limits. We also study the complementarity of this with dedicated LLP detectors, like MATHUSLA and CODEX-b detectors, for long-lived mediator particles arising in the scenario that we discussed above. 

We take a step in the future to study the prospect of the muon spectrometer of the future circular hadron collider (FCC-hh) experiment, which is proposed to have a centre of mass energy of 100\,TeV, in the search for scalar mediator particles having longer lifetimes and coming from the Higgs boson decay. It is to be noted that the muon spectrometer of the FCC-hh detector is different from the one at the CMS detector, i.e. it has a forward muon spectrometer in addition to the barrel and endcap detectors. Therefore, a dedicated study is required to understand how this design affects the search of LLPs coming from the decay of the SM Higgs boson.

Since the design of the 100\,TeV collider and detector are all currently under development, we take this opportunity to perform some studies of dedicated LLP detector designs in the periphery of the 100\,TeV collider, which can aid in increasing the sensitivity of the long-lived mediator particles that we are concerned with in this paper. We propose a dedicated LLP detector at the 100\,TeV collider experiment, DELIGHT (\textbf{De}tector for \textbf{l}ong-l\textbf{i}ved particles at hi\textbf{gh} energy of 100 \textbf{T}eV), with a few variations of dimensions, and study the prospect of these detectors in enhancing the sensitivities of long-lived particles produced at the 100\,TeV collider from the SM Higgs boson decay, in the three dominant production modes of the Higgs boson.

The rest of the paper is organised as follows: In section\,\ref{sec:model}, we provide a brief discussion on the production as well as decay of a light scalar mediator particle, well-motivated from the standpoint of the DM problem, and the motivation behind focusing on the long-lived parameter space of the mediator particle, and also a description of the minimal model. Next, in section\,\ref{sec:higgs-lhc}, we begin by outlining our simulation and computational setup along with the analysis strategies for finding displaced activities using the CMS muon spectrometer for each of the decay modes of the mediator particle in a model-independent way, along with a combination of them following the minimal model. In the latter part of this section, we present the results of MATHUSLA and CODEX-b analyses, and their complementarity with the CMS muon spectrometer analysis. Section\,\ref{sec:higgs-100TeV} discusses the sensitivity of the 100\,TeV collider experiment and the proposed DELIGHT LLP detectors accompanying the FCC-hh detectors for a light scalar mediator particle from the Higgs boson decay. Finally, we summarise and conclude the findings of our study in section\,\ref{sec:summary}.

\section{The scalar mediator}
\label{sec:model}

A mediator particle which is singlet under the standard model (SM) gauge symmetries is predicted in many dark sector scenarios. It plays a role in connecting dark sector particles to those of the SM. Among various possibilities for the mediator particle, in this paper, we focus on the (real) scalar mediator whose mass is smaller enough than the electroweak scale and decay width is small enough to make it long-lived. Some properties of the mediator particle, how it could be produced at high energy colliders and how it could decay into SM particles, are summarized below.

\subsection{Production of the mediator particle at high energy colliders}
\label{ssec:prod-med}

From the dimensional analysis, lower dimensional interactions between the scalar mediator and SM particles are from $\Phi |H|^2$ and $\Phi^2 |H|^2$ with $\Phi$ and $H$ being fields describing the SM singlet mediator and the Higgs doublet in the SM, respectively.

After the electroweak symmetry breaking, the former interactions cause mixing between $\Phi$ and $H$, with the mass eigenstates and hence, the physical states, being the mediator particle ($\phi$), and the SM-like discovered Higgs boson ($h$). Thus, it induces various processes for the mediator production production at high energy collider experiments: the gluon fusion process ($g g \to \phi$), the vector boson fusion process ($q q \to q q \phi$, $e e \to e e \phi$, etc.) and the mediator-strahlung process ($q \bar{q} \to V \phi$, $e^- e^+ \to V \phi$) with $q$, $e$ and $V$ being the quark, electron and vector boson, respectively. Moreover, the mediator particle can also be produced by the decay of various mesons if it is very light, such as $\Upsilon \to \gamma \phi$, $B \to K \phi$, $K \to \pi \phi$, etc., with $\Upsilon$, $B$, $K$ and $\pi$ being Upsilon, B, K, and $\pi$ mesons, respectively. On the other hand, the mixing has already been severely constrained by various experiments, including those of the Higgs precision measurement at the LHC experiment\,\cite{Bechtle:2014ewa}, and hence the production of the mediator particle via the above processes is not very much efficient.

On the other hand, the latter interaction, $\Phi^2 |H|^2$, is not severely constrained so far, as it must be accompanied by an on-shell Higgs boson to probe it sensitively. Indeed, when the mass of the mediator particle is lighter than half of the Higgs boson mass, the interaction induces the Higgs boson to decay into a pair of mediator particles, i.e. $h \to \phi \phi$. Though the decay is currently not severely constrained, it is expected to be intensively searched for in the (near) future at the HL-LHC, 100\,TeV collider and future lepton collider experiments. Hence, in this paper, we focus on the search of the scalar mediator particle at the hadron collider experiments via the $h \to \phi \phi$ process, assuming that the hidden sector particles and the mediator particle do not alter the Higgs boson production process significantly, as most of the hidden dark sector scenarios predict. We therefore consider usual SM processes for the Higgs boson production, i.e., the gluon fusion process (ggF, $g g \to h$), the vector boson fusion process (VBF, $q q \to q q h$) and the Higgs-strahlung process (Vh, $q \bar{q} \to V h$), at the HL-LHC as well as the 100\,TeV collider experiments\,\cite{higgsxs}.

\subsubsection*{The mediator particle from Higgs boson decay at the colliders:}

Here, we discuss the production of the mediator particle at the 14\,TeV HL-LHC and the 100\,TeV collider experiments in more details, i.e., the transverse momentum and pseudo-rapidity distributions, when it is produced from the decay of the SM Higgs boson. The SM Higgs boson, as discussed above, has multiple production channels at the hadron colliders, of which we study the ggF, VBF, and Vh production channels in this paper. The cross-sections of these production channels for the 14 TeV and 100 TeV colliders, as taken from Ref.\,\cite{higgsxs}, has been summarised in Table\,\ref{tab:cs}.

\begin{table}[t]
    \centering
    \begin{tabular}{c|c|c|}
    $\sqrt{s}$\,[TeV] & Process & Cross section\,[pb] \\
    \hline
    \multirow{3}{*}{14} & ggF & 50.35 \\
					    & VBF & 4.172 \\
					    &  Vh & 2.387 (Wh:1.504, Zh:0.8830) \\
    \hline
    \multirow{3}{*}{100} & ggF & 740.3 \\
					     & VBF & 82.00 \\
					     &  Vh & 27.16 (Wh:15.90, Zh:11.26) \\
    \hline 
    \end{tabular}
    \caption{\sl \small The cross sections of the SM Higgs boson production for the center of mass energies $\sqrt{s} =$ 14\,TeV and $\sqrt{s} =$ 100\,TeV. These numbers have been taken from Ref.\,\cite{higgsxs}.}
    \label{tab:cs}
\end{table}

\begin{figure}[t]
    \centering
    \includegraphics[width=\textwidth]{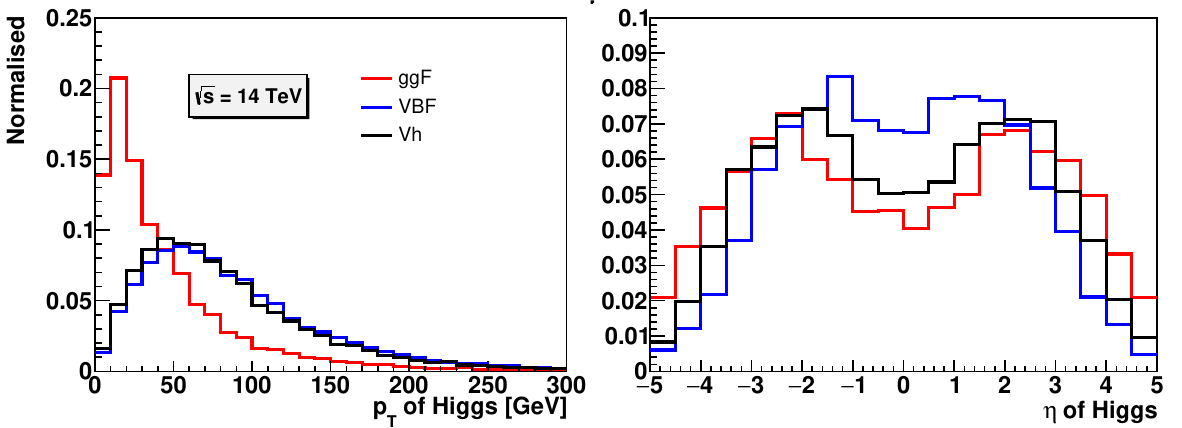}
    \caption{\small \sl Normalised distributions of $p_T$ (left panel) and $\eta$ (right panel) of the SM Higgs boson produced through the ggF, VBF and Vh production channels at $\sqrt{s} =$ 14\,TeV.}
\label{fig:higgs-dist-14TeV}
\end{figure}

\begin{figure}[t]
    \centering
    \includegraphics[width=\textwidth]{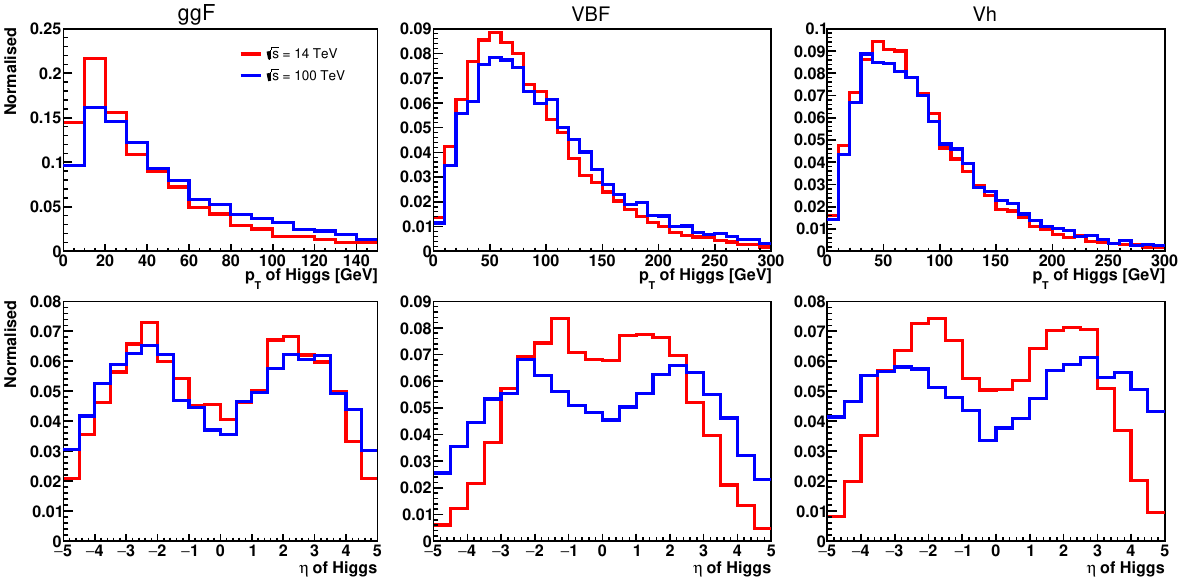}
    \caption{\small \sl Comparison of the $p_T$ (top panels) and $\eta$ (bottom panels) distributions of the SM Higgs boson produced via the ggF (left panels), VBF (centre panels) and Vh (right panels) production channels at the $\sqrt{s} =$ 14\,TeV and 100\,TeV collider experiments.}
    \label{fig:higgs-dist-14TeV-100TeV}
\end{figure}

A discussion on the transverse momentum, $p_T$, and the pseudo-rapidity, $\eta$, distributions of the SM Higgs boson in the three production channels mentioned above is a good starting point; it affects the boost of the scalar mediator particle in the laboratory frame and its detector acceptance. For generating the Higgs boson in various production channels and its subsequent decay into mediator particles with varying masses, $m_{\phi}$, we have used the \texttt{PYTHIA6} code\,\cite{Sjostrand:2006za}. Fig.\,\ref{fig:higgs-dist-14TeV} shows the normalised $p_T$ (left panel) and $\eta$ (right panel) distributions of the SM Higgs boson produced via the ggF, VBF and Vh production channels at $\sqrt{s} = 14$\,TeV. We observe that the Higgs boson has the least transverse momentum and the most forward production in the ggF production channel, and it is followed by the Vh and VBF channels, respectively. On comparing the distributions of the SM Higgs boson at the 14 TeV and 100 TeV colliders, in Fig.\,\ref{fig:higgs-dist-14TeV-100TeV}, we find that the Higgs boson has slightly higher $p_T$ values and more forward $\eta$ values for the 100 TeV collider in all production channels.

\begin{figure}[t]
    \centering
    \includegraphics[width=\textwidth]{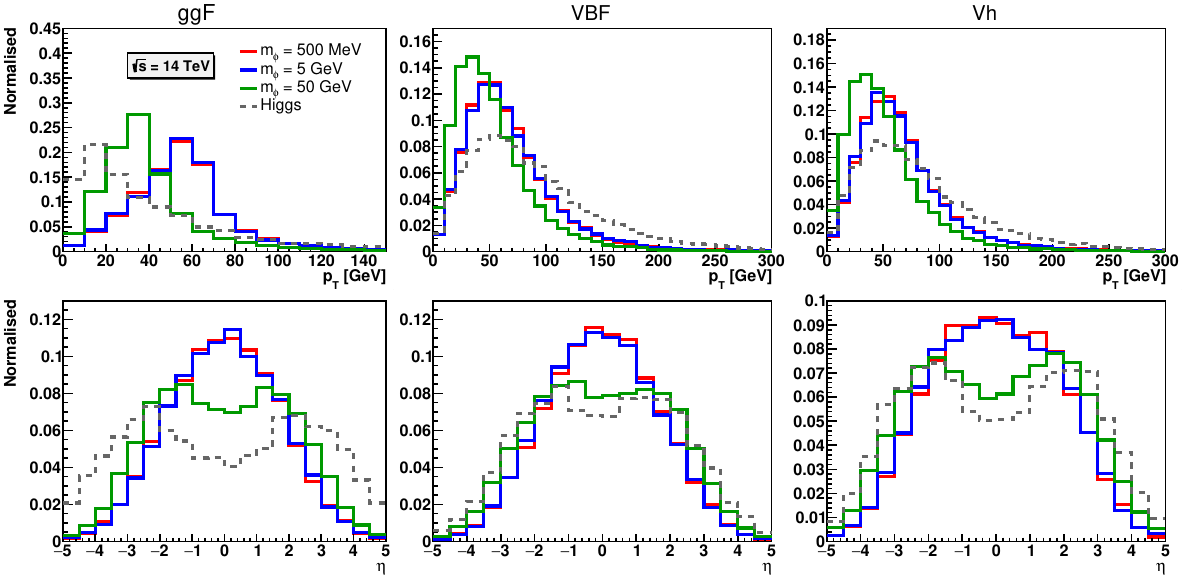}
    \caption{\small \sl Normalised $p_T$ (top panels) and $\eta$ (bottom panels) distributions of the mediator particle from the decay of the Higgs boson, for three values of the mediator mass ($m_\phi =$ 500\,MeV, 5\,GeV and 50\,GeV), along with the latter for comparison in the ggF (left panels), VBF (centre panels) and Vh (right panels) production channels at $\sqrt{s} =$ 14\,TeV.}
    \label{fig:LLPs-dist-14TeV}
\end{figure}

We now look at the distributions of the scalar mediator particle, which is produced from the decay of the Higgs boson. Fig.\,\ref{fig:LLPs-dist-14TeV} shows the normalised distributions of $p_T$ (top panels) and $\eta$ (bottom panels) of the mediator particle from the decay of the Higgs boson, for three values of the mediator mass ($m_\phi =$ 500\,MeV, 5\,GeV and 50\,GeV) in the ggF (left panels), VBF (centre panels) and Vh (right panels) production channels at $\sqrt{s} = 14$\,TeV. The corresponding distributions for the Higgs boson are also shown (dashed gray lines) for comparison. With the increasing mass of the mediator particle, its transverse momentum decreases and produces more in the forward direction; it follows the SM Higgs boson distributions more closely than the lighter mediator particles due to the smaller phase-space availability in the former.

\subsection{Decay of the mediator particle}
\label{ssec:decay-med}

Once the mediator particle is produced at the hadron colliders, it could decay into SM particles. Assuming that its flavor-changing decays and those into hidden sector particles are suppressed and/or forbidden, decay processes into SM particles that could dominate the decay are $\phi \to b \bar{b}$, $c \bar{c}$, $s \bar{s}$, $d \bar{d}$, $u \bar{u}$, $e^- e^+$, $\mu^- \mu^+$, $\tau^- \tau^+$, $g g$ and $\gamma \gamma$, with $b$, $c$, $s$, $d$, $u$, $e$, $\mu$, $\tau$, $g$ and $\gamma$ being $b$, $c$, $s$, $d$, $u$ quarks, electron, muon, tau lepton, gluon and photon, respectively.\footnote{Since we focus on the $h \to \phi \phi$ process for the production of the mediator particle, the mass of the mediator particle is assumed to be less than half of the Higgs boson mass, i.e. $m_\phi \leq m_h/2$.} The use of the decay processes into light quarks and gluons are valid if the mediator mass ($m_\phi$) is larger enough than the hadronization scale, say $m_\phi \gtrsim 2$\,GeV. If the mediator mass is less than this value, the decay processes should be expressed by those into mesons: $\phi \to \pi \pi$, $KK$ and other hadronic ($\eta$ and $\rho$, etc.) processes such as $\phi \to \eta \eta$, $\rho \rho$ and $4\pi$ (four pions).

The above decay processes are described by several higher-dimensional interactions from the viewpoint of the effective theory including the mediator and SM particles: $\Phi\,\bar{Q}_i H^c U_i/\Lambda_{u_i}$, $\Phi\,\bar{Q}_i H D_i/\Lambda_{d_i}$, $\Phi\,\bar{L}_i H E_i/\Lambda_{e_i}$, $\Phi\,G^a_{\mu \nu} G^{a \mu \nu}/\Lambda_g$ and $\Phi\,F_{\mu \nu} F^{\mu \nu}/\Lambda_\gamma$ with $Q_i$, $U_i$, $D_i$, $L_i$, $E_i$ being the quark doublet, up-type quark singlet, down-type quark singlet, lepton doublet, charged lepton singlet fields of the $i$-th generation, respectively, while $G^a_{\mu \nu}$ and $F_{\mu \nu}$ are gluon and photon field strength tensors.\footnote{Though it is better to use the interaction $\Phi\,B_{\mu \nu} B^{\mu \nu}$ instead of $\Phi\,F_{\mu \nu} F^{\mu \nu}$ with $B_{\mu \nu}$ being the field strength tensor of the hyper-charged gauge boson to make the effective theory invariant under the SM gauge symmetries, we have used the latter interaction $\Phi\,F_{\mu \nu} F^{\mu \nu}$ in the text for it is more directly related to the physical quantity, i.e. the decay width of the mediator into two photons.}  Then, it turns out that the mediator becomes a long-lived particle, if the (effective) cut-off scales of the interactions ($\Lambda$s) are large. In particular, when we impose a condition inspired by the minimal flavor violation on the interactions that flip the chirality of SM fermions\,\cite{DAmbrosio:2002vsn, Colangelo:2008qp}, the cutoff scales are inversely proportional to corresponding Yukawa couplings, i.e. $\Lambda_{u_i} \propto y_{u_i}^{-1}$, $\Lambda_{d_i} \propto y_{d_i}^{-1}$ and $\Lambda_{e_i} \propto y_{e_i}^{-1}$. It makes the mediator particle further long-lived when it decays mainly into the light SM fermions.

We consider the cases that the mediator particle decays into a pair of muons, light mesons, quarks and gluons ($\pi\pi$, $KK$, $s \bar{s}$ and $g g$), as well as charm quarks ($c \bar{c}$), tau leptons and bottom quarks ($b \bar{b}$).\footnote{We do not consider the cases that $\phi$ decays mainly into a pair of photons ($\gamma \gamma$), electrons ($e^- e^+$) and $u/d$ quarks ($u \bar{u}$, $d \bar{d}$), for the processes tend to be sub-dominant in most of dark sector scenarios when $\phi$ is not very light, such as $m_\phi \geq 2 m_\mu$ with $m_\mu$ being the muon mass.} We analyse each case individually with the mass and the decay length of the mediator particle ($m_\phi$ and $c \tau_\phi$) being free parameters assuming that the mediator decays into each specific final state with 100\,\% branching fraction. We present the result of the analysis quantitatively based on this strategy; for example, we will show the parameter region in the $(c \tau_\phi,\,m_\phi)$-plane that can be probed at the future hadron collider experiments for all the above cases.

\subsubsection*{Minimal model of the scalar mediator:}

In addition to the above strategy, we also consider the minimal model of the (real) scalar mediator to present the result of the analysis. This model is often referred for the scalar mediator particle and allows us to predict a branching fraction for each decay mode mentioned above. The model is defined by the following Lagrangian:
\begin{equation}
    {\cal L} = {\cal L}_{\rm SM} + \frac{1}{2} (\partial_\mu \Phi)^2
    - A_{\Phi H} \Phi |H|^2 - \frac{\lambda_{\Phi H}}{2} \Phi^2 |H|^2
	- \mu_1^3 \Phi - \frac{\mu_\Phi^2}{2} \Phi^2 - \frac{\mu_3}{3!} \Phi^3 - \frac{\lambda_\Phi}{4!} \Phi^4 + {\cal L}_{\rm DS},
	\label{eq: minimal model}
\end{equation}
where ${\cal L}_{\rm SM}$ is the SM Lagrangian, while ${\cal L}_{\rm DS}$ is the Lagrangian for dark sector particles including their interactions with $\Phi$. After the electroweak symmetry breaking and taking the unitary gauge, i.e. $H = (0, v_H + h')^T/\sqrt{2}$ with $v_H \simeq 246$\,GeV being the vacuum expectation value of $H$, while $\Phi = v_\Phi + \phi'$ with $v_\Phi = 0$ being that of $\Phi$,\footnote{We can take the vacuum expectation value of the mediator field to be zero without the loss of generality. Then, this condition gives the relation among the parameters; $2\mu_1^3 + \mu_{\Phi H} v_H^2 = 0$.} $H$ is mixed with $\Phi$. Then, the quadratic terms of $h'$ and $\phi'$ in the Lagrangian is\footnote{The contribution from the dark sector particles to the mixing is assumed to be negligibly small.}
\begin{align}
    {\cal L} \supset
	-\frac{1}{2}
	\begin{pmatrix} h' & \phi' \end{pmatrix}
	\begin{pmatrix} m^2_{h' h'} & m^2_{h' \phi'} \\ m^2_{h' \phi'} & m^2_{\phi' \phi'} \end{pmatrix}
	\begin{pmatrix} h' \\ \phi' \end{pmatrix}
	=
	-\frac{1}{2}
	\begin{pmatrix} h & \phi \end{pmatrix}
	\begin{pmatrix} m_h^2 & 0 \\ 0 & m_\phi^2 \end{pmatrix}
	\begin{pmatrix} h \\ \phi \end{pmatrix},
\end{align}
where $m_{h' h'}^2 = \lambda_H v_H^2$, $m_{h' \phi'}^2 = A_{\Phi H} v_H$ and $m_{\phi' \phi'}^2 = \mu_\Phi^2 + \lambda_{\Phi H} v_H^2/2$, respectively, with $\lambda_H$ being the quartic Higgs coupling in the SM, i.e. $-(\lambda_H/2)(|H|^2 - v_H^2/2)^2 \subset {\cal L}_{\rm SM}$. The mixing matrix diagonalizing the mass matrix and connecting the states is
\begin{align}
	\begin{pmatrix} h \\ \phi \end{pmatrix} =
	\begin{pmatrix} \cos\theta & -\sin\theta \\ \sin\theta & \cos\theta \end{pmatrix}
	\begin{pmatrix} h' \\ \phi' \end{pmatrix}.
\end{align}
Mass eigenstates and mixing angle are given by  $m_h^2,\,m_\phi^2 = [(m_{h' h'}^2 + m_{\phi' \phi'}^2) \pm \{(m_{h' h'}^2 - m_{\phi' \phi'}^2)^2 + 4m_{h' \phi'}^2\}^{1/2}]/2$ and $\tan (2\theta) = 2m_{h' \phi'}^2/(m_{\phi' \phi'}^2 - m_{h' h'}^2)$. Then, the mediator and the Higgs boson interacts with SM fermions and gauge bosons as follows:\footnote{Other interactions among SM fermions and gauge bosons, i.e. gauge interactions of SM fermions and self-interactions of SM gauge bosons, are not altered from the SM. In addition to the interactions addressed so far, there are various scalar interactions among the mediator and the Higgs boson.}
\begin{equation}
    {\cal L}_{\rm int}
	=
	-\frac{s_\theta \phi + c_\theta h}{v_H} \sum_f m_f \bar{f}f
	+[\frac{s_\theta \phi + c_\theta h}{v_H} + \frac{(s_\theta \phi + c_\theta h)^2}{2v_H^2}] (2m_W^2 W_\mu^\dagger W^\mu + m_Z^2 Z_\mu Z^\mu),
	\label{eq: interactions}
\end{equation}
where $c_\theta\equiv \cos\theta$, $s_\theta\equiv \sin\theta$. Here, SM fermions (quarks and leptons) are denoted by $f$ with $m_f$ being its mass, while $W_\mu$ and $Z_\mu$ are charged and neutral weak bosons ($W$ and $Z$ bosons) in the SM with $m_W$ and $m_Z$ being their masses, respectively.

As addressed in the previous subsection, the mixing angle $\theta$ is constrained to be enough less than one, so that the interactions of the Higgs boson with SM fermions and gauge bosons shown in eq.\,(\ref{eq: interactions}) are not sizeably affected by the mixing. Hence, the productions of the Higgs boson at  high energy colliders are not significantly altered compared to those predicted in the SM. On the other hand, the same mixing angle does suppress the interactions of the mediator particle with SM fermions and gauge bosons as also shown in eq.\,(\ref{eq: interactions}), and in addition to the existing prompt mediator phase space, it can extend the parameter space to include a long-lived mediator, where it has a lifetime greater than $\mathcal{O}$(100\,ps), which corresponds to a decay length greater than $\sim$ 1\,cm. In particular, the mediator particle must decay into light SM fermions when the mediator is light, so that it becomes further long-lived as the interactions are also proportional to the small fermion masses.

\begin{figure}[t]
    \centering
    \includegraphics[width=0.46\textwidth]{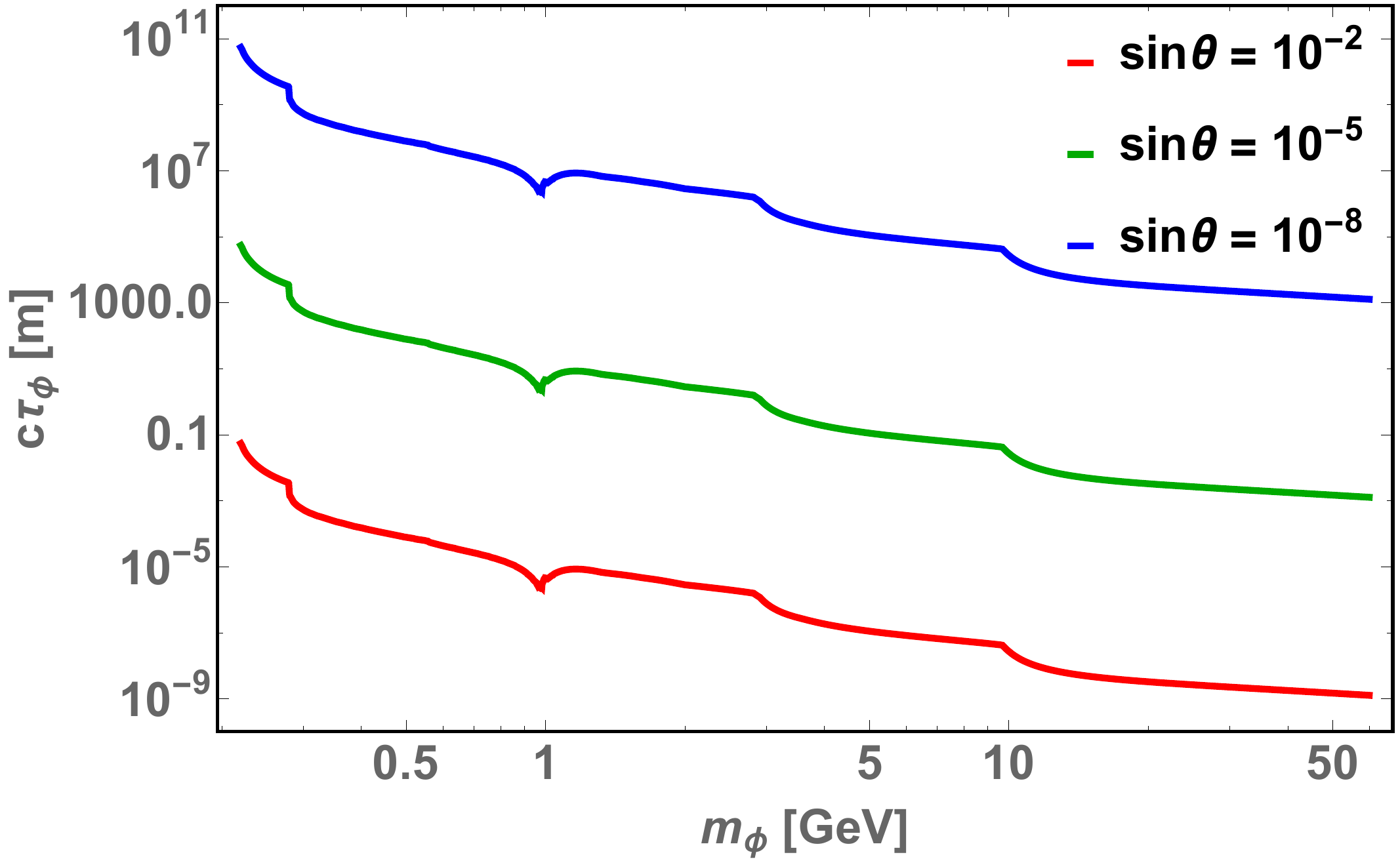}
    \qquad
    \includegraphics[width=0.46\textwidth]{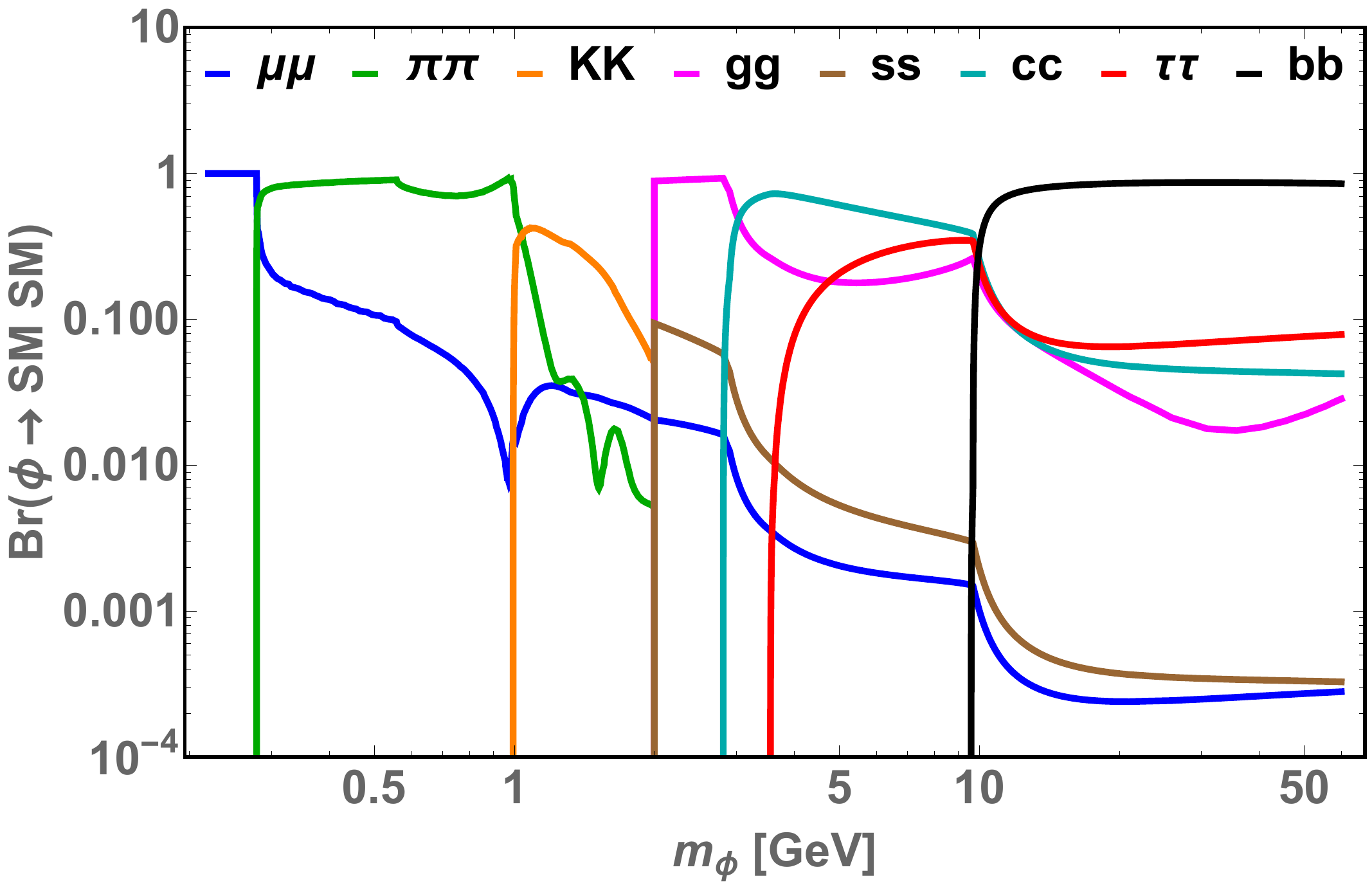}
    \caption{\small \sl The mean proper decay length of the mediator particle, $c \tau_\phi$, predicted in the minimal model in eq.\,(\ref{eq: minimal model}) and the branching fraction of each decay mode, ${\rm Br}(\phi \to {\rm SM}\,{\rm SM})$, contributing to the length are show in the left and right panels, respectively, assuming that the mediator does not decay into dark sector particles. In the left panel, the decay lengths with the mixing angles $\sin \theta = 10^{-2}$, $10^{-5}$ and $10^{-8}$ are shown. In the right panel, only the branching fractions that are analysed in the following sections are shown.}
    \label{fig:width_branching}
\end{figure}

The mean proper decay length of the mediator particle, $c \tau_\phi$, predicted in the minimal model and the branching fraction of each decay mode, ${\rm Br}(\phi \to {\rm SM}\,{\rm SM})$, contributing to the length are show in the left and right panels of Fig.\,\ref{fig:width_branching}, respectively, assuming that the mediator particle does not decay into dark sector particles. We have adapted the method developed in Ref.\,\cite{Winkler:2018qyg} in order to calculate the mean proper decay length and the branching fractions. Concerning partial decay widths into hadrons or light quarks/gluons, the mediator particle is postulated to decay into hadrons when $m_\phi \leq 2$\,GeV, while it decays into light quarks and gluons when $m_\phi \geq 2$\,GeV within this method. Moreover, the total decay width into all the hadrons is matched with that into light quarks and gluons at $m_\phi = 2$\,GeV by adjusting the normalization of the partial decay width of the process into $4\pi$, $\eta\eta$ and $\rho\rho$.\footnote{The partial decay widths has been computed at leading order in Ref.\,\cite{Winkler:2018qyg} when $m_\phi \geq 2$\,GeV, while we used the result of the \texttt{HDECAY} code which involves some next to leading order corrections. Hence, the normalization that we have used is slightly different from that used in the reference.}

The left panel of the figure shows that decreasing mass and sin$\theta$ of the mediator particle increases its $c\tau_\phi$, and a mediator particle with $c \tau_\phi$ greater than $\sim$ 1\,cm is considered to be long-lived. For a mixing angle of $10^{-5}$, the mean proper decay length, $c\tau_\phi$, ranges from about 77\,m to 0.11\,m to 0.15\,cm for mediator masses of 0.5, 5 and 50\,GeV, respectively. In the right panel of the figure, we show the variation of branching fractions of the various decay modes that we consider in this work ($\mu\mu$, $\pi\pi$, $KK$, $gg$, $ss$, $cc$, $\tau\tau$ and $bb$) with the mass of the mediator particle.

\section{The long-lived mediator particle at the HL-LHC}
\label{sec:higgs-lhc}

As motivated in the previous section where we have discussed the properties of the mediator particle, we study the case where the mediator is produced from the decay of the SM Higgs boson, assuming that its production is not altered by new physics, i.e., dark sector particles and the mediator particle. Then, there are three dominant components depending on the production mode of the SM Higgs boson at the LHC experiment; gluon-gluon fusion (ggF), vector boson fusion (VBF), and Higgs-strahlung (Vh). These production modes, along with the mediator mass, $m_{\phi}$, affect the boost ($\beta \gamma$) of the mediator particle, which in turn controls its decay length in the laboratory frame ($d=\beta \gamma c \tau_\phi$) in addition to its mean proper lifetime, $\tau_{\phi}$.\footnote{Hereafter for simplicity, we will refer to the mean proper lifetime ($\tau_{\phi}$) or the mean proper decay length ($c\tau_{\phi}$) of the mediator as just the lifetime or the decay length, unless otherwise stated.}

\begin{figure}[t]
    \centering
    \includegraphics[width=\textwidth]{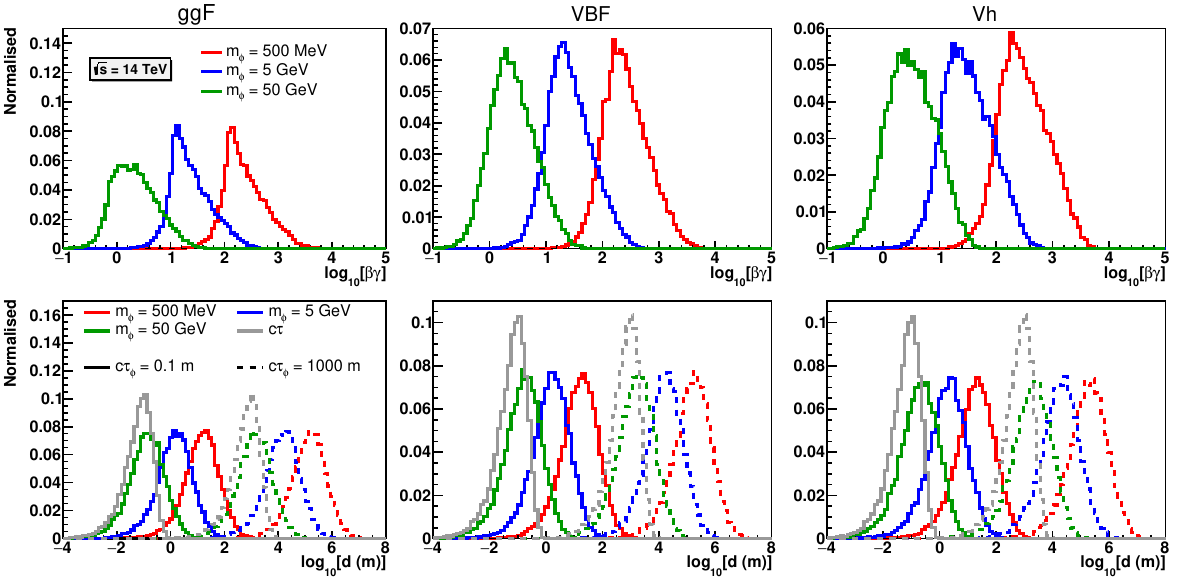}
    \caption{\small \sl Normalised histograms of the boost factor distribution\,(top panels) and the decay length in the laboratory frame\,(bottom panels) of the mediator particle from the SM Higgs boson decay, for $m_{\phi} =$ 500\,MeV, 5\,GeV and 50\,GeV, and $c \tau_\phi = 0.1$\,m (solid lines) and 1000\,m (dashed lines), in the ggF (left), VBF (centre) and Vh (right) production modes of the Higgs boson production at $\sqrt{s} =$ 14\,TeV. The histogram of the exponential distribution, ${\rm exp(-d/c\tau_\phi)}$, for the two different decay lengths are shown in gray.}
    \label{fig:LLPs-dist-14TeV-boost}
\end{figure}

We have used the \texttt{PYTHIA6} code\,\cite{Sjostrand:2006za} to generate the Higgs boson in various production channels and its further decay into a pair of mediator particles with varying masses, $m_{\phi}$, and decay lengths, $c\tau_{\phi}$. Fig.\,\ref{fig:LLPs-dist-14TeV-boost} shows the variation of the histogram of the boost factor distribution (top three panels) as well as that of the decay length in the laboratory frame (bottom three panels) with three values of mediator masses, $m_\phi = 500$\,MeV, 5\,GeV, 50\,GeV, where each has two different decay lengths, $c \tau_\phi = 0.1$\,m, 1000\,m, for the ggF, VBF and Vh production channels of the SM Higgs boson at $\sqrt{s} = 14$\,TeV. We also show the histogram of the distribution of the proper decay length, $\exp(-d/c \tau_\phi)$, for the two lifetimes mentioned above as a reference to compare how the boost factor changes the decay lengths in the laboratory frame. The boost factor increases for lighter mediator particles, and therefore, shifts the decay length $d$ to very large values. Out of the three production channels, the ggF process imparts the lowest boost to the mediator particle since the Higgs boson itself tends to have lower transverse momentum, $p_T$, as we have seen in section\,\ref{ssec:prod-med}.

Fig.\,\ref{fig:LLPs-dist-14TeV-boost} indicates that the general purpose detectors at the LHC experiment, i.e. the ATLAS and CMS detectors which have sizes of $\mathcal{O}(10)$\,m, can be sensitive for a range of mediator masses greater than 100\,MeV and decay lengths with $d < 1000$\,m. Other proposed dedicated experiments for LLPs, such as the FASER, CODEX-b, and MATHUSLA experiments, will be sensitive to LLPs which are lighter and have greater decay lengths, i.e., $d > \mathcal{O}(10)$\,m, because they will be/are planned to be situated at larger distances from the interaction point\,(IP). Since these experiments have a much cleaner environment than those at the ATLAS and CMS detectors owing to their increased distance from the IP and therefore, increased shielding, even a small number of LLP decays within the detectors can signal new physics. Unlike the CODEX-b and MATHUSLA experiments, both of which have transverse detectors, the FASER experiment has a forward detector, so that limits from the former experiments are stronger than the FASER experiment for LLPs from the Higgs boson decay, as has already been studied in Ref\,\cite{Boiarska:2019vid}. This fact can be understood from the pseudo-rapidity distribution of the mediator particle when it is produced from the Higgs boson decay, as shown in Fig.\,\ref{fig:LLPs-dist-14TeV}, which is more central for lighter mediators. The FASER experiment is sensitive only for the mediator particle heavier than $\sim$ 40\,GeV when they are produced more in the forward direction. We therefore focus on the experiments with transverse LLP detectors in this paper.

\subsection{Mediator decays within CMS muon spectrometer}
\label{ssec:CMS-MS}

High luminosity runs of the LHC (HL-LHC) experiment is proposed to have a peak luminosity of $5 \times 10^{34}$\,cm$^{-2}$\,s$^{-1}$\,($7.5\times 10^{34}$\,cm$^{-2}$\,s$^{-1}$) which corresponds to around an average of 140\,(200) pile-up (PU) vertices per bunch crossing at the start\,(end) of the runs\,\cite{Apollinari:2017cqg}. This huge amount of pile-up poses a severe challenge, and therefore dedicated efforts are being made to maintain the physics reach of the experiment. For instance, the CMS collaboration has designed a level-1 trigger menu in Ref.\,\cite{CERN-LHCC-2020-004} to be used for its Phase-II runs which can control PU rates. Triggering LLPs in such a high PU scenario is studied in Ref.\,\cite{Bhattacherjee:2020nno}, where Phase-II upgrades like L1 tracking has been used. Also, the HLT and offline analyses for LLPs are not yet fully available for the HL-LHC experiment. Dedicated studies for the prospects of the LLP search in each sub-detector, i.e., tracker, electromagnetic calorimeter (ECal), as well as hadronic calorimeter (HCal), is hence required for the HL-LHC experiment.

The inner parts of the ATLAS and CMS detectors, i.e., trackers and calorimeters, will be affected the most due to the increased amount of activity. This motivates us to study LLP decays inside the detector which lies farthest from the IP, and is therefore the least affected by the increased PU rate, i.e., the muon spectrometer (MS). It also has the largest coverage in the radial and the $z$-direction (both $\sim$ 3\,m for CMS) compared to all others in ATLAS and CMS detectors. This large decay volume compensates for its distance from the IP, and the MS can still have a significant acceptance for LLP decays. To demonstrate this, we present the ratio of efficiency of the mediator decaying inside the MS ($d_T > 4$\,m or $|d_z| > 7$\,m, and, $d_T < 7$\,m and $|d_z| < 10$\,m) to that within the tracker ($d_T < 1.29$\,m and $|d_z| < 3$\,m), $\epsilon_{\rm MS}/\epsilon_{\rm Tracker}$, for some benchmark mediator masses and decay lengths in Table\,\ref{tab:T-MS-ratio}.\footnote{Here, $d_T$ and $d_z$ are the transverse and the longitudinal components of the decay length in the laboratory frame, defined as $d_T = \beta_T \gamma c \tau$ and $d_z = \beta_z \gamma c \tau$ with $\beta_T = p_T/m_\phi$ and $\beta_z = p_z/m_\phi$.} We find that for all the three mediator masses, 0.5\,GeV, 5\,GeV and 50\,GeV, there exists a range of decay lengths where this ratio is equal to or greater than one, and it means there are equal or more decays within the MS than within the tracker.

\begin{table}[t]
    \centering
    \begin{tabular}{r|c|c|c|}
    \backslashbox{$c\tau_{\phi}$}{$m_{\phi}$} & 0.5\,GeV & 5\,GeV &\,50 GeV \\
    \hline
      0.01\,m & 0.09 & 0.00 & 0.00 \\
       0.1\,m & 1.10 & 0.09 & 0.00 \\
       1.0\,m & 1.68 & 1.07 & 0.07 \\
      10.0\,m & 2.04 & 1.67 & 0.85 \\
     100.0\,m &    - & 1.59 & 1.53 \\
    1000.0\,m &    - &    - & 1.52 \\
    \hline
    \end{tabular}
    \caption{\small \sl The ratio of efficiencies for the LLP (the mediator particle) which decays inside the muon spectrometer and the tracker of the CMS detector, refereed as $\epsilon_{\rm MS}/\epsilon_{\rm Tracker}$ in the text, for some benchmark masses and decay lengths of the mediator particle.}
    \label{tab:T-MS-ratio}
\end{table}

In addition, the MS has the capability to detect various final states from the mediator decay other than muons, as studied in Ref.\,\cite{CMS:2021juv}. The CMS muon stations, consisting of resistive plate chambers (RPCs), drift tubes (DTs) and cathode strip chambers (CSCs), are interleaved with return yokes made of iron plates. Electrons, photons and hadrons can interact with the yokes and make a shower producing multiple hits in the muon chambers, giving a cluster of hits in the MS\,\cite{CMS:2021juv}.

Focusing on the MS for LLP searches has already been explored by experimental collaborations such as CMS and ATLAS, where the CMS collaboration has studied the prospects of using the CMS end-cap MS for LLP decays into a pair of quarks and tau leptons in Ref.\,\cite{CMS:2021juv}, whereas the ATLAS collaboration has performed searches using both the barrel and end-cap parts of the MS for displaced jets in Refs.\,\cite{ATLAS:2018tup,ATLAS:2019jcm,ATLAS-CONF-2021-032} as well as collimated muons from dark photons in Ref.\,\cite{ATL-PHYS-PUB-2019-002}. We take inspirations from these analyses and extend them to include various production channels of the SM Higgs boson and a variety of decay channels of the mediator particle which are not very well studied in the literature for the HL-LHC as well as future colliders. Before discussing our analysis strategy in details, we give a brief description of the simulation and computational setup used in this paper in the next section.

\subsubsection{Simulation and computational setup}
\label{sssec:sim}

As mentioned previously, we use the \texttt{PYTHIA6} code to generate the process:
\begin{align}  
    p p \rightarrow h X, \qquad h \rightarrow \phi \phi, \qquad \phi \rightarrow {\rm SM}\,{\rm SM},
    \nonumber
\end{align}
where $X$ is associated objects which are prompt and depends on the production channel, $\phi$ is the long-lived mediator particle, and ``SM" denotes various SM decay products of the mediator particle. As outlined before, we concentrate on the ggF, VBF and Vh production channels of the Higgs boson, and among ``SM" decay modes of $\phi$ we consider its decays into $\mu^+ \mu^-$, $\pi^+ \pi^-$, $K^+ K^-$, $g g$, $s \bar{s}$, $c \bar{c}$, $\tau^+ \tau^-$ and $b \bar{b}$. We use the \texttt{cteq6l1} PDF set from \texttt{LHAPDF6}\,\cite{Buckley:2014ana} to generate all the processes.

We use the \texttt{Delphes-3.4.2} code\,\cite{deFavereau:2013fsa} with some modifications for fast-detector simulation and the \texttt{ROOT6}\,\cite{BRUN199781} framework for our analyses. Since our benchmarks include light LLPs which will have high boosts at the $\sqrt{s} = 14$\,TeV LHC experiment, as observed in Fig.\,\ref{fig:LLPs-dist-14TeV-boost}, their decay products are expected to be highly collimated, and have very small opening angles between them. Moreover, when the decay products are highly displaced, they will have a very small physical distance between them in the $\eta$-$\phi$ plane, as has been identified in Ref.\,\cite{Bhattacherjee:2019fpt}. Smaller opening angles between the decay products will affect their isolation,  and it also makes reconstructing the secondary vertex difficult. Hence, the magnetic field present in the detector plays an important role in bending the charged tracks and affects the opening angle between the decay products. We, therefore, have to take into account the appropriate magnetic field present in the detector for correct simulation of such boosted LLP scenarios, otherwise we lose most of them due to the isolation. Since the CMS magnetic field is solenoidal, it is easier to implement in the \texttt{Delphes} code, and therefore from this point onwards we focus on the CMS detector geometry, perform our analyses and present results using the CMS MS. Charged particles are propagated in a magnetic field of $B_z = 3.8$\,T till the tracker edge of the CMS detector in the \texttt{Delphes-3.4.2} code,
whereas further propagation till the HCal detector as well as propagation in the MS under the influence of the opposite magnetic field are absent.
The magnetic field strength in the MS after the solenoid reduces significantly.
From Ref.\,\cite{CMS:2009moq}, we use a conservative value of $B_z = -0.5$\,T in the MS.
We have added these parts of the propagation in the \texttt{Delphes} modules to take into account the full range of the magnetic field in the detector.

\begin{figure}[t]
    \centering
    \includegraphics[width=0.6\textwidth]{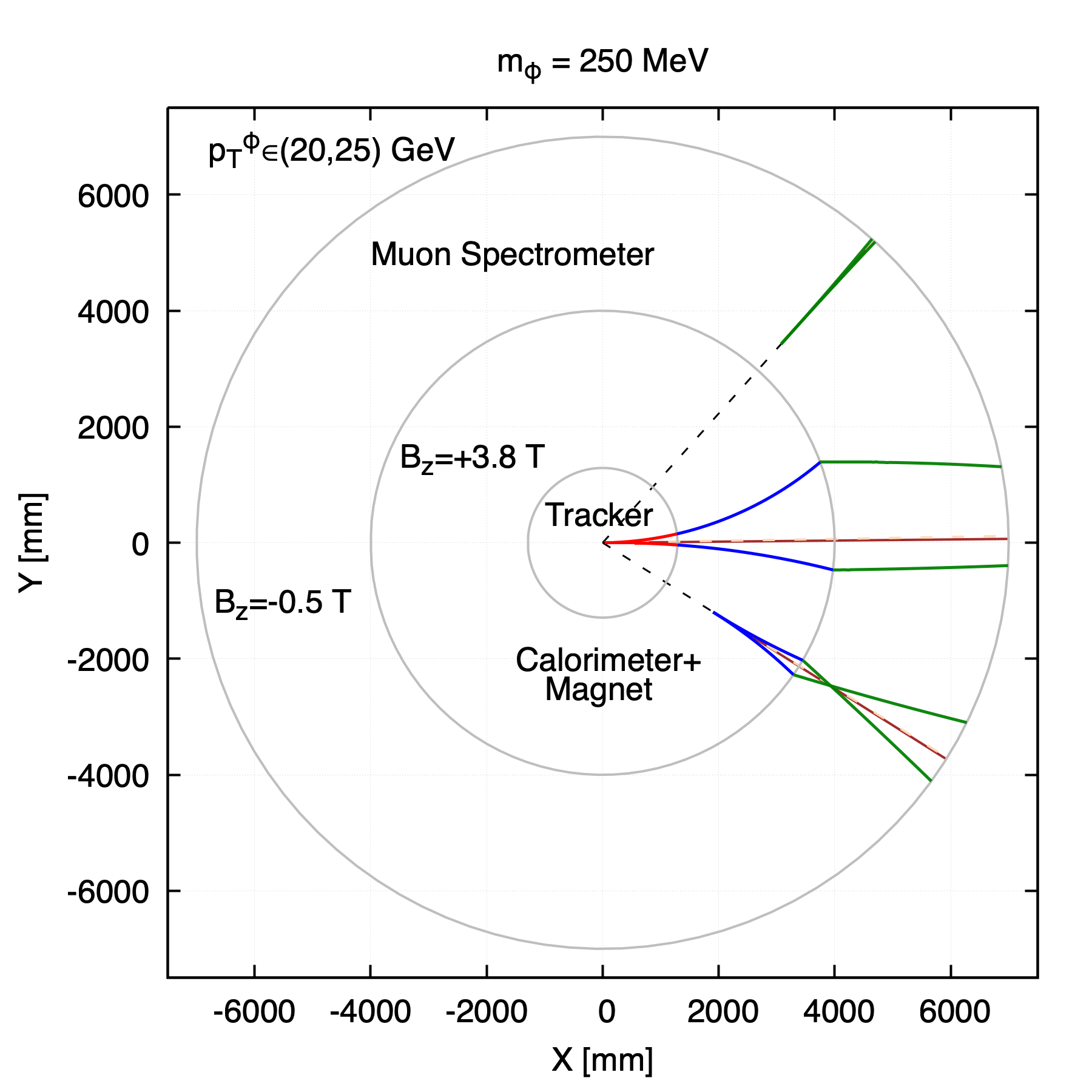}
    \caption{\small \sl Transverse projection of the propagation of a muon pair in the magnetic field of the CMS detector, where the pair is assumed to be from the decay of the LLP  ($m_\phi =$ 250\,MeV) that is pair-produced from the Higgs boson decay. See text for more details.}
    \label{fig:magnet}
\end{figure}

Fig.\,\ref{fig:magnet} shows the propagation of a muon pair from the LLP (mediator particle $\phi$) decay in the full CMS magnetic field using our modified \texttt{Delphes} setup. The mass of the mediator is assumed to be 250\,MeV, and three different events are displayed where $\phi$ decays within the tracker, inside the calorimeter or Magnet and inside the MS; in all these three cases, the mediator $p_T$ lies between 20-25 GeV. The lines in solid dark brown and dashed light brown represent muon trajectories in the absence of the magnetic field.
We find that our implementation of the full magnetic field helps in isolating the muon tracks when the LLP decays within the Tracker or in the `Calorimeter+Magnet' region as compared to the scenario with no magnetic field. When the LLP decays inside the muon spectrometer, we find that the reduced magnetic field strength outside the solenoid is not very efficient in isolating the tracks.

\begin{figure}[t]
    \centering
    \includegraphics[width=0.46\textwidth]{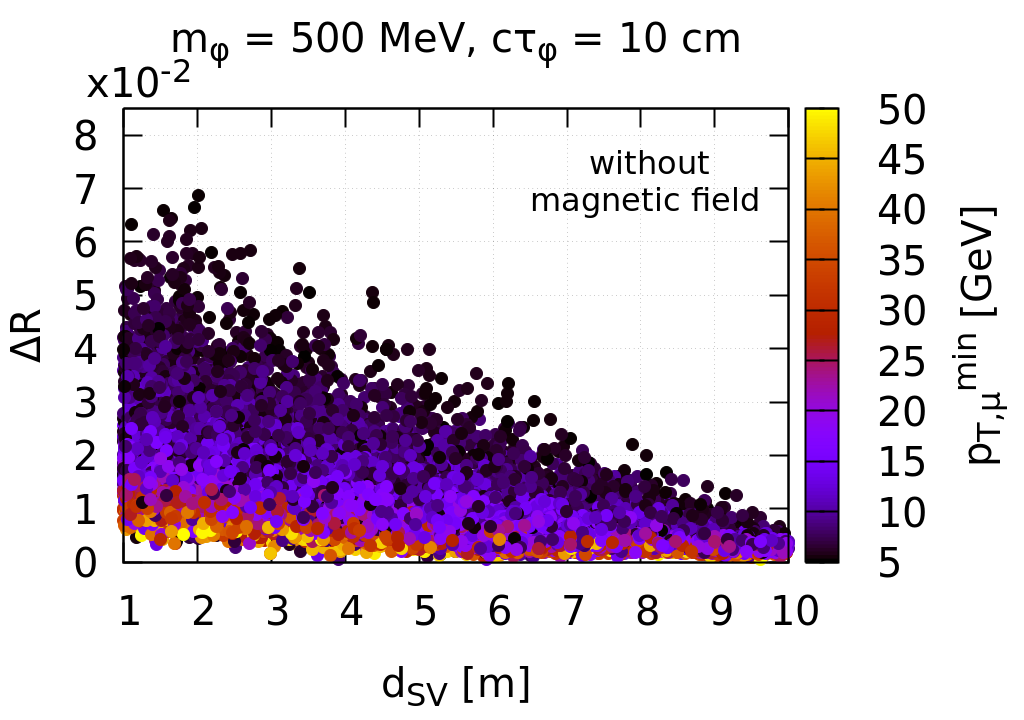} \qquad
    \includegraphics[width=0.46\textwidth]{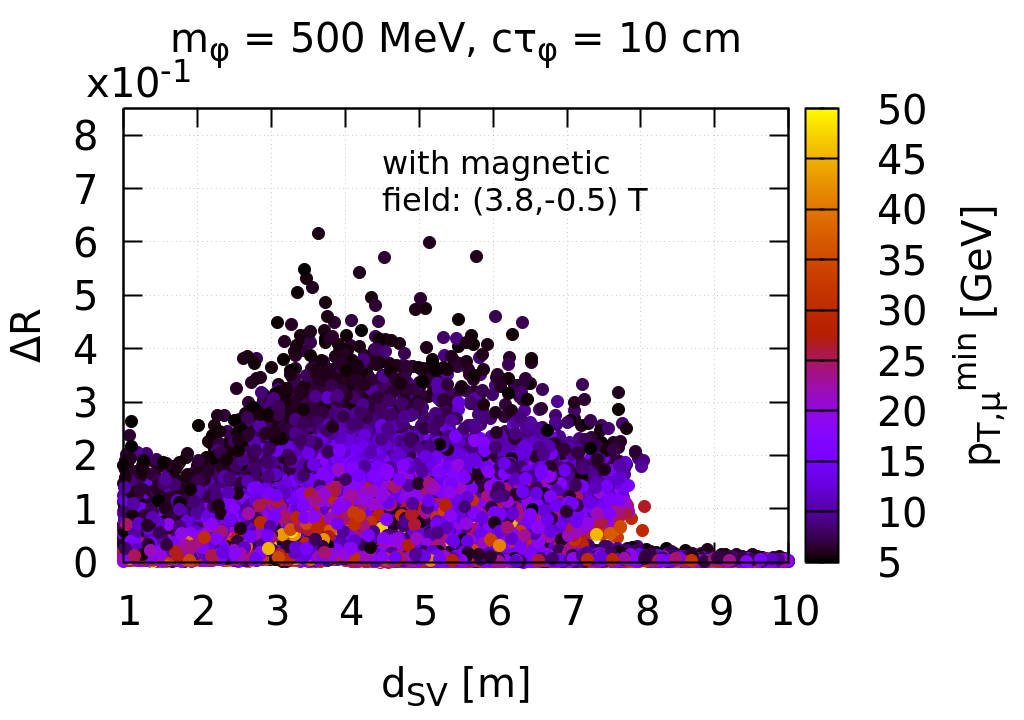} \\
    \includegraphics[width=0.46\textwidth]{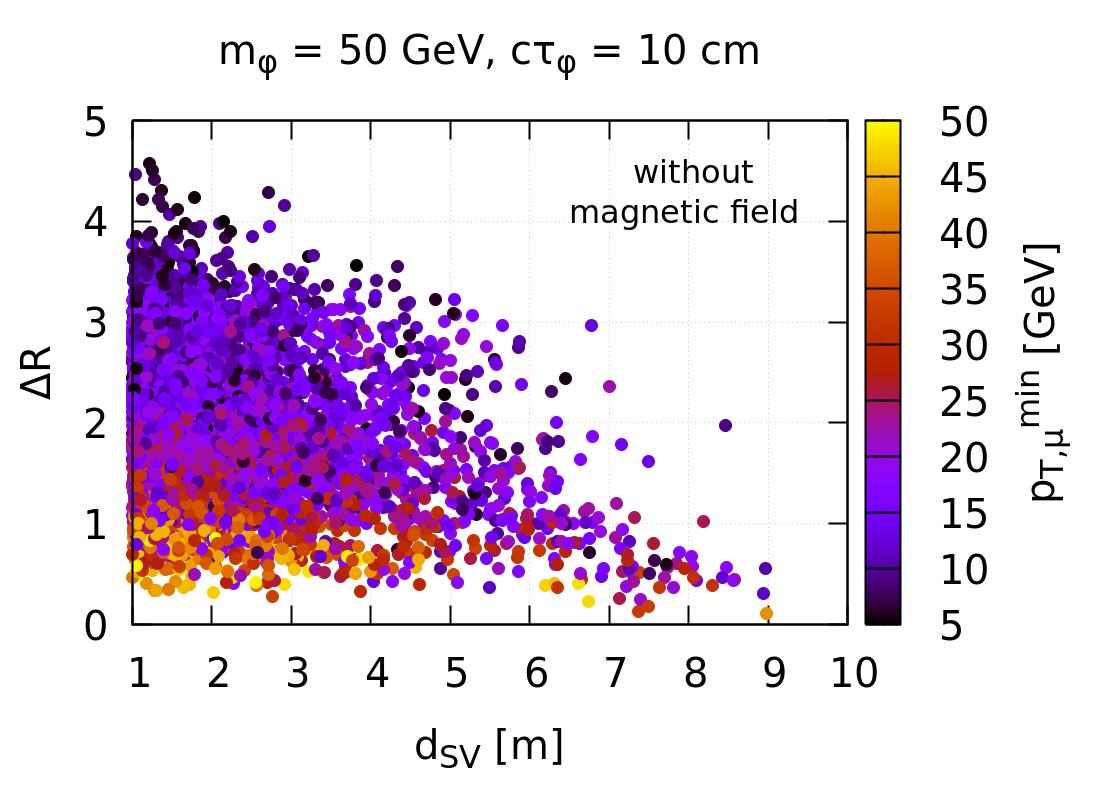} \qquad
    \includegraphics[width=0.46\textwidth]{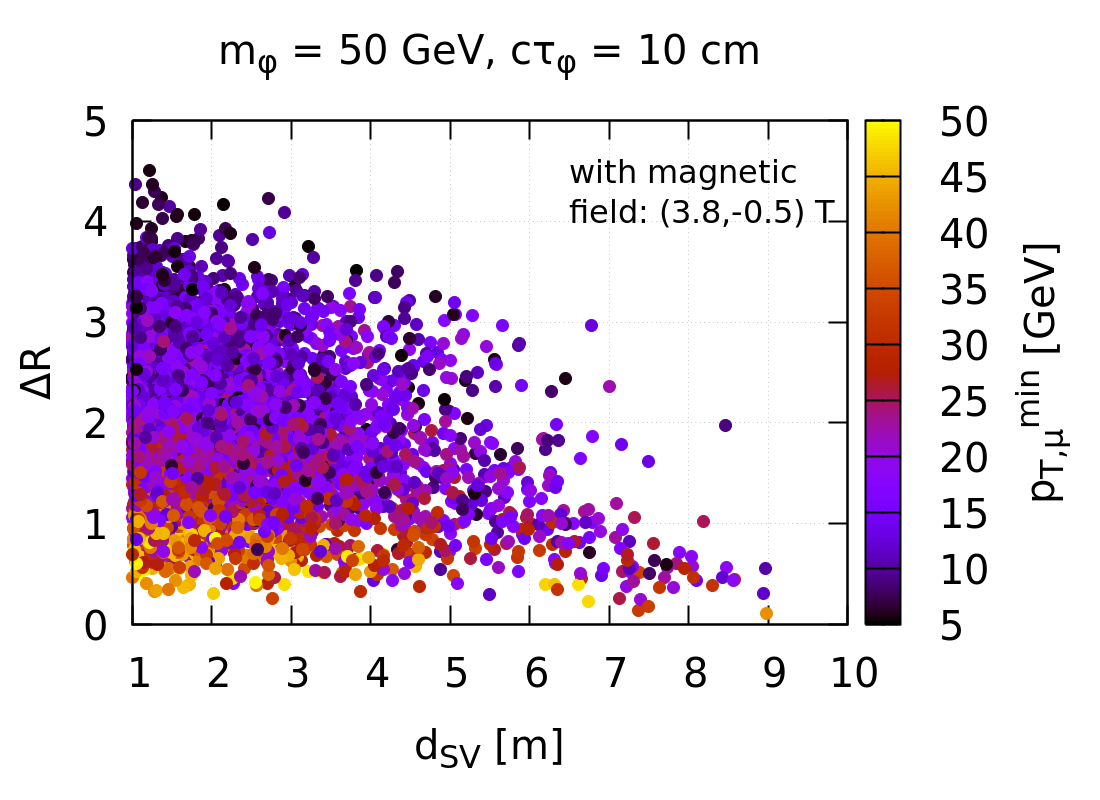}
    \caption{\small \sl Variation of the $\Delta R$ values between the muons coming from the decay of the boosted long-lived LLP (mediator particle $\phi$) as a function of the distance between the secondary vertex and the IP, $d_{SV}$, with (right panels) and without (left panes) 3.8\,T solenoidal magnetic field. The color bar shows the minimum $p_T$ from the muon pair.}
    \label{fig:delR}
\end{figure}

Fig.\,\ref{fig:delR} shows the scatter plot of $\Delta R$ (which is calculated using the $\eta$ and $\phi$ of the particles at the detector edge) between the muons from the decay of the boosted long-lived mediator particle as a function of the distance between the secondary vertex and the IP, $d_{SV}$, with the color bar showing the minimum $p_T$ from each muon pair, in the absence (left panels) and the presence (right panels) of the (3.8\,T,-0.5\,T) solenoidal magnetic field of the CMS detector. The top panels are for a light mediator particle of 500\,MeV mass, while the bottom panels are for a heavier mediator particle of 50\,GeV mass. We find that the $\Delta R$ values for the light mediator particle are extremely small without the magnetic field and keep decreasing with increasing $d_{SV}$, as is expected. In the presence of the magnetic field, the $\Delta R$ values increase to reasonable ones. We observe a turnover at around 4\,m, which marks the change in the direction of the magnetic field. Before 4\,m, the particles experience magnetic fields in both directions, as we have seen in Fig.\,\ref{fig:magnet}, and these two counter each other to give smaller $\Delta R$. This effect of magnetic fields of opposite directions countering each other is more evident for particles with lower $d_{SV}$ values, hence, smaller opening angles. After 4\,m, the particles only experience magnetic field in one direction and follow the usual trend of decreasing $\Delta R$ with increasing $d_{SV}$. On the other hand, for the heavier mediator particle, the effects addressed above are much smaller. For each case, mediator particles with smaller $p_T$ have comparatively larger $\Delta R$. 

Now that we have a proper implementation of the magnetic field, we start discussing physics objects used in our analyses. We broadly categorise the physics objects into ``prompt'' and ``displaced'' objects, and they consist of the following:
\begin{itemize}
    \item \textbf{Prompt:} Since we study various production channels of the Higgs boson which differ in the type of prompt objects produced in association with the Higgs boson, we store information of all prompt electrons, muons and jets with the resolutions, efficiencies and scaling present in the default \texttt{Delphes} CMS card. For muons, we also apply an additional cut of $d_T <$ 1\,cm to distinguish them from those coming from the displaced decay of the mediator particle. We will apply standard trigger cuts, like the single jet, di-jet, VBF jets, single and double electrons, and muons, all described in the next section, on these prompt objects to select the event based on the production channel of the Higgs boson.

    \item \textbf{Displaced:} Muons having $d_T >$ 1\,cm are stored as displaced muons which are used for the displaced muons analysis. As we are concentrating only on activities in the MS, all the other particles, i.e.,  neutral and charged hadrons, electrons and photons are stored if their production vertices lie within the MS. This is because, if these particles are produced before the MS, they will deposit their energies in the calorimeters. This is elaborately discussed in the next section. For the displaced particles, we do not apply any efficiencies as of yet, since these are not very well-documented for the HL-LHC experiment.
\end{itemize}
We will make use of prompt objects in our analyses in addition to displaced objects, in order to make easier triggering and further background reduction. We are now ready to discuss the detail of the analyses performed for each of the production channels of the Higgs boson and decay channels of the long-lived mediator particle. The possible backgrounds for our analyses are discussed later in section\,\ref{ssec:background}.

\subsubsection{Analyses for various production and decay channels}
\label{sssec:analysis}

The crucial part of selecting any kind of new physics signals at the LHC experiment is triggering. This is especially true for long-lived signatures because they need dedicated triggers. The production channel of the Higgs boson plays a key role in triggering the event. We can trigger the event using {\bf the prompt associated objects}, like jets, electrons, or muons.  For the ggF channel, we have ISR jets, and two forward jets with high invariant mass for the VBF one. We divide the Vh channel into two parts, i.e., Vh-jet and Vh-lep, depending on the decay mode of the vector boson into jets and leptons, respectively. Table\,\ref{tab:prod-cuts} shows prompt triggers which we use for the various production channels. We use the single jet, di-jet as well as VBF jet triggers for the ggF, VBF, and Vh-jet channels, and a single electron, double electron, single muon and double muon triggers for the Vh-lep one. The set of cuts is denoted by $P_{\rm Mode}^{H/S}$, where the subscript denotes the production mode, i.e., ggF, VBF, Vh-jet, and Vh-lep, while the superscript refers to a hard set of cuts ($P_{\rm Mode}^H$) motivated from the CMS Phase-II L1 trigger TDR\,\cite{CERN-LHCC-2020-004} or to a softer set of cuts ($P_{\rm Mode}^S$), obtained from reducing most of the thresholds by half. The fact that we also observe a displaced activity in addition to the prompt associated objects motivates the second set of relaxed cuts. Since each decay mode can be triggered with more than one of the available prompt objects, we define the sets of cuts in terms of the production modes rather than the individual triggers, to make our study more inclusive. To further clarify, $P^H_{\rm Mode}$ and $P^S_{\rm Mode}$ for each mode are defined as follows:
\begin{itemize}
    \item $P^{H/S}_{\rm ggF/VBF/Vh-jet}$ are sets of cuts constituting the single jet, di-jet, and VBF jet triggers applied on events that the Higgs boson is produced from the gluon-gluon fusion process, vector boson fusion process and Higgs-strahlung process with an associated weak gauge boson which decays into jets, respectively.

    \item $P^{H/S}_\text{Vh-lep}$ are sets of cuts constituting the single electron, double electron, single muon,  and double muon triggers applied on events that the Higgs boson is produced along with a vector boson which decays to leptons ($e,\mu,\tau$).
\end{itemize}
This implies that $P^{H/S}_{\rm ggF}$, $P^{H/S}_{\rm VBF}$ and $P^{H/S}_{\rm Vh-jet}$ are the same sets of cuts, only that they are applied to Higgs boson processes produced in the three different modes. Usually,  the thresholds at the high-level trigger (HLT) are higher than the L1 trigger to reduce the background rates. We assume that the HLT thresholds can be lowered and made comparable to those at L1 due to the observation of additional displaced activities.

\begin{table}[t]
    \centering
    \resizebox{\textwidth}{!}{
        \begin{tabular}{r|l|l|l|}
            Trigger & \qquad\qquad\qquad In $P_{\rm Mode}^H$ & \qquad\qquad\qquad In $P_{\rm Mode}^S$ & Mode \\
            \hline
            Single jet & $p_T^j >$ 180\,GeV, $|\eta_j| < 2.4$.
                       & $p_T^j >$  90\,GeV, $|\eta_j| < 2.4$.
                       & \\
            \cline{1-3}
                 Di-jet & $p_T^j >$ 112\,GeV, $|\eta_j| < 2.4$, $\Delta\eta < 1.6$.
                       & $p_T^j >$  90\,GeV, $|\eta_j| < 2.4$, $\Delta\eta < 1.6$.
                       & \\
            \cline{1-3}
                       & $p_T >$ 70\,GeV for Leading jet,
                       & $p_T >$ 60\,GeV for Leading jet,
                       & ggF, \\
                       & $p_T >$ 40\,GeV for Sub-leading jet,
                       & $p_T >$ 30\,GeV for Sub-leading jet,
                       & VBF, \\
            VBF jet & $|\eta_j| < 5$, $\eta_{j_1} \times \eta_{j_2} < 0$, $\Delta\eta > 4.0$,
                    & $|\eta_j| < 5$, $\eta_{j_1} \times \eta_{j_2} < 0$, $\Delta\eta > 4.0$,
                    & Vh-jet. \\
                    & $\Delta\phi < 2.0$,
                    & $\Delta\phi < 2.0$,
                    & \\
                    & $m_{jj} >$ 1000\,GeV. & $m_{jj} >$ 500\,GeV.
                    & \\
            \hline
            Single electron & $p_T^e >$ 36\,GeV, $|\eta| < 2.4$.
                            & $p_T^e >$ 18\,GeV, $|\eta| < 2.4$.
                            & \multirow{4}{*}{Vh-lep.} \\
            \cline{1-3}
            Double electron & $p_T^{e_1} >$ 25\,GeV, $p_T^{e_2} >$ 12\,GeV, $|\eta| < 2.4$.
                            & $p_T^{e_1} >$ 12\,GeV, $p_T^{e_2} >$ 12\,GeV, $|\eta| < 2.4$.
                            & \\
            \cline{1-3}
            Single muon & $p_T^\mu >$ 22\,GeV, $|\eta| < 2.4$.
                        & $p_T^\mu >$ 11\,GeV, $|\eta| < 2.4$.
                        & \\
            \cline{1-3}
            Double muon & $p_T^{\mu_1} >$ 15\,GeV, $p_T^{\mu_2} >$ 7\,GeV, $|\eta| < 2.4$.
                        & $p_T^{\mu_1} >$  7\,GeV, $p_T^{\mu_2} >$ 7\,GeV, $|\eta| < 2.4$.
                        & \\
            \hline
        \end{tabular}
    }
    \caption{\small \sl The prompt triggers used and the selection cuts applied in our analysis on the signal based on the production mode of the Higgs boson at the HL-LHC experiment.}
    \label{tab:prod-cuts}
\end{table}

The selection of the event based on {\bf the displaced decay products} using objects reconstructed in the CMS MS depends on the final state that the long-lived mediator particle decays into. As mentioned in section\,\ref{ssec:decay-med}, we focus on the decay of the mediator particle into a pair of muons, pions, Kaons, gluons, strange quarks, charm quarks, taus, and bottom quarks, depending on its mass. Analyses for these final states differ due to various factors, such as whether the final state consists of muons or other particles, and the charged particle multiplicity associated with a displaced secondary vertex\,(dSV). As we have already discussed, particles apart from muons will look different in the MS due to their interactions with the iron yokes, i.e., they shower and give rise to a cluster of hits. Questions like how they exactly look in the MS, whether these hits can be reconstructed into tracks and whether the position of the dSV can be identified with such clusters of hits are all purely experimental in nature that we cannot address properly in a phenomenological study such as the one in this paper. We just devise our cuts to ensure that a cluster with a high multiplicity of hits can be detected in the MS for various final states other than muons. 

In addition, if the mediator particle decays into muons, we will have hits in the MS irrespective of where it decays within the detector. However, if it decays into electrons, mesons, or jets before (inside) the calorimeters, these decay products will shower and deposit all (part) of their energies in the calorimeters, and there will not be any hits (bad quality hits) in the MS. Therefore, for such final states, which do not involve at least two muons that can reconstruct the dSV, we consider only those events where at least one mediator particle decays after the calorimeters (and magnet), namely well inside the MS. This requirement translates to a cut of $d_T >$ 4\,m or $|d_z| >$ 7\,m on the position of the dSV. Moreover, if the mediator particle decays near the MS outer edge, there won't be enough hits associated with the MS cluster, and therefore, we put an upper cut on the position of the dSV such that the charged decay products of the mediator particle have hits in at least two layers of the MS, which then restricts the fiducial decay volume to be $d_T <$ 6\,m and $|d_z| <$ 9\,m.

Also, because charged pions and Kaons are detector stable, in case the mediator particle decays into such particles, the number of charged particles associated with each vertex will be just two and a smaller number of hits might be associated with the MS cluster, unless they have higher energies. When the mediator particle decays into displaced jets, the multiplicity increases, and it implies that the energy threshold for the individual particles is brought down, maintaining similar number of hits in a MS cluster. When the mediator particle decays into taus, we apply a hybrid analysis; the displaced muons analysis is used when both taus from the same mediator particle decay into muons, while the displaced jets analysis is used otherwise.

Based on the above discussion, we divide the final states into four groups:
\begin{itemize}
    \item[(A)] {\bf Muons:} The mediator particle (LLP) decays anywhere within the detector; hits in the muon spectrometer (MS) can be reconstructed as muon tracks.

    \item[(B)] {\bf Pions, Kaons:} The mediator particle decays within the MS; due to the iron yoke, they can shower and produce a cluster of hits in the MS. On the other hand, because the initial multiplicity of charged particles associated with a dSV is low, the particles should be energetic to produce enough hits in the MS.

    \item[(C)] {\bf Jets from $gg$, $s\bar{s}$, $c\bar{c}$, $b\bar{b}$:} Decay of the mediator particle shows up as a cluster of hits in the MS; the energy threshold for individual particles can be reduced due to the high multiplicity of charged particles associated with the dSV.

    \item[(D)] {\bf Taus:} If both the taus from a single mediator particle decay into $\mu +$ anything, we follow the analysis for muons, otherwise we follow the analysis of jets.
\end{itemize}

We present in detail model-independent analyses for displaced muons, displaced mesons, displaced jets and displaced taus in the MS, where in all final states, apart from those that we have at least two displaced muons from a mediator particle (LLP), we have a cluster of hits in the MS. We present our results with four sets of combinations for the cuts on the prompt as well as the displaced objects. We describe the combinations below along with a {\it motivation} for each of them:

\begin{itemize}
    \item $D_{\rm Mode}^H\,(\geq 1\,{\rm vtx})$: Selection based only on at least one displaced activity with hard cuts. {\it Since we make a selection using at least one displaced activity, the harder set of cuts is used in order to ensure efficient background rejection.}

    \item $D_{\rm Mode}^H\,(= 2 {\rm~vtx})$: Selection based on two displaced activities with the hard cuts mentioned above. {\it Events with two identified displaced activities satisfying the harder set of cuts are expected to have significantly low background rates.}

    \item $P_{\rm Mode}^H \times D_{\rm 
    Mode}^S\,(\geq 1\,{\rm vtx})$: Selection based on the set of harder cuts on the prompt objects and at least one displaced activity with the set of the soft cuts. {\it When we are using the harder set of cuts on the prompt objects, the cuts on displaced objects can be relaxed without increasing the backgrounds much.}

    \item $P_{\rm Mode}^S \times D_{\rm Mode}^S\,(\geq 1\,{\rm vtx})$: Selection based on the set of softer cuts on the prompt objects along with at least one displaced activity with again the set of softer cuts. {\it This is an optimistic version of the previous set, $P_{\rm Mode}^H \times D_{\rm Mode}^S\,(\geq 1\,{\rm vtx})$, where we also relax the cuts on the prompt particles as well, expecting that the combination of prompt and displaced cuts keep the backgrounds in control.}
\end{itemize}
Here, the subscript of the set of cuts, $D^{H/S}_{\rm Mode}$, denotes the final state particles, i.e., muons, pions, Kaons, and jets. We explore such possibilities of combining cuts in this paper, for the exact analyses for the HL-LHC experiment are not yet finalised, and detailed studies of projected limits with varying sets of cuts can guide such analyses. The concrete definition of each $D^{H/S}_{\rm Mode}$ is given in the subsequent sections.

In the sections, we also describe our analysis strategies for all the four groups of the final states, and present their projected sensitivities on the branching fraction, Br($h \to \phi \phi$), in the $(m_\phi,\,c \tau_\phi)$-plane, assuming 100\% branching fraction on the mediator particle decays. For each case, we present the limit presuming an observation of a minimum of 50 events. On the other hand, in section\,\ref{ssec:background} where we discuss briefly regarding possible backgrounds, we also present a brief discussion on whether or not an observation of 50 events at the HL-LHC experiment might be enough for exclusion or claiming discovery. In the end, we combine these results according to the branching ratios predicted by the minimal model, as shown in Fig.\,\ref{fig:width_branching}; we translate our results in the plane of the mediator mass and its mixing angle with the SM Higgs boson, assuming particular branching of the latter into the mediator particles.

\subsubsection{Mediator particle decaying into muons}

For the mediator particles (LLPs) decaying into a pair of muons, we can use the displaced di-muon trigger proposed to be added in the HL-LHC runs of the CMS experiment\,\cite{CERN-LHCC-2020-004} which does not require matching between the track segments in the muon detectors and the inner tracker, and therefore increases sensitivity to larger displacements. The CMS displaced di-muon trigger requires two muons with transverse momentum, $p_T$, greater than 20\,GeV and 15\,GeV. For our harder set of cuts, we use a $p_T$ cut of 20\,GeV and reduce this $p_T$ threshold by half to 10\,GeV for the softer set of cuts\footnote{In case that the 10\,GeV cut is too low for the standard analyses at the HL-LHC experiment, the sensitivity of scouting analyses with such low $p_T$ thresholds for displaced muons can be studied\,\cite{CMS:2019buh}.}. At the trigger level, it might be difficult to reconstruct the dSV. We can use the transverse impact parameter ($d_0$) of tracks that have been reconstructed using the hits in the MS to discriminate these muons from prompt ones. We follow Ref.\,\cite{ATLAS:2019fwx} and perform our analysis with a $|d_0|>2$\,mm cut.

In section\,\ref{sssec:sim}, we have discussed that for lighter mediator masses having high boosts or for decays with higher displacements, the decay products could be very close in the $\eta$-$\phi$ plane, which will affect the multiplicity associated with the dSV. The implementation of the CMS magnetic field brings the simulated separation between the decay products, depending on its boost and decay position, close to what will be observed in the experiment. We need to demand a minimum cut on this separation to ensure that the detector can identify both the muons coming from the same dSV separately, and they are not too close to be misidentified as a single muon. In the latter case, $d_0$ of the reconstructed track will be very small, since this track will point towards the mediator particle's direction, which comes from the collision vertex, and looks like a prompt single muon\footnote{For very displaced decays, the absence of hits in the inner tracker compatible with the muon track reconstructed in the MS provides some discrimination compared to prompt single muons.}. In addition, if the two muons are very close, it makes the reconstruction of the dSV difficult as well. This motivates a $\Delta\phi$ cut between the pair of muons from the same dSV. The ATLAS HL-LHC proposed analysis for dark photons\,\cite{ATL-PHYS-PUB-2019-002}, which also faces a similar problem of being highly boosted, thereby making the muons coming from their decay very collimated, looks for a secondary pattern within a single Muon Region of Interest (RoI) to deal with such collimated cases and puts a $\Delta\phi >$ 0.01 cut between the secondary and the primary patterns in the same RoI. For the CMS analysis, we put a similar cut. The $\phi$ values of displaced particles, as observed in the detector, change with the radial distance due to its mismatch with the detector's $\phi$ segments\,\cite{Bhattacherjee:2019fpt}, and therefore, it is important to specify the radial value at which we are computing this $\Delta\phi$, like specified in Chapter 7 (section 7.1.4.1) of the TDR for the Phase-II upgrade of the CMS Muon Detectors\,\cite{CERN-LHCC-2017-012}, where they discuss trigger algorithm for displaced muons. We have used the $\phi$ values at the outermost edge of the CMS detector, i.e., $R=7$\,m, $Z=10$\,m. One could use the standard way of calculating the $\Delta\phi$ or $\Delta R$ using the momentum of the muons from their tracks ($\Delta\phi_{\mu\mu,{\rm mom}}$), and later we have compared the two.  

Once the event passes the L1 trigger, at the High-Level Trigger (HLT) and offline analyses, it is a lot easier to find out the dSV due to the availability of more information as well as more time. Then, we can apply a cut on the transverse distance of the dSV from the interaction point in order to further reduce any prompt di-muon background. Table\,\ref{tab:muon_cuts} shows the various sets of cuts applied on the signal when the mediator particle (LLP) decays into a pair of displaced muons, denoted as $D^i_{\mu \mu}$, where $i$ indicates either a harder set of cuts\,($D^H_{\mu \mu}$) or a softer set of cuts ($D^S_{\mu \mu}$). These cuts can be combined with the selection cuts based on the various production channels denoted by $P^{H/S}_\text{Mode}$, as discussed in section\,\ref{sssec:analysis}, which can let us relax both the sets of cuts, allowing for better signal efficiencies with lower backgrounds. 

\begin{table}[t]
    \centering
    \begin{tabular}{c|l|l|}
        & \qquad\qquad $D^H_\mu$ & \qquad\qquad $D^S_\mu$ \\
        \hline
        \multirow{4}{*}{Muons} & $p_T^\mu >$ 20\,GeV & $p_T^\mu >$ 10\,GeV \\
        \cline{2-3}
                               & $n_\mu \geq$ 2 & $n_\mu \geq$ 2 \\
        \cline{2-3}
                               & $|\eta^\mu| <$ 2.8 & $|\eta^\mu| <$ 2.8 \\
        \cline{2-3}
                               & $|d_0^\mu| >$ 2\,mm & $|d_0^\mu| >$ 2\,mm \\
        \hline
        \multirow{2}{*}{Muon pair from} & $d_T >$ 1\,cm & $d_T >$ 1\,cm \\
        \cline{2-3}
        \multirow{2}{*}{the same dSV} & $d_T <$ 6\,m \& $|d_z| <$ 9\,m & $d_T <$ 6\,m \& $|d_z| <$ 9\,m \\
        \cline{2-3}
                                      & $\Delta\phi_{\mu\mu} >$ 0.01 & $\Delta\phi_{\mu\mu} >$ 0.01 \\
        \hline
        Event & $n_{vtx} \geq 1$ or $n_{vtx} = 2$ &  $n_{vtx} \geq 1$ or $n_{vtx} = 2$ \\
        \hline
    \end{tabular}
    \caption{\small \sl Harder and softer sets of cuts, $D^H_\mu$ and $D^S_\mu$, applied for the displaced muons.}
    \label{tab:muon_cuts}
\end{table}

\begin{figure}[t]
    \centering
    \includegraphics[width=0.46\textwidth]{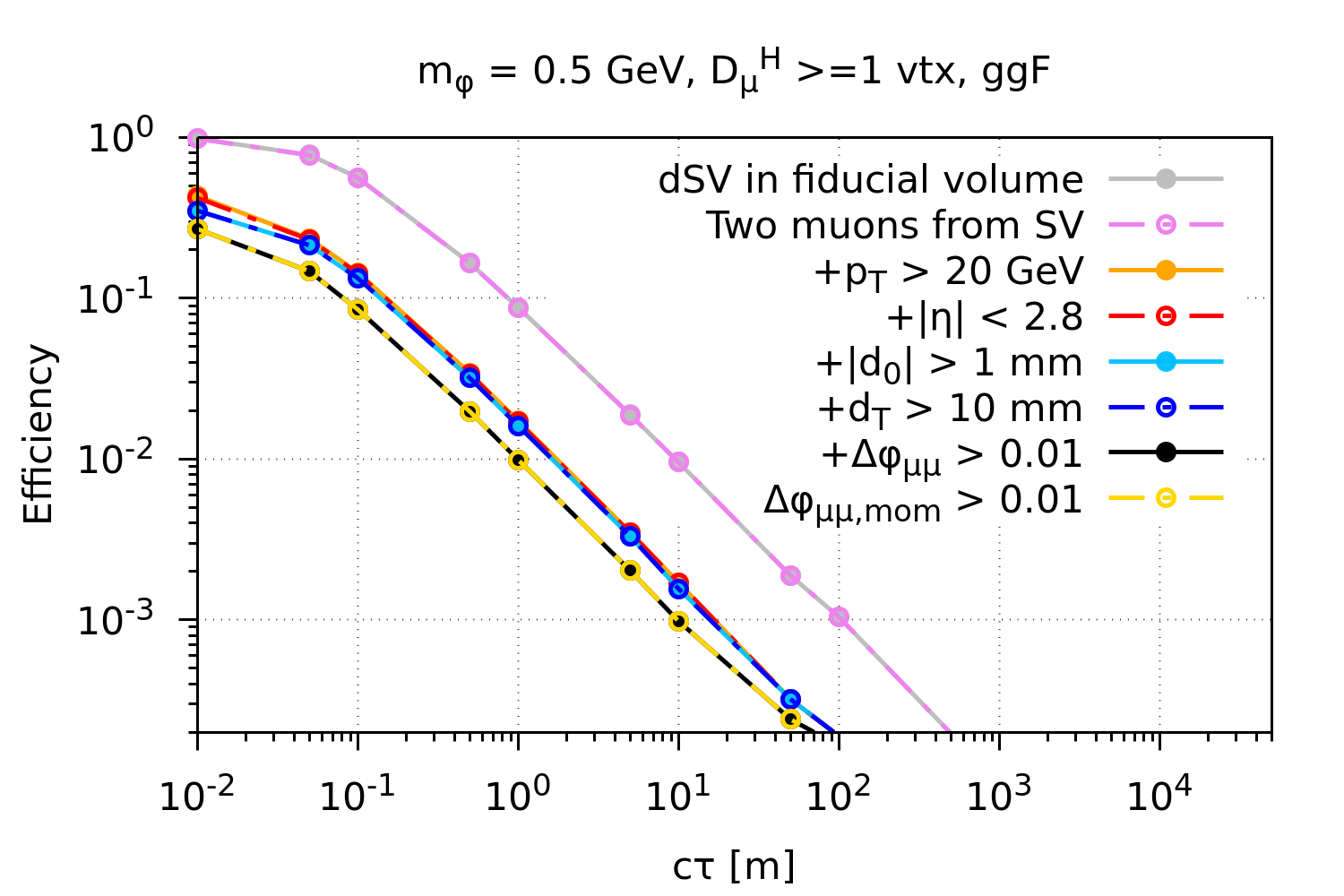} \qquad
    \includegraphics[width=0.46\textwidth]{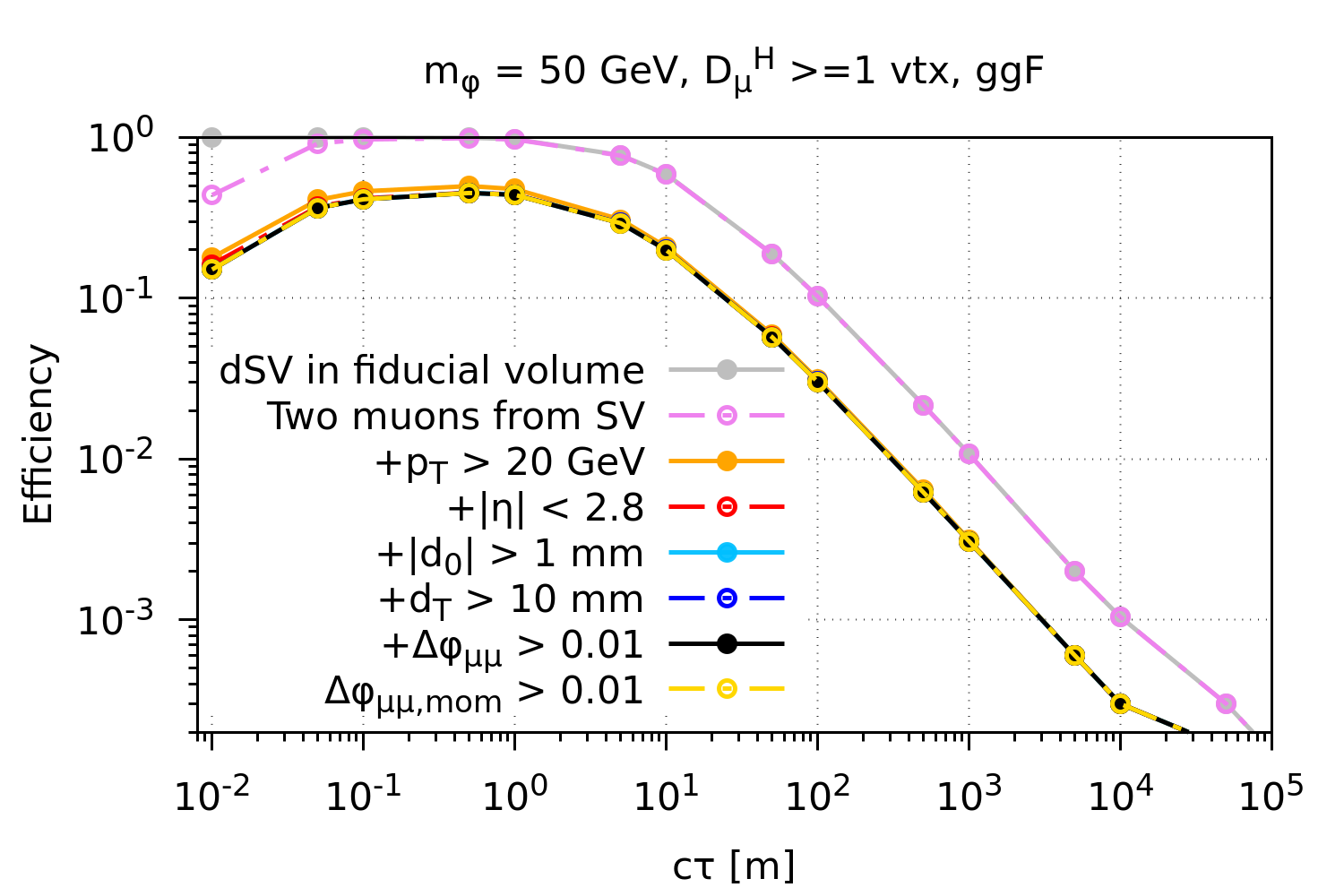}
    \caption{\small \sl Efficiencies of the various cuts on mediator particles of mass 0.5\,GeV (left panel) and 50\,GeV (right panel) for varying lifetimes when it decays to a pair of muons.}
    \label{fig:eff-mumu-14TeV}
\end{figure}

Fig.\,\ref{fig:eff-mumu-14TeV} shows the effect of the various sets of cuts on the signal efficiency as a function of decay length of the mediator particle $\phi$ for two benchmark masses of 0.5\,GeV (left panel) and 50\,GeV (right panel). For both the masses, the $p_T$ cut on the muons affects the efficiency the most. On the other hand, all the other cuts do not have a  significant effect on the efficiency. For the lighter benchmark, the $\Delta\phi_{\mu\mu}$ cut (solid black line) has slightly more effect as compared to the heavier benchmark, which is due to a larger boost of the former. We discuss this in more detail in the case where the mediator decays into a pair of $b$-quarks in section \ref{sssec:bb}.
We observe that in both these benchmarks, the efficiency is not affected much if we apply a cut on $\Delta\phi_{\mu\mu,{\rm mom}}>0.01$ instead of using the separation at the detector edge ($\Delta\phi_{\mu\mu}$).

\begin{figure}[t]
    \centering
    \includegraphics[width=0.46\textwidth]{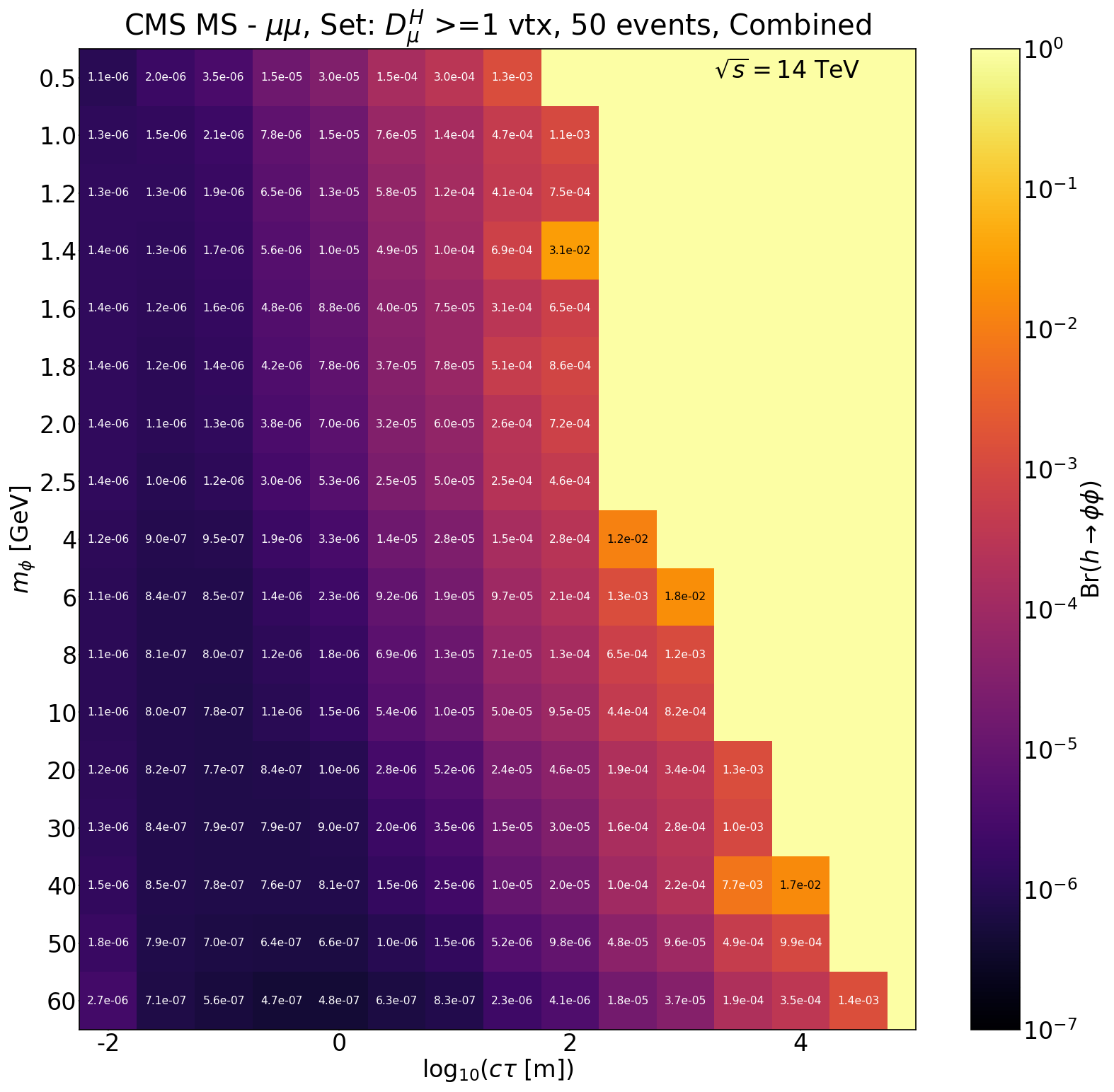} \qquad
    \includegraphics[width=0.46\textwidth]{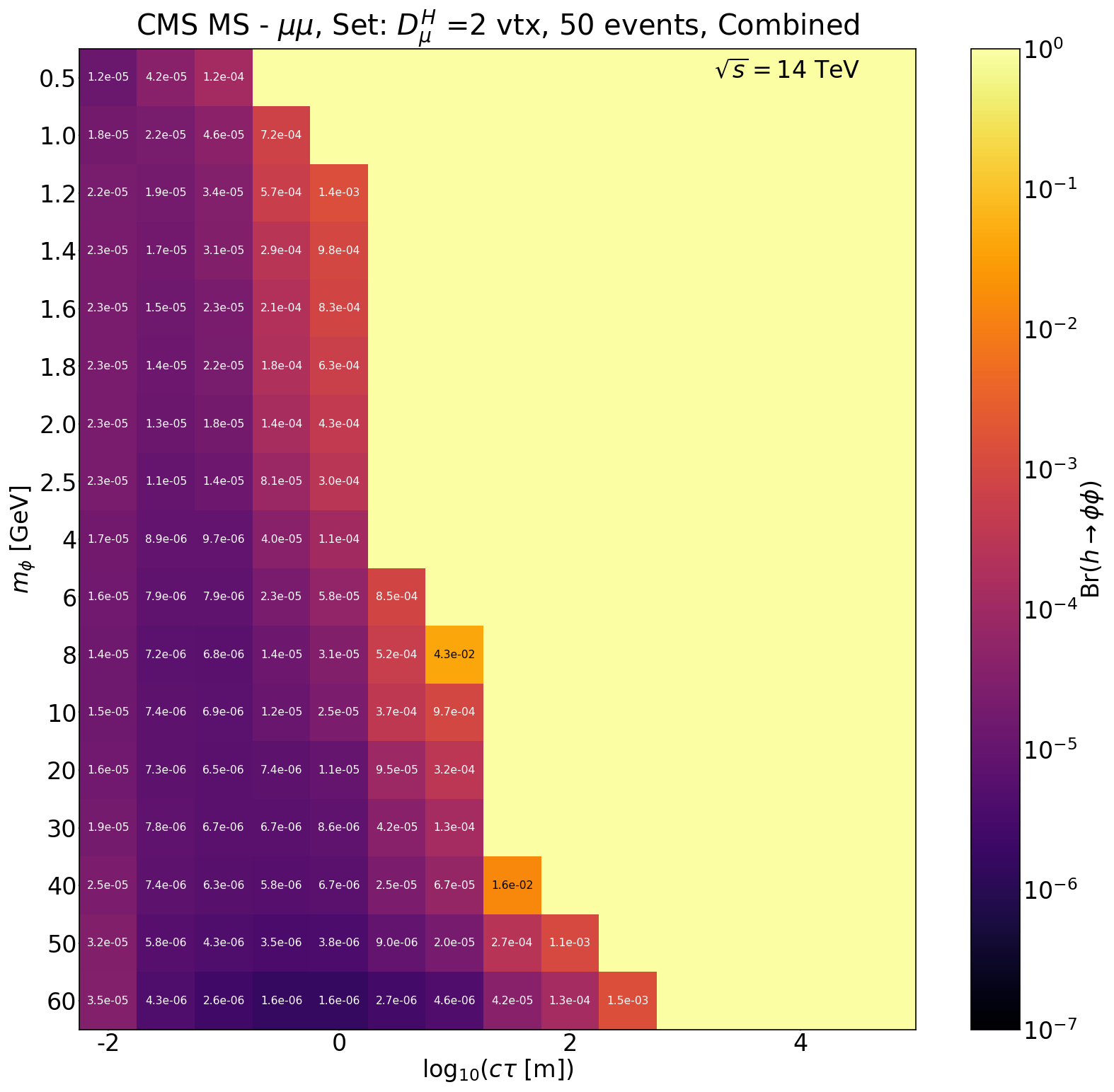} \\
    \includegraphics[width=0.46\textwidth]{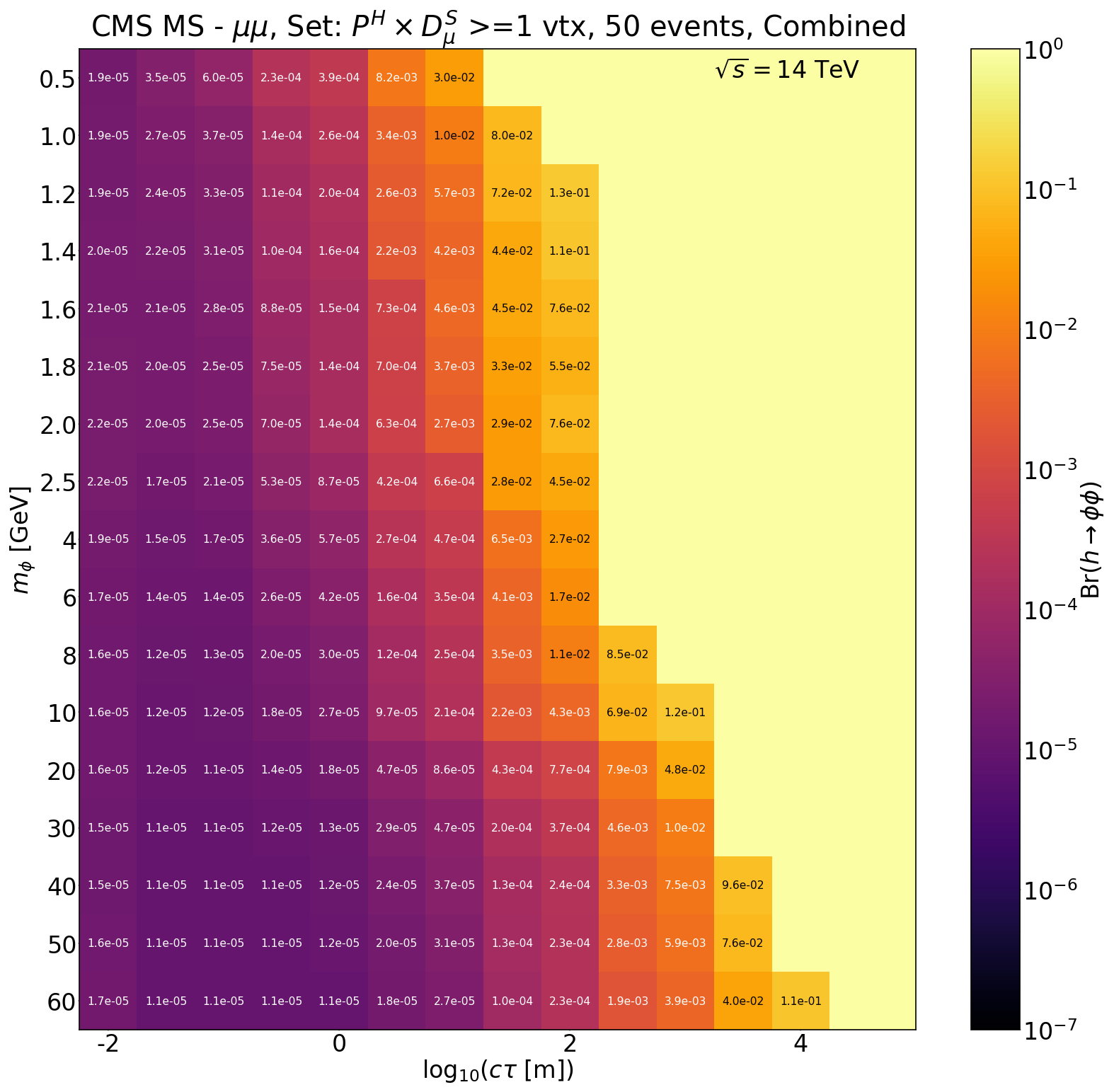} \qquad
    \includegraphics[width=0.46\textwidth]{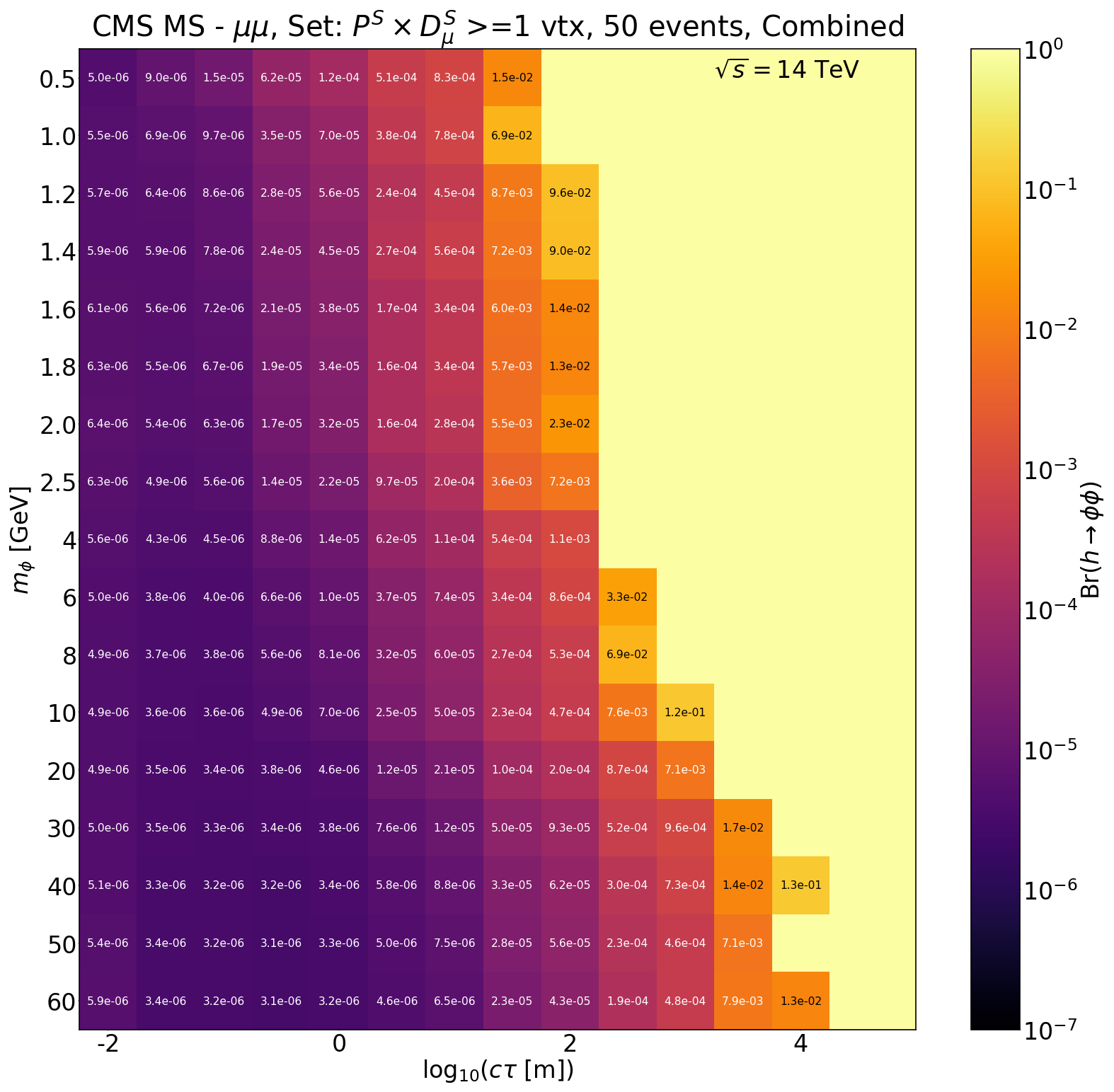}
    \caption{\small \sl Projected upper limits on the branching fraction of the Higgs boson decaying into a pair of the mediator particles, Br($h \to \phi\phi$), for 50 observed decays of the mediator particles to muons for sets of cuts applied only on the displaced muons (top panels) and those combining the prompt and displaced activities (bottom panels). The shown limits are obtained by combining the ggF, VBF and Vh productions for the Higgs boson.}
    \label{fig:mumu-14TeV-combo}
\end{figure}

\begin{figure}[t]
    \centering
    \includegraphics[width=0.46\textwidth]{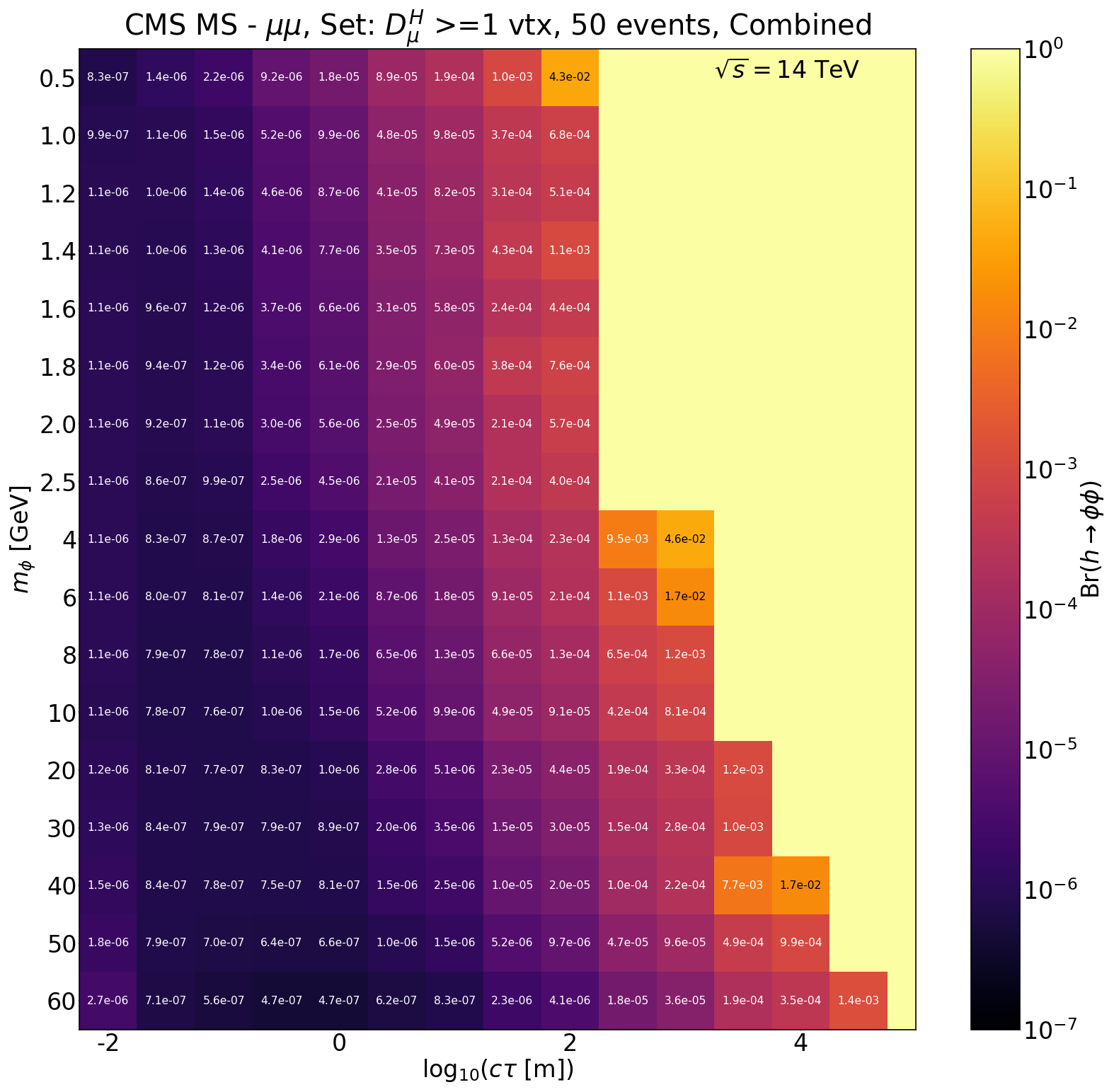} \qquad
    \includegraphics[width=0.46\textwidth]{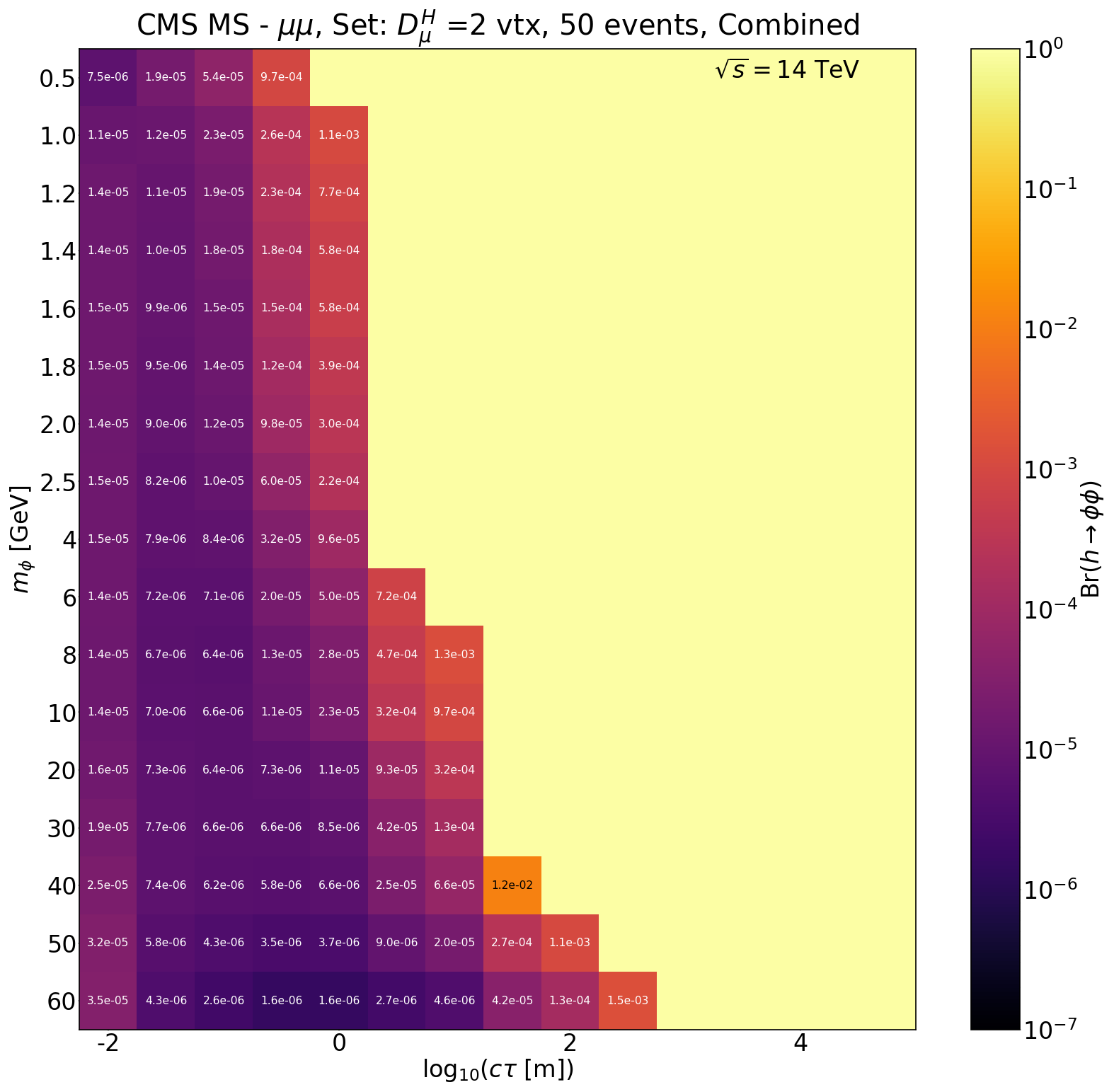} \\
    \includegraphics[width=0.46\textwidth]{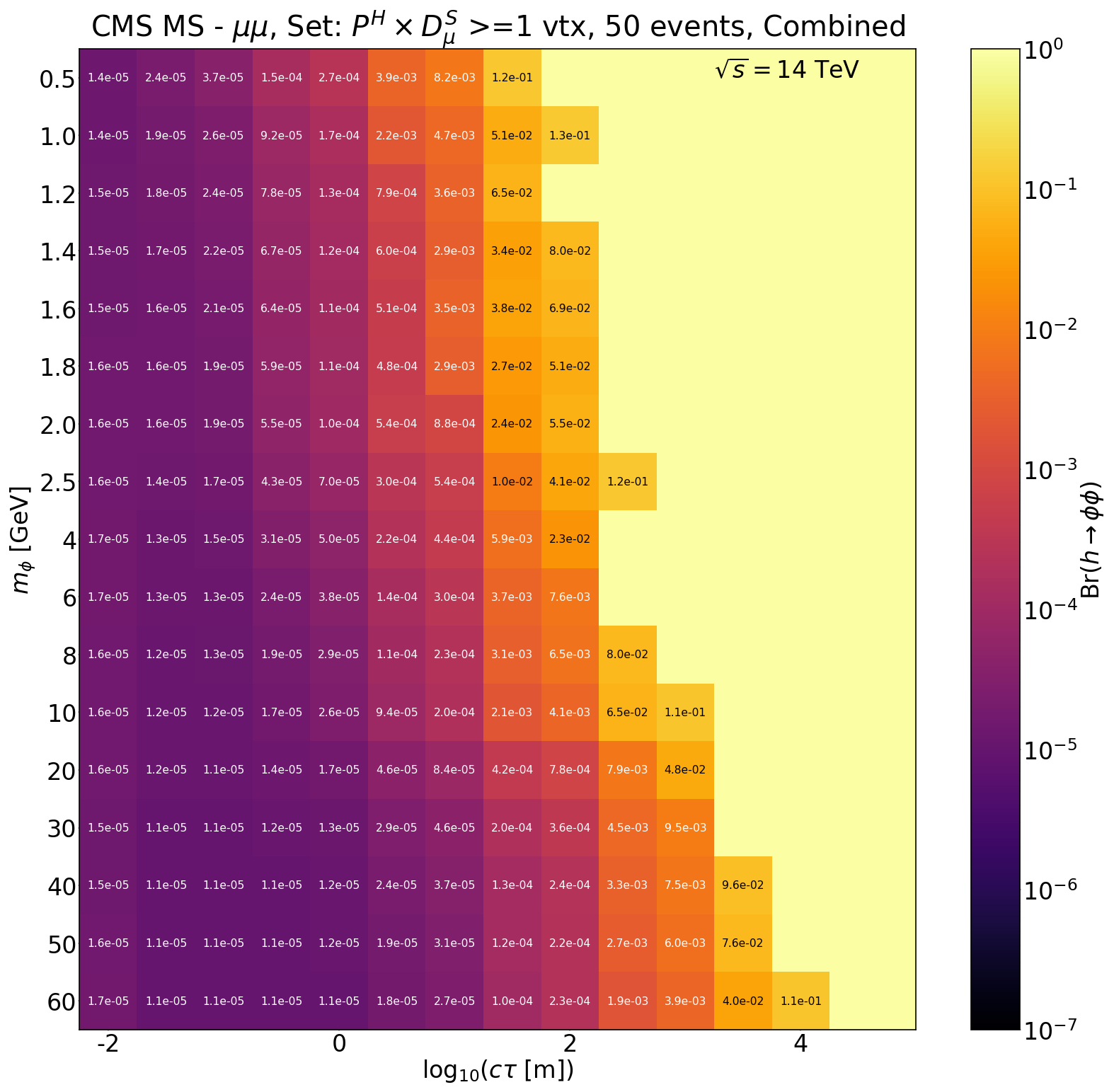} \qquad
    \includegraphics[width=0.46\textwidth]{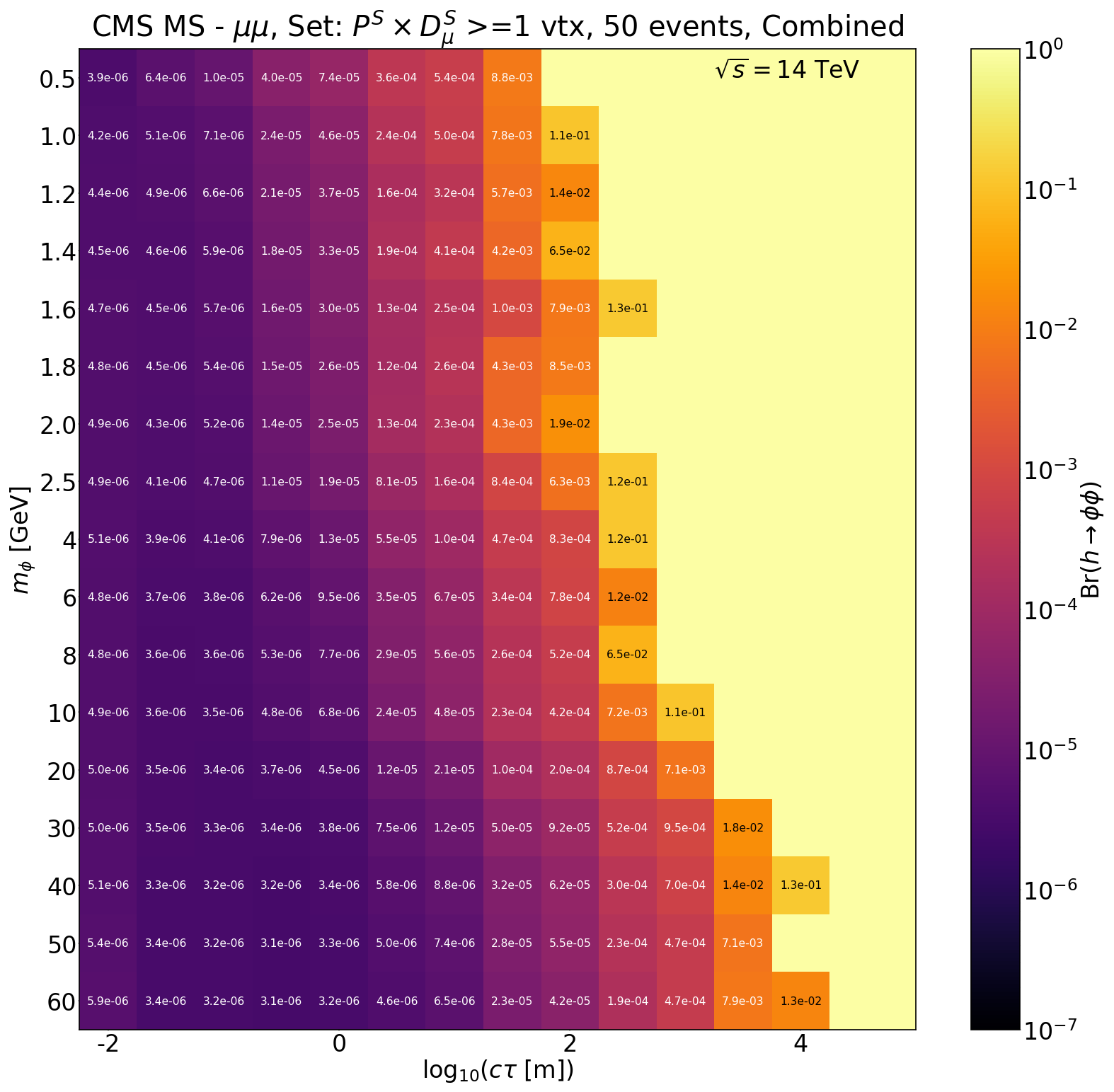}
    \caption{\small \sl Projected upper limits on the branching fraction of the Higgs boson decaying into a pair of the mediator particles, Br($h \to \phi\phi$), for 50 observed decays of the mediator particles to muons 
    The shown limits are obtained by combining the ggF, VBF and Vh productions for the Higgs boson.}
    \label{fig:mumu-14TeV-sv-combo}
\end{figure}

Fig.\,\ref{fig:mumu-14TeV-combo} shows the projected upper limits on the branching fraction of the Higgs boson decaying to a pair of the mediator particles with masses ranging from 0.5\,GeV to 60\,GeV, and a wide range of decay lengths starting from 1\,cm at the 14\,TeV HL-LHC experiment with a luminosity of 3000\,fb$^{-1}$, when we apply the cuts on displaced muons outlined in Table\,\ref{tab:muon_cuts} along with the cuts on the various prompt particles coming from the production channel described in Table\,\ref{tab:prod-cuts}, combined into four sets discussed earlier. The results obtained from production channels of ggF, VBF, Vh-jet, and Vh-lep have been combined in this figure.
In Fig.\,\ref{fig:mumu-14TeV-sv-combo}, we show similar limit grid plots with the isolation cut applied on $\Delta\phi_{\mu\mu,{\rm mom}}$ instead of the physical separation at the detector edge. We find that for lighter LLP masses, the $\Delta\phi_{\mu\mu,{\rm mom}}$ cut performs better than the latter, and for heavier mediators the limits from the two approaches are very close. 
Since we are demanding the displaced muons above some $p_T$ threshold, cases where the $\Delta\phi_{\mu\mu,{\rm mom}}$ is below some required threshold, however they are physically well separated due to bending in the magnetic field, do not occur. 
However, there are cases where the muon pair is isolated as per $\Delta\phi_{\mu\mu,{\rm mom}}$, but not physically separated due to the large displacement of the muons. 
In Fig.\,\ref{fig:eff-mumu-14TeV}, where we plot the variation of efficiency of the $D_\mu^H \geq 1$ vtx set of cuts with $c\tau$, we observe that in going from a mediator of mass 0.5\,GeV to 50\,GeV, the most sensitive decay length shifts from 1\,cm to 0.5\,m. This is also reflected in the top left plot of Fig.\,\ref{fig:mumu-14TeV-combo}, where for a 0.5\,GeV mediator decaying to muons, the most sensitive limit achieved is Br($h \to \phi \phi) <$ 
1.1$\times 10^{-6}$ at $c \tau_\phi =$ 1\,cm decay, whereas, for $m_\phi =$ 50\,GeV, Br($h \to \phi \phi) <$ $6.4 \times 10^{-7}$ at $c \tau_\phi =$ 0.5\,m. With an increasing mediator mass, the range of decay lengths sensitive to the analysis also increases since they have lower boosts and, even for higher lifetimes, decay within the detector, unlike lighter particles whose probability to decay within the detector decreases much rapidly with increasing $c\tau$.

Comparing the other set of cuts, we find that the best limits for the 0.5\,GeV mediator particle remain at $c \tau_\phi =$ 1\,cm, however, increase by factors of 
11, 17, and 4.5, respectively for the $D_\mu^H =$2\,vtx, $P^H \times D_\mu^S \geq$ 1\,vtx and $P^S \times D_\mu^S \geq$ 1\,vtx sets of cuts. Similar numbers for the 50\,GeV mediator particle are 
5.5, 17, and 4.8, respectively. Out of these four sets, the $D_\mu^H \geq$ 1\,vtx cuts is the most sensitive set since it demands only a single displaced di-muon pair. It will be feasible if it is possible to trigger the events at L1 of the HL-LHC experiment using only the displaced di-muons. In case that is not possible, we can demand two such displaced pairs of muons ($D_\mu^H =$ 2\,vtx), or select the event using the L1 menu for prompt objects and then identify the displaced muons at HLT ($P^H \times D_\mu^S \geq$ 1\,vtx). Another hybrid set that we have considered with reduced thresholds of both the prompt and displaced objects ($P^S \times D_\mu^S \geq$ 1\,vtx) may be used to improve the limits maintaining reasonable background. 
For the $D_\mu^H \geq$ 1\,vtx ($P^H \times D_\mu^S \geq$ 1\,vtx), sensitivity falls off around 50\,m (1-5\,m) and 10$^4$\,m (500\,m) for mediator masses of 0.5\,GeV and 50\,GeV respectively. The $D_\mu^H =$ 2\,vtx set of cuts loses its sensitivity at around 100\,m for the 50\,GeV mediator particle.
Among the production modes, although the ggF mode dominates the limits due to its higher cross section (Table\,\ref{tab:cs}), the VBF and Vh production modes have similar or even higher efficiencies for varying decay lengths over a range of mediator masses.

\subsubsection{Mediator particle decaying into pions/Kaons}

We now look for the mediator particle decaying into a pair of charged mesons, such as pions and Kaons. As already discussed in section\,\ref{ssec:decay-med}, when the mass of the mediator particle is below 2 GeV, it is better to consider that the mediator particle is decaying into mesons rather than into light quarks and gluons. We, therefore, consider mediator masses starting from $2\times m_{\pi/K}$ till 2\,GeV decaying to two pions and two Kaons in this section. The charged particle multiplicity associated with the mediator decay in these cases is two, to begin with. However, as discussed earlier, the charged hadrons can interact with the iron yoke to shower and produce further particles. Since the \texttt{PYTHIA6} code treats both $\pi^\pm$ and $K^\pm$ as detector stable particles, and we do not have a full detector simulation, we cannot quantify the final charged particle multiplicity for a mediator particle decaying into these particles.

We, therefore, closely follow the analysis for muons, the only difference being that we now only consider decays happening inside the muon spectrometer. The motivation to follow the muon analysis is that, if the pions and Kaons are as energetic as the muons, they might lead to an amount of activity in the MS, which will be enough to make it possible to trigger on. Table\,\ref{tab:piK_cuts} summarises the cuts used for the analysis of the mediator particle decaying to a pair of charged pions and Kaons. We do not apply any $d_0$ cut here since it might not be possible to measure this quantity as the mesons will shower and, therefore, might not have well-reconstructed tracks like the muons. Also, the possibility of the event passing the selection might improve when we can trigger on some prompt objects from the associated production mode along with this displaced activity. We also discuss the results of such combined sets of cuts on prompt and displaced activities, similar to the case of the muons.

\begin{table}[t]
    \centering
    \begin{tabular}{c|l|l|}
        & \qquad\qquad $D^H_\pi$/$D^H_K$ & \qquad\qquad $D^S_\pi$/$D^S_K$ \\
        \hline
        \multirow{3}{*}{Pions/Kaons} & $p_T^{\pi/K} >$ 20\,GeV & $p_T^{\pi/K} >$ 10\,GeV \\
        \cline{2-3}
        & $n_{\pi/K} \geq$ 2 & $n_{\pi/K} \geq$ 2 \\
        \cline{2-3}
        & $|\eta^{\pi/K}| <$ 2.8 & $|\eta^{\pi/K}| <$ 2.8 \\
        \hline
        \multirow{3}{*}{$\pi$ pair/$K$ pair} & $d_T >$ 4\,m or $|d_z| >$ 7\,m & $d_T >$ 4\,m or $|d_z| >$ 7\,m \\
        & $d_T <$ 6\,m and $|d_z| <$ 9\,m & $d_T <$ 6\,m and $|d_z| <$ 9\,m \\
        \cline{2-3}
        & $\Delta\phi_{\pi\pi/KK} >$ 0.01 & $\Delta\phi_{\pi\pi/KK} >$ 0.01 \\
        \hline
        Event & $n_{vtx} \geq$ 1 or $n_{vtx} =$ 2 & $n_{vtx} \geq$ 1 or $n_{vtx} =$ 2 \\
        \hline
    \end{tabular}
    \caption{\small \sl Harder and softer sets of cuts applied for the displaced pions or Kaons.}
    \label{tab:piK_cuts}
\end{table}

\begin{figure}[t]
    \centering
    \includegraphics[width=0.46\textwidth]{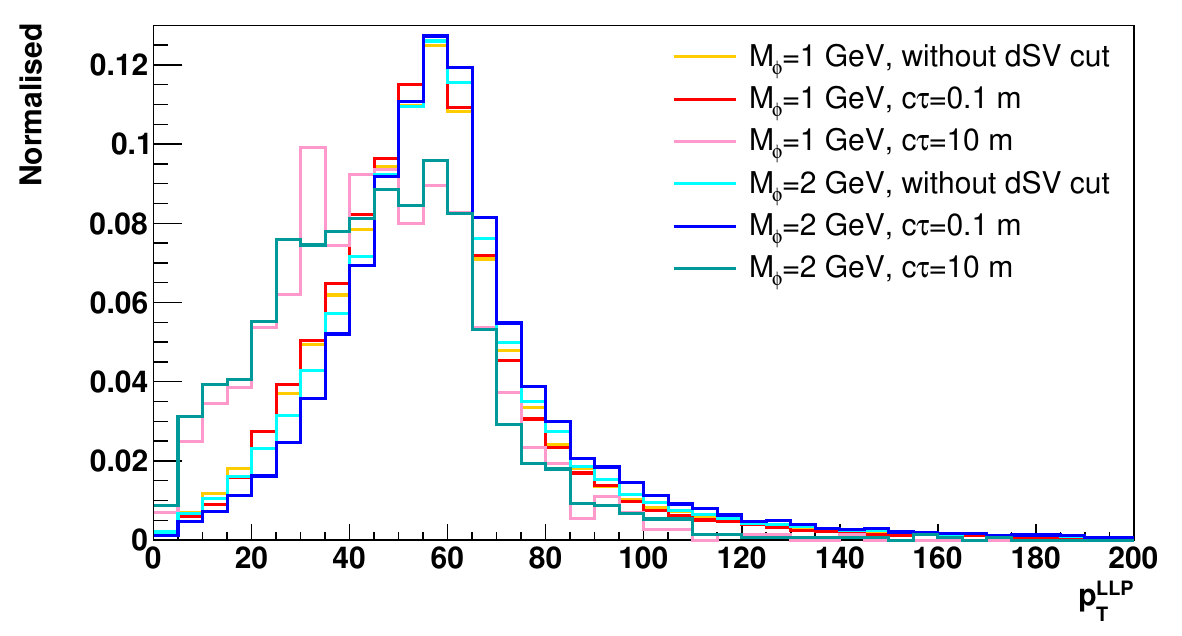} \qquad
    \includegraphics[width=0.46\textwidth]{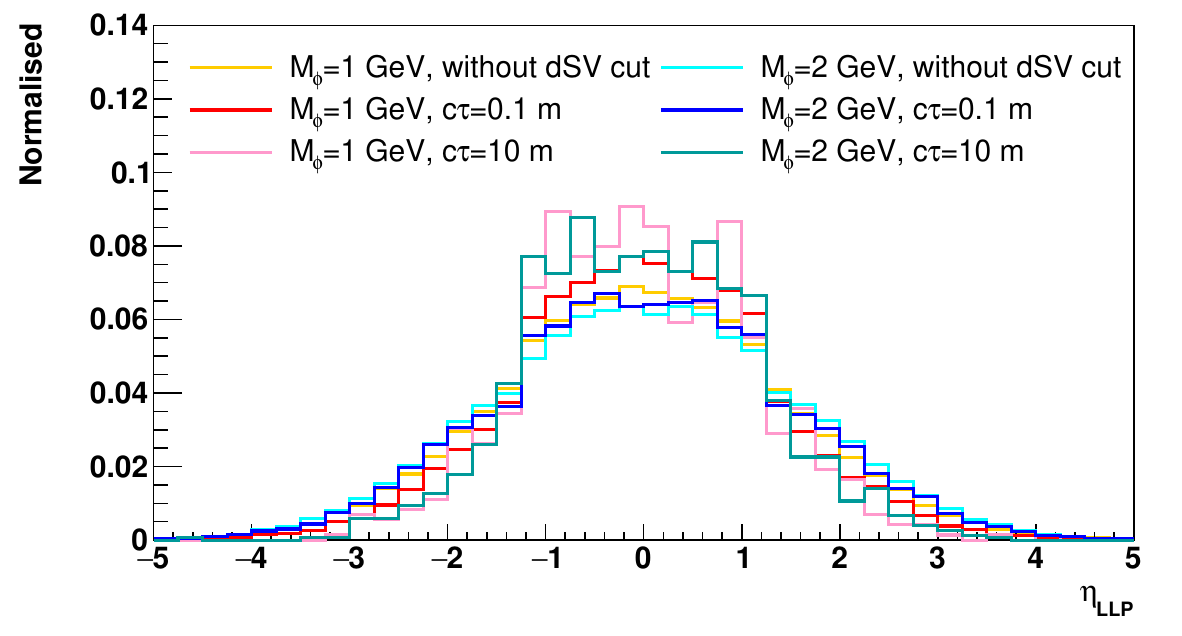}
    \caption{\small \sl Normalised histograms of the transverse momentum (left panel) and the pseudo-rapidity (right panel) distributions of the long-lived mediator particle for two benchmark masses ($m_\phi =$ 1\,GeV and 2\,GeV) and decay lengths ($c \tau_\phi =$ 0.1\,m and 10\,m) before (dashed lines) and after (solid lines) applying the required cut on the dSV.}
    \label{fig:pt-eta-LLP-smallmass}
\end{figure}

As we now constrain the decays within the MS, let us have a look at the distributions of transverse momentum and pseudo-rapidity of the mediator particles that decay inside the MS for some benchmark masses and decay lengths. Fig.\,\ref{fig:pt-eta-LLP-smallmass} shows the normalised histograms of $p_T$ and $\eta$ distributions of the long-lived mediator particles when it has a mass of 1\,GeV and 2\,GeV and how these distributions get biased when we apply a cut on the dSV position for two decay lengths of 0.1\,m, and 10\,m. We observe that for both the 1\,GeV and 2\,GeV mediator particles with $c \tau_\phi =$ 0.1\,m, the $p_T$ distributions do not change much on applying the dSV cut, however, for the higher decay length of 10\,m, the dSV cut selects low $p_T$ mediators, otherwise high boost leads to decay outside the detector. For the $\eta$ distributions, the difference made by the dSV cuts is not significant, however, it makes the distributions slightly more central compared to the original ones for both the benchmark masses and decay lengths. We will revisit this plot for higher mass benchmarks in the next section, as there we discuss the decay of the mediator particle into a pair of $b$-quarks.

\begin{figure}[t]
    \centering
    \includegraphics[width=0.46\textwidth]{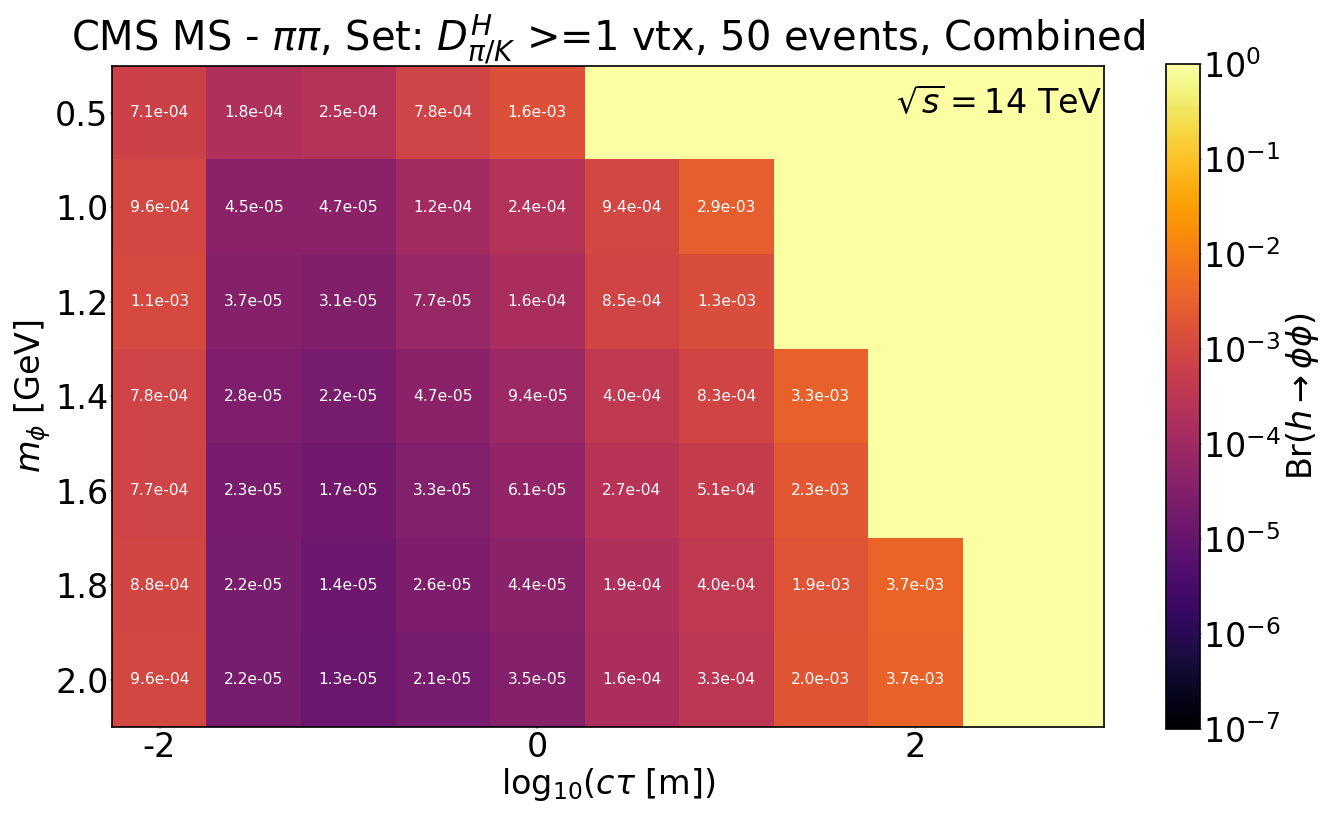} \qquad
    \includegraphics[width=0.46\textwidth]{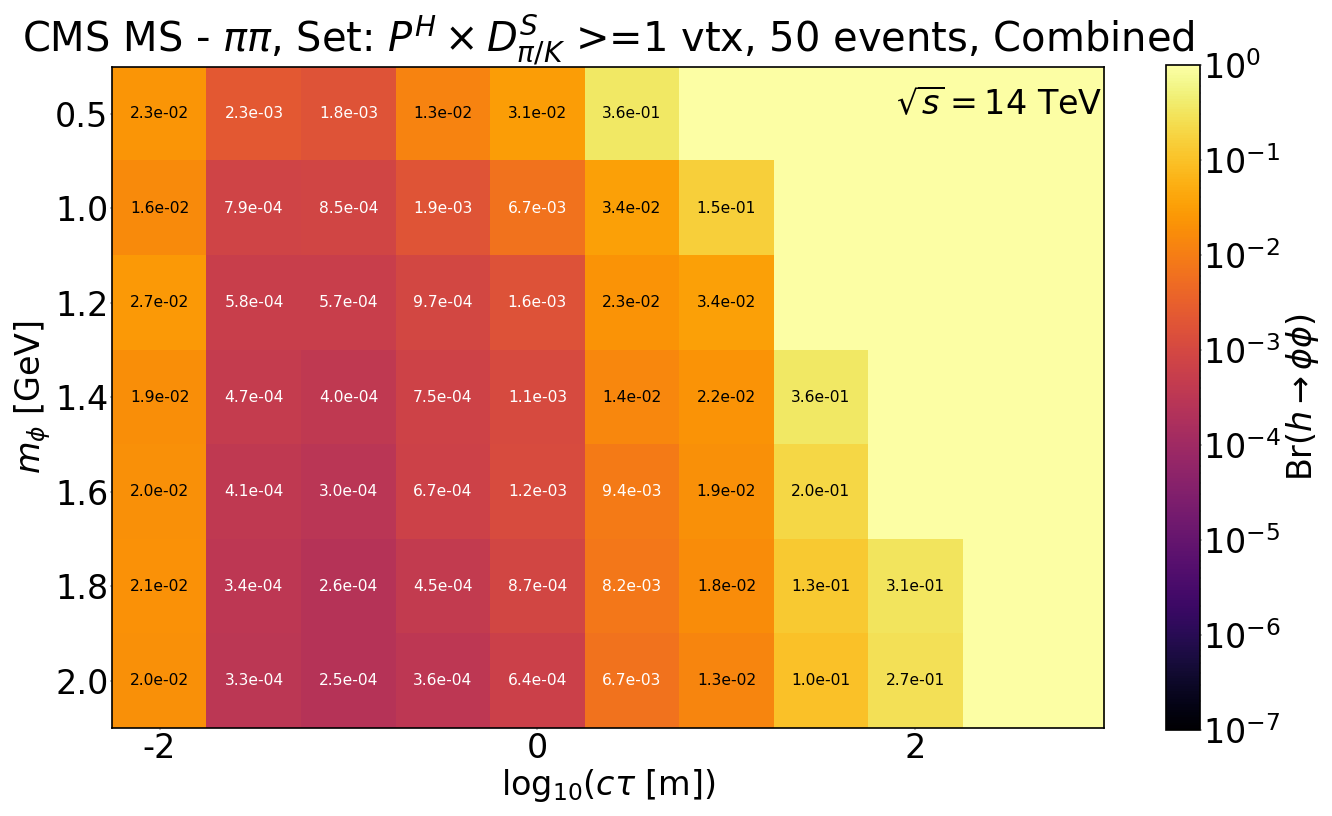} \\
    \includegraphics[width=0.46\textwidth]{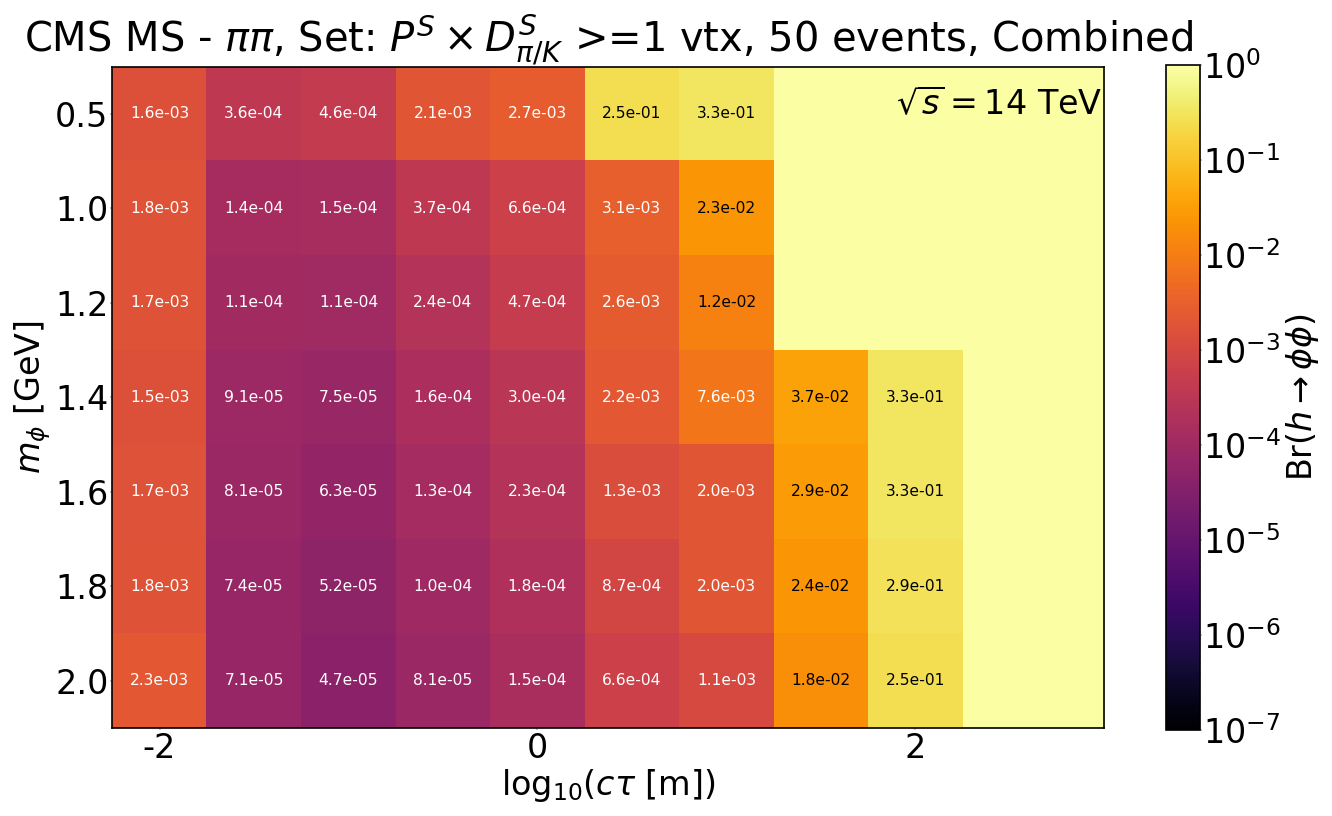} \qquad
    \caption{\small \sl Projected upper limits on Br($h \to \phi \phi$) for 50 observed decays of the mediator particles into pions for three sets of cuts explained in the text. The shown limits are obtained by combining the ggF, VBF and Vh production modes of the Higgs boson.}
    \label{fig:pipi-14TeV-combo}
\end{figure}

\begin{figure}[t]
    \centering
    \includegraphics[width=0.46\textwidth]{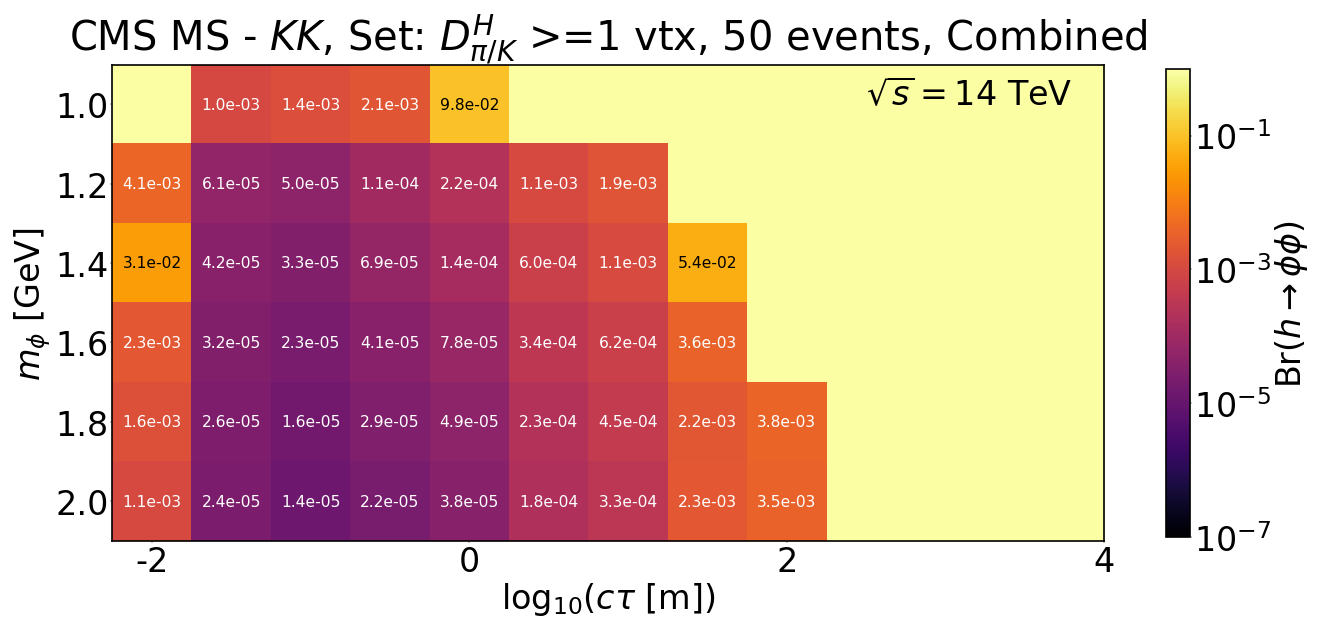} \qquad
    \includegraphics[width=0.46\textwidth]{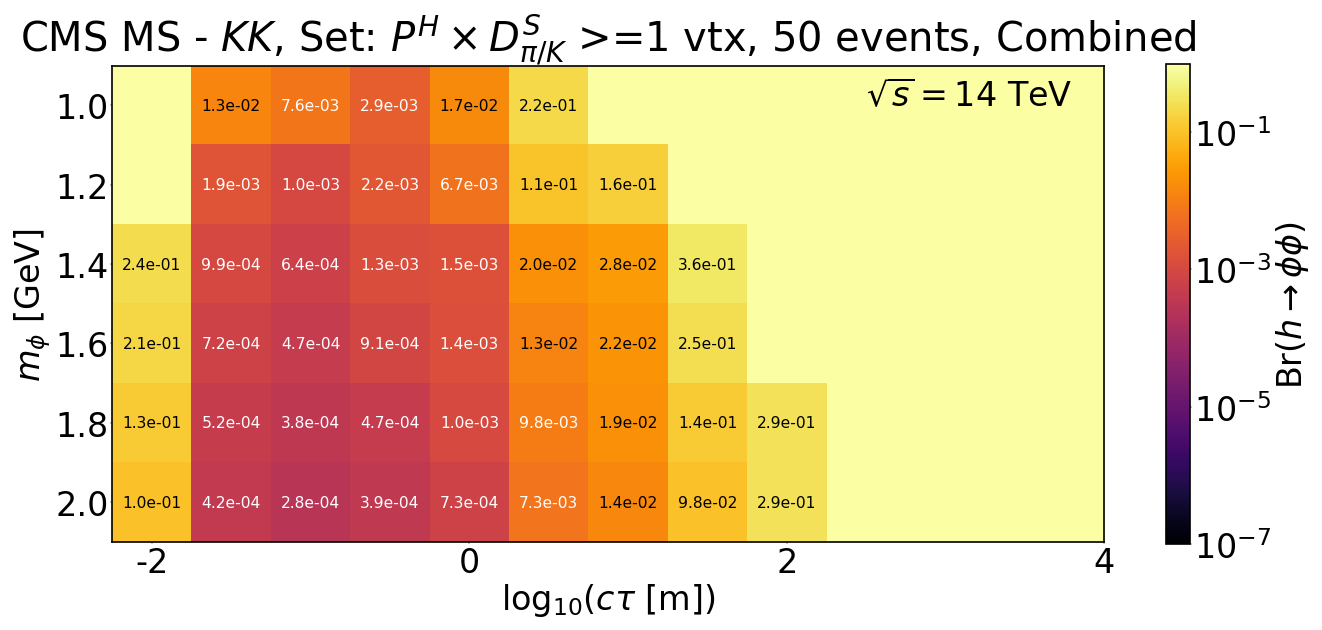} \\
    \includegraphics[width=0.46\textwidth]{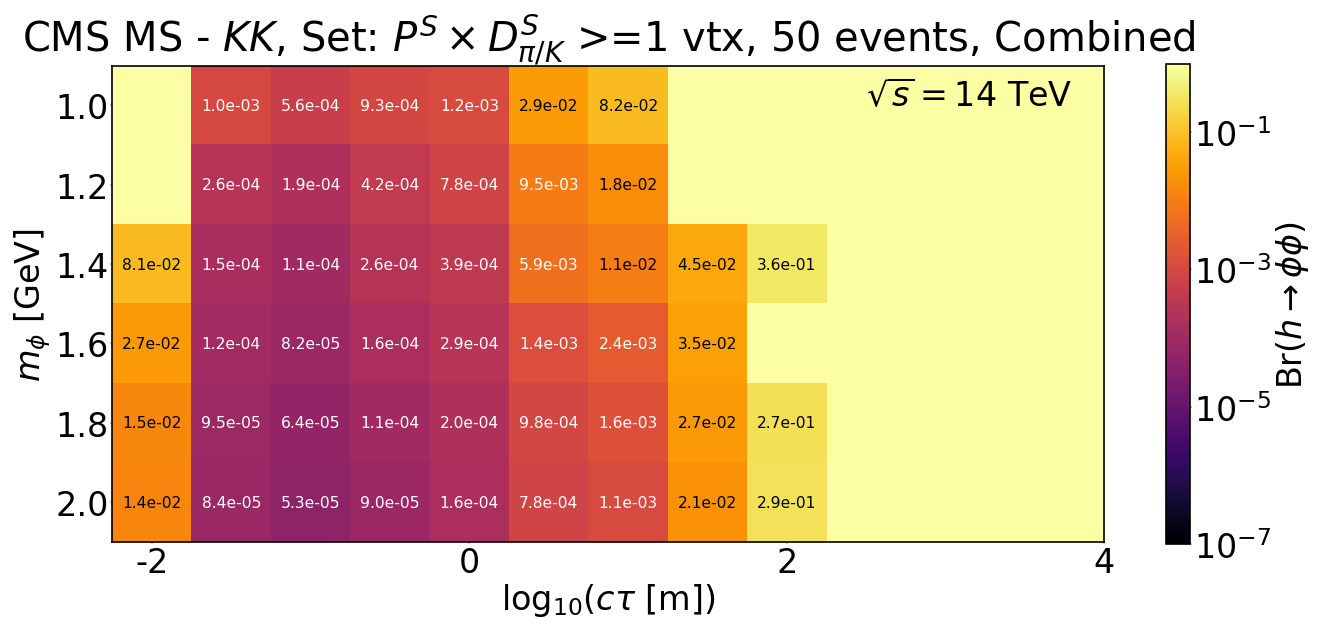}
    \caption{\small \sl Projected upper limits on Br($h \to \phi \phi$) for 50 observed decays of the mediator particles into Kaons for three sets of cuts explained in the text. The shown limits are obtained by combining the ggF, VBF and Vh production modes of the Higgs boson.}
    \label{fig:KK-14TeV-combo}
\end{figure}

Figs.\ref{fig:pipi-14TeV-combo} and \ref{fig:KK-14TeV-combo} show the projected upper limit on the Br$(h \to \phi\phi)$ for the set of cuts $D^H \geq$ 1\,vtx, $P^H \times D^S \geq$ 1\,vtx, and $P^S \times D^S \geq$ 1\,vtx, for some benchmark masses and a range of lifetimes assuming 100\,\% decay to pions and Kaons, respectively, combined over ggF, VBF, Vh-jet and Vh-lep production modes at 14\,TeV HL-LHC experiment with an integrated luminosity of 3000\,fb$^{-1}$. For the set of cuts outlined in Table\,\ref{tab:piK_cuts}, we get negligibly small efficiencies for events where both the mediator particles decay inside the MS and the mesons coming from their decay satisfy the harder set of cuts, $D^H_{\pi/K} =$ 2\,vtx, and therefore we do not show these results. We observe that the $P^S\times D_{\pi/K}^S \geq$ 1\,vtx set of cuts has the best sensitivity when the mediator decays to pions and Kaons, and the best limits with this set of cuts for masses 1\,GeV and 2\,GeV are 
Br$(h \to \phi\phi)<3.6\times10^{-4}$, $c\tau=5$\,cm ($5.6\times10^{-4}$, $c\tau=10$\,cm) and $4.7\times10^{-5}$, $c\tau=10$\,cm ($5.3\times10^{-5}$, $c\tau=10$\,cm)
achieved respectively for pions (Kaons). The limits achieved for pions and Kaons are comparable, and both degrade by a factor of 20 as compared to the case where the mediator decays to muons. The mass range of mediators studied in this section varies over a small interval, and therefore, we do not see much difference in the variation of limits with decay length over the masses. 
The $D^H \geq$ 1\,vtx and $P^S \times D^S \geq$ 1\,vtx set of cuts lose their sensitivity around decay lengths of 5-10\,m, whereas the $P^H \times D^S \geq$ 1\,vtx analysis loses sensitivity around 1\,m.

\subsubsection{Mediator particle decaying into $b \bar{b}$/$g g$/$s \bar{s}$/$c \bar{c}$}
\label{sssec:bb}

The mediator particle can decay into gluons and quarks when its mass exceeds a few GeV as discussed in section\,\ref{sec:model}, and these will lead to displaced jets in the final state. The mediator particle tends to dominantly decay into two $b$-quarks when its mass exceeds 10\,GeV as we have also discussed in section\,\ref{ssec:decay-med}. We, therefore, outline the analysis for displaced jets in the CMS MS by considering $b$-jets first, and then apply this analysis to cases for jets from gluons, strange quarks and charm quarks.

Jets are objects with high particle multiplicity. Moreover, jets consist of both charged and neutral hadrons, photons, and even electrons. All of these particles can interact with the iron yokes in the MS and give rise to multiple hits, as we have previously discussed. Due to the absence of a full detector simulation that takes into account the particles' interactions with the detector elements, our analysis makes use of some phenomenological cuts motivated by what might be achieved in the real HL-LHC experiment. As already mentioned earlier, the first criterion is that the decay vertex needs to be present within the CMS MS, i.e., we select only those events where the mediator particle decays after a transverse distance of 4\,m or half-length 7\,m, and before the second last MS station which is located at $d_T =$ 6\,m and $|d_z| =$ 9\,m. We, therefore, put this cut on the position of the dSV in our analysis.

\begin{figure}[t]
    \centering
    \includegraphics[width=0.46\textwidth]{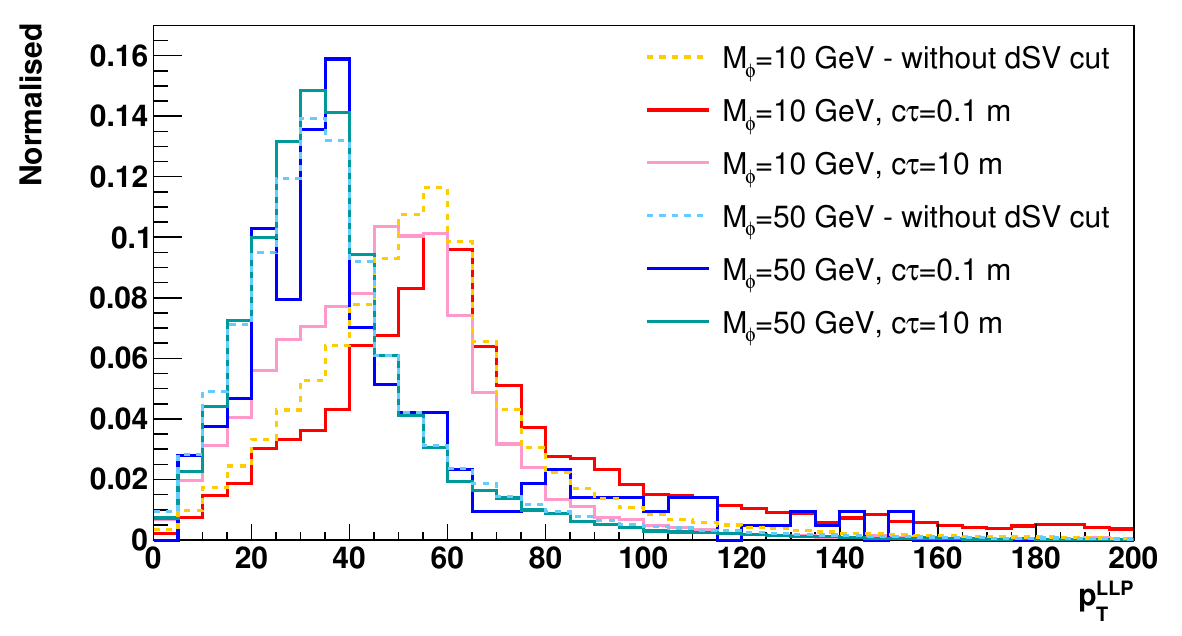} \qquad
    \includegraphics[width=0.46\textwidth]{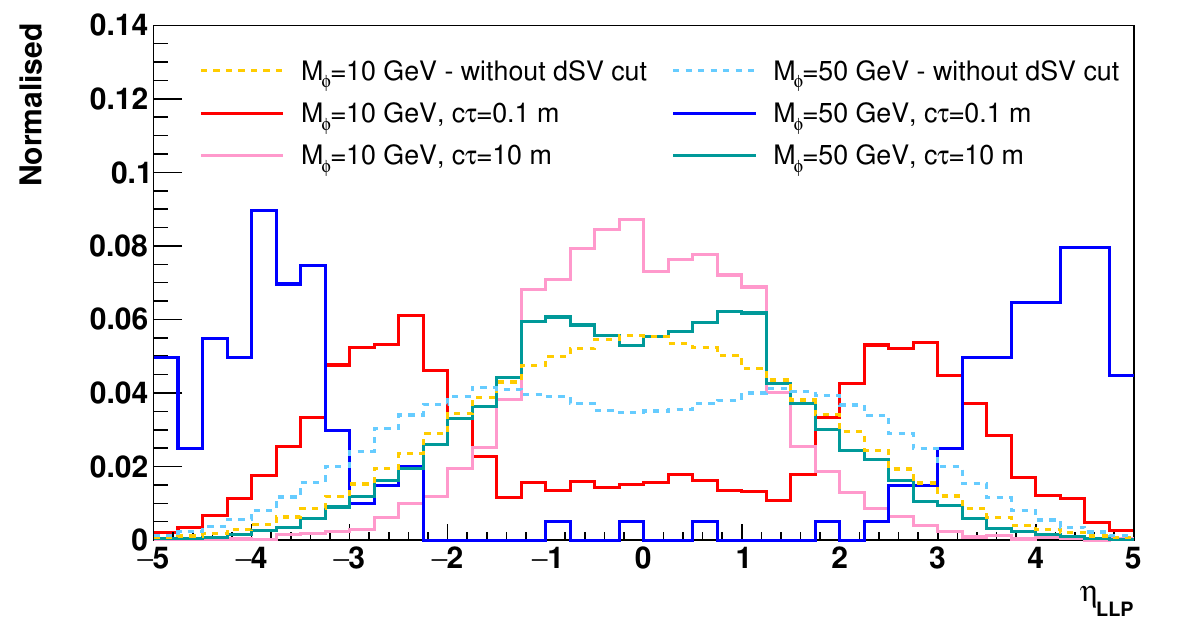}
    \caption{\small \sl Normalised histograms of the transverse momentum (left panel) and the pseudo-rapidity (right panel) distributions of the long-lived mediator particle for two benchmark masses ($m_\phi =$ 10\,GeV and 50\,GeV) and decay lengths ($c \tau_\phi =$ 0.1\,m and 10\,m) before (dashed lines) and after (solid lines) applying the required cut on the dSV.}
    \label{fig:pt-eta-LLP}
\end{figure}

Fig.\,\ref{fig:pt-eta-LLP} shows the $p_T$ and $\eta$ distributions of the mediator particles, coming from the decay of the Higgs boson produced in the ggF mode, with two benchmark masses of 10\,GeV and 50\,GeV, each having a decay length of 0.1\,m and 10\,m after putting the cut on dSV compared to the initial distributions without any cut on the position of the dSV. This plot is the extension of Fig.\ref{fig:pt-eta-LLP-smallmass} for mediators with higher masses, and we find that the effect of the dSV cut is more evident here. The first observation is that the long-lived particles of mass 10\,GeV have higher $p_T$ than the 50\,GeV's, and the former is more central than the latter. This is due to the fact that the lighter mediator particle has more boost than the latter. Also, since these come from the decay of the Higgs boson, which itself has mostly low $p_T$ and high $\eta$ values in the ggF production mode, the heavier mediator particles follow these distributions owing to their lower boosts. The second observation is that mediator particles with a decay length of 0.1\,m have harder $p_T$ distributions and more forward $\eta$ distributions for both 10\,GeV and 50\,GeV mass compared to the respective larger decay length (10\,m) counterparts. For low values of decay length, the mediator particle needs higher boost to reach the MS, and the cut on dSV position captures mostly the tail of the $p_T$ and $p_z$ distributions for such decay lengths, which explains the observed trend with decreasing decay length in Fig.\,\ref{fig:pt-eta-LLP}. For the mediator particle with lifetime 10\,m, the dSV cut biases the distribution towards lower $p_T$ and more central $\eta$ when compared to the actual distributions without requiring the decay to be in the MS, opposite to what we observed for the benchmarks with smaller decay length. This is on account of the fact that a higher boost will make the mediator particle with a larger decay length to decay outside the fiducial MS decay volume that we have considered. Thus, for each mass and decay length, we capture a different range of boosts inside the CMS MS (or any other detector volume kept away from the IP).

\begin{figure}[t]
    \centering
    \includegraphics[width=0.46\textwidth]{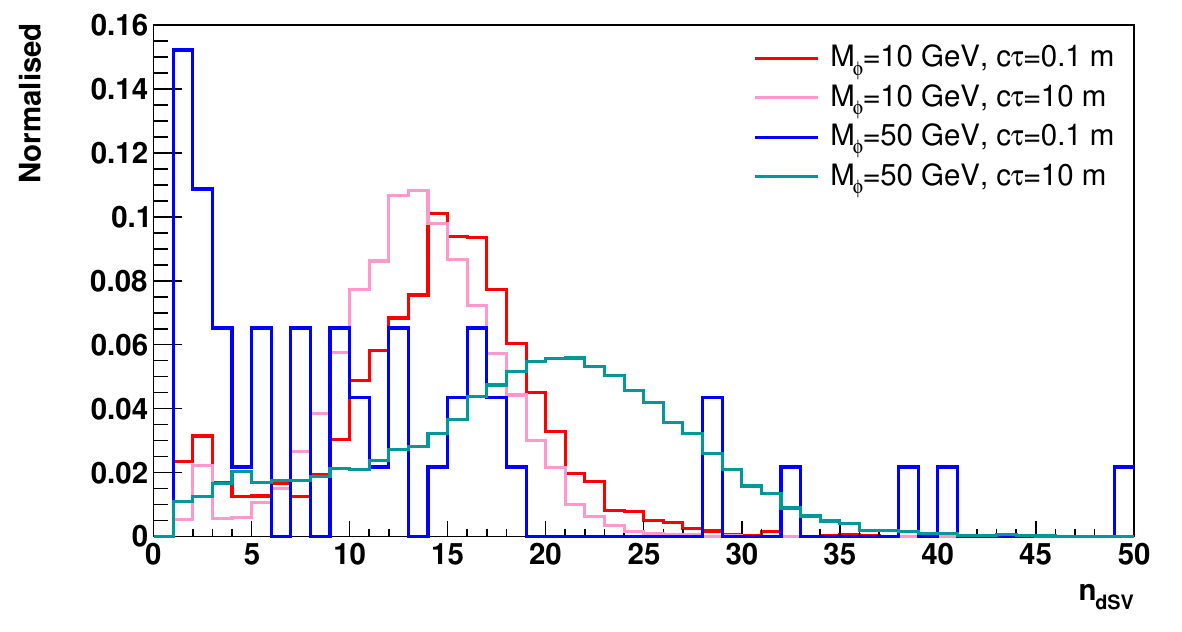} \qquad
    \includegraphics[width=0.46\textwidth]{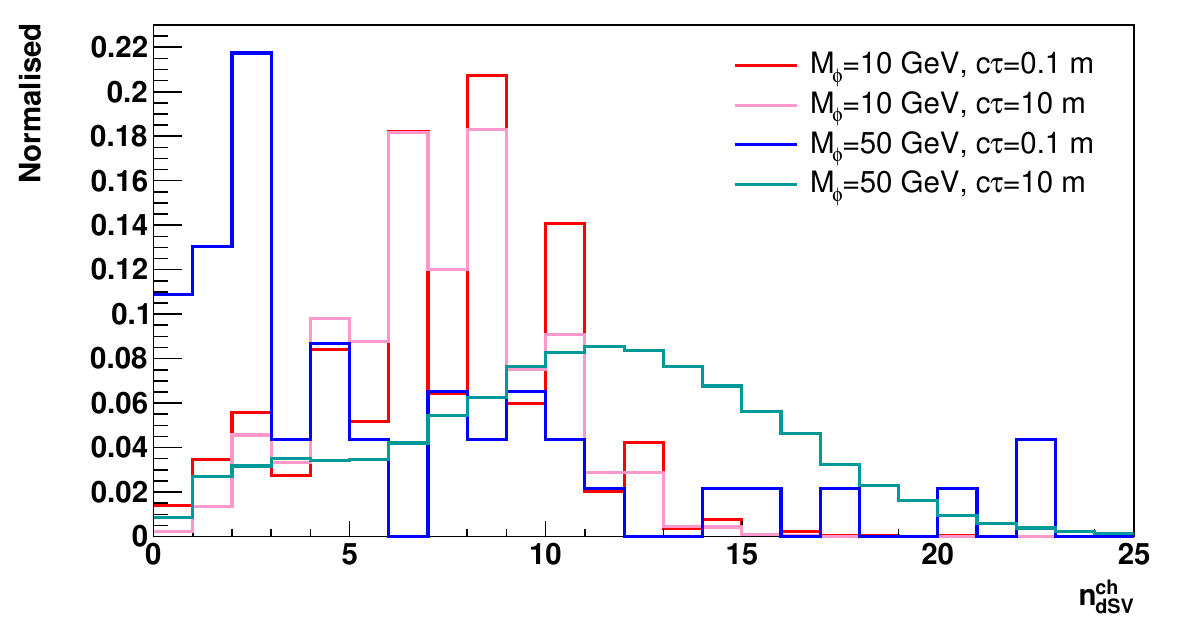}
    \caption{\small \sl Normalised histograms of multiplicity distributions of all particles associated with a dSV, ($n_{\rm dSV}$, (left panel) and charged particles associated with a dSV, $n_{\rm dSV}^{\rm ch}$, (right panel) for long-lived mediator particles having masses of $m_\phi =$ 10\,GeV and 50\, GeV, and the decay lengths of $c \tau_\phi =$ 0.1\,m and 10\,m after applying the required cut on the dSV. Multiplicity is calculated with particles having $p_T >$ 0.5\,GeV and $|\eta| < 2.8$.}
    \label{fig:bb_14_ndsv}
\end{figure}

Once we select decays of mediator particles into a pair of $b$-jets inside the CMS MS, we need to ensure that they can be triggered efficiently with a reasonably low background. Particle multiplicity from the same dSV, i.e. $n_{\rm dSV}$, is an important factor, because it will eventually control the number of hits in a cluster detected in the CMS MS, namely, a higher multiplicity ensures that the cluster of MS hits can be easily distinguished from the SM punch-through background. Also, if we demand a minimum cut on the number of charged particles associated with the same dSV, i.e. $n_{\rm dSV}^{\rm ch}$, for instance, we demand at least 3 or 5 charged particles, we can significantly reduce the background from the decay of long-lived particles in the SM, such as $K_S$ and $\Lambda$, which have a charged multiplicity of two\footnote{Other SM long-lived hadrons which might decay after punching through the calorimeter are, $\Sigma^+$, $\Sigma^-$, $\Xi$, $\Xi^-$ and, $\Omega^-$. Some of these, like $\Xi^-$ or $\Omega^-$, might result in $n_{\rm dSV}^{\rm ch} \geq 3$, however, these will be rare with low $p_T$ and our harder set of cuts with $n_{\rm dSV}^{\rm ch} \geq 5$ can get rid of such backgrounds.}. We calculate the multiplicities including hadrons, electrons and photons coming from the same dSV with $p_T >$ 0.5\,GeV and $|\eta| < 2.8$. Our definition of the same dSV includes tolerance of 1\,cm in each $x$-, $y$- and $z$-direction, in order to take into account the displaced decays of $b$-mesons. Some mesons can still have longer decay lengths in the laboratory frame, and hence we will lose them in this analysis.  Fig.\,\ref{fig:bb_14_ndsv} shows the histograms of the $n_{\rm dSV}$ distribution (left panel) as well as the $n_{\rm dSV}^{\rm ch}$ distribution (right panel).

In actual experiments, the number of hits associated with a MS cluster will also depend on the momentum or energy of all the particles from the same vertex, i.e., more energy leads to more showering, and hence increase the hit multiplicity. We, therefore, require the sum of the transverse momentum of all the particles associated with a mediator particle decay ($\sum p_{T,\,dSV}$) to be greater than some threshold, which we choose to be 20\,GeV and 50\,GeV as the softer and harder cuts respectively, to guarantee that these particles can shower to leave enough number of hits in the MS. As we are demanding higher $\sum p_{T,\, dSV}$ requirement, we do not apply any cut on $n_{\rm dSV}$, because both have a similar goal of increasing the hits in an MS cluster. Another feature of this cluster is its spread in the $\phi$ direction which we quantify by finding the maximum of the $\Delta\phi$ between all possible pairs of particles with $p_T >$ 2\,GeV\footnote{We are only using particles with $p_T>$ 2\,GeV to calculate $\Delta\phi_{\rm max}$, because some stray low $p_T$ particles are possible to give high $\Delta\phi$ values, while they will not add many hits to the MS cluster due to lower $p_T$. Hence, it does not determine the actual spread of the MS cluster.}, $|\eta| < 2.8$ from the same dSV ($\Delta\phi_{\rm max}$)\footnote{$\Delta\phi$ is calculated at the outermost edge of the detector, i.e. using \texttt{PhiOuter} from \texttt{Delphes}.}. Identifying the cluster of hits will be easier if it is more spread out, and therefore the particles are more separated from each other. Motivated from this, we also apply a minimum cut on $\Delta\phi_{\rm max}$ as well.

\begin{figure}[t]
    \centering
    \includegraphics[width=0.46\textwidth]{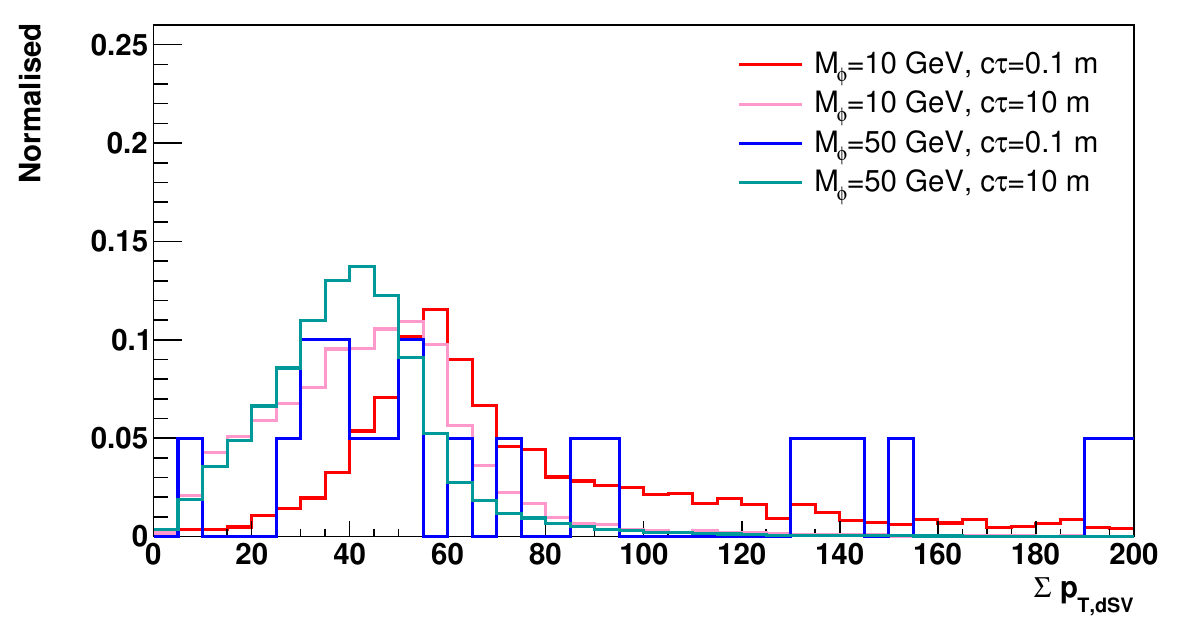} \qquad
    \includegraphics[width=0.46\textwidth]{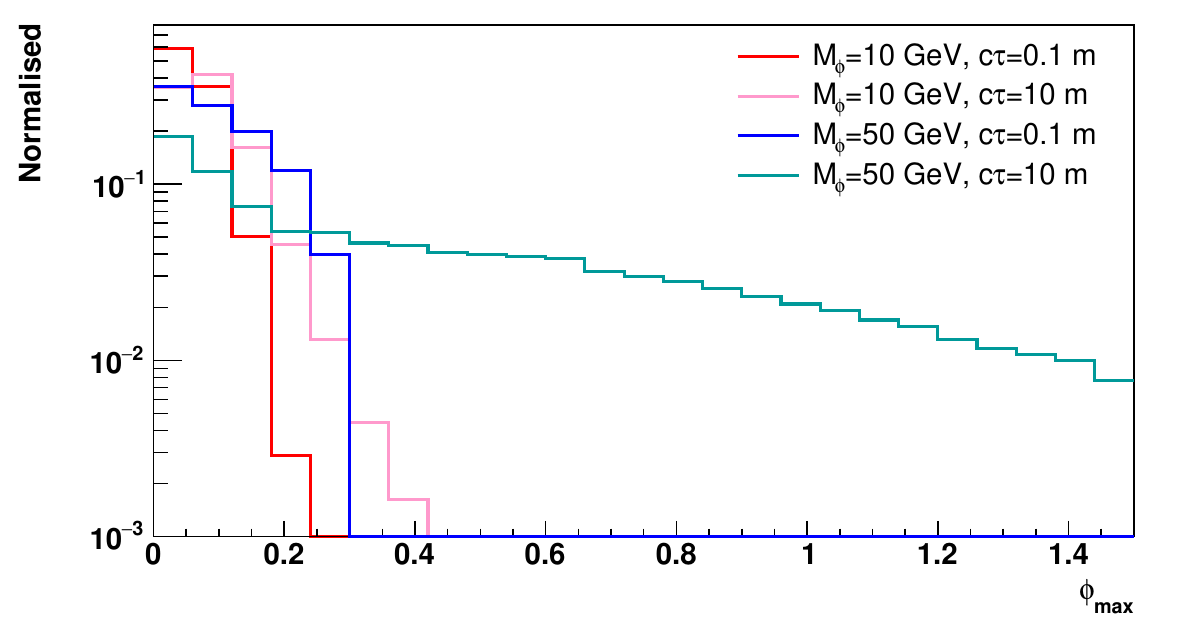}
    \caption{\small \sl Normalised histograms of the $\sum p_{T,\,dSV}$ (left panel) and $\Delta\phi_{max}$ (right panel) distributions for long-lived mediator particles having masses of $m_\phi =$ 10\,GeV, and 50\,GeV and decay lengths of $c \tau_\phi =$ 0.1\,m and 10\,m after applying the required cut on the position of the dSV and $n_{\rm dSV}^{\rm ch} \geq$ 3.}
    \label{fig:bb_14_ptsum}
\end{figure}

Fig.\,\ref{fig:bb_14_ptsum} shows the histograms of $\sum p_{T,\,dSV}$ and $\Delta\phi_{\rm max}$ distributions for the four benchmarks with masses of 10\,GeV and, 50\,GeV and decay lengths of 0.1\,m and 10\,m, for the ggF production of the Higgs boson. These distributions are drawn for only those vertices where $n_{\rm dSV}^{\rm ch} \geq$ 3. The benchmark with the 50\,GeV and $c \tau_\phi =$ 0.1\,m mediator particle has very few events left after the $n_{\rm dSV}^{\rm ch} \geq$ 3 and $|\eta| < 2.8$ cuts, therefore the distributions for this benchmark are not very continuous. We find that the $\sum p_{T,\,dSV}$ distributions show similar trends as the $p_T^{LLP}$ distributions, which is expected because $p_T$ sum associated with a displaced vertex, modulo the detector effects, is equivalent to the $p_T$ of the mediator particle. The value of $\Delta\phi_{\rm max}$ is controlled by two factors; boost of the mediator particle which affects the $p_T$ of its decay products as well as the decay position. A higher value of both indicates lower $\Delta\phi_{\rm max}$. 
We find that the distribution sharply falls around $\Delta\phi_{\rm max}$ value of 0.2-0.4 for a mediator of mass 10\,GeV with decay length 0.1\,m and 10\,m, and also the 50\,GeV mediator having a lower decay length (0.1\,m), whereas the $\Delta\phi_{\rm max}$ distribution extends to much higher values for the heavier mediator with higher decay length (50\,GeV with $c\tau=10$\,m). 
We put a softer cut of 0.1 and a harder cut of 0.2 on this variable.

\begin{figure}[t]
    \centering
    \includegraphics[width=0.46\textwidth]{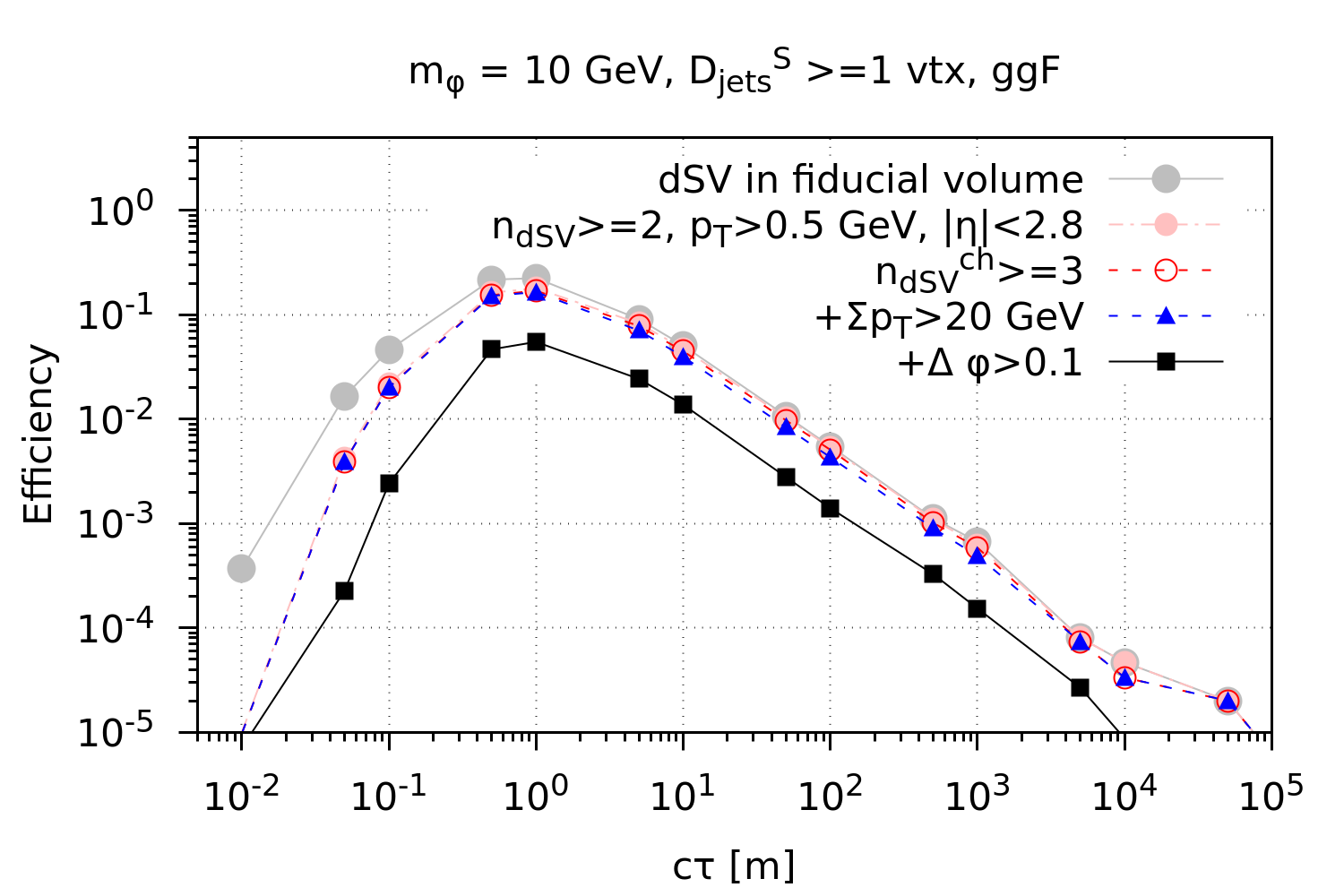} \qquad
    \includegraphics[width=0.46\textwidth]{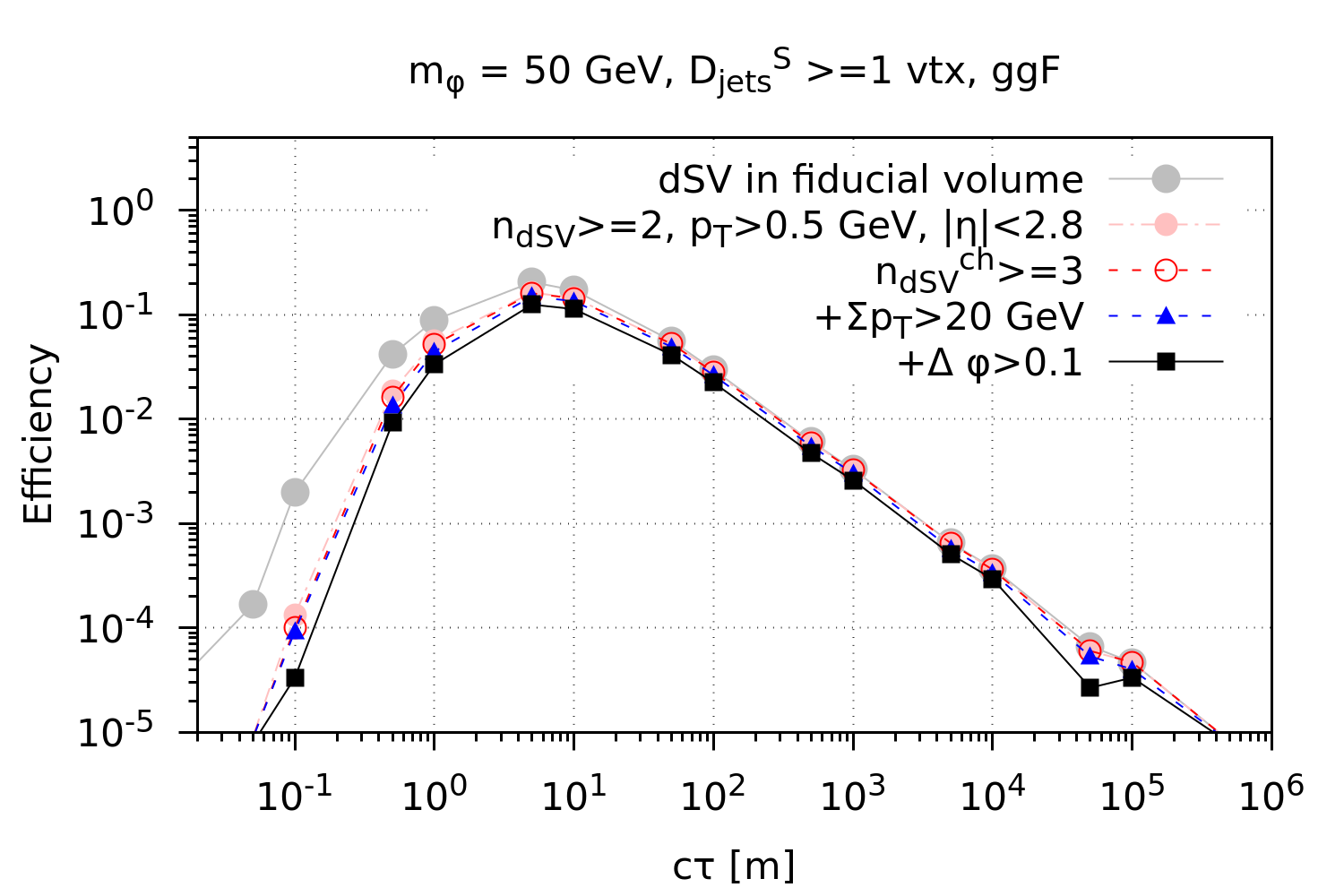} 
    \caption{\small \sl Efficiencies of the $n_{\rm dSV}^{\rm ch}$, $\sum p_{T,\,dSV}$ and $\Delta\phi_{\rm max}$ cuts on mediator particles of 10\,GeV (left panel), and 50\,GeV (right panel) masses as a function of the decay length. }
    \label{fig:cuts-bb-1}
\end{figure}

\begin{figure}[t]
    \centering    
    \includegraphics[width=0.46\textwidth]{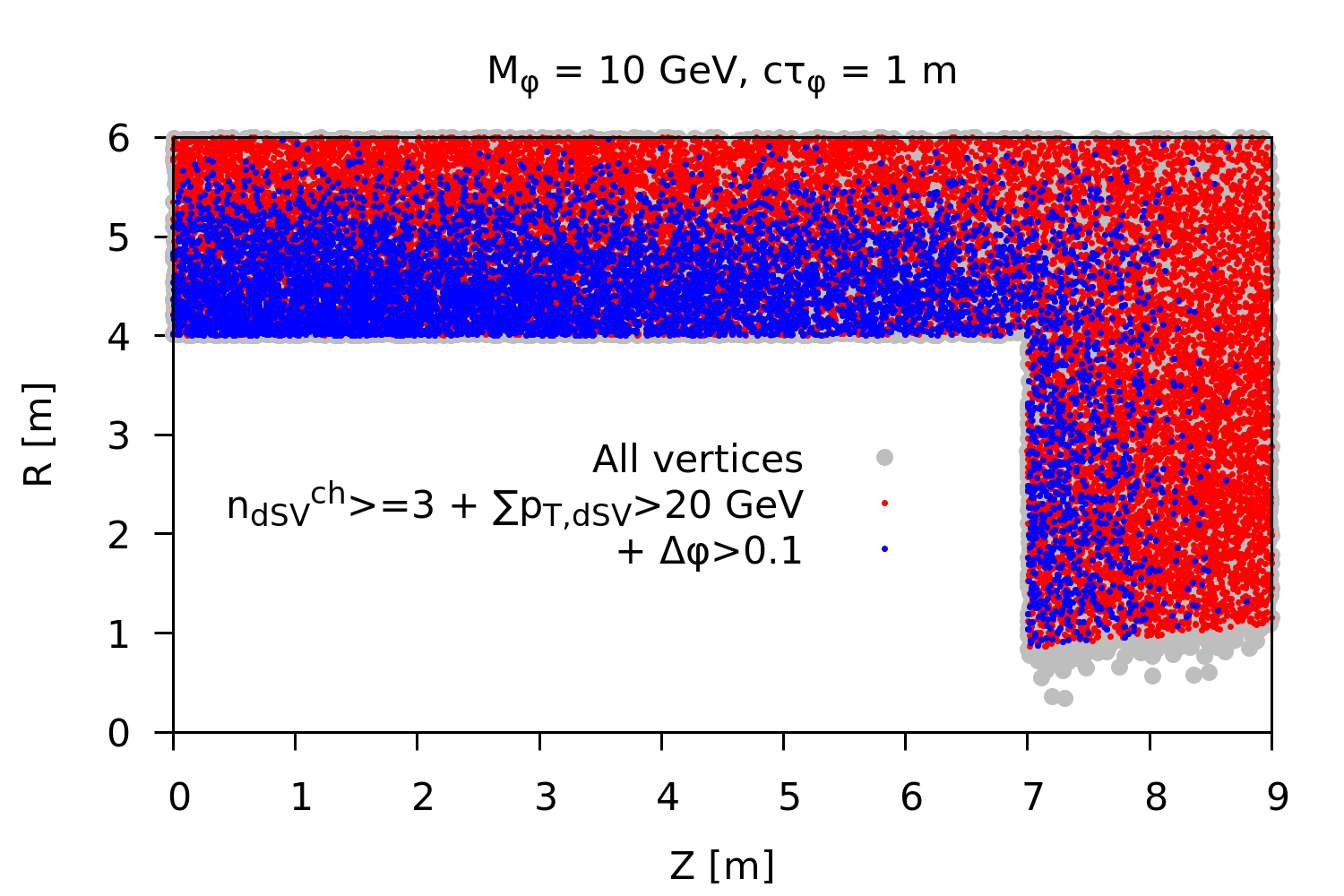} \qquad
    \includegraphics[width=0.46\textwidth]{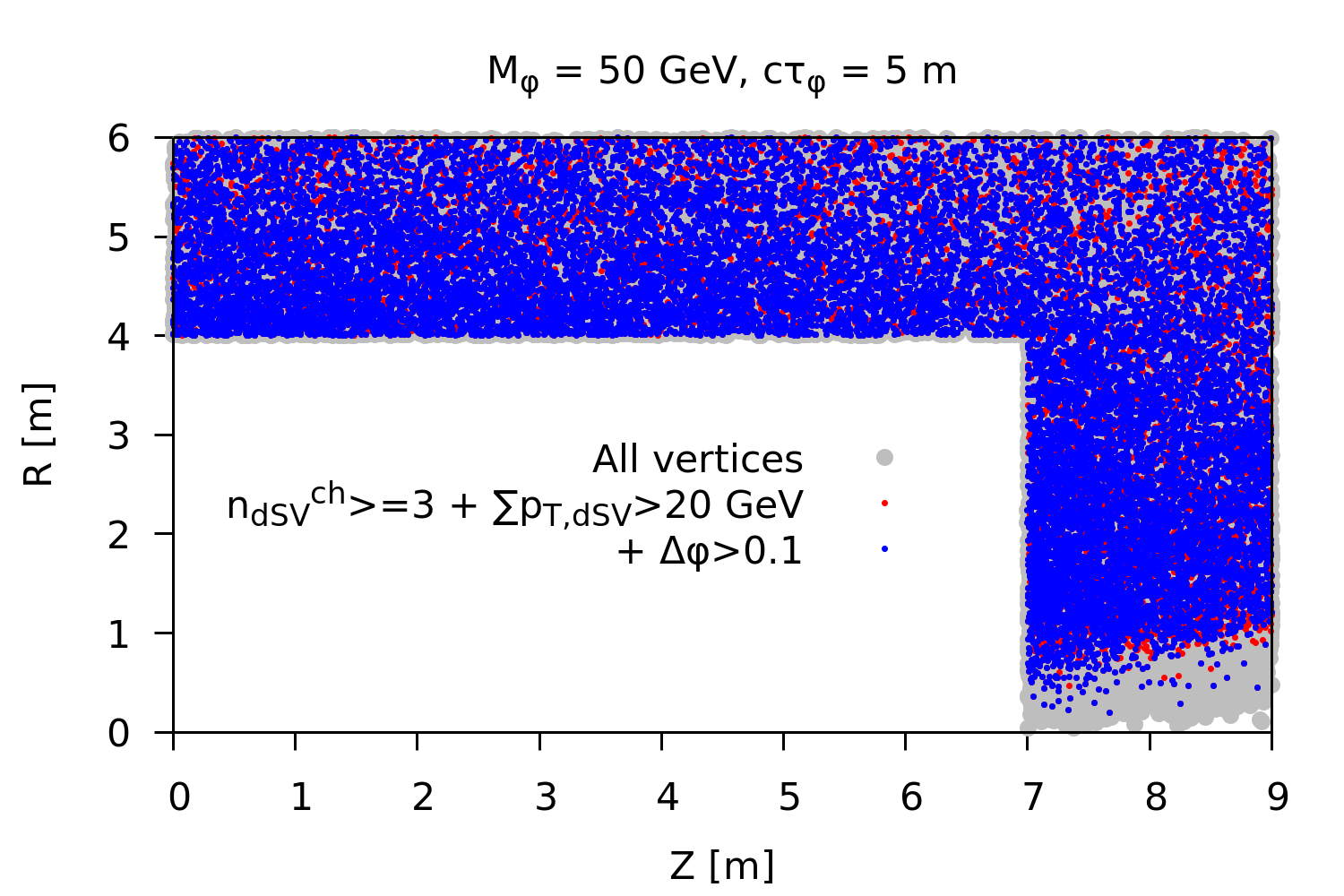}
    \caption{\small \sl Distribution of displaced vertices in the $(R,\,Z)$-plane of the detector for mediator particles of 10\,GeV (left panel), and 50\,GeV (right panel) masses having $c \tau_\phi =$ 1\,m and 5\,m, respectively, and how it is affected by the selection cuts.}
    \label{fig:cuts-bb-2}
\end{figure}

The panels of Fig.\,\ref{fig:cuts-bb-1} show the efficiencies of selecting the long-lived mediator particle signal that comes from the decay of the Higgs boson produced in the ggF mode with the cuts discussed above when it has a mass of 10\,GeV, and 50\,GeV and varying decay lengths. We find that the $\Delta\phi_{\rm max}$ cut affects the 10\,GeV benchmark much more than that of 50\,GeV, the efficiency drops by about half on applying this cut compared to that before applying it. We have already seen a similar effect for displaced muons as well. To dig a little deeper, we take a look into the distribution of these vertices in the $R-Z$ plane of the detector. For both the mediator particle masses, we have shown in the panels of Fig.\,\ref{fig:cuts-bb-2} the vertex distribution for the decay length which has the best sensitivity for that particular mass; 1\,m and 5\,m for the 10\,GeV, and 50\,GeV mediator particles, respectively. We can observe that the $\Delta\phi_{\rm max}$ cut selects only events with a much lesser displacement of the dSV for the 10\,GeV mediator particle than the 50\,GeV one. This could be understood from the fact that the 10\,GeV mediator particle has more boost and that combined with longer displacements, make the cluster of particles from the mediator particle decay more collimated. The 50\,GeV mediator particle having less boost can satisfy the $\Delta\phi_{\rm max}$ cut even for longer displacements and thus, is not affected much by this cut.

\begin{figure}[t]
    \centering
    \includegraphics[width=0.66\textwidth]{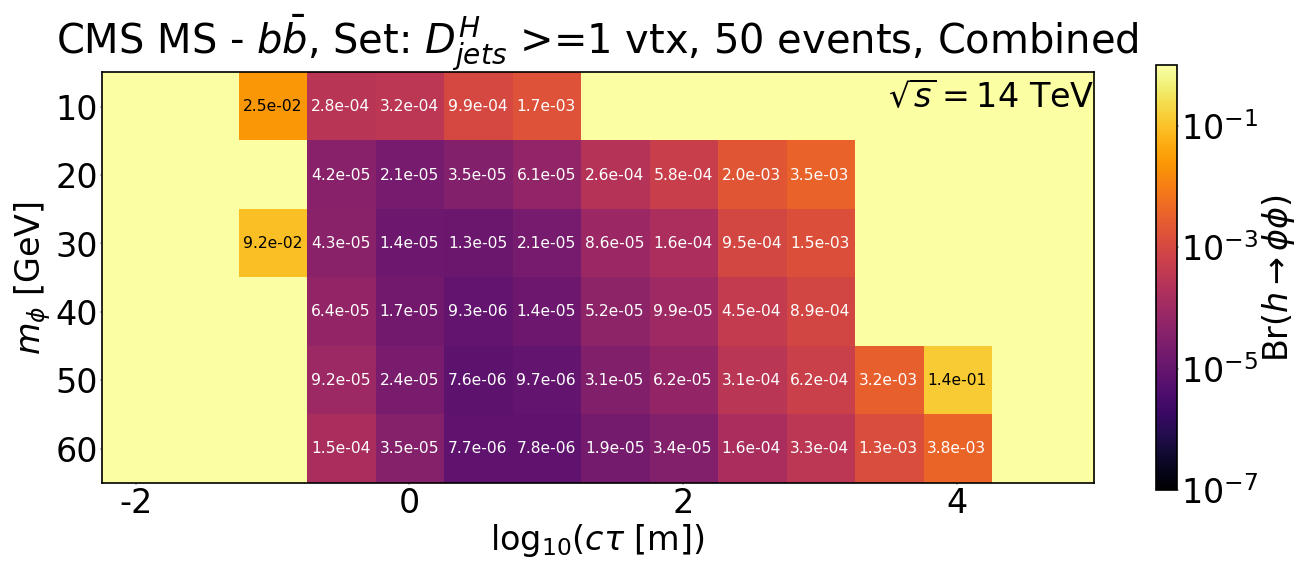} \\
    \includegraphics[width=0.66\textwidth]{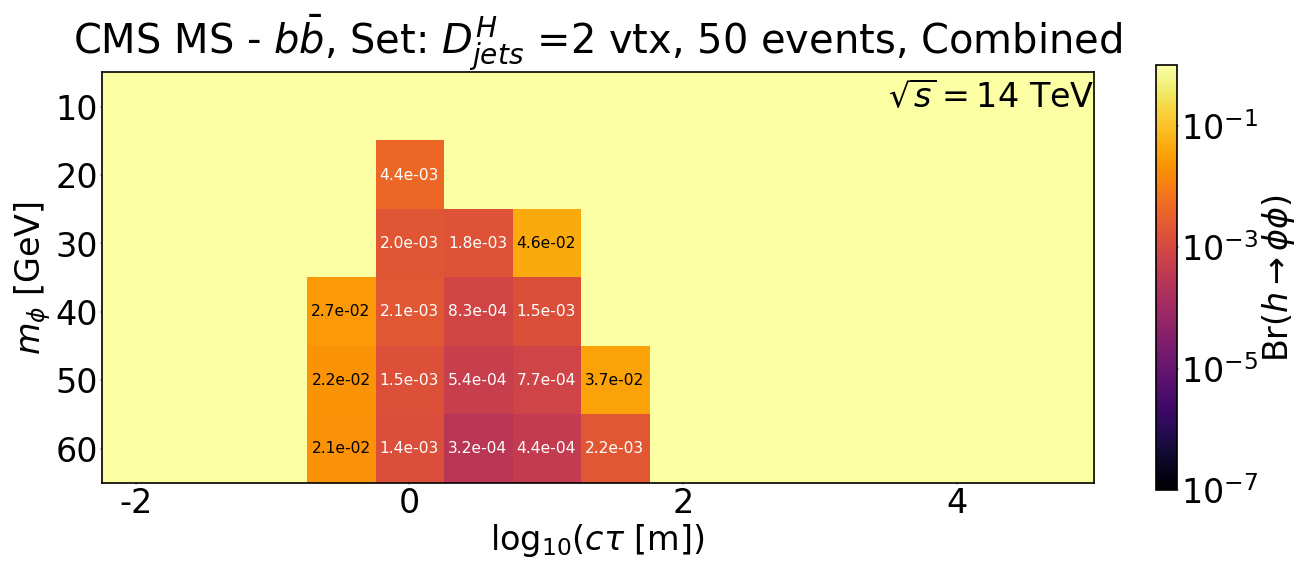} \\
    \includegraphics[width=0.66\textwidth]{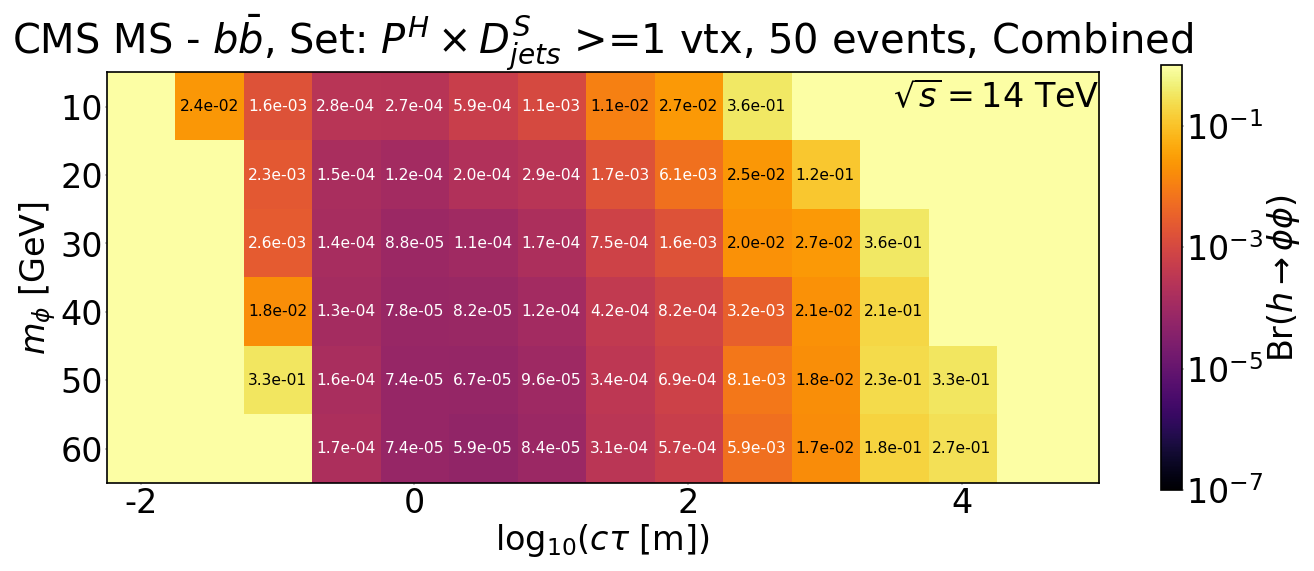} \\
    \includegraphics[width=0.66\textwidth]{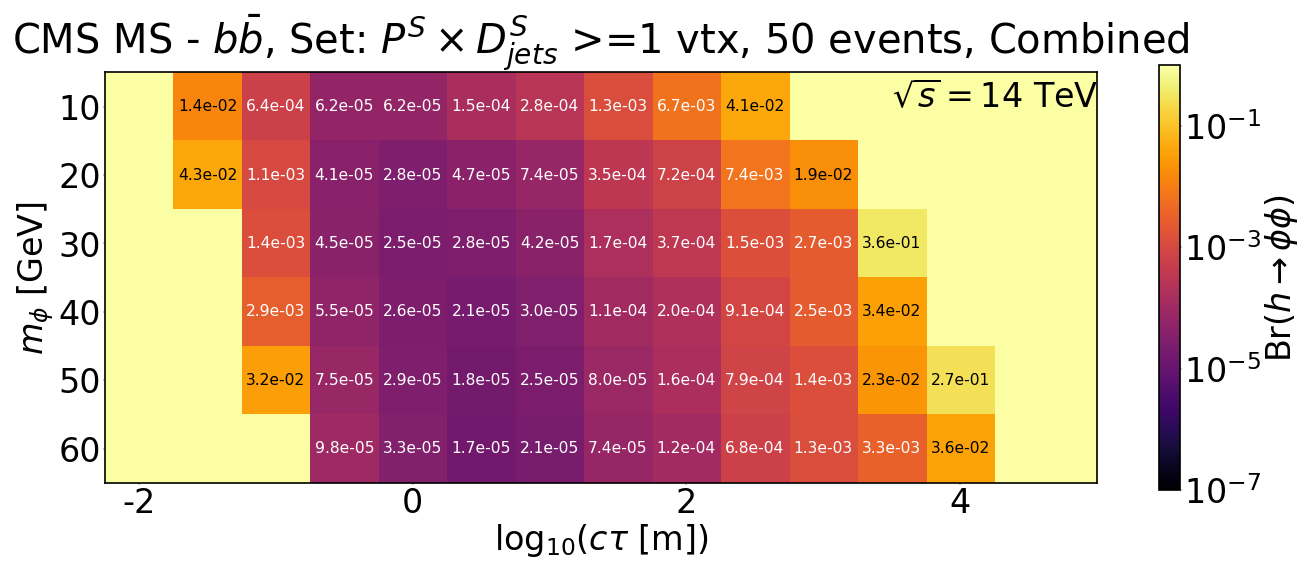}
    \caption{\small \sl Projected upper limits on Br$(h \to \phi \phi)$ for 50 observed decays of long-lived mediator particles decaying into a pair of $b$-quarks within the CMS MS for four sets of cuts explained in the text. The shown limits are obtained by combining the ggF, VBF and Vh modes of the Higgs boson production at the 14\,TeV HL-LHC experiment (3000\,fb$^{-1}$).}
    \label{fig:bb-14TeV-combo}
\end{figure}

Table\,\ref{tab:jets_cuts} summarises the various sets of cuts applied when the mediator particle decays into a pair of light quarks ($u$, $d$ or $s$ quarks), $c$ quarks, $b$ quarks or gluons giving displaced jets, denoted as $D_{jets}^i$, where $i$ stands for either a harder set of cuts\,($D^H_{jets}$) or a softer set of cuts\,($D^S_{jets}$). On the other hand, Fig.\,\ref{fig:bb-14TeV-combo} shows the projected upper limit on the branching fraction, Br$(h \to \phi \phi)$, for the four combinations of cuts as described in the beginning of this section for six benchmark masses starting from 10\,GeV and a range of decay lengths, assuming 100\,\% decay branching fraction into a pair of $b$-quark pair. We have combined the results of ggF, VBF, Vh-jet, and Vh-lep production modes of the Higgs boson for the 14\,TeV HL-LHC experiment with an integrated luminosity of 3000\,fb$^{-1}$, which is the amount of data expected to be collected by the end of the experiment. 
When the mediator decays into a pair of $b$-quarks, we observe that the 
the best limit of Br$(h \to \phi \phi)<7.5\times10^{-6}$ is obtained with the $D_{jets}^H \geq$ 1\,vtx set of cuts for the 50\,GeV mediator at $c\tau=5$\,m and Br$(h \to \phi \phi)<6.2\times10^{-5}$ is obtained with the $P^S\times D_{jets}^S \geq$ 1\,vtx set of cuts for the 10\,GeV mediator at $c\tau=1$\,m. The analysis with the $P^S\times D_{jets}^S \geq$ 1\,vtx ($D_{jets}^H \geq$ 1\,vtx) set of cuts is sensitive corresponding to branching $\sim10^{-3}$ till a decay length of around 5000\,m for the 10\,GeV (50\,GeV) mediator.
Compared to our displaced di-muon analysis, the best limit for the 50\,GeV mediator degrades by factor of $\sim$ 12 and the most sensitive $c\tau$ value shifts from 50\,cm to 5\,m when we constrain the decay to the CMS MS, and the mediator decays to $b$-jets.

\begin{table}[t]
    \centering
    \begin{tabular}{c|l|l|c|}
        & \qquad\qquad $D^H_{jets}$ & \qquad\qquad $D^S_{jets}$ \\
        \hline
        Electrons, photons, & $p_T >$ 0.5\,GeV & $p_T >$ 0.5\,GeV \\
        \cline{2-3}
        hadrons & $|\eta| <$ 2.8 & $|\eta| <$ 2.8 \\
        \cline{2-3}
        \hline
        \multirow{4}{*}{MS cluster from} & $d_T >$ 4\,m or $|d_z| >$ 7\,m & $d_T >$ 4\,m or $|d_z| >$ 7\,m \\
        & $d_T <$ 6\,m and $|d_z| <$ 9\,m & $d_T <$ 6\,m and $|d_z| <$ 9\,m \\
        \cline{2-3}
        \multirow{2}{*}{same dSV ($<$ 1\,cm)} & $n_{\rm dSV}^{\rm ch} \geq$ 5 & $n_{\rm dSV}^{\rm ch} \geq$ 3 \\
        \cline{2-3}
        & $\sum p_{T,\,dSV} >$ 50\,GeV & $\sum p_{T,\,dSV} >$ 20\,GeV \\
        \cline{2-3}
        & $\Delta\phi_{\rm max} >$ 0.2 & $\Delta\phi_{\rm max} >$ 0.1 \\
        \hline
        Event & $n_{\rm cluster} \geq$ 1, $n_{\rm cluster}$ =2 & $n_{\rm cluster} \geq$ 1, $n_{\rm cluster} =$ 2 \\
        \hline
    \end{tabular}
    \caption{\small \sl Harder and softer sets of cuts applied for the displaced jets.}
    \label{tab:jets_cuts}
\end{table}

\begin{figure}[t]
    \centering
    \includegraphics[width=0.46\textwidth]{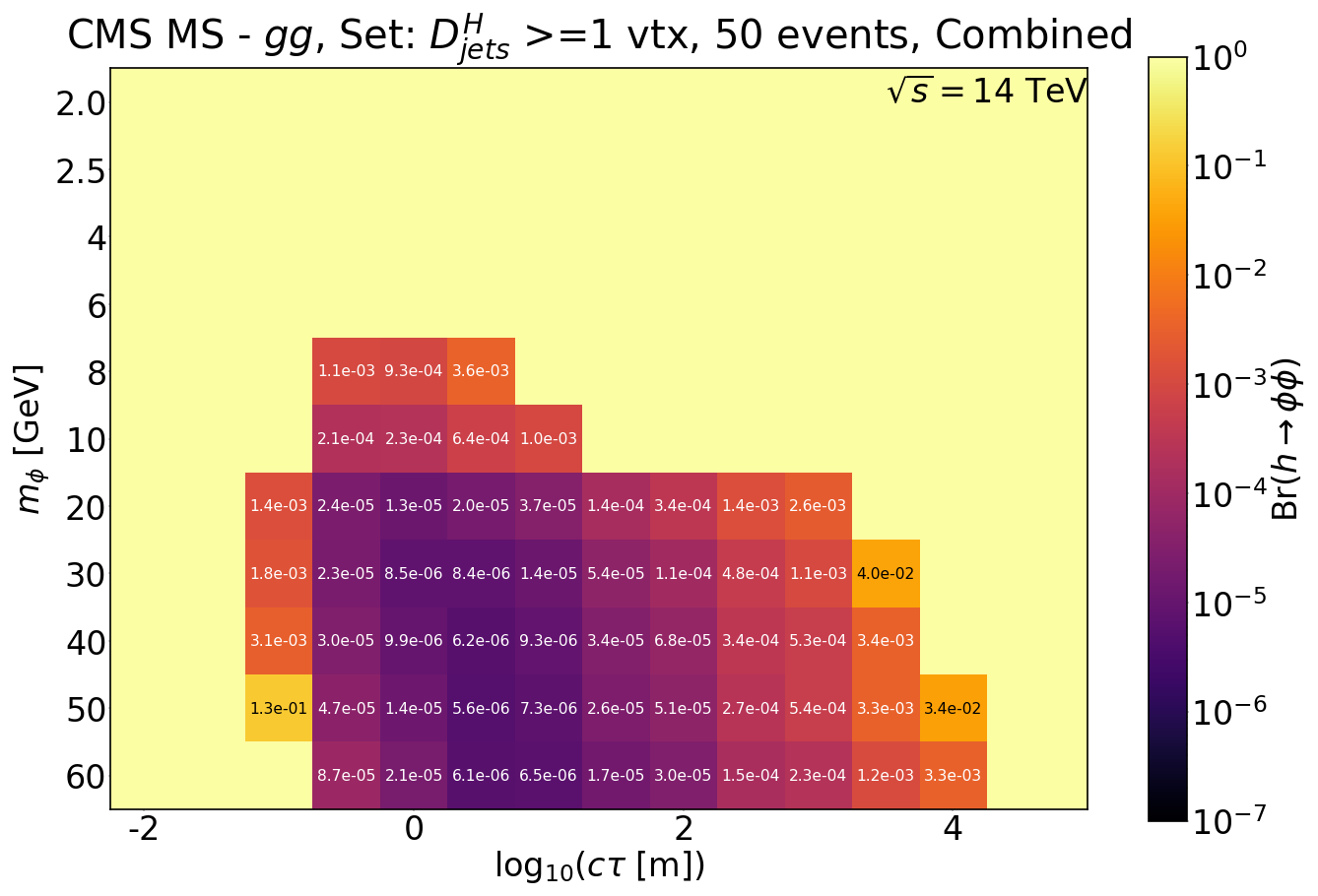}   \qquad \includegraphics[width=0.46\textwidth]{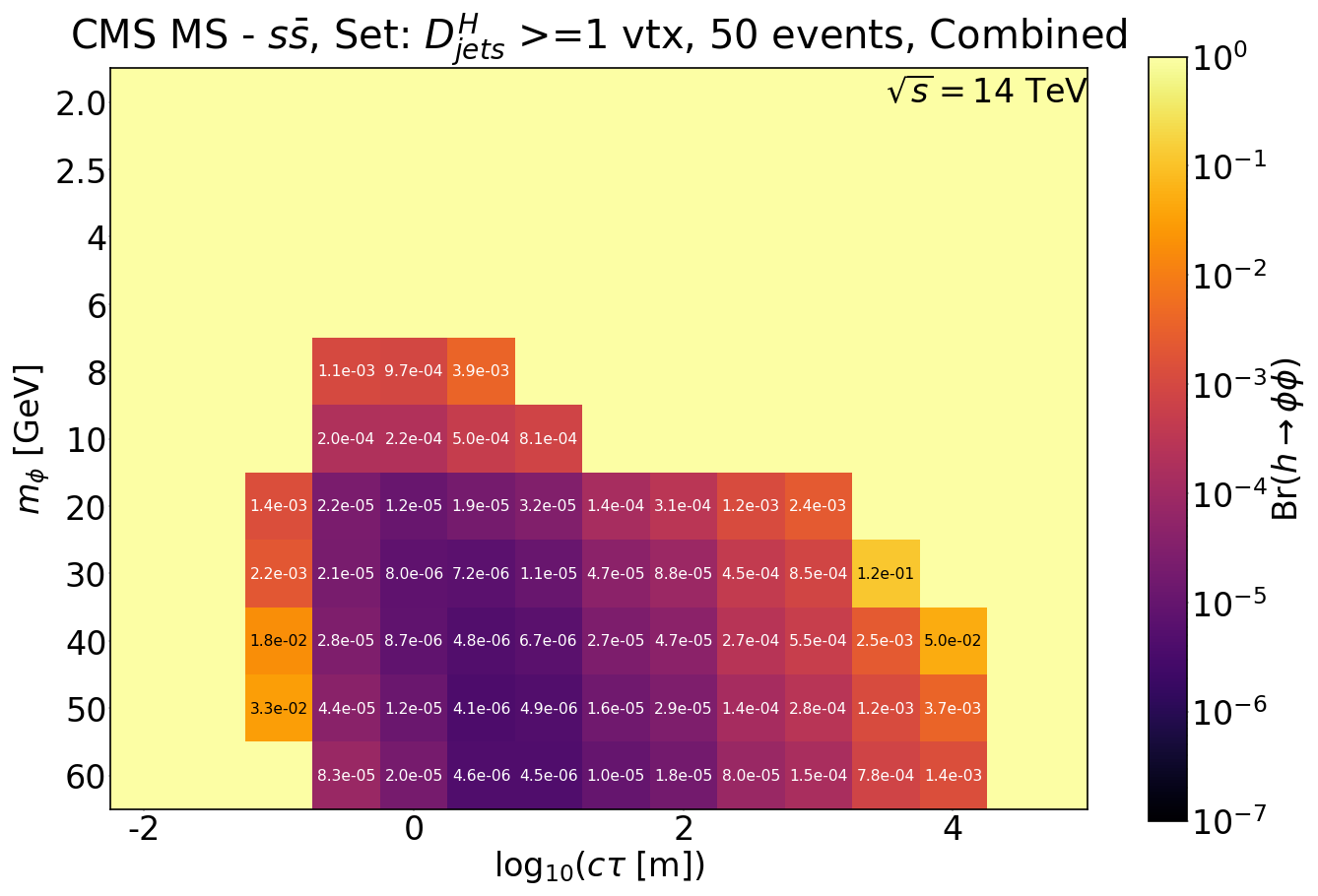} \\
    \includegraphics[width=0.46\textwidth]{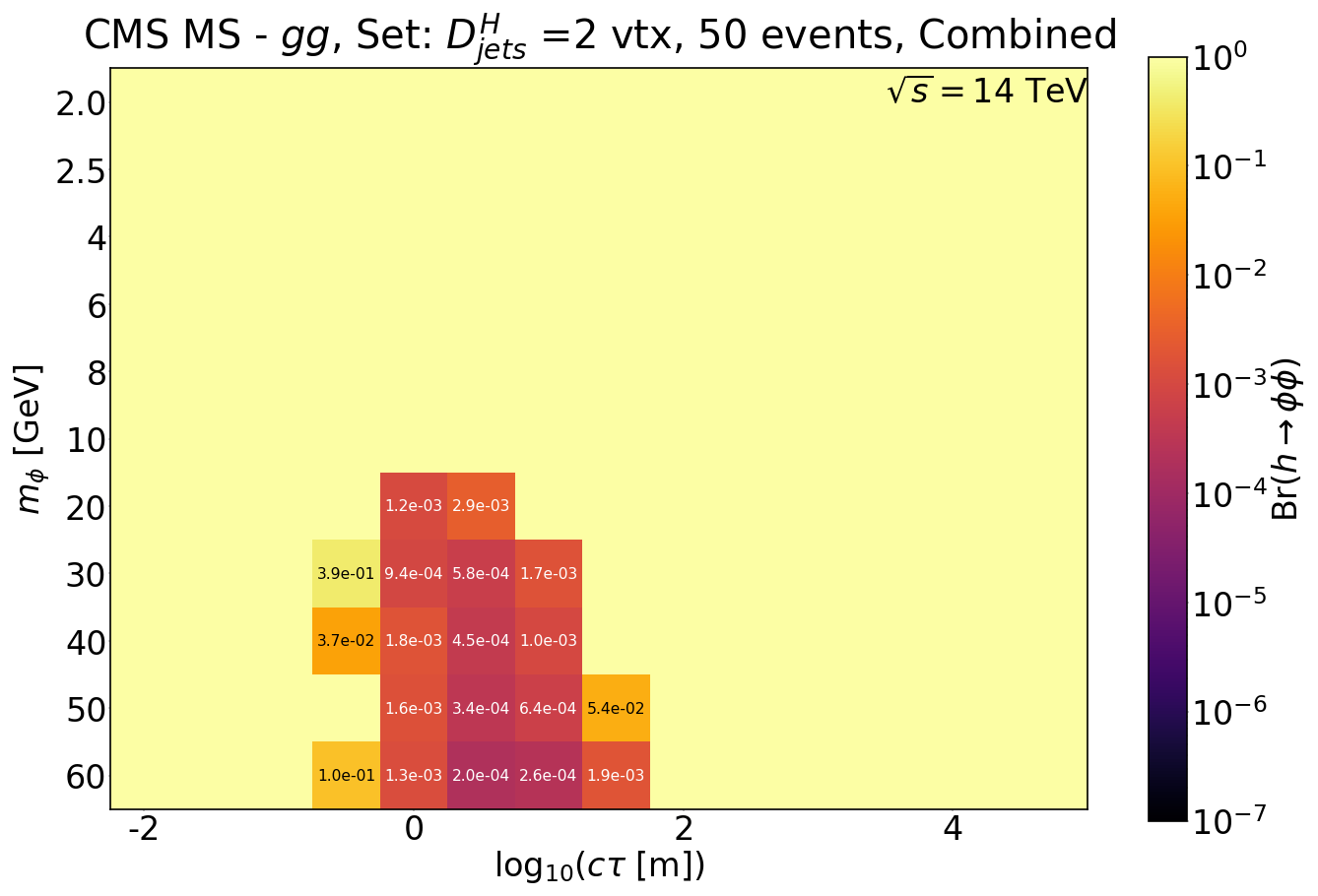}   \qquad \includegraphics[width=0.46\textwidth]{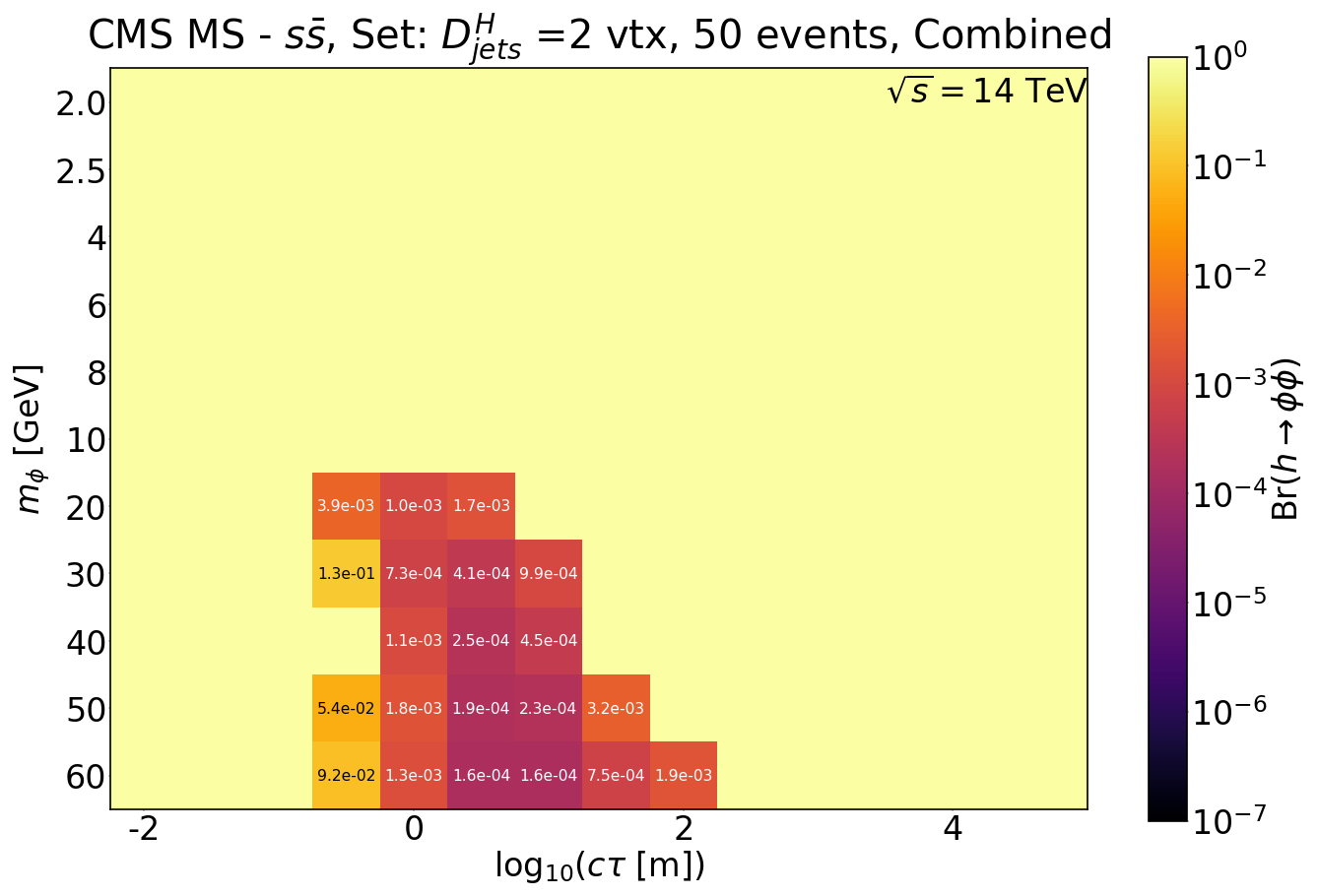} \\
    \includegraphics[width=0.46\textwidth]{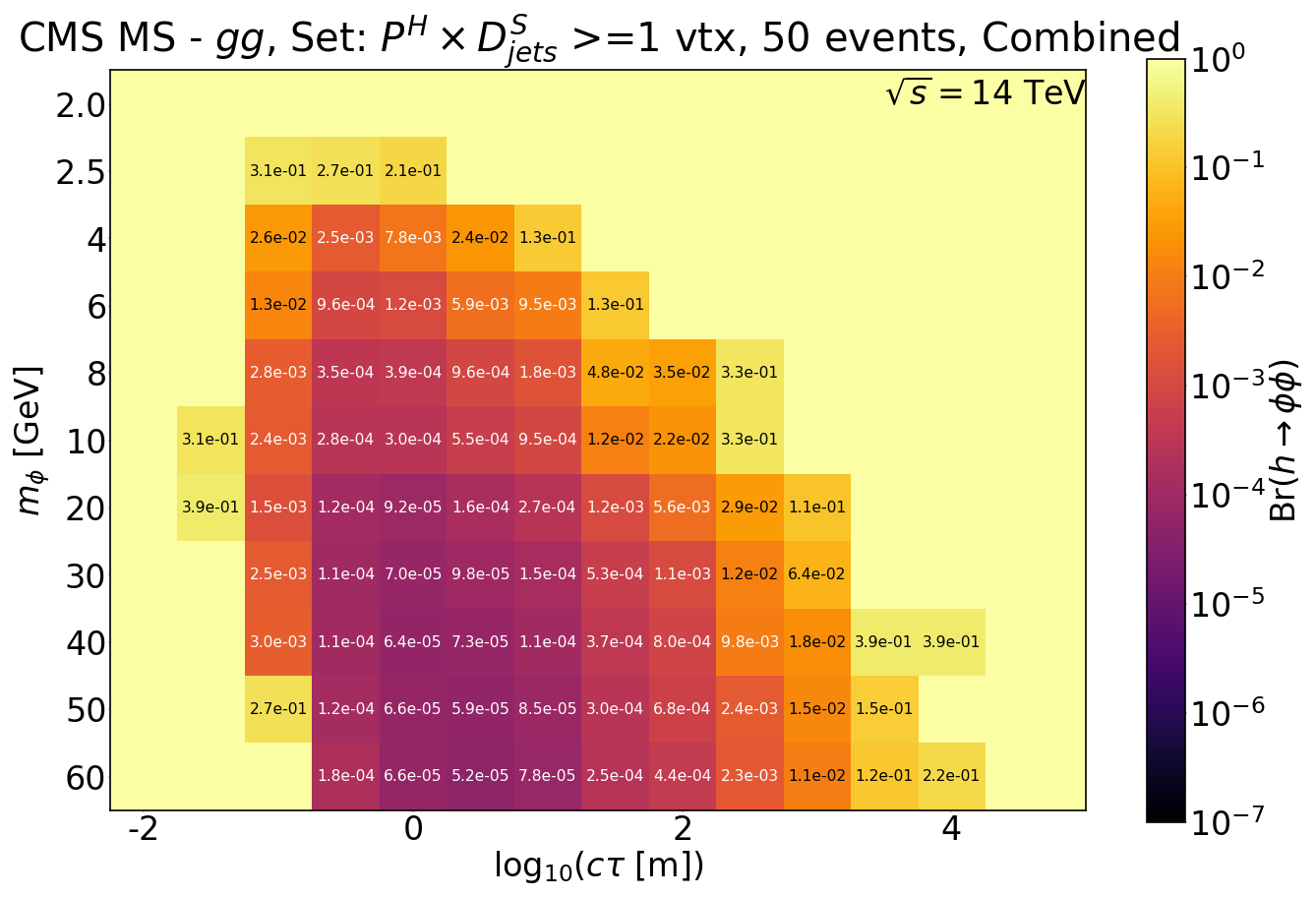} \qquad \includegraphics[width=0.46\textwidth]{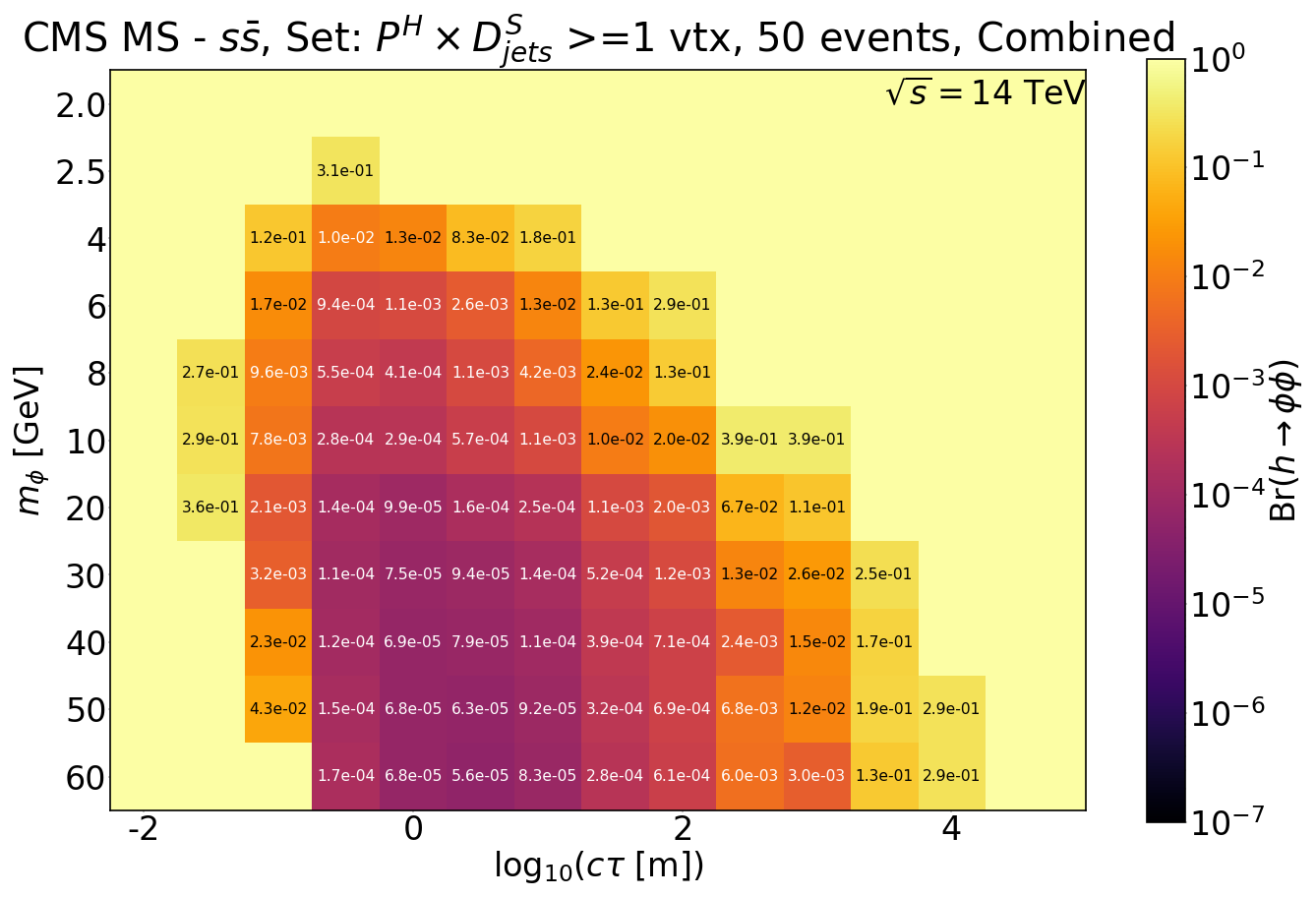} \\
    \includegraphics[width=0.46\textwidth]{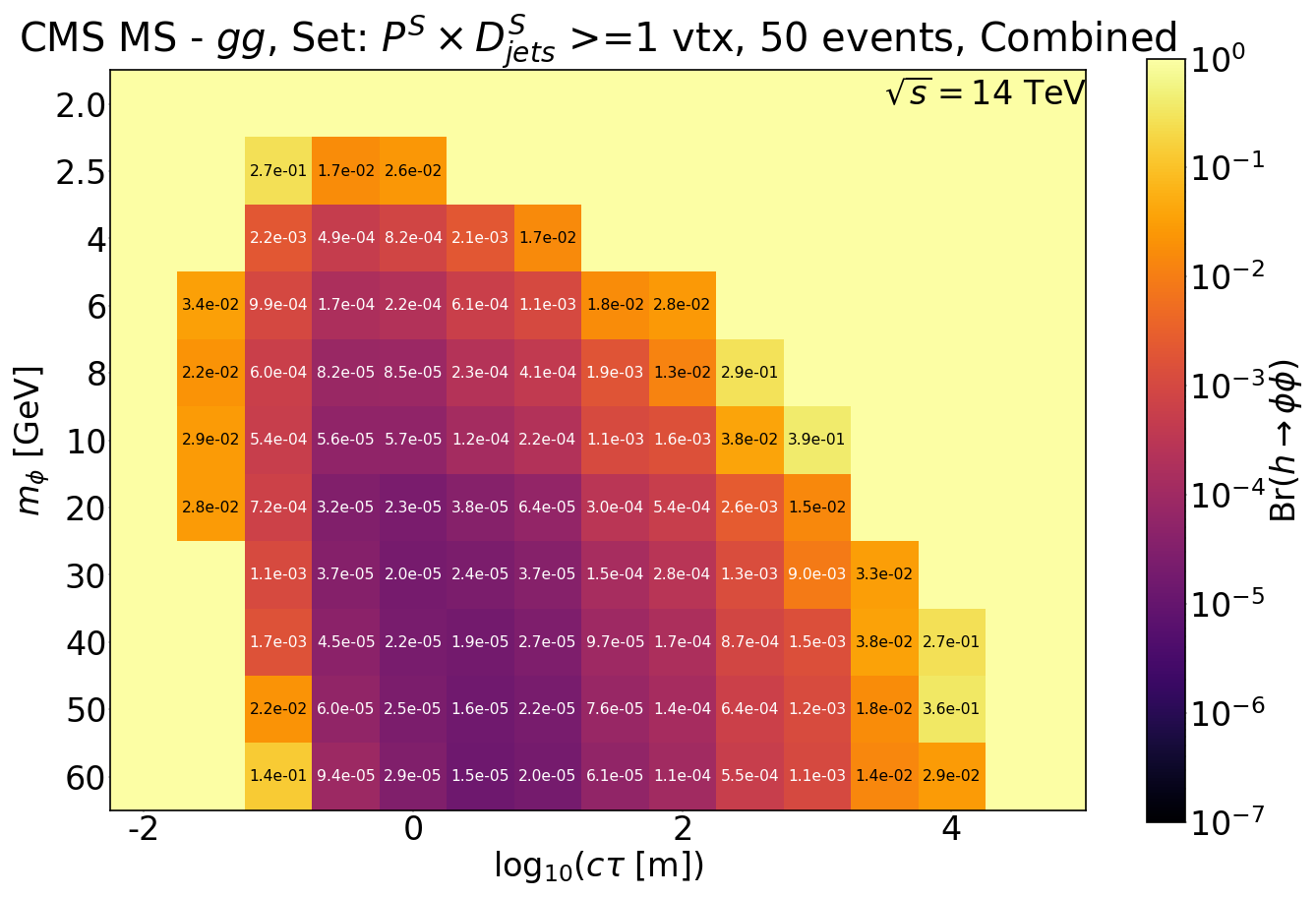} \qquad \includegraphics[width=0.46\textwidth]{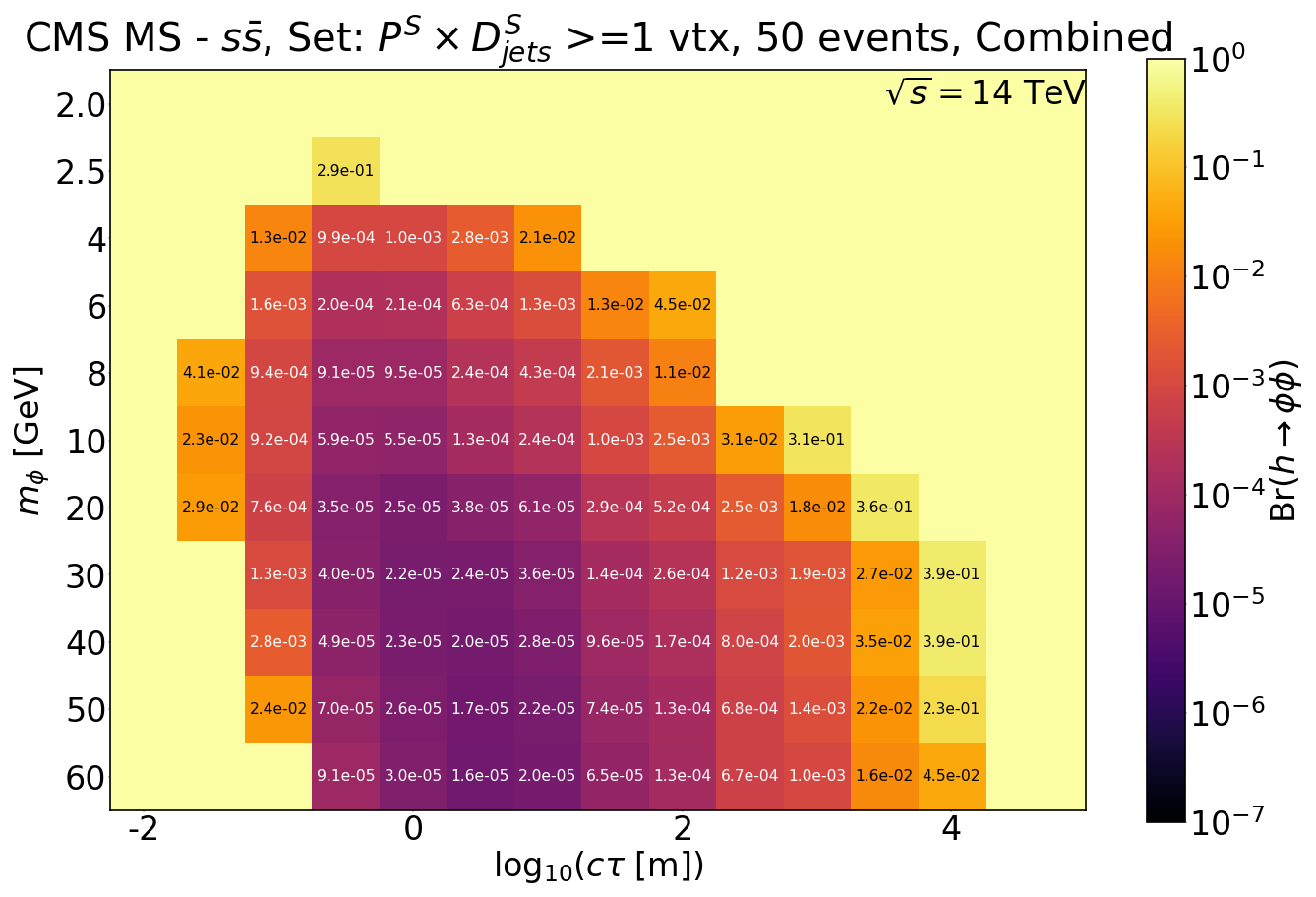}
    \caption{\small \sl Projected upper limits on the branching fraction, Br$(h \to \phi \phi)$, for 50 observed decays of long-lived mediator particles decaying into a pair of gluons (left four panels) and $s$-quarks (right four panels) within the CMS MS for four sets of cuts explained in the text. The shown limits are obtained by combining the ggF, VBF and Vh channels of the Higgs boson production at the 14\,TeV HL-LHC experiment with 3000\,fb$^{-1}$ integrated luminosity.}
    \label{fig:ggss-14TeV-combo}
\end{figure}

\begin{figure}[t]
    \centering
    \includegraphics[width=0.46\textwidth]{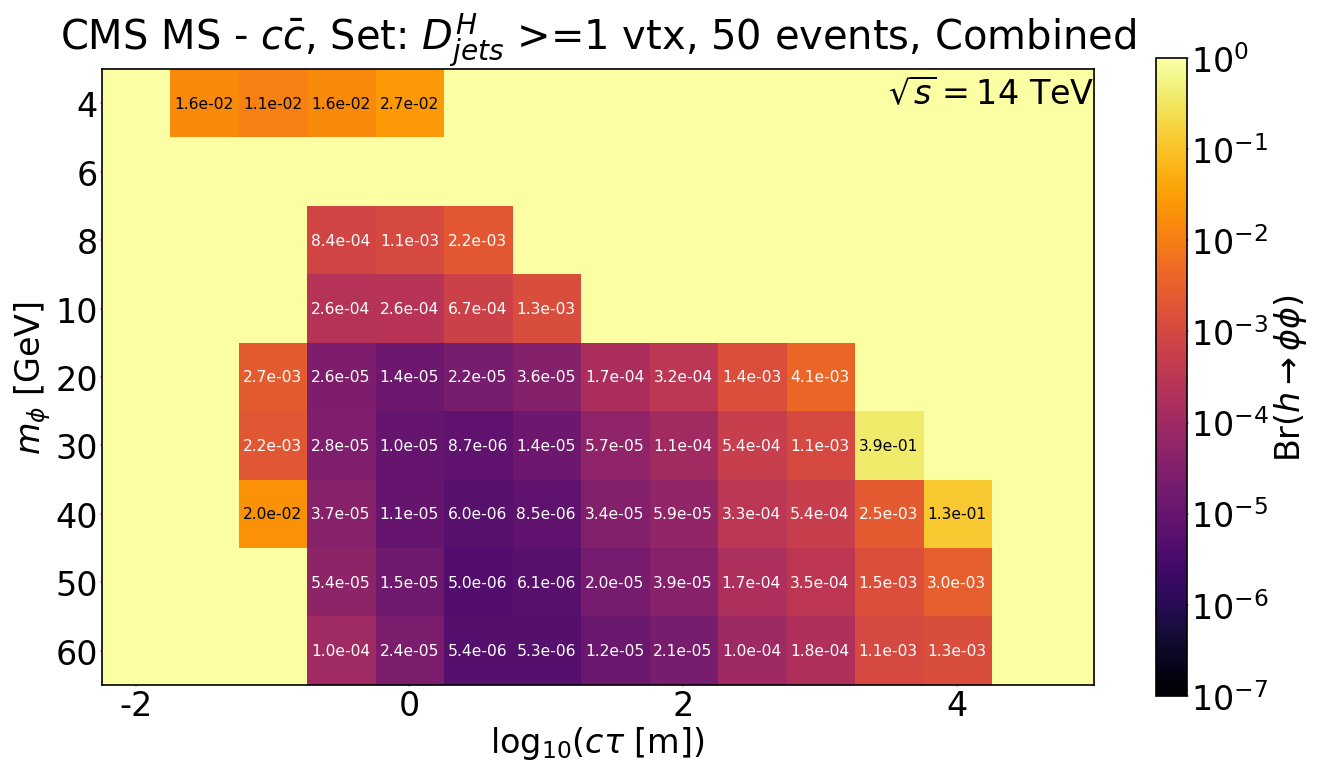} \qquad
    \includegraphics[width=0.46\textwidth]{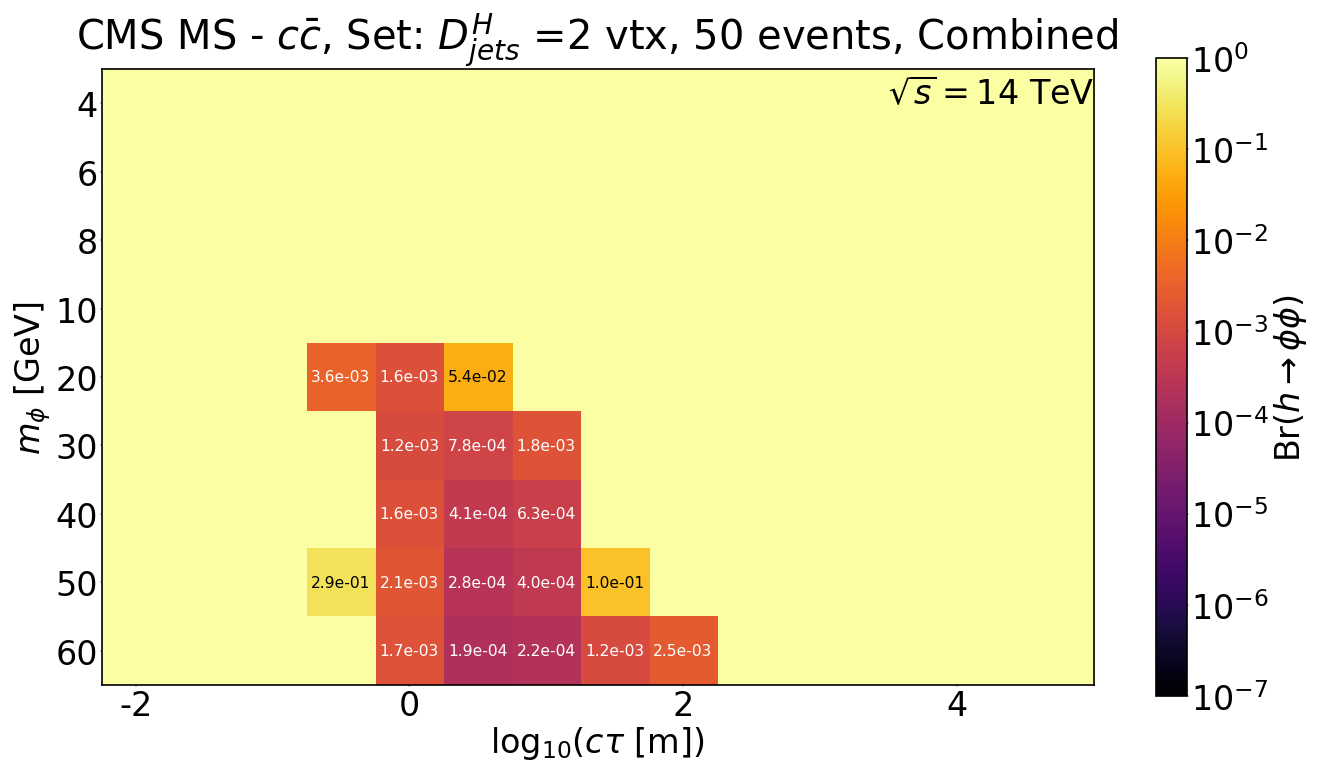} \\
    \includegraphics[width=0.46\textwidth]{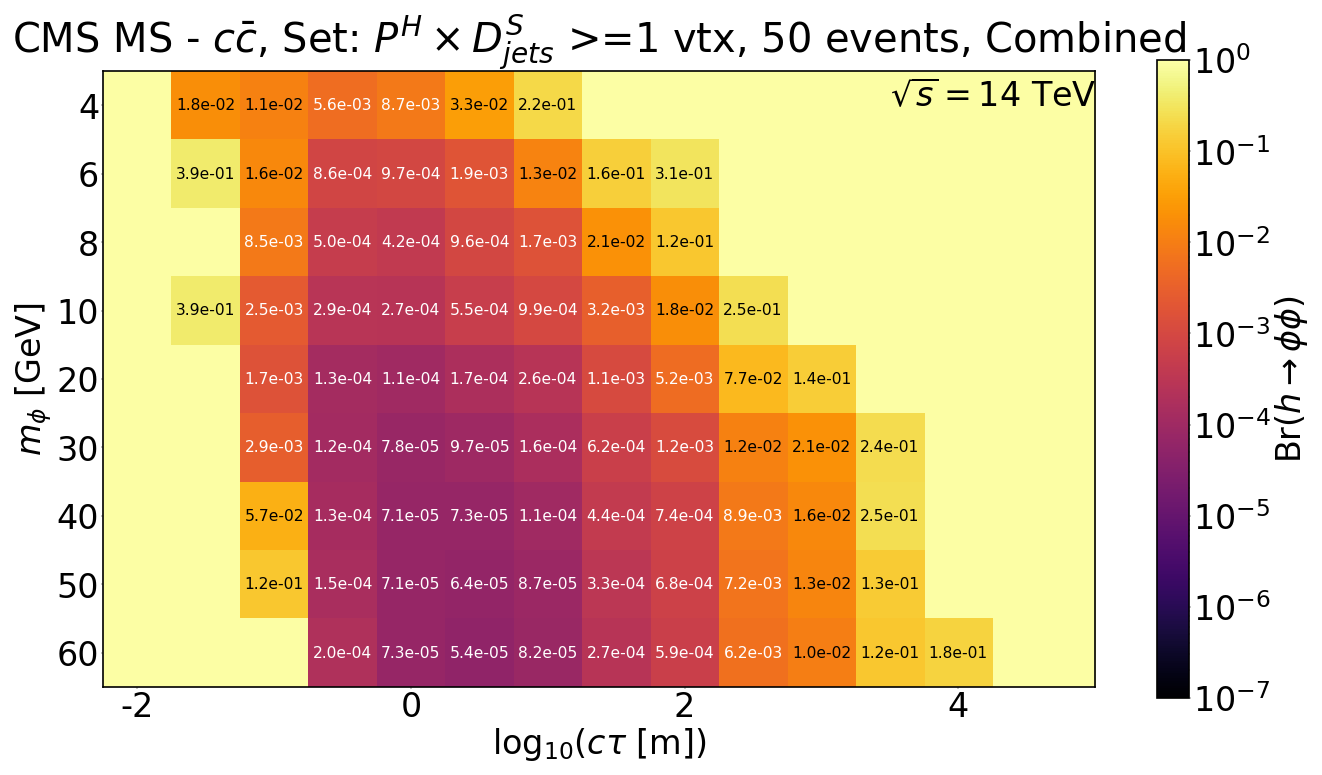} \qquad
    \includegraphics[width=0.46\textwidth]{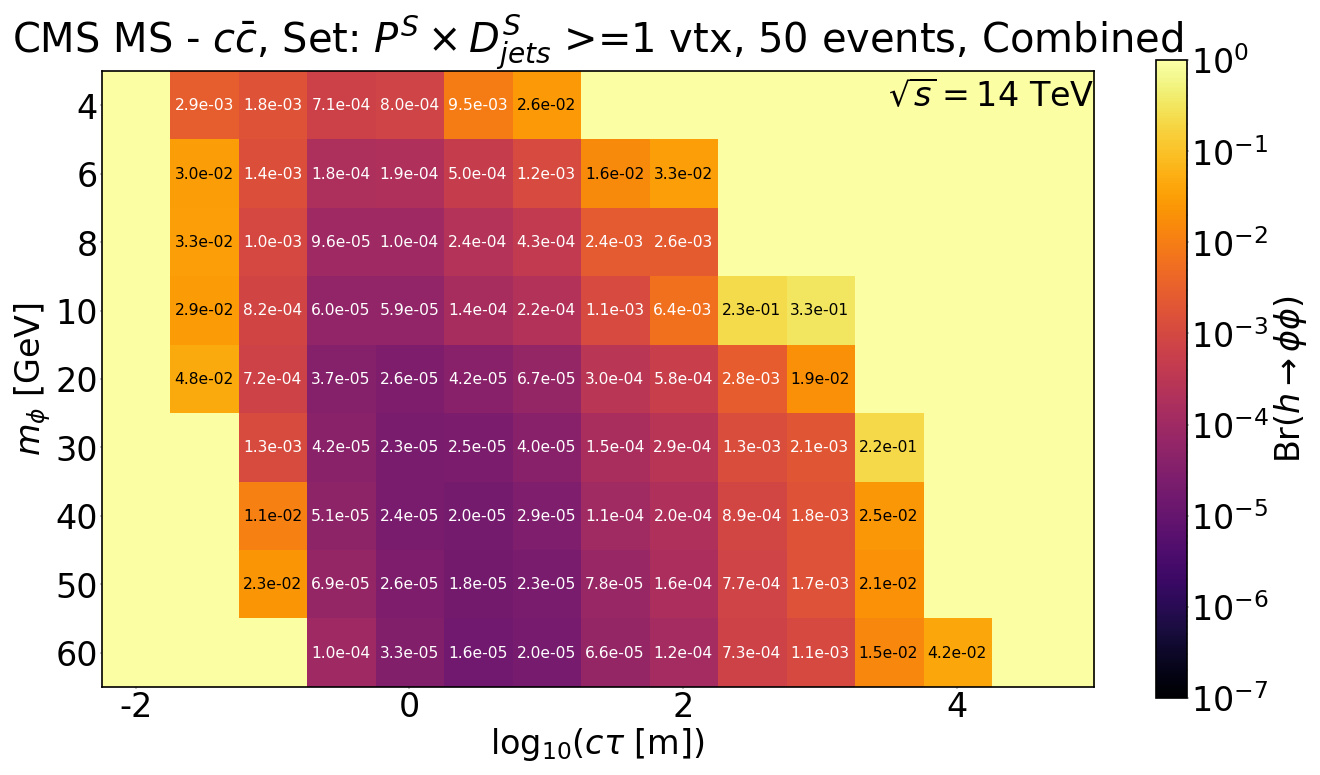} 
    \caption{\small \sl Projected upper limits on Br$(h \to \phi \phi)$ for 50 observed decays of long-lived mediator particles decaying into a pair of $c$-quarks within the CMS MS for four sets of cuts explained in the text. The shown limits are obtained by combining the ggF, VBF and Vh modes of the Higgs boson production at the 14\,TeV HL-LHC experiment (3000\,fb$^{-1}$).}
    \label{fig:cc-14TeV-combo}
\end{figure}

Next, we apply the same analysis for displaced jets when the mediator particle decays into gluons, strange quarks, and charm quarks. Figs.\,\ref{fig:ggss-14TeV-combo} and \ref{fig:cc-14TeV-combo} show the projected upper limit on the Br$(h \to \phi\phi)$ for the four combinations of cuts at some benchmark masses and a range of decay lengths assuming 100\,\% decay into gluons, $s$-quarks and $c$-quarks, respectively, combined over ggF, VBF, Vh-jet and Vh-lep production modes for the Higgs boson at 14\,TeV HL-LHC experiment with an integrated luminosity of 3000\,fb$^{-1}$. The $gg$ and $s\bar{s}$ decay modes open up from a mediator mass of a few GeV, as seen from Fig.\,\ref{fig:width_branching}, and therefore we have selected mass points starting from 2\,GeV. The $c\bar{c}$ channel, however, opens up around 3 GeV, and we have shown the result starting from a mediator mass of 4\,GeV for this channel. We find that our analyses, especially the ones with harder cuts on the displaced objects, loses sensitivity as the mediator mass decreases. This is due to their high boost which leads to high displacement, and the magnetic field in the MS is not enough to bend the tracks to satisfy the condition on $\Delta\phi_{\rm max}$.

\begin{figure}[t]
    \centering
    \includegraphics[width=0.7\textwidth]{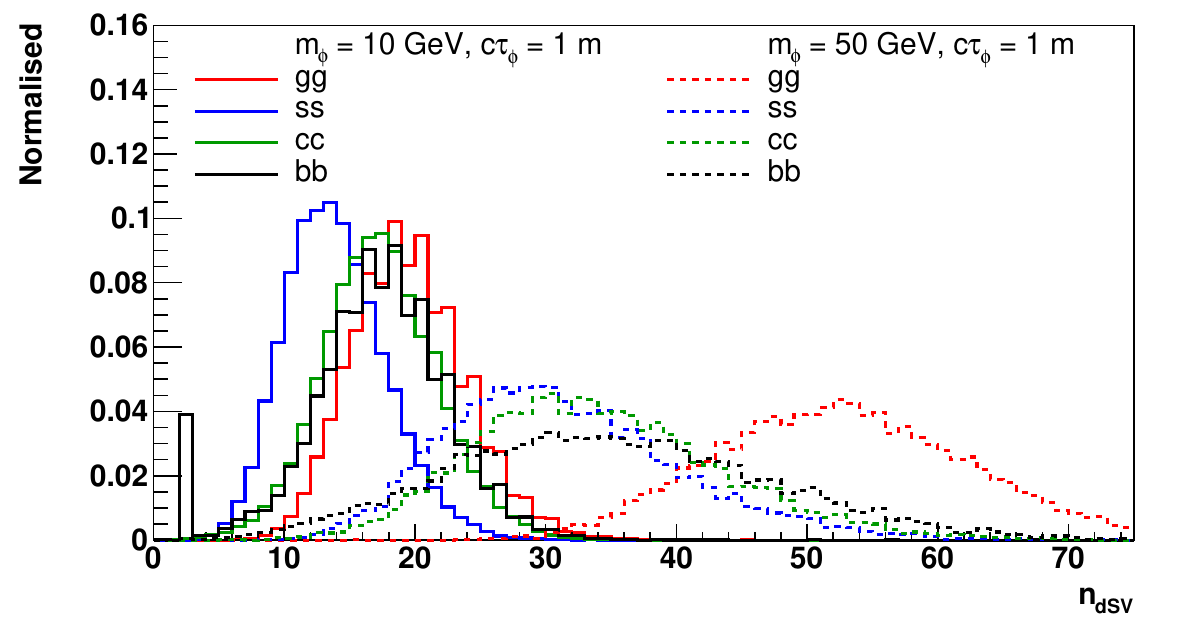}
    \caption{\small \sl Multiplicity of stable particles produced from the displaced secondary vertex (with a tolerance of 1\,cm in each of the $x$, $y$ and $z$ directions) due to the decay of mediator particles (LLPs) of 10\,GeV and 50\,GeV masses, each having a decay length of 1\,m, when it decays into a pair of gluons, strange quarks, charm quarks and bottom quarks.}
    \label{fig:particle-multiplicity}
\end{figure}

The performance remains similar for the decays into $g g$, $s \bar{s}$, $c \bar{c}$ and $b \bar{b}$ with the $g g$ case being slightly better, followed by $b \bar{b}$, $c \bar{c}$ and $s \bar{s}$, respectively, which is accounted by the decreasing particle multiplicity in going down this list, which is reflected in Fig.\,\ref{fig:particle-multiplicity} which shows the distribution of the number of particles associated with a dSV when the mediator decays to gluons, and $s\,,c\,,b$ quarks for two benchmark LLP (mediator) masses of 10 GeV and 50 GeV. We observe that with increasing LLP mass, the particle multiplicity also increases, and the effect is more for gluons than the three flavors of quarks. The smaller peak at $n_{dSV}=2$ for mediator mass of 10 GeV when it decays to a pair of $b$-quarks is due to the production of $\Upsilon$ since the mediator mass is close to the $\Upsilon$ resonance, and its eventual decay to a pair of electrons or muons. Since we are selecting events with $n_{dSV}^{ch}\geq3$, these events do not get selected in our analysis. We also observe that the $D_{jets}^H =$ 2\,vtx cut is not sensitive for lighter mediators below 6\,GeV. This cut demands two MS clusters satisfying the harder set of cuts with $\sum p_{T,dSV}>50$\,GeV and $\Delta\phi_{max}>0.3$, and for lighter LLPs with larger boosts, it is difficult to get two events within the MS decay volume, and then extra suppression comes from the $\Delta\phi_{max}$ cut. In experiments, if the $\Delta\phi_{max}$ cut can be reduced or removed, then we can gain in sensitivity for lighter LLPs.

\subsubsection{Mediator particle decaying into tau leptons}

When the mediator particle decays into $\tau$ leptons, the analysis that we apply depends on what the $\tau$ lepton decays into. If both the $\tau$ leptons from a mediator particle decays into muons, then we can apply the displaced muons analysis where we do not restrict the decay of the mediator particle inside the MS. Otherwise, we apply the displaced jets analysis, which we discussed previously. In the former case, where we have two displaced muons associated with the same mediator particle, we need to relax the same vertex condition than what we were using for the displaced muons case. This is due to the fact that the $\tau$ lepton has a lifetime of around $2.9 \times 10^{-13}$\,s, i.e., it does not decay promptly. We hence increase the tolerance of the distance between the two muons from the decay of the two $\tau$ leptons to 1\,cm in each of the $x$, $y$, $z$ directions, which is the same as what we have considered for the displaced jets analysis. For all other decays of the $\tau$ lepton, we follow the analysis developed for displaced jets. However, the multiplicity of charged particles from $\tau$ leptons will be smaller than jets, and therefore, we reduce the $n_{\rm dSV}^{\rm ch}$ cut to 3 and 2 for the harder and softer set of cuts, respectively. The important thing to note here is that in addition to the particle multiplicity, the energy of the $\tau$ leptons and its decay products are equally important to ensure a high multiplicity of hits in the MS cluster, which can distinguish it from the punch-through background, discussed in section \ref{ssec:background}.

\begin{figure}[t]
    \centering
    \includegraphics[width=0.46\textwidth]{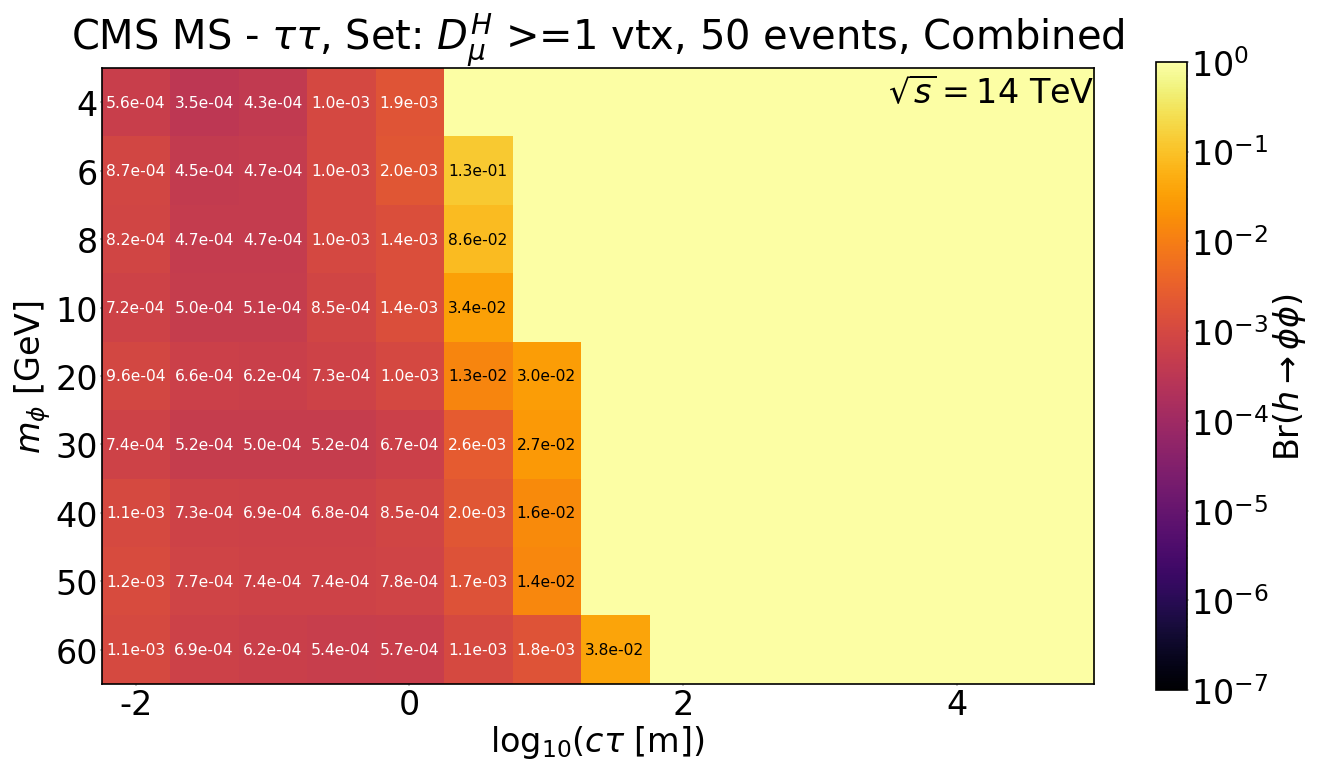} \qquad
    \includegraphics[width=0.46\textwidth]{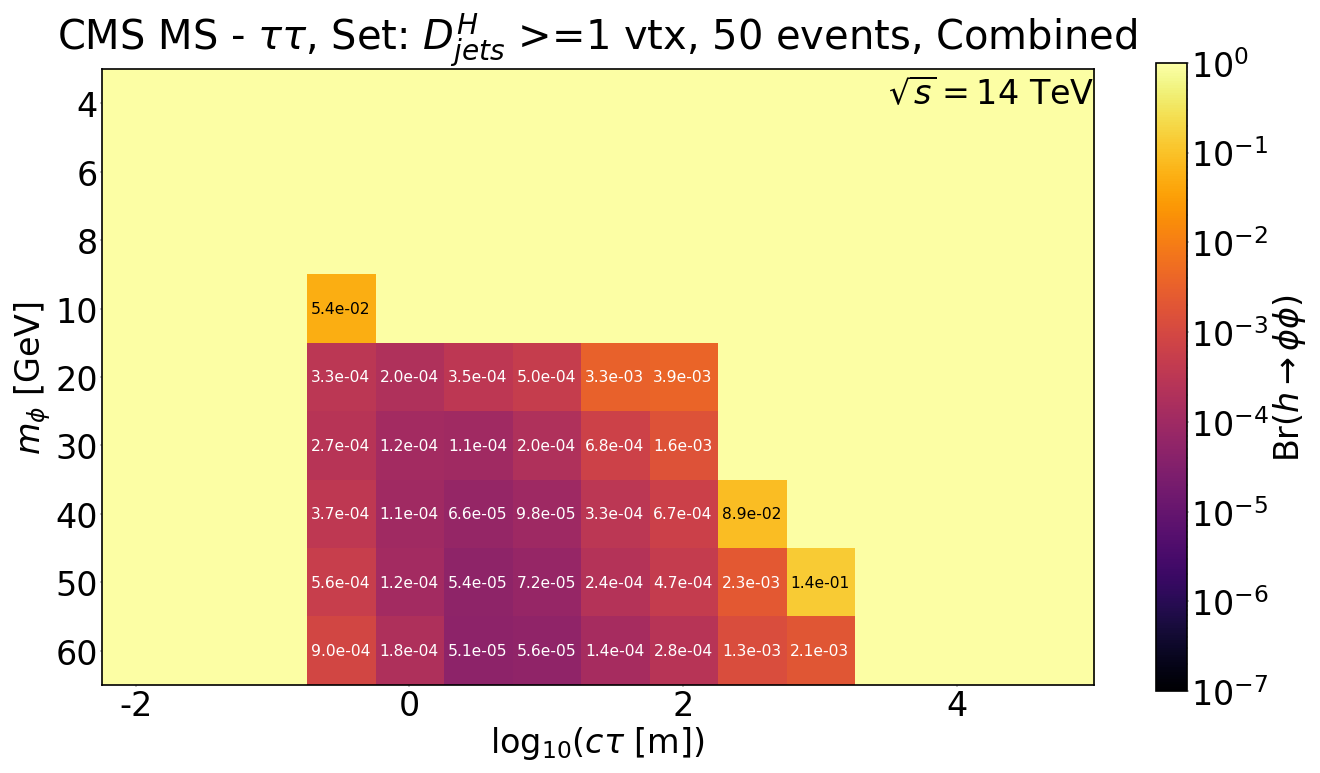} \\
    \includegraphics[width=0.46\textwidth]{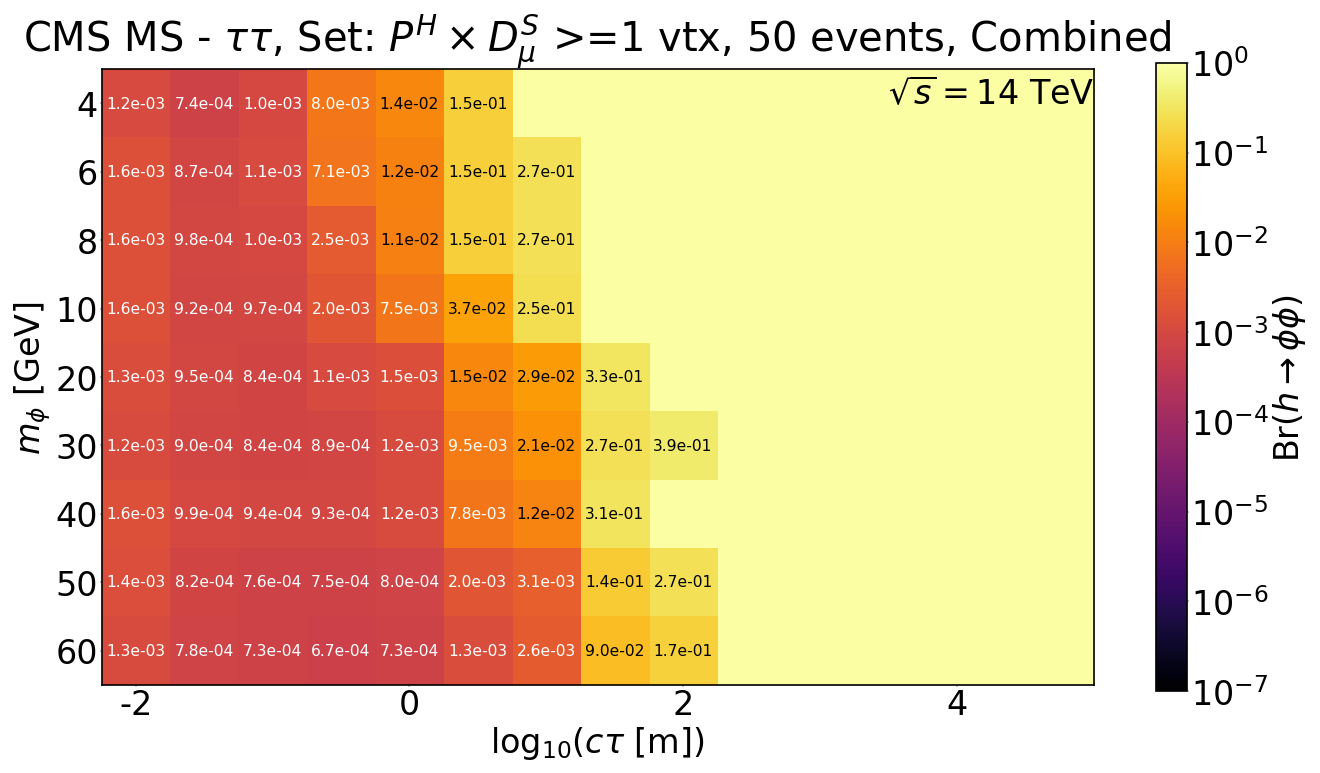} \qquad
    \includegraphics[width=0.46\textwidth]{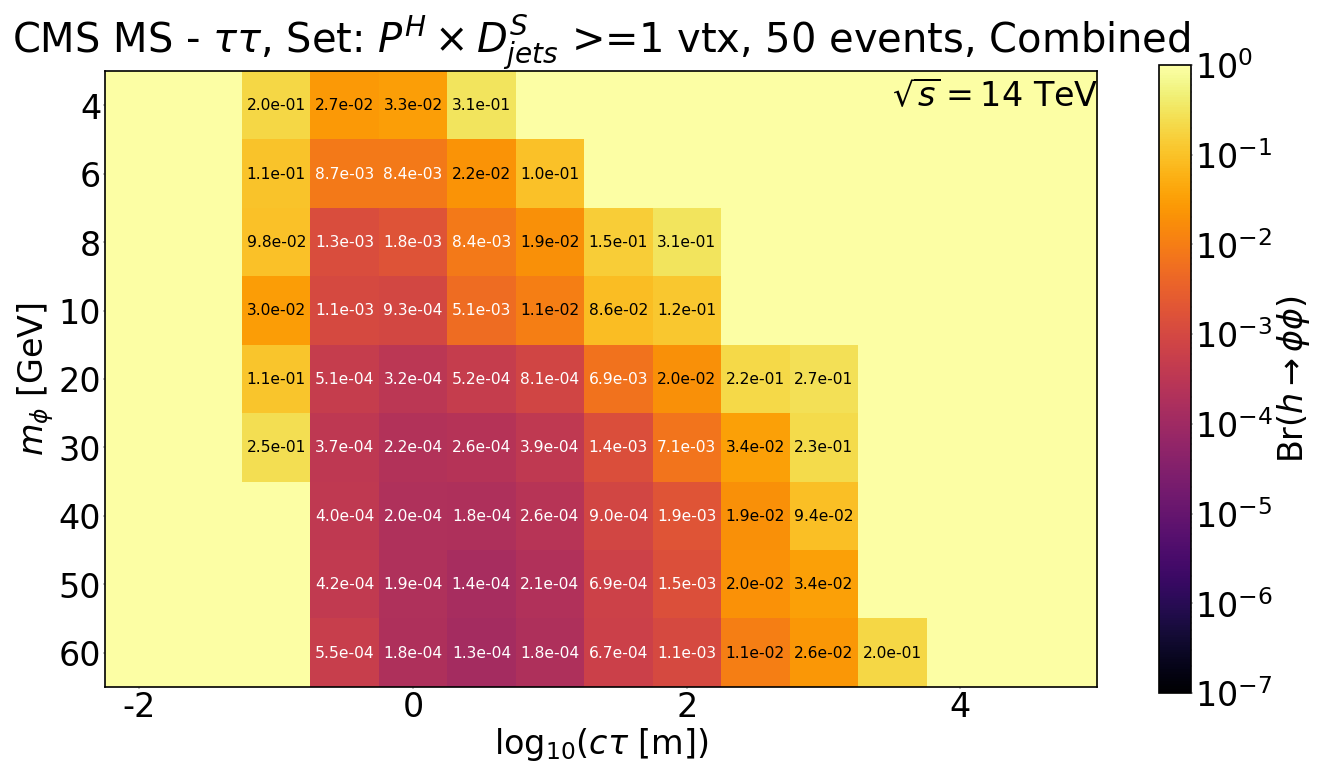} \\
    \includegraphics[width=0.46\textwidth]{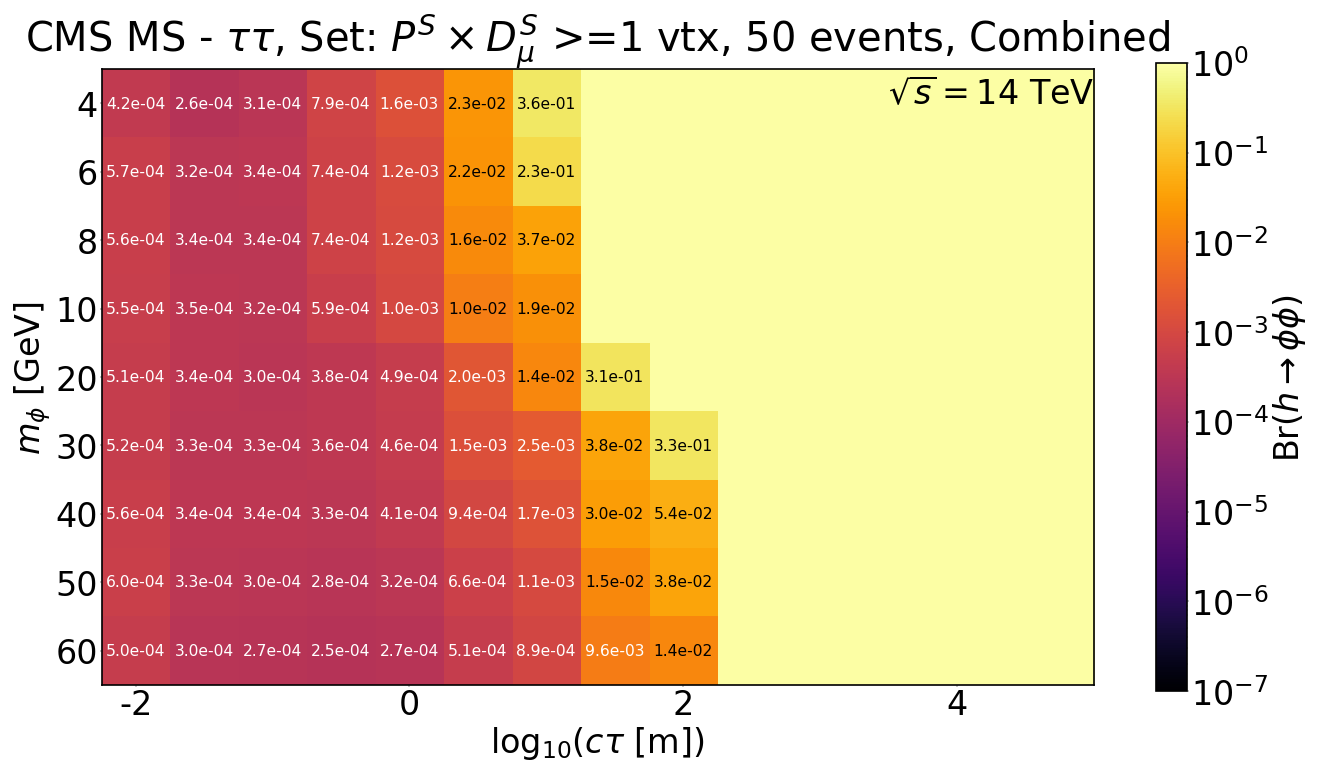} \qquad
    \includegraphics[width=0.46\textwidth]{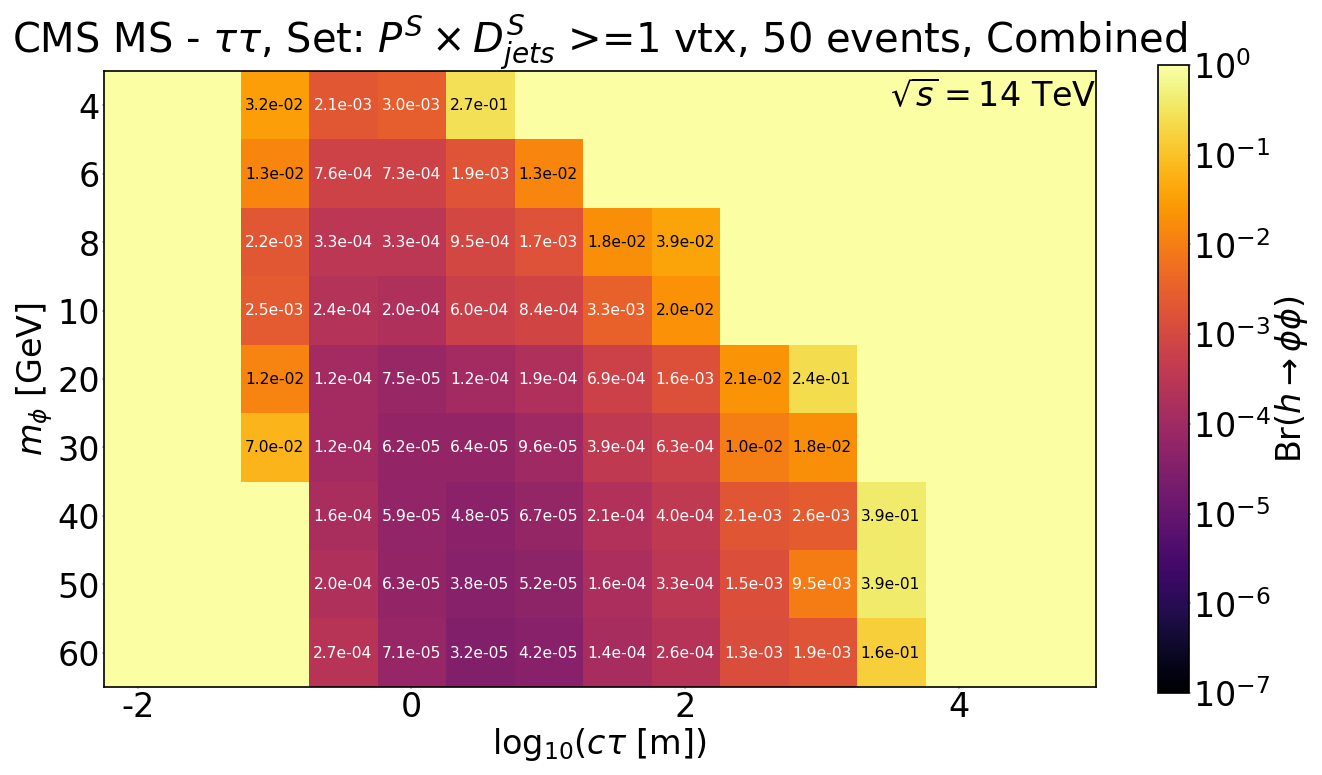}
    \caption{\small \sl Projected upper limits on Br$(h \to \phi \phi)$ for 50 observed decays of long-lived mediator particles decaying into $\tau$ leptons, when both the $\tau$s from a mediator particle decay into muons (left panel), otherwise within the CMS MS (right panel), for three sets of cuts explained in the text. The shown limits are obtained by combining the ggF, VBF and Vh modes for the Higgs boson production at the 14\,TeV HL-LHC experiment (3000\,fb$^{-1}$).}
    \label{fig:tautau-14TeV-combo}
\end{figure}

Fig.\,\ref{fig:tautau-14TeV-combo} shows the projected upper limit on the branching fraction, Br$(h\rightarrow\phi\phi)$, for three combinations of cuts, $D^H \geq$ 1\,vtx, $P^H\times D^S \geq$ 1\,vtx and $P^S \times D^S \geq$ 1\,vtx, for some benchmark masses and a range of decay lengths assuming 100\,\% decay of the mediator particle into $\tau$ leptons, combined over ggF, VBF, Vh-jet and Vh-lep modes for the Higgs boson production at the 14\,TeV HL-LHC experiment with an integrated luminosity of 3000\,fb$^{-1}$. The mediator particle decaying into $\tau$ leptons is feasible only when $m_{\phi} > 2 \times m_\tau$, and therefore, we present the results starting from a mediator mass of 4\,GeV. We get very low efficiencies for events where both the mediator particles, which come from the Higgs boson, decay inside the MS and for that, the decay products of the $\tau$ leptons, muons or otherwise, satisfy the harder set of cuts; $D^H_\mu =$ 2\,vtx or $D^H_{jets} =$ 2\,vtx, which imply that these sets of cuts do not have sensitivity for this decay mode. We,  therefore, do not include these in the figure.

The displaced di-muons analysis is most sensitive at $c\tau=5$\,cm and 0.5\,m for mediator masses of 4\,GeV and 50\,GeV with Br$(h \to \phi \phi) < 2.6 \times 10^{-4}$ and $2.8 \times 10^{-4}$, respectively, with the $P^S \times D_\mu^S \geq$ 1\,vtx set of cuts. 
For the MS cluster analysis, we get the strongest limits from the $P^S \times D_{jets}^S \geq$ 1\,vtx set of cuts, with Br$(h \to \phi \phi) < 2.1\times 10^{-3}$ and $3.8 \times 10^{-5}$ at $c \tau_\phi = 0.5$\,m and 5\,m for mediator masses of 4\,GeV and 50\,GeV, respectively. The limit for the displaced di-muons analysis has a suppression of $\sim 0.06\,(= 0.17 \times 0.17 \times 2)$ for both $\tau$s from a mediator particle to decay to muons, because the branching of $\tau \to \mu \nu_\mu \nu_\tau$ is around 17.41\,\%, where $\nu_\mu$ and $\nu_\tau$ are the muon and tau neutrinos. In addition to the suppression due to branching, the muons coming from $\tau$ lepton decay are mostly not energetic enough to pass the required $p_T$ thresholds, and we get around an order of suppression from that.

\subsubsection{Mediator particle in the minimal model}

We have so far discussed the analysis strategy for eight possible decay modes of the mediator particle, i.e. $\phi \to \mu^+ \mu^-$, $\pi^+ \pi^-$, $K^+ K^-$, $g g$, $s \bar{s}$, $c \bar{c}$, $\tau^+ \tau^-$ and $b \bar{b}$, and presented results assuming 100\% decay to each of these modes separately. We now combine all these decay modes taking into account the branching ratios predicted in the minimal model, i.e. Table\,\ref{fig:width_branching}. We span mediator masses from 0.5\,GeV to 60\,GeV, and for each benchmark mass, take the following decay modes into account:
\begin{center}
\begin{tabular}{c r l}
    $\bullet$ & 0.5\,GeV: & $\mu^+ \mu^-$, $\pi^+ \pi^-$. \\
    $\bullet$ & 1.0, 1.2, 1.4, 1.6, 1.8\,GeV: & $\mu^+ \mu^-$, $\pi^+ \pi^-$, $K^+ K^-$. \\
    $\bullet$ & 2.0\,GeV: & $\mu^+ \mu^-$, $\pi^+ \pi^-$, $K^+ K^-$, $g g$, $s \bar{s}$. \\
    $\bullet$ & 2.5\,GeV: & $\mu^+ \mu^-$, $g g$, $s \bar{s}$. \\
    $\bullet$ & 4.0, 6.0, 8.0\,GeV: & $\mu^+ \mu^-$, $g g$, $s \bar{s}$, $c \bar{c}$, $\tau^+ \tau^-$. \\
    $\bullet$ & 10, 20, 30, 40, 50, 60\,GeV: & $\mu^+ \mu^-$, $g g$, $s \bar{s}$, $c \bar{c}$, $\tau^+ \tau^-$, $b \bar{b}$. \\
\end{tabular}
\end{center}
From Fig.\,\ref{fig:width_branching} we can see that in the minimal model, for mediator masses up to 1\,GeV, limits from pions dominate followed by the muons. At 1\,GeV decay to Kaons open up and it dominates till 2\,GeV mediator mass, from where the mediator can decay to $s$-quarks and gluons, with gluons dominating the branching fractions. Decay to $c$-quarks and $\tau$ leptons are possible after 3-4\,GeV and branching to $c$-quarks dominates followed by $\tau$ leptons and gluons. From 10\,GeV onwards, the dominant branching of the mediator is to a pair of $b$-quarks, with all the other modes having branching less than or around 10\%. The combined limits will be dominated by the individual decay modes according to their branching fractions, that we discussed above. For the displaced muons, we use the result where the physical separation at the detector edge is used for isolation to keep our results conservative.

\begin{figure}[t]
    \centering
    \includegraphics[width=0.46\textwidth]{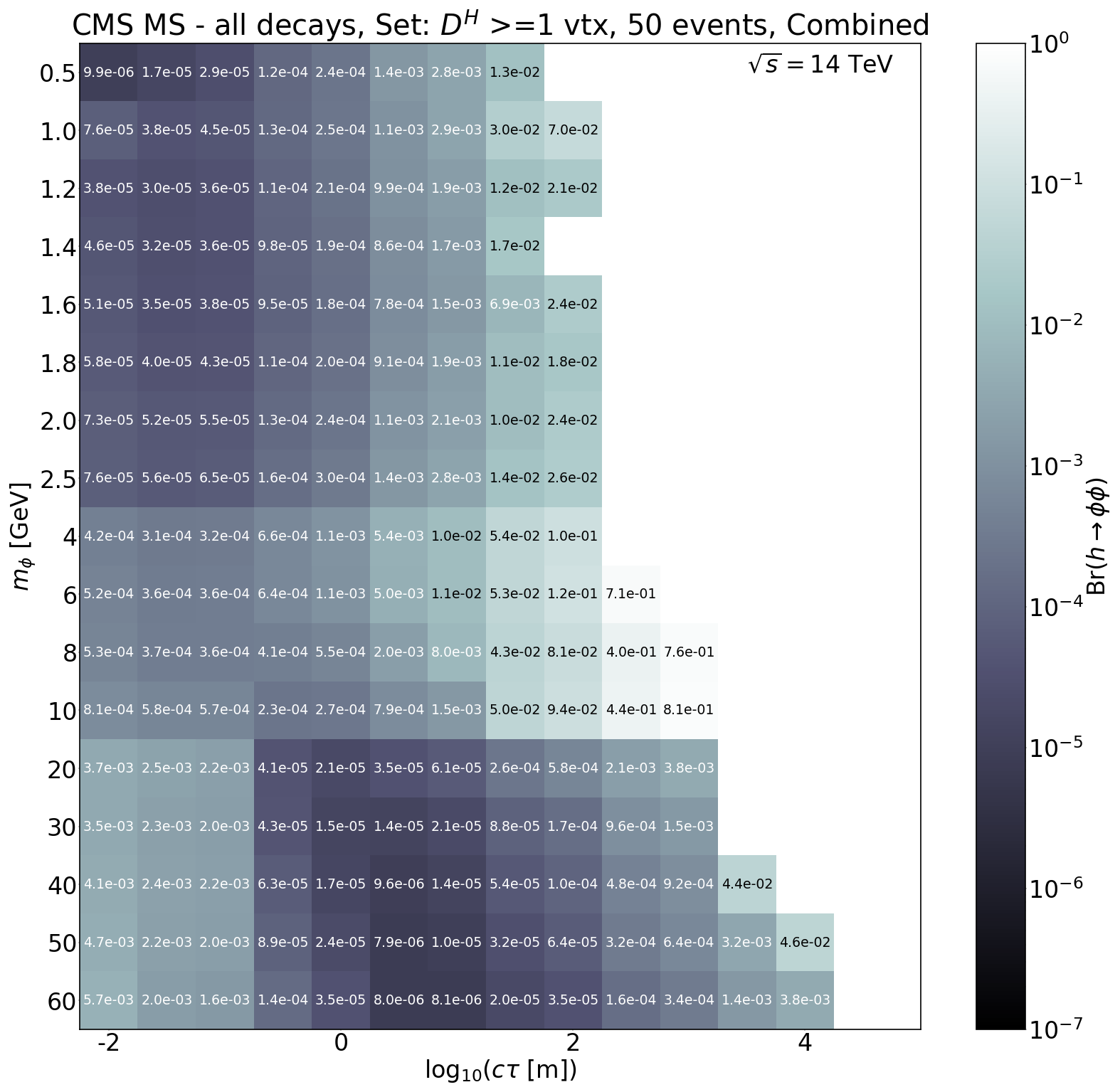} \qquad
    \includegraphics[width=0.46\textwidth]{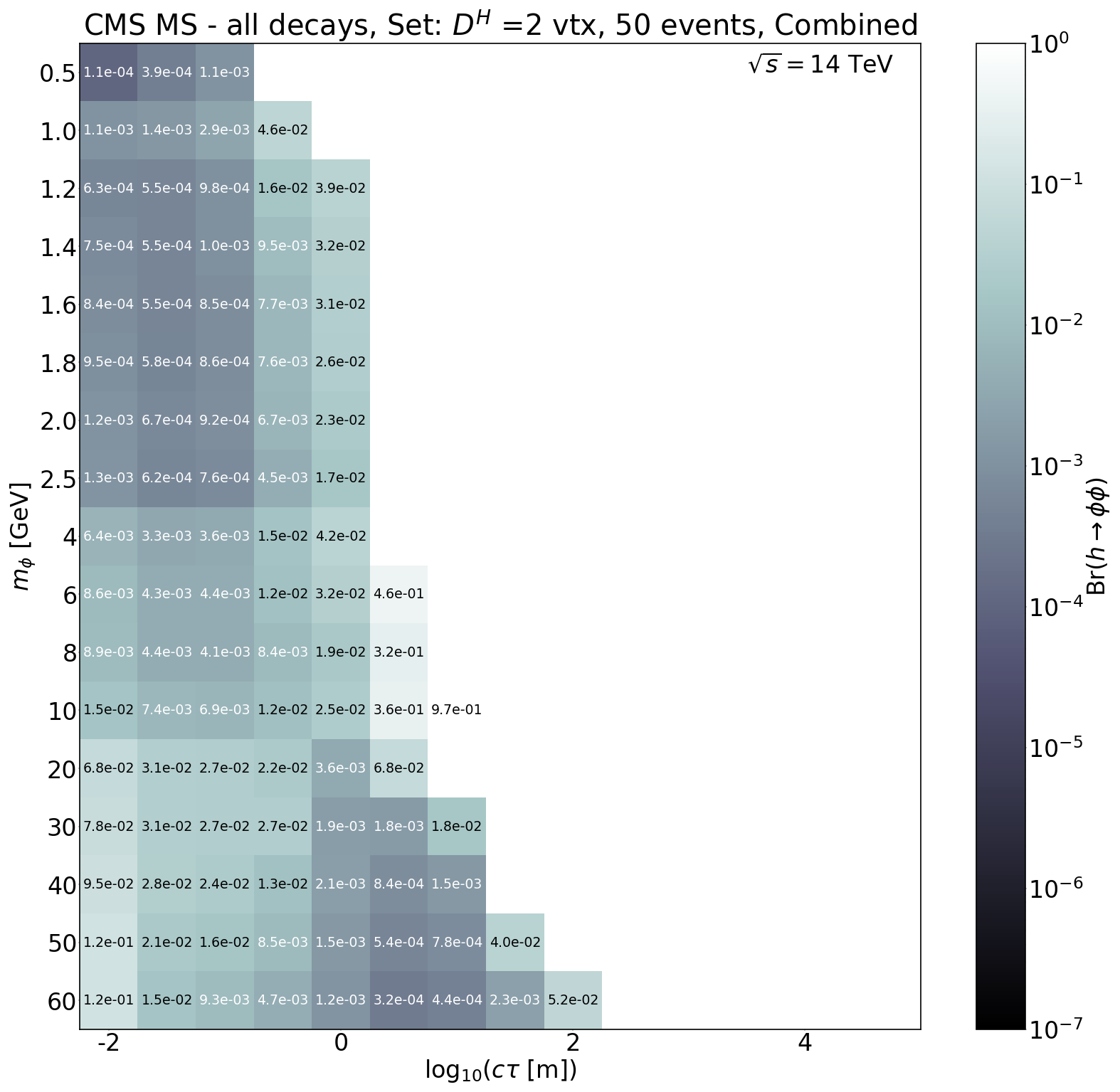} \\
    \includegraphics[width=0.46\textwidth]{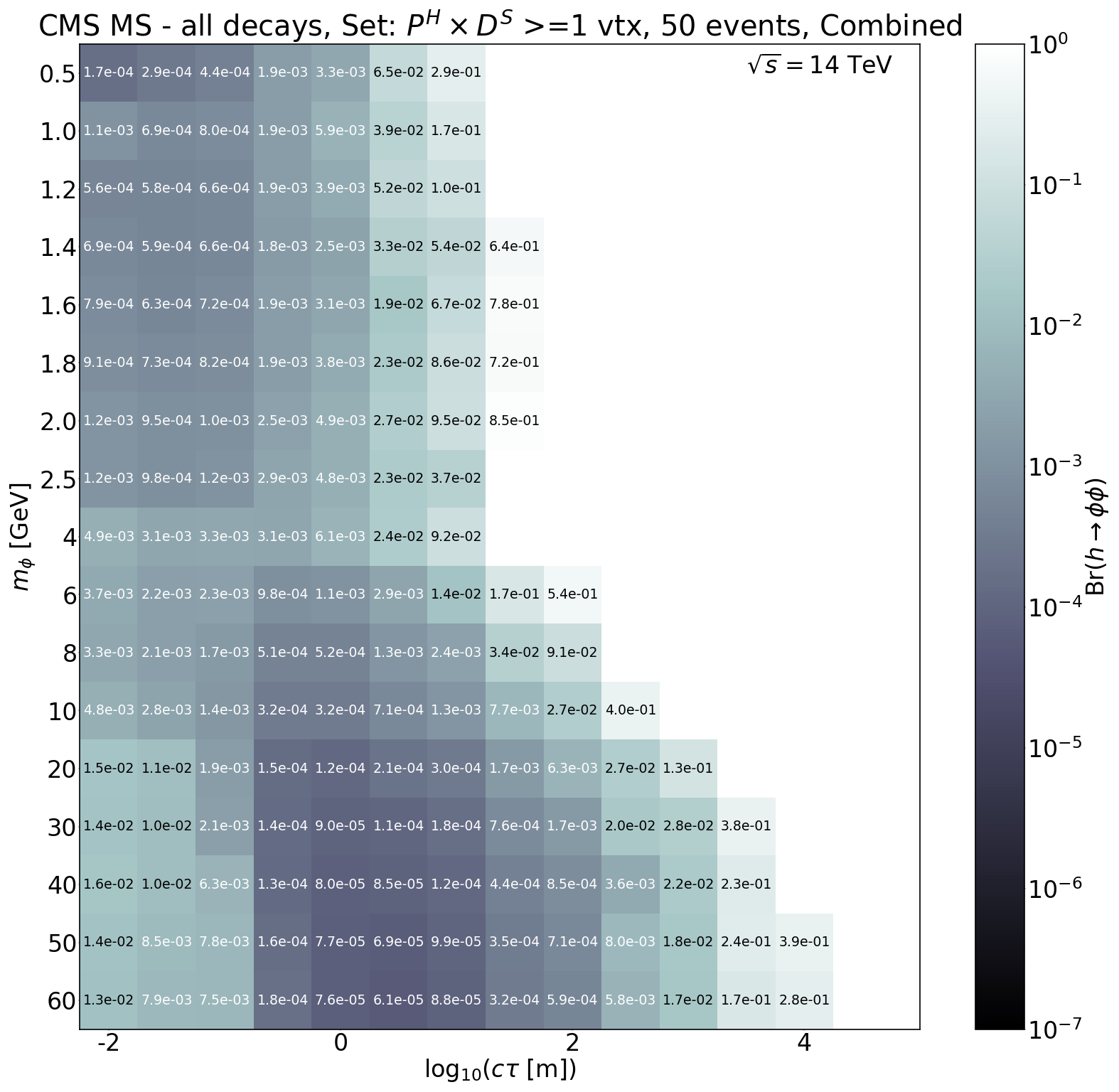} \qquad
    \includegraphics[width=0.46\textwidth]{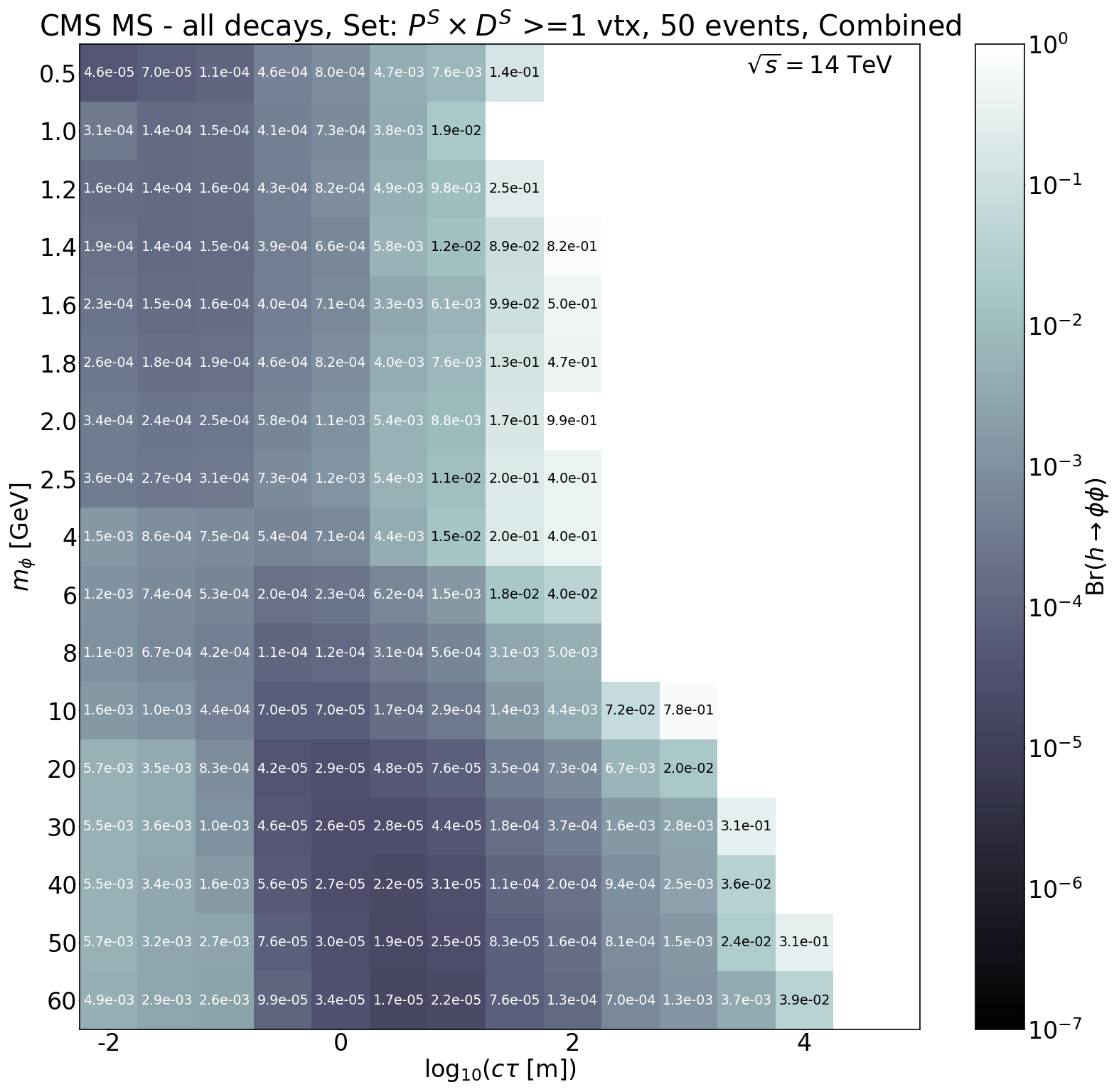}
    \caption{\small \sl Projected upper limits on the branching fraction, Br$(h \to \phi \phi)$, for 50 observed decays of long-lived mediator particles using the CMS MS when events are selected by applying the four sets of cuts described in the beginning of section\,\ref{sssec:analysis}. The shown limits are obtained by combining the ggF, VBF and Vh channels for the Higgs boson production at the 14\,TeV HL-LHC experiment (3000\,fb$^{-1}$) and the decay modes of the mediator particle $\phi \to \mu^+ \mu^-$, $\pi^+ \pi^-$, $K^+ K^-$, $c \bar{c}$, $\tau^+ \tau^-$ and $b \bar{b}$ according to the branching ratios in Fig.\,\ref{fig:width_branching}.}
    \label{fig:combined-CMS-14-1}
\end{figure}

\begin{figure}[t]
    \centering
    \includegraphics[width=0.46\textwidth]{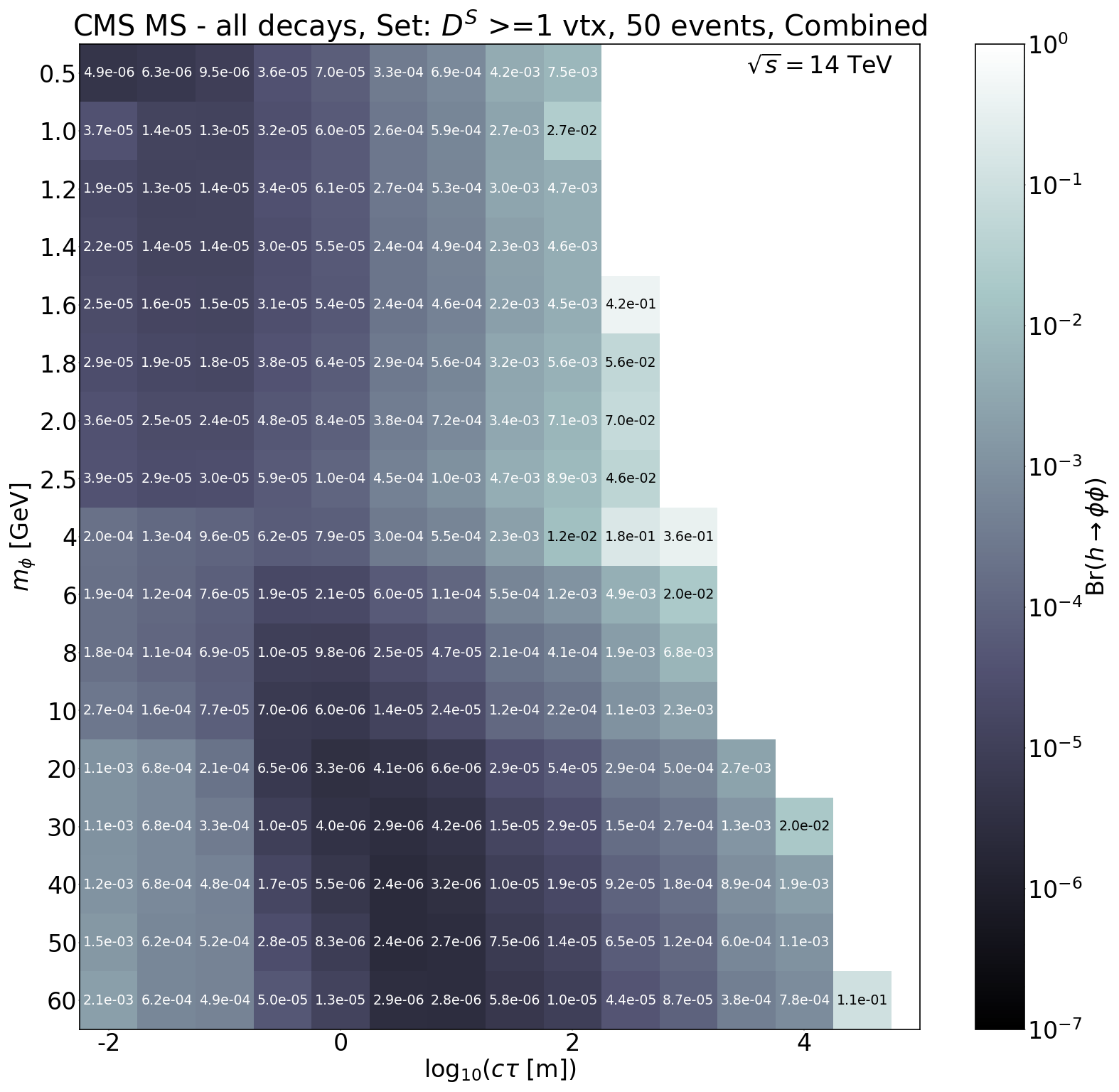} \qquad
    \includegraphics[width=0.46\textwidth]{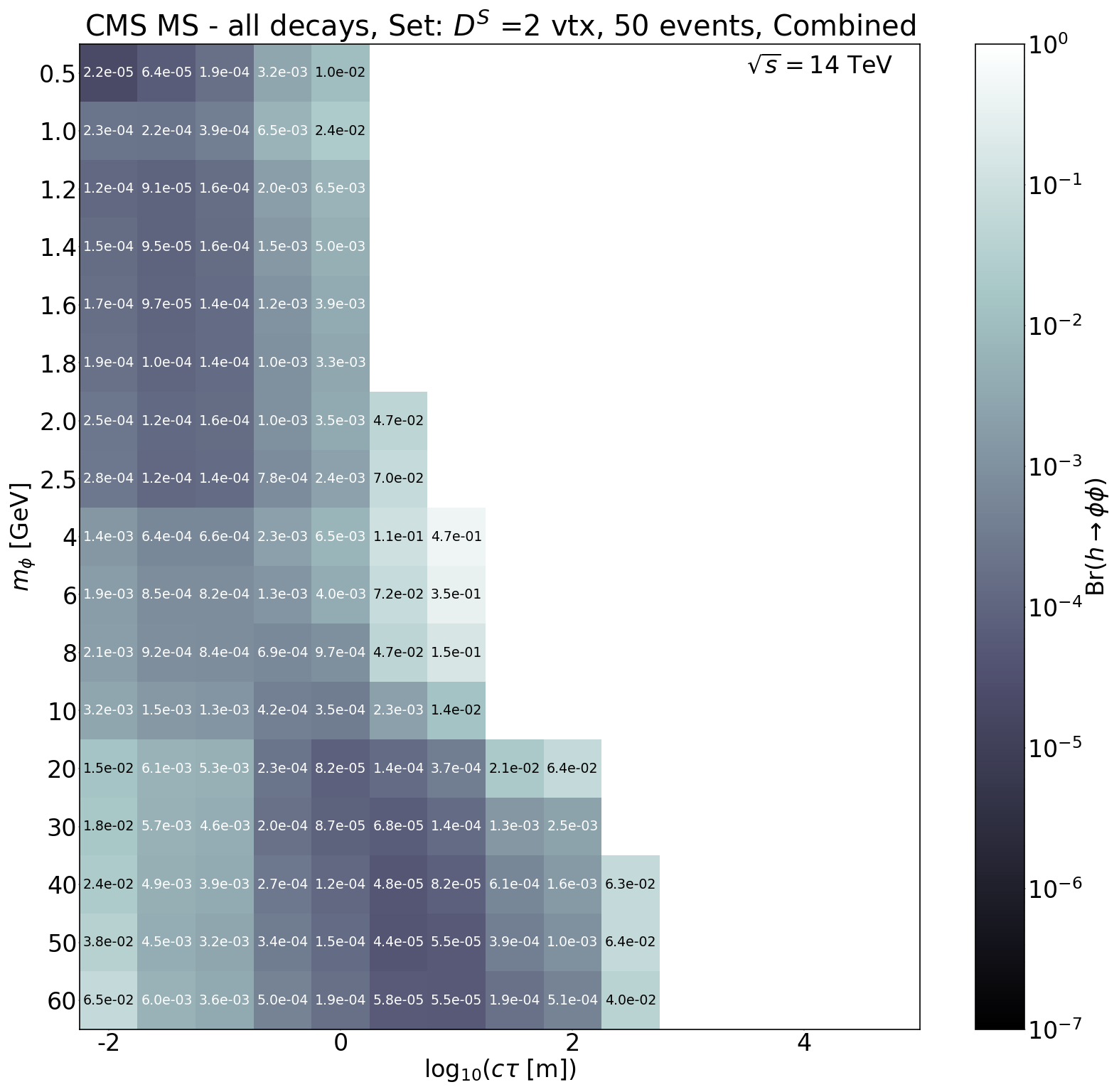}
    \caption{\small \sl Projected upper limits on the branching fraction, Br$(h \to \phi \phi)$, for 50 observed decays of long-lived mediator particles using the CMS MS when events are selected by applying softer cuts on the displaced particles only. The shown limits are obtained by combining the ggF, VBF and Vh modes for the Higgs boson production at the 14\,TeV HL-LHC experiment with an integrated luminosity of 3000\,fb$^{-1}$ and the decay modes of $\phi \to \mu^+\mu^-$, $\pi^+\pi^-$, $K^+K^-$, $c\bar{c}$, $\tau^+\tau^-$ and $b\bar{b}$ according to the branching ratios in Fig.\,\ref{fig:width_branching}.}
    \label{fig:combined-CMS-14-2}
\end{figure}

Figs.\,\ref{fig:combined-CMS-14-1} and \ref{fig:combined-CMS-14-2} shows the projected upper limit on the branching fraction, Br$(h \to \phi \phi)$, for the six combinations of cuts for the benchmark masses mentioned above over a range of decay lengths from 0.01\,m to $10^4$\,m, assuming the minimal model defined in eq.\,(\ref{eq: minimal model}), combined over ggF, VBF, Vh-jet and Vh-lep modes for the Higgs boson production at the 14\,TeV HL-LHC experiment with an integrated luminosity of 3000\,fb$^{-1}$. In addition to the results (shown in Fig.\,\ref{fig:combined-CMS-14-1}) which use the four sets of cuts described in the beginning of section\,\ref{sssec:analysis}, we also show results of $D^S \geq$ 1\,vtx and $D^S =$ 2\,vtx set of cuts (in Fig.\,\ref{fig:combined-CMS-14-2}), where we trigger on the displaced activity alone with a softer set of cuts, demanding at least one vertex and two vertices, respectively. The best limits that we can achieve are reflected by the result of $D^S \geq$ 1\,vtx cuts in Fig.\,\ref{fig:combined-CMS-14-2}. 
We can probe up to a branching fraction of the Higgs boson to the mediator particle of around $4.6 \times 10^{-5}$ ($4.9 \times 10^{-6}$) for the mediator mass of 0.5\,GeV at $c \tau_\phi = 1$\,cm and $5.4 \times 10^{-4}$ ($6.2 \times 10^{-5}$) for the 4\,GeV mediator at $c \tau_\phi = 0.5$\,m and $1.9 \times 10^{-5}$ ($2.4 \times 10^{-6}$) for the 50\,GeV mediator mass at $c \tau_\phi = 5$\,m for $P^S\times S^S\geq1$\,vtx ($D^S\geq1$\,vtx) set of cuts\,\footnote{
The CMS LLP search results at 13\,TeV for 137\,fb$^{-1}$ using the endcap MS in Ref.\cite{CMS:2021juv} gives an upper limit of $3\times10^{-3}$ on Br($h \to \phi \phi$) for 40\,GeV mediator at around $c \tau_\phi = 1$\,m. Our limits from the MS cluster analysis for the 40\,GeV mediator at 3000\,fb$^{-1}$ is around $8\times10^{-5}$ at $c \tau_\phi = 1$\,m for the $P^H\times D^S\geq1$\,vtx set of cuts, where we select the event based on the prompt objects at L1, and assume that in HLT, the displaced activity in MS maintain the L1 thresholds. We achieve an improvement by a factor of $\sim 37$, of which a factor of $\sim 22$ is accounted by the increase in luminosity, and the extra enhancement might be due to using both the barrel and endcap MS detectors.}. For the lower $c \tau_\phi$ values, the limit is dominantly coming from displaced muons analyses which are applied to the mediator decay into a pair of muons, even at higher mediator masses of 50-60\,GeV.

We observe that the combination of various production modes of Higgs boson as well as the decay modes of the mediator particle contribute non-trivially to the limits. For example, for the 50\,GeV mediator particle with $c \tau_\phi = 1$\,m, the ggF production mode where the mediator particle decays into a pair of $b$ quarks contribute 65\,\% to the total limit, while the rest comes from other production modes and decay of the mediator particle into a pair of muons and $c$ quarks for the $D^H = 2$\,vtx set of cuts. For the LLP benchmark with the same mass and decay length, we observe that the various decay and production modes have some contribution other than the dominant $b\bar{b}$ decay mode and ggF production mode. This highlights the importance of such a comprehensive study

\begin{figure}[t]
    \centering
    \includegraphics[width=0.46\textwidth]{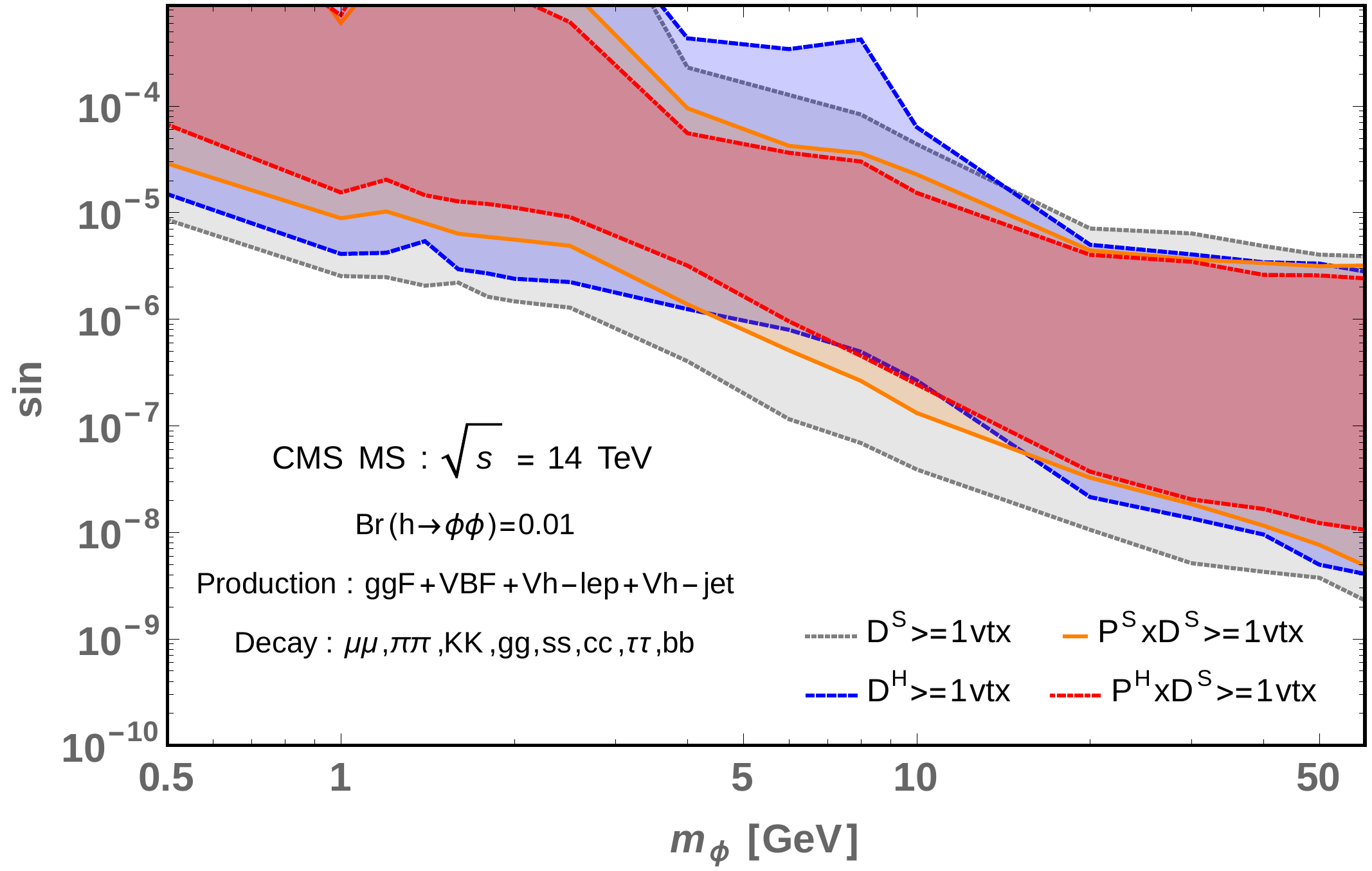} \qquad
    \includegraphics[width=0.46\textwidth]{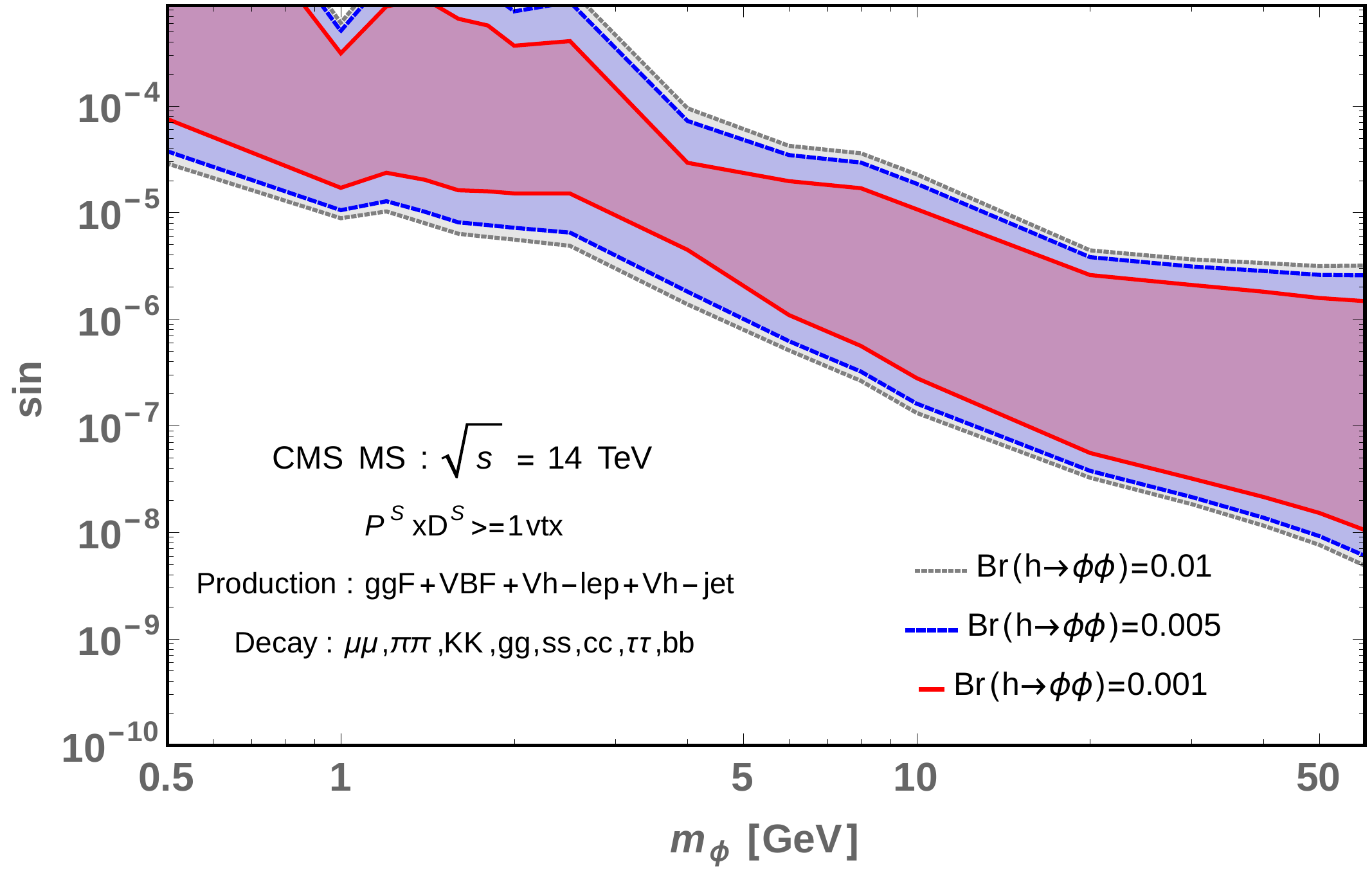}
    \caption{\small \sl Projected sensitivity on the $(sin \theta,\,m_\phi)$-plane for the mediator particle in the minimal model with various sets of cuts assuming Br$(h \to \phi \phi) =$ 0.01 (left panel) and for different values of Br$(h \to \phi\phi)$ with the $P^S \times D^S \geq$ 1\,vtx set of cuts (right panel).}
    \label{fig:sintheta-mass-CMS}
\end{figure}

Fig.\,\ref{fig:sintheta-mass-CMS} shows the projected sensitivity of our HL-LHC analysis with the CMS MS on the $(\sin \theta,\,m_\phi)$-plane of the minimal model which is defined in eq.\,(\ref{eq: minimal model}). The left plot shows the sensitivity of the various sets of cuts: $D^S \geq$ 1\,vtx, $D^H \geq$ 1\,vtx, $P^H \times D^S \geq$ 1\,vtx, and $P^S \times D^S \geq$ 1\,vtx, with the branching fraction of the Higgs boson decaying into a pair of the mediator particles, Br$(h \to \phi \phi)$, being equal to 0.01. Since we start all our analyses from 1\,cm decay length, the largest mixing angle that we can probe corresponds to the angle which makes the 0.5\,GeV mediator particle have a decay length of 1\,cm, and is $8.8\times10^{-4}$. We, therefore, start our $\sin \theta$ axis from this value onward. 
For a 1\% branching of the Higgs boson to the mediator particle, we can probe mixing angles to $8\times 10^{-6}$ and $3\times 10^{-5}$ for 0.5\,GeV mediator particle with the $D^S \geq 1$\,vtx and $P^S \times D^S \geq1$\,vtx cuts respectively. For the 50\,GeV mediator, we can probe sin$\theta$ from $4\times10^{-6}$ ($3\times10^{-6}$) to $4\times10^{-9}$ ($7\times10^{-9}$) with the $D^S\geq1$\,vtx ($P^S\times D^S\geq1$\,vtx) cuts assuming 1\% branching of Higgs boson to the mediator.
On the other hand, the right plot shows the sensitivity for the $P^S \times D^S \geq$ 1\,vtx set of cuts assuming the branching fraction of the Higgs boson decaying into a pair of the mediator particles, Br$(h \to \phi \phi)$, is equal to 0.01, 0.005 and 0.001.

\subsection{Possible backgrounds in the CMS MS}
\label{ssec:background}

Any analysis at the CMS is not complete without a word on possible backgrounds affecting the signals. Signatures of long-lived particles are quite clean compared to other prompt new physics signals, because most of the SM processes are indeed prompt. However, along with this bright side, the hindrance comes from non-standard background processes, which are challenging to simulate. In this paper, we discuss the possible backgrounds that can arise in our various analyses. We discuss, qualitatively, how these backgrounds can be handled and reduced in our analyses, though  quantification of the same is outside the scope of the present work.

As seen in the discussion so far, it is possible to broadly classify our analyses into two categories. We, therefore, discuss possible backgrounds for each of them below.
\begin{itemize}
    \item For displaced di-muons produced anywhere within the CMS detector:

    \begin{enumerate}
        \item There are SM particles decaying into di-muons with a displaced vertex, such as $J/\psi$ or $\Upsilon$. These, however, have very small decay lengths ($\sim$ few pm) and therefore, can be separated from signals with the $d_0$ or $d_T$ cuts. Also, the invariant mass of the two muons can be used to distinguish the SM particles by applying masks near the $J/\psi$ and $\Upsilon$ resonances.

        \item Cosmic muons can be misidentified as displaced muons in the detector. Those usually appear back-to-back in the detector, and this background can be suppressed by rejecting back-to-back muon pairs with a $\Delta \phi$ cut. In Ref.\,\cite{CMS-PAS-FTR-18-002}, they estimate a suppression factor of $10^{-9}$ for cosmic muon events in the absence of $pp$ collisions, which will be valid even in the HL-LHC runs, for this is independent of the conditions of the collider.

        \item Muons from the beam halo can have large displacements and, therefore, can mimic long-lived mediator particle signals. They, however, have very low transverse momentum\,\cite{Liu:2007ad}), and a cut of $p_T >$ 15\,GeV on displaced muons can completely suppress the beam halo background\,\cite{CMS-PAS-FTR-18-002}. 
    \end{enumerate}

    \item For clusters of electrons, photons, and hadrons within the CMS MS:

    \begin{enumerate}
        \item Several long-lived hadrons in the SM can punch through the calorimeter from the transition regions and then decay, such as $K_S \to \pi^+ \pi^-$, $\Lambda \to p \pi^-$, $\Sigma^+ \to p \pi^0 / n \pi^+$, $\Sigma^- \to n \pi^-$, $\Xi \to \Lambda \pi^0$, $\Xi^- \to \Lambda \pi^-$ and $\Omega^- \to \Lambda K^- / \Xi \pi^- / \Xi^- \pi^0$. Demanding a minimum of 3-5 charged particles associated with a displaced vertex can minimise such backgrounds.

        \item The HL-LHC experiment will have an increased amount of pile-up (PU), and then a fraction of the PU jets can punch through and produce activities inside the MS. These activities will also have high particle multiplicity, which is in contrast to the previous case of long-lived SM hadron decays. These can be suppressed either by vetoing events from the transition regions, or by checking for activities inside the calorimeters as well as the trackers within a cone of some radius around the activity in the MS, as has already been discussed in Ref.\,\cite{ATLAS:2015xit} for the ATLAS detector.
    \end{enumerate}
\end{itemize}

Since we are not quantifying any of the backgrounds, we keep our limits conservative by demanding a minimum of 50 observed events, as mentioned in section\,\ref{sssec:analysis}. For our limits to correspond to at least a significance ($S/\sqrt{B}$ with $S$ and $B$ denoting the number of signal and background events, respectively) of $2\sigma$, the number of background events at the HL-LHC experiment that can be accommodated is around 625. The recent CMS analysis,  which makes use of the end-cap MS gives a background estimate of around $\sim 3$ events with 137\,fb$^{-1}$ of data, which, if scaled only by luminosity, corresponds to around 66 events at the HL-LHC experiment. Our limits can, therefore, accommodate about ten times more background events, which can make the results robust against the increasing amount of PU at the HL-LHC experiment. With the proper simulation of the backgrounds using the HL-LHC conditions and a correct estimation of the background at the HL-LHC experiment, one can scale our limits accordingly. In the next subsection, we study the prospects of dedicated LLP detectors for long-lived mediator particles produced from the Higgs boson decay.

\subsection{Dedicated LLP detectors}
\label{ssec:dedicated_detectors}

In section\,\ref{sec:higgs-lhc}, the distributions of decay length in the lab frame (Fig.\,\ref{fig:LLPs-dist-14TeV-boost}) for a range of mediator masses and lifetimes motivated the importance of dedicated LLP detectors for capturing lighter mediator particles with longer lifetimes, and it was evident from these distributions that such detectors placed far away from the IP might be sensitive to a range of lifetimes which is complementary to the CMS MS. In this section, we compute the limits obtained from two such detectors, which are proposed to collect data during the HL-LHC runs; MATHUSLA and CODEX-b. We emphasize once again on the fact that, since these detectors are proposed to be placed a few tens of meters away from the IP of the $pp$ collision, there is enough shielding of rock or concrete to guarantee very little or almost no backgrounds. Therefore, observation of even a few events ($\sim 4$) can be claimed as a discovery of displaced decays of LLP particles. Here, we study the potential of such detectors for a long-lived scalar mediator particle produced from the decay of the Higgs boson, for the ggF, VBF, and Vh modes of the Higgs boson production, which impart different boosts to the mediator particle, and therefore need to be studied separately. To the best of our knowledge, such a study combining all the production modes of Higgs boson is not available in the literature for such dedicated LLP detectors. We start with brief descriptions of these detectors, along with their planned position and size specifications.

\subsubsection{The CODEX-b detector}

The CODEX-b detector is proposed to be placed near the so-called interaction point 8 (IP8) around which the LHCb detector is located, specifically in the UX85 cavern. The position of the CODEX-b detector is currently proposed to be at
\begin{eqnarray}
    26 < x < 36\,{\rm m} \nonumber\\
    -7 < y < 3\,{\rm m} \nonumber\\
    5 < z < 15\,{\rm m},
\end{eqnarray}
where $z$ is the direction along the beam line and IP8 is the origin. This has a decay volume of $10 \times 10 \times 10$\,m$^3$\,\cite{Aielli:2019ivi}. Since the detector will be placed near the LHCb detector, the data is collected at an integrated luminosity of 300\,fb$^{-1}$, about 10 times lower than ATLAS and CMS. We have also added more optimistic results with a bigger decay volume of $20 \times 10 \times 10$\,m$^3$, in case that the DELPHI detector is removed, and with an integrated luminosity of 1\,ab$^{-1}$, motivated from Ref.\,\cite{Gligorov:2017nwh}.

\subsubsection{The MATHUSLA detector}

The MATHUSLA detector is another proposal of a large decay volume to be placed in the transverse direction of any of the $pp$ collision points at the HL-LHC experiment. The detector is currently planned to have a $25 \times 100 \times 100$\,m$^3$ decay volume fat
\begin{eqnarray}
    60 < x < 85\,{\rm m} \nonumber \\
    -50 < y < 50\,{\rm m}\nonumber \\
    68 < z < 168\,{\rm~m},
\end{eqnarray}
where the origin is at the interaction point and the beam line is along $z$\,\cite{Alpigiani:2020iam}. MATHUSLA is proposed to be placed near the CMS detector, and expected to collect an integrated luminosity of 3\,ab$^{-1}$ by the end of the HL-LHC experiment.

\subsubsection{The long-lived mediator particle at dedicated LLP detectors}

Both of these detectors mentioned above have multiple layers of RPCs (Resistive Plate Chambers) to trace the hits of the charged decay products of the LLP to ensure that the decay vertex lies within the fiducial decay volume. As long as at least two of the decay products from a LLP are charged, the exact nature of the decay mode is not very relevant for these detectors. Since all of the decay modes considered in this paper include a minimum of two charged particles, the results of these detectors do not depend on the decay products of the mediator particle, unlike the case of the CMS MS analyses. We have validated our implementation of these detectors by comparing the projected upper limits on Br$(h \to \phi \phi)$ at the CODEX-b and MATHUSLA detectors as a function of $c \tau_\phi$ for some benchmark masses by demanding at least four observed events within these detectors, with those provided by these experimental collaborations, as shown in Fig.\,\ref{fig:val} in Appendix\,\ref{app:LLP-detectors}.

\begin{figure}[t]
    \centering
    \includegraphics[width=0.46\textwidth]{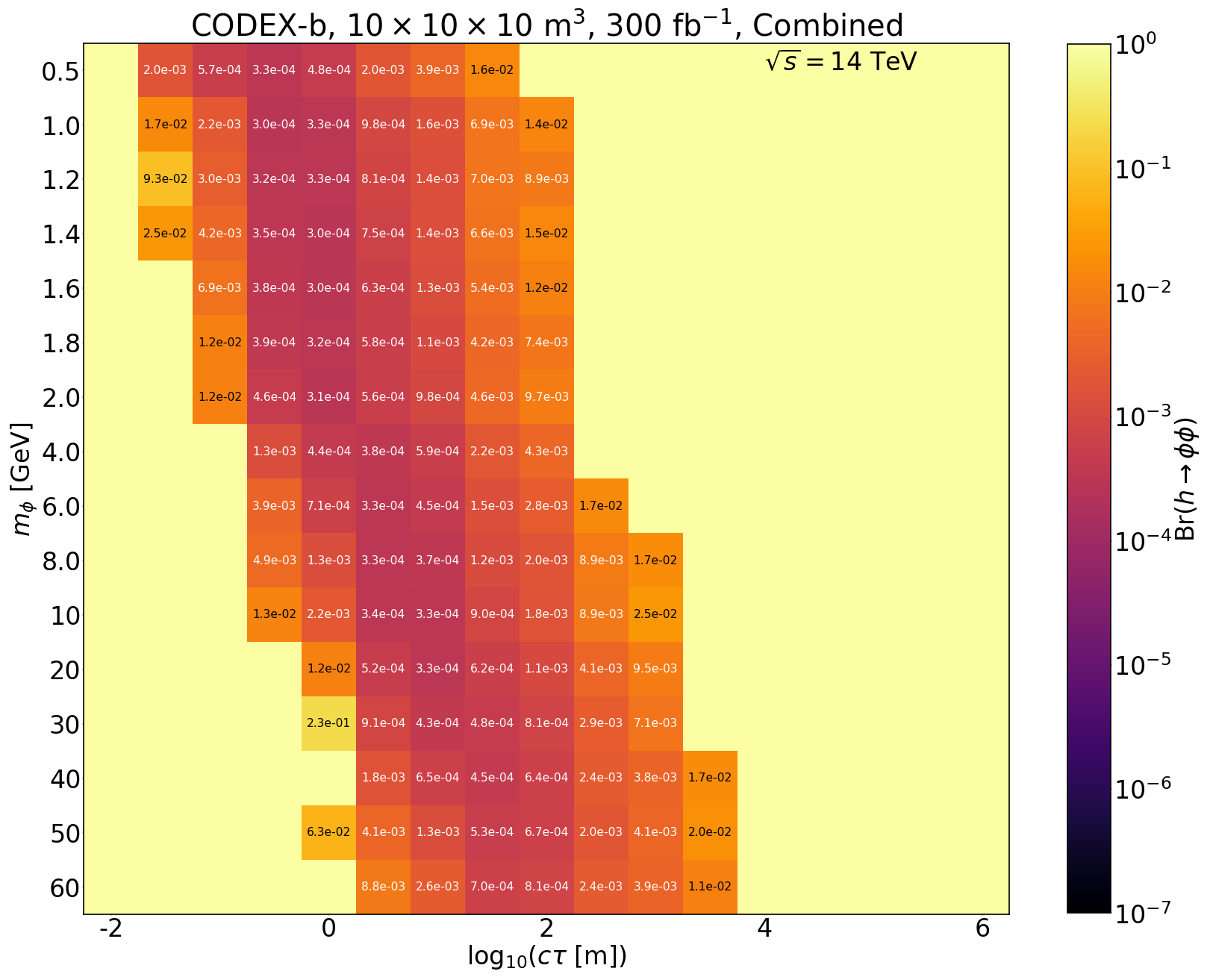} \qquad
    \includegraphics[width=0.46\textwidth]{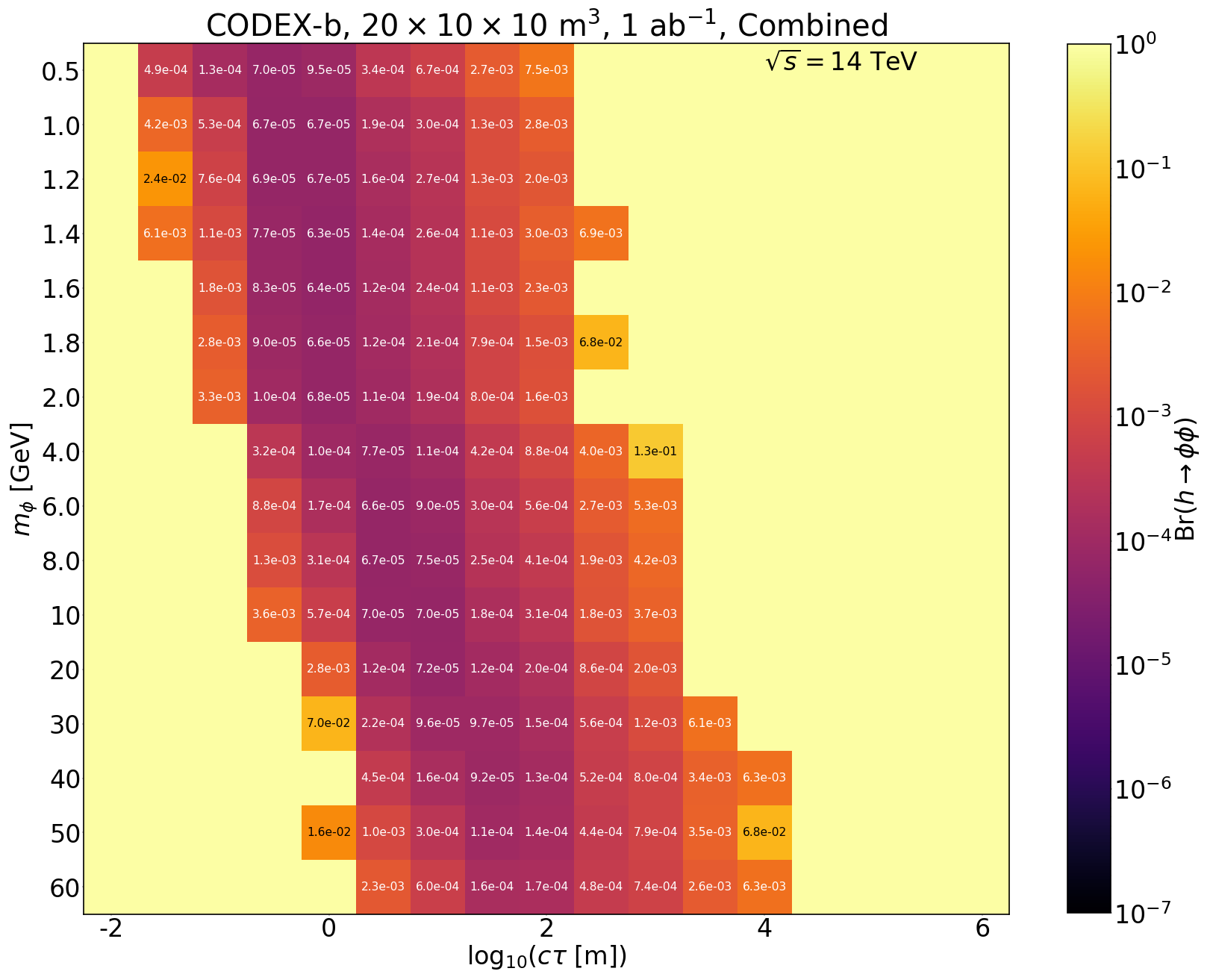} \\
    \includegraphics[width=0.46\textwidth]{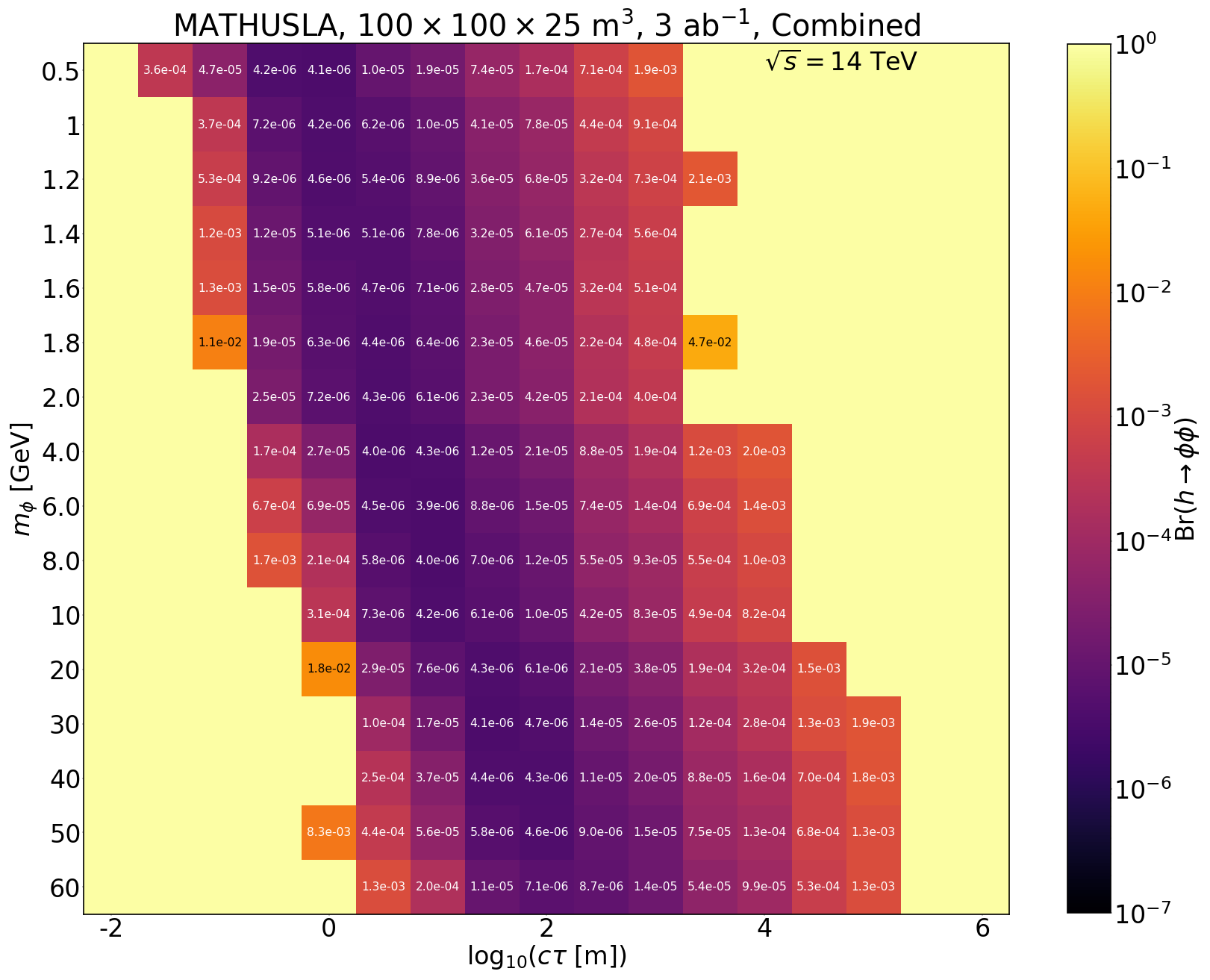}
    \caption{\small \sl Projected upper limits on the branching fraction, Br$(h \to \phi \phi)$, for 4 observed decays of long-lived mediator particles within the CODEX-b (top panels) and MATHUSLA (bottom panel) detectors. The shown limits are obtained by combining the ggF, VBF and Vh channels for the production of the Higgs boson. See text for more details.}
    \label{fig:limits-codex}
\end{figure}

The validated setups are used to extend the same exercise for a range of $m_{\phi}$ and $c \tau_\phi$, and the corresponding projected upper limits on Br$(h \to \phi \phi)$ are shown in Fig.\,\ref{fig:limits-codex} for both the designs of the CODEX-b (top panels) and MATHUSLA (bottom panel) detectors, combined over the ggF, VBF and Vh modes for the Higgs boson production. We also present the corresponding efficiency maps in Figs.\,\ref{fig:eff-maps-codex} and \ref{fig:eff-maps-mathu} in Appendix\,\ref{app:LLP-detectors}) for each production mode, which can be more useful for further analyses. The ggF production mode dominates the limits among all other production modes due to its larger cross-section. We observe that the MATHUSLA detector is sensitive over a larger range of lifetimes and gives stronger limits compared to CODEX-b, owing to its large decay volume and increased integrated luminosity of 3000\,fb$^{-1}$. The strongest limits obtained from MATHUSLA for a mediator mass of 0.5\,GeV is $4.1\times10^{-6}$ at 1\,m, whereas it is $3.3\times10^{-4}$ and $7.0\times10^{-5}$ at $c\tau=0.5$\,m for CODEX-b $10\times10\times10$\,m$^3$ (300\,fb$^{-1}$) and $20\times10\times10$\,m$^3$ (1000\,fb$^{-1}$) detectors respectively. For the 50\,GeV mediator, the most sensitive limit from MATHUSLA is Br$(h \to \phi \phi)<4.6\times10^{-6}$ at $c\tau=100$\,m and from CODEX-b $10\times10\times10$\,m$^3$, 300\,fb$^{-1}$ ($20\times10\times10$\,m$^3$, 1000\,fb$^{-1}$) detector, it is Br$(h \to \phi \phi)<5.3\times10^{-4}$ ($1.1\times10^{-4}$) at $c\tau=50$\,m. The most sensitive decay length for CODEX-b detectors are always smaller than that of MATHUSLA, since CODEX-b is nearer to the IP.

\begin{figure}[t]
    \centering
    \includegraphics[width=0.6\textwidth]{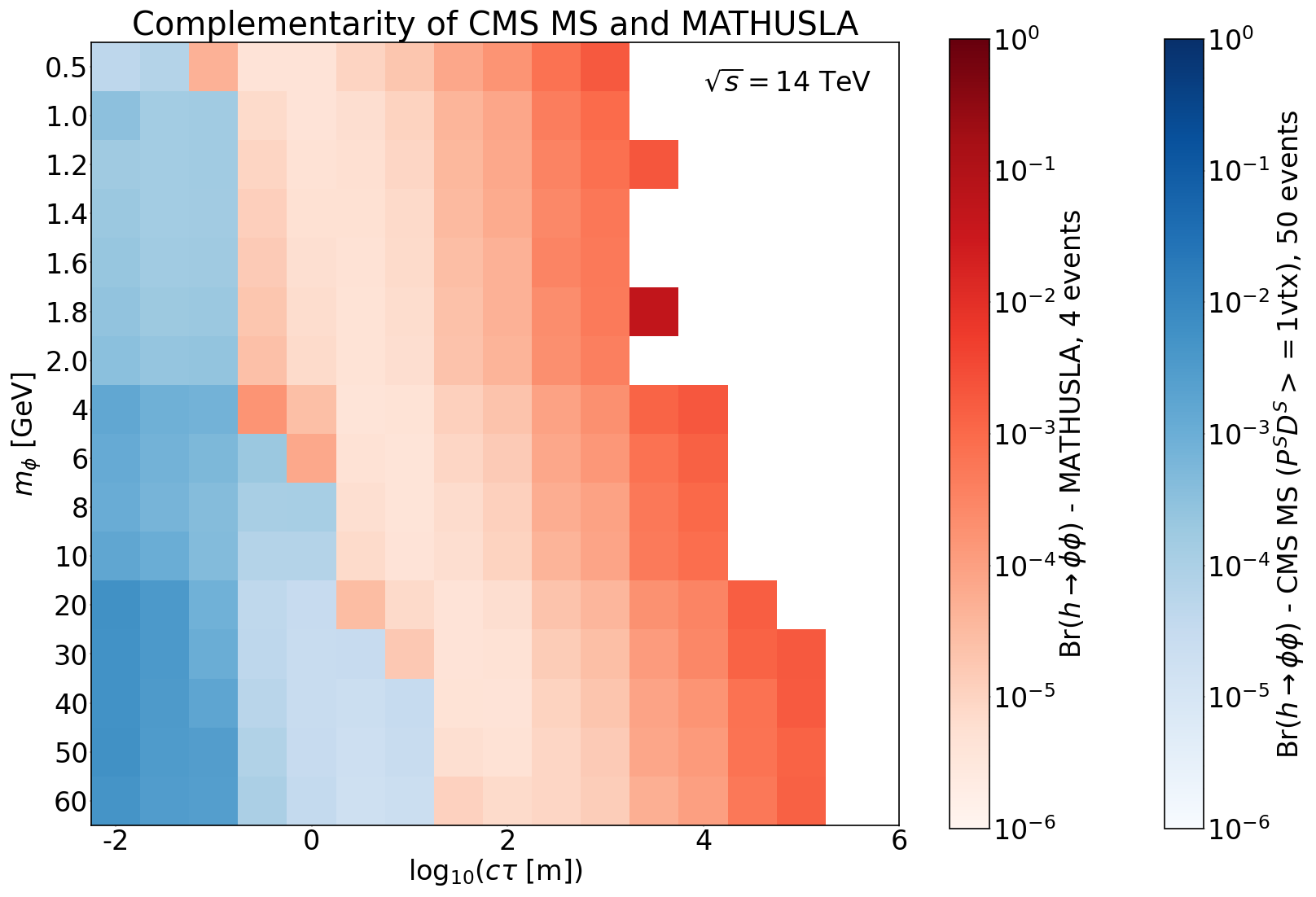}
    \caption{\small \sl Complementarity of the CMS MS analysis and the MATHUSLA LLP detector analysis at 14\,TeV HL-LHC experiment with an integrated luminosity of 3000\,fb$^{-1}$.}
    \label{fig:CMS-MATHU}
\end{figure}

In order to compare these limits with those from the CMS MS analyses, we plot the strongest limits obtained for each of the mediator masses and decay lengths from CMS MS analysis, with the $P^{S}\times D^{S}\geq1$\,vtx set of cuts, and MATHUSLA in Fig.\,\ref{fig:CMS-MATHU}. We have used a blue color bar if it is coming from the CMS MS analysis, and a red color bar if it is coming from MATHUSLA experiment. We have shown results from CMS MS and MATHUSLA analyses together in Fig.\ref{fig:CMS-MATHU}, because they are expected to collect the same integrated luminosity of 3000\,fb$^{-1}$. We can see that the regions of decay lengths where these two experiments are the most sensitive are complementary to each other, and together they probe a large range of lifetimes. 
The best limit from CMS ($P^S \times D^S \geq 1$\,vtx set of cuts) is $1.7\times 10^{-5}$ for a 60\,GeV mediator particle with 5\,m decay length, and from MATHUSLA is $3.9\times10^{-6}$ for a mediator particle of 6\,GeV with decay length 10\,m. Together they can probe decay lengths as high as 1000\,m for 0.5\,GeV mediator particle and 10$^5$\,m for 50\,GeV mediator particle.

\begin{figure}[t]
    \centering
    \includegraphics[width=0.65\textwidth]{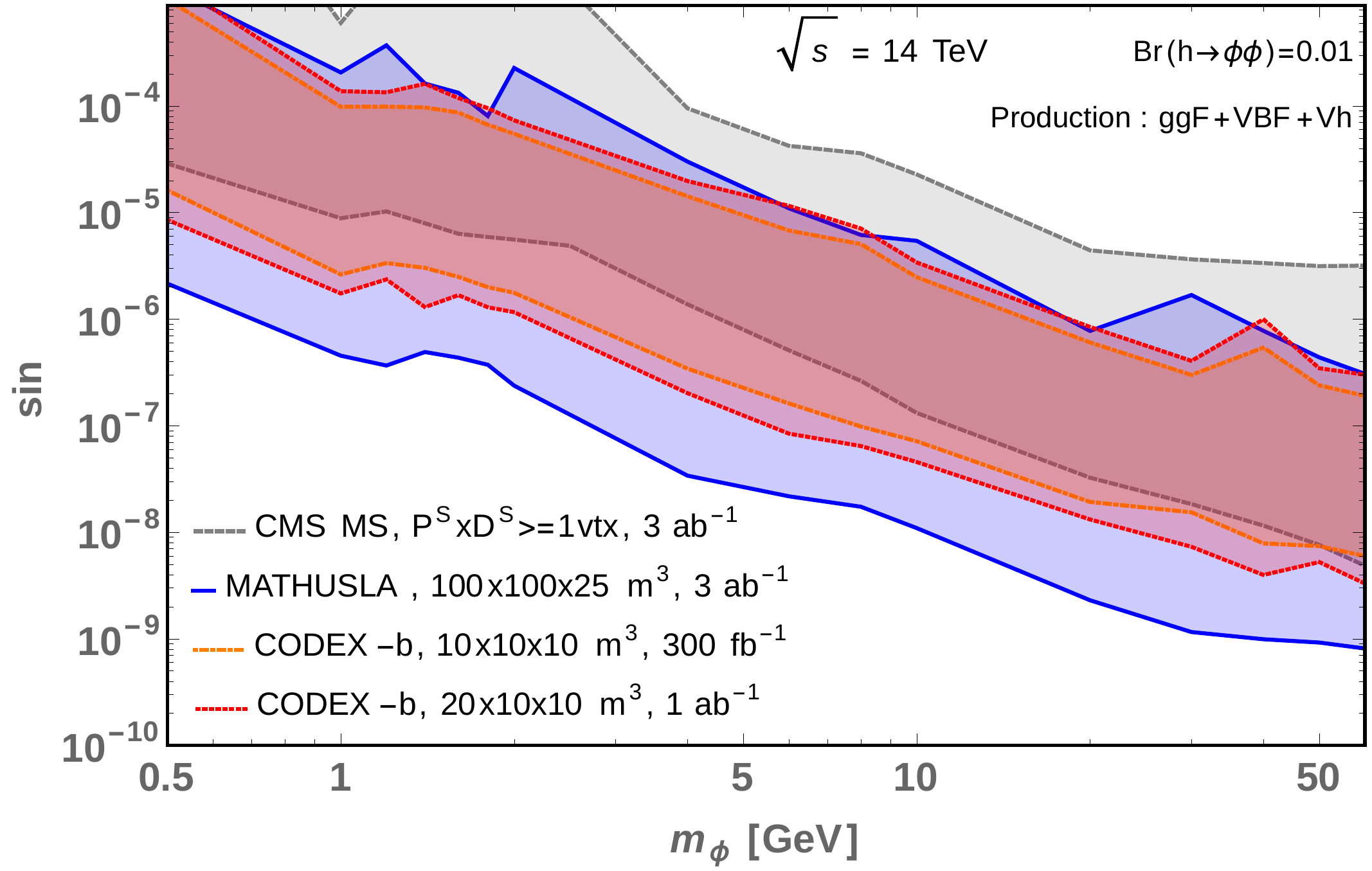}
    \caption{\small \sl Projected sensitivity on the $(\sin \theta,\, m_\phi)$-plane for the mediator particle in the minimal model achieved with the CMS MS with the $P^S \times D^S \geq$ 1\,vtx set of cuts, MATHUSLA and both the designs of CODEX-b assuming Br$(h \to \phi \phi) =$ 0.01.}
    \label{fig:14TeV-all-final}
\end{figure}

Fig.\,\ref{fig:14TeV-all-final} shows the translated limits from CODEX-b and MATHUSLA in the $(\sin \theta,\, m_\phi)$-plane of the mediator particle in the minimal model defined in eq.\,(\ref{eq: minimal model}), assuming a branching fraction of 1\,\% from the Higgs boson to a pair of the long-lived mediator particles. It also shows the limits achieved by the CMS MS search with the $P^S \times D^S \geq$ 1\,vtx set of cuts. As was expected from the boost distributions of the mediator particles, the complementarity of the CMS MS analyses with dedicated LLP detectors, i.e.,  the CODEX-b and MATHUSLA detectors, is seen in their projected sensitivities as well. MATHUSLA and CODEX-b, being situated farther from the IP than the CMS MS are sensitive to mediator particles with higher decay lengths, or lower mixing angles, than the latter. The larger decay volume and luminosity of MATHUSLA extend its reach to lower values of $\sin \theta$ than CODEX-b. For a mediator mass of 0.5\,GeV, our CMS MS analysis with the $P^S\times D^S\geq1$\,vtx cuts probes mixing angles larger than $8.8\times10^{-4}$ (corresponding to 1\,cm decay length) and MATHUSLA can extend the sensitivity to mixing angles as low as $2\times10^{-6}$. For 50\,GeV mediator mass, sensitivity of the CMS analysis ($P^S\times D^S\geq1$\,vtx) starts from a mixing angle of $3 \times 10^{-6}$, and MATHUSLA extends the sensitivity till $8 \times 10^{-10}$.

\section{The long-lived mediator particle at the 100\,TeV collider}
\label{sec:higgs-100TeV}

Future Circular Collider (FCC) is a proposal of a 100\,TeV collider, with three different possibilities of collisions of particles: hadron-hadron (FCC-hh), electron-electron (FCC-ee), and electron-proton (FCC-eh). We are studying the prospects of searching for long-lived mediator particles from the Higgs boson decay in hadron colliders, and therefore focus on FCC-hh in this paper. The 100\,TeV $pp$ collider has a rich physics program, and the reader is referred to Ref.\,\cite{Mangano:2017tke}. The motivation for exploring the prospect of 100\,TeV collider in this paper, apart from the increased cross-sections as well as luminosity, is the fact that it is still in its designing phase and a study of the sensitivity of the features of the present detector design for scalar mediator particles will be able to guide further improvements in the design for enhancing its sensitivities for such long-lived scenarios. There is also a lack of studies understanding the prospects of LLPs at the 100 TeV collider. Such a study is, therefore, timely and a kick-off to strengthen the physics case of LLPs at the FCC-hh.

\begin{figure}[t]
    \centering
    \includegraphics[width=\textwidth]{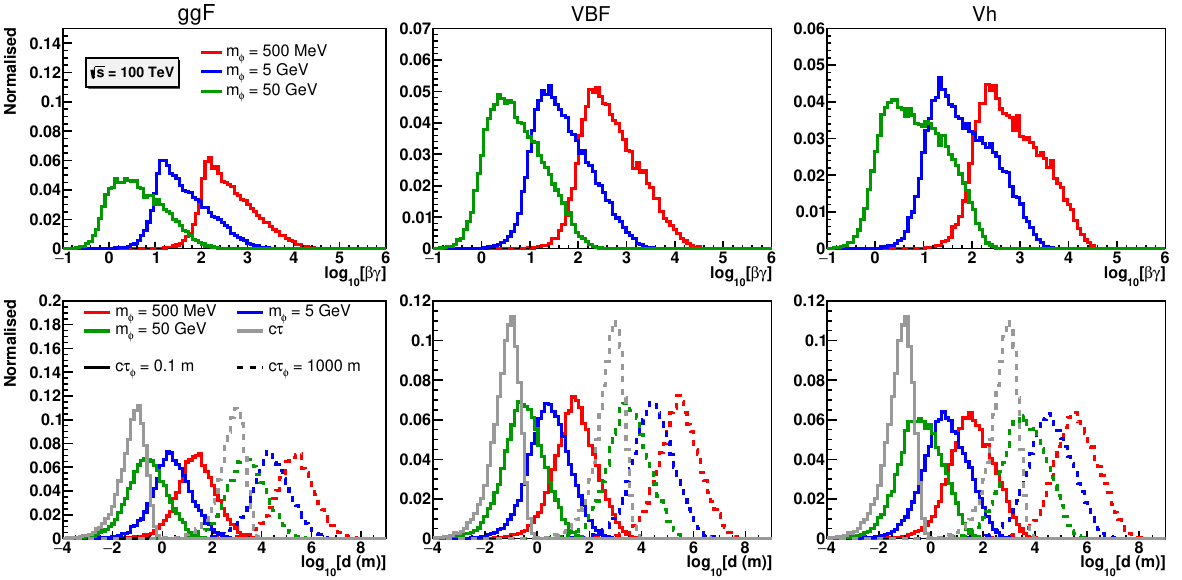}
    \caption{\small \sl Normalised histograms of the boost factor distribution (top panels) and the decay length distribution in lab frame (bottom panels) of the mediator particle from the Higgs boson decay, for $m_\phi =$ 100\,MeV, 5\,GeV and 50\,GeV, and $c \tau_\phi$ = 0.1\,m (solid lines) and 1000\,m (dashed lines), in the ggF (left panels), VBF (center panels) and Vh (right panels) Higgs boson production modes at $\sqrt{s} =$ 100\,TeV. The histogram of the exponential distribution, $\exp[-d/(c \tau_\phi)]$, for the two different decay lengths are shown in gray.}
    \label{fig:LLPs-dist-100TeV-boost}
\end{figure}

In this section we study the sensitivity of the future hadron collider with $\sqrt{s} =$ 100\,TeV. Due to the high centre-of-mass energy, the long-lived mediator particles will have a harder boost distribution, which also increases their decay lengths in the lab frame. Fig.\,\ref{fig:LLPs-dist-100TeV-boost} shows the $\beta\gamma$ and $d$ distributions of the mediator particle from the Higgs boson produced at the 100\,TeV collider in the ggF, VBF and Vh production modes. Comparing these to the corresponding distributions for $\sqrt{s} =$ 14\,TeV in Fig.\,\ref{fig:LLPs-dist-14TeV-boost}, we observe that the distributions for mediator particles produced in the 100\,TeV collider extends to larger $\beta\gamma$ and $d$ values due to higher centre of mass energy. We follow similar sets of analyses for the various decay modes of the mediator particle using the CMS MS detector at the HL-LHC experiment for the MS of a detector at the future 100\,TeV collider experiment. In addition, to increase the sensitivity to higher decay lengths, we propose some future LLP detectors, like MATHUSLA and CODEX-b, near the IP of the 100\,TeV collider detector.

\subsection{Extending the CMS MS analyses}

We begin by extending the previous CMS MS analyses at the HL-LHC experiment performed for the various final states to the 100\,TeV collider experiment. We choose some benchmarks for the size and magnetic field of the detector motivated from the \texttt{Delphes} card for the FCC-hh experiment and the detector geometry shown in Ref.\,\cite{FCC-hh}. We take the strength of the magnetic field to be 4\,T in the $z$-direction till the beginning of the MS, and after that the magnitude of the strength is reduced to 0.5\,T and the direction reversed. The FCC-hh detector is proposed to be designed as almost a combination of the CMS/ATLAS geometry for covering the transverse physics and the LHCb geometry for capturing the forward physics, because the latter will be an important part of the 100\,TeV collider experiment. We deal with the forward MS of FCC-hh separately, and take the dimension and pseudo-rapidity of the detector as follows:
\begin{align}
    \textbf{Tracker:} \qquad & R \leq 1.5\,{\rm m}, \quad |Z| \leq 5\,{\rm m}, \nonumber \\
    \textbf{Barrel + Endcap MS:} \qquad & 6\,{\rm m} \leq R \leq 9\,{\rm m}, \quad 9\,{\rm m} \leq |Z| \leq 12\,{\rm m}, \quad \eta \leq 2.5, \nonumber \\
    \textbf{Forward MS:} \qquad & 12\,{\rm m} \leq |Z| \leq 23\,{\rm m}, \quad 2.5 \leq \eta \leq 5.0.
\end{align}
We now revisit the displaced muons as well as displaced jets (for a pair of $b$ quarks) analyses in light of the FCC-hh MS detector at the 100\,TeV collider experiment.


\subsubsection{Mediator particle decaying into muons}

When the mediator particle decays into a pair of muons, we do not restrict the decay position within the MS of the FCC-hh detector, just like we did in case of the MS of the CMS detector. We use the same cuts as we used for the CMS analysis of displaced muons, as given in Table\,\ref{tab:muon_cuts}. The only change we make is in the cut on the position of the dSV to contain it within the FCC-hh detector, which now becomes $d_T <$ 9\,m and $|d_z| <$ 12\,m for the barrel and endcap MS and $|d_z| <$ 23\,m for the forward MS, with the corresponding appropriate pseudo-rapidity cuts of the 100\,TeV collider discussed in the beginning of this section. Since we are not yet sure exactly how many MS stations will be there inside the FCC-hh detector and their positions and performances, we do not restrict the decays within the second last MS layer.

\begin{figure}[t]
    \centering
    \includegraphics[width=0.65\textwidth]{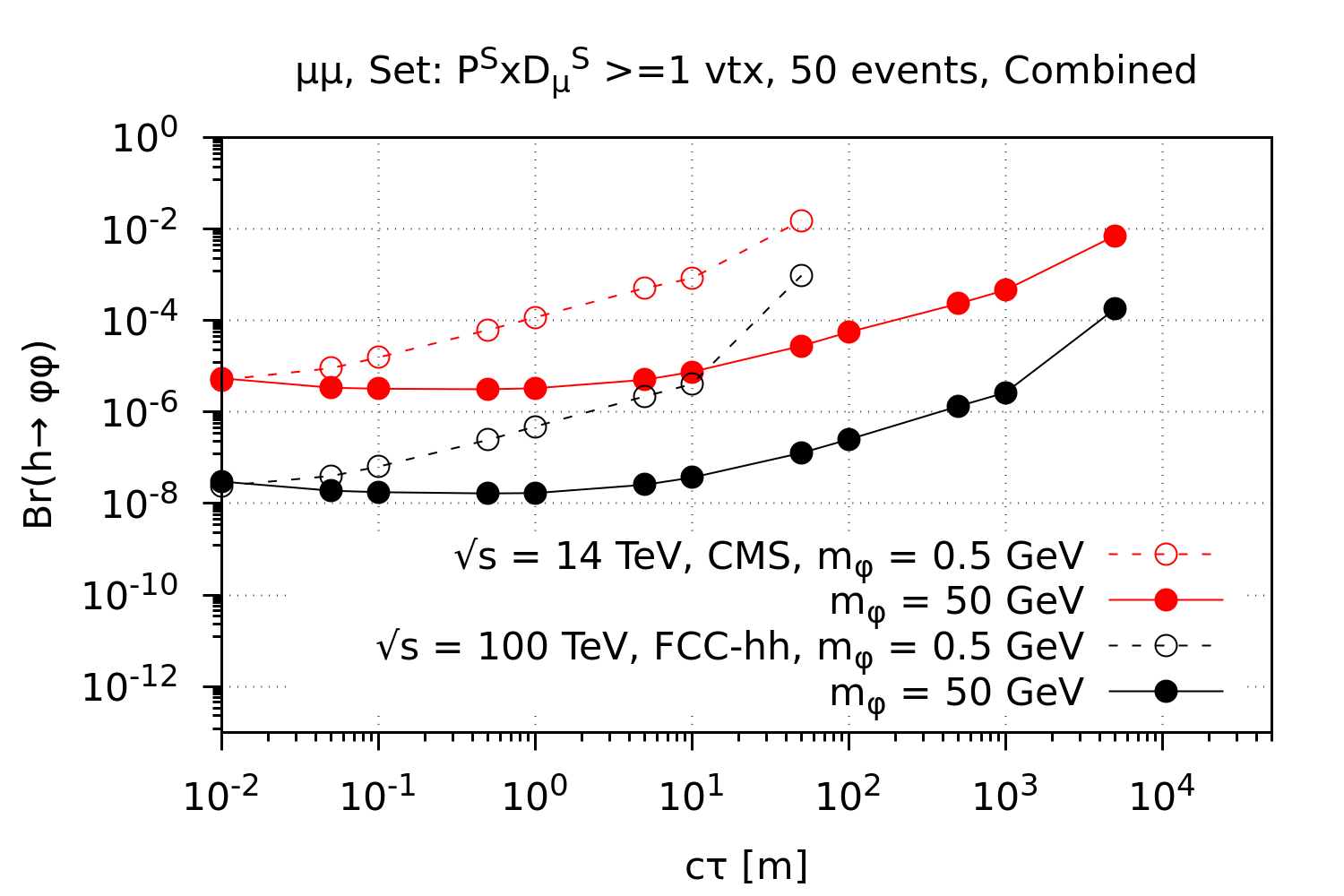}
    \caption{\small \sl Comparison of the projected upper limits on the branching fraction, Br$(h \to \phi \phi)$, for 50 observed decays of long-lived mediator particles into a pair of muons in the CMS and the FCC-hh detectors for the $P^S \times D^S_\mu \geq$ 1\,vtx set of cuts. The shown limits are obtained by combining the ggF, VBF and Vh modes for the Higgs boson production.}
    \label{fig:mumu-14vs100-combo}
\end{figure}

Fig.\,\ref{fig:mumu-14vs100-combo} compares the projected upper limits on the branching fraction of the Higgs boson into a pair of mediator particles obtained with the $P^S \times D^S_\mu \geq$ 1\,vtx set of cuts, between the 14\,TeV CMS MS and the 100\,TeV FCC-hh MS (barrel-endcap and forward MS combined) for two mediator masses of 0.5\,GeV and\,50 GeV. For both the benchmark masses, we have an improvement by a factor of around 200 in the FCC-hh detector compared to the CMS detector. Increase in the cross-section and the luminosity alone accounts for a factor of $\sim$ 150, and an increase in the efficiency by a factor of around 1.4 in going from the CMS to the FCC-hh detectors.

\begin{figure}[t]
    \centering
    \includegraphics[width=0.7\textwidth]{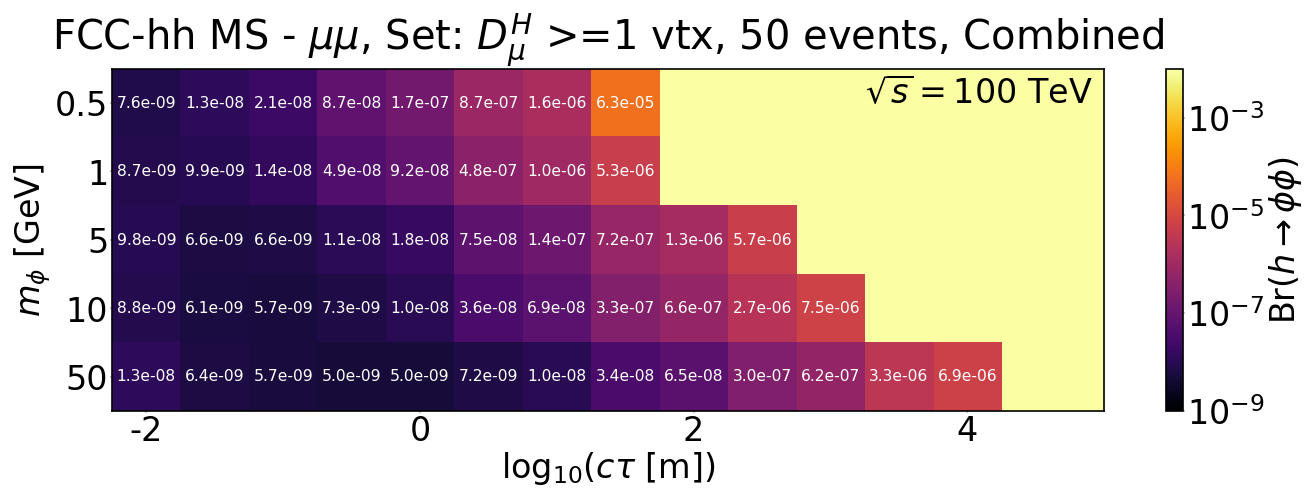} \\
    \includegraphics[width=0.7\textwidth]{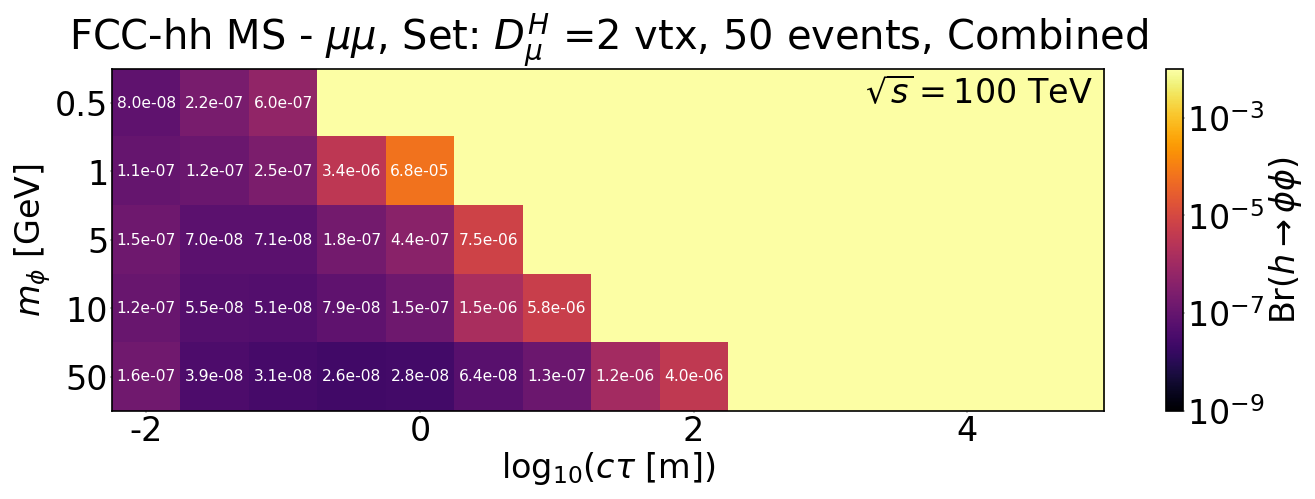} \\
    \includegraphics[width=0.7\textwidth]{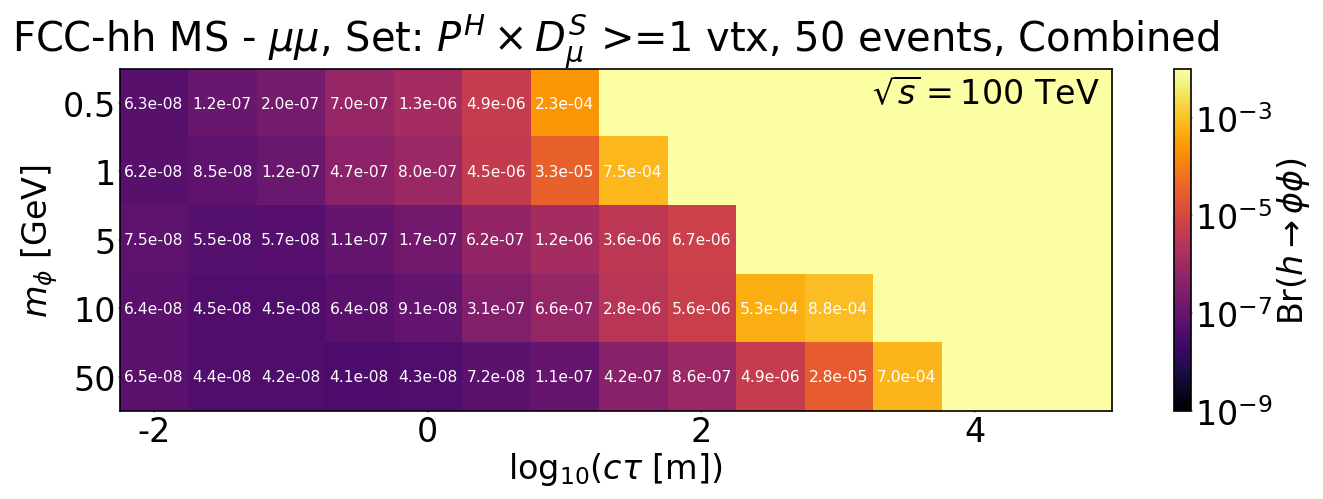} \\
    \includegraphics[width=0.7\textwidth]{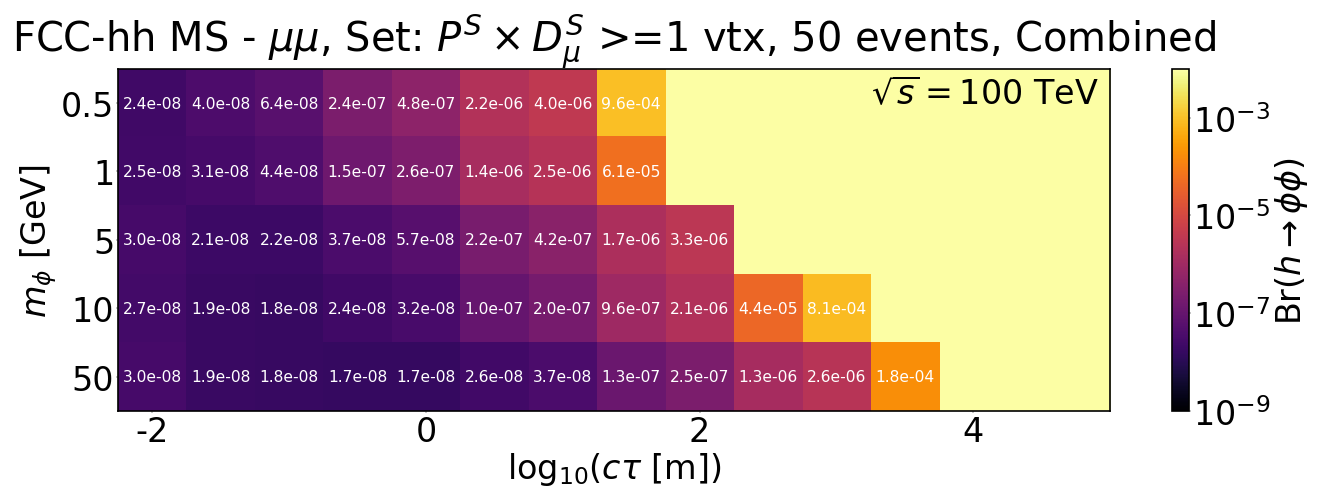}
    \caption{\small \sl Projected upper limits on the branching fraction, Br$(h \to \phi \phi)$, for 50 observed decays of long-lived mediator particles into muons within the FCC-hh MS at the 100\,TeV collider experiment for the four sets of cuts explained in the text. The shown limits are obtained by combining the ggF, VBF and Vh production modes of the Higgs boson.}
    \label{fig:mumu-100TeV-combo}
\end{figure}

\begin{figure}[t]
    \centering
    \includegraphics[width=0.46\textwidth]{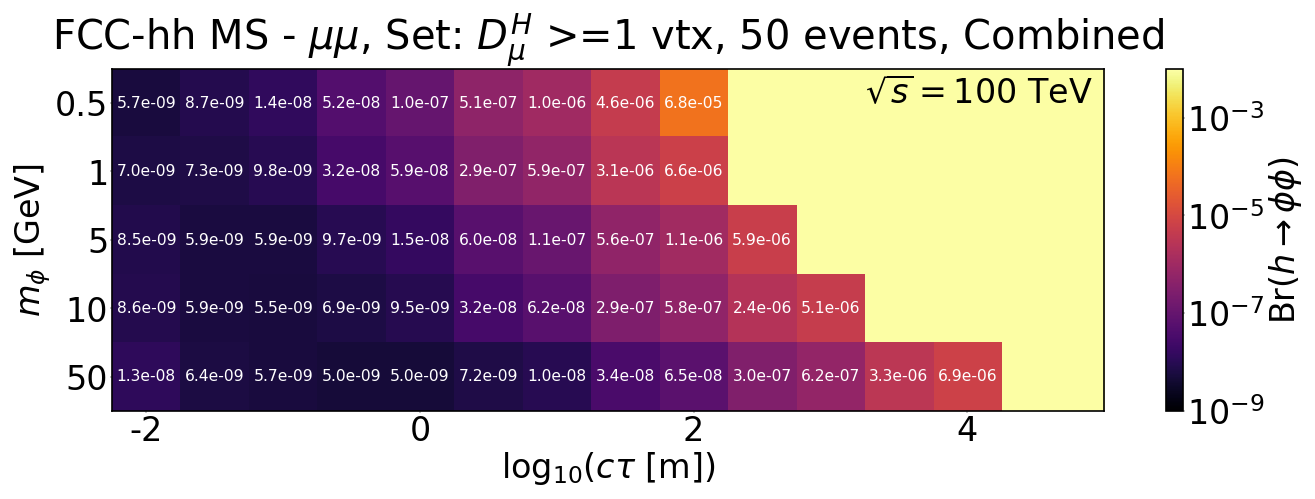} \qquad
    \includegraphics[width=0.46\textwidth]{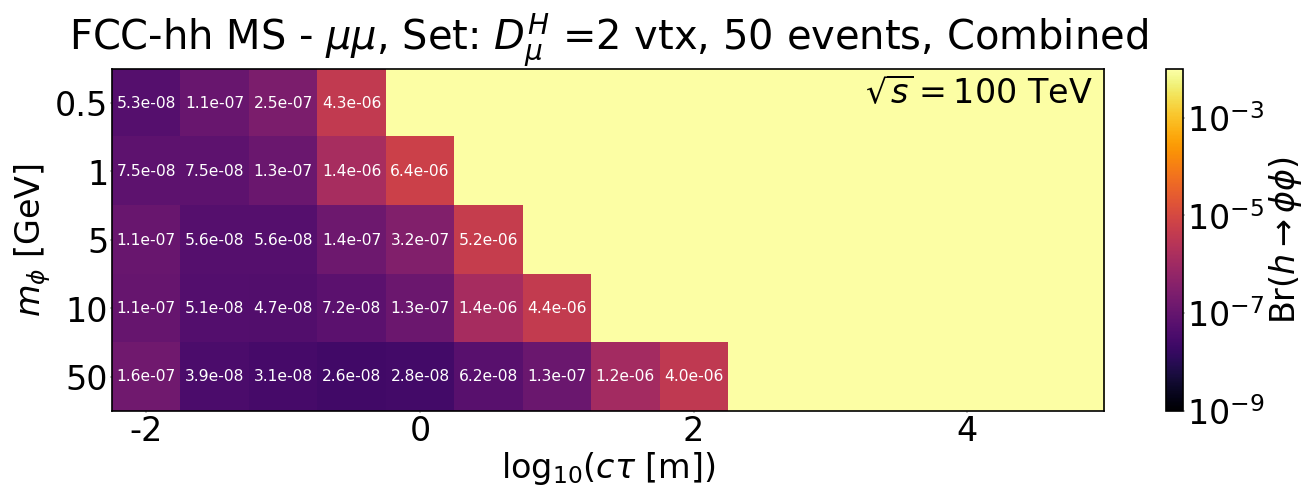} \\
    \includegraphics[width=0.46\textwidth]{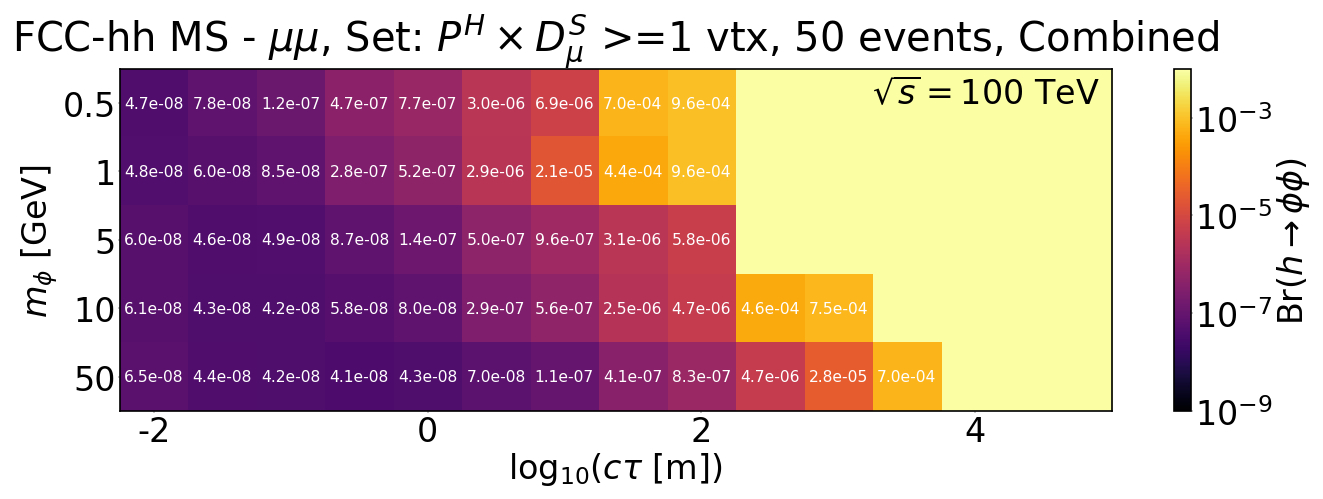} \qquad
    \includegraphics[width=0.46\textwidth]{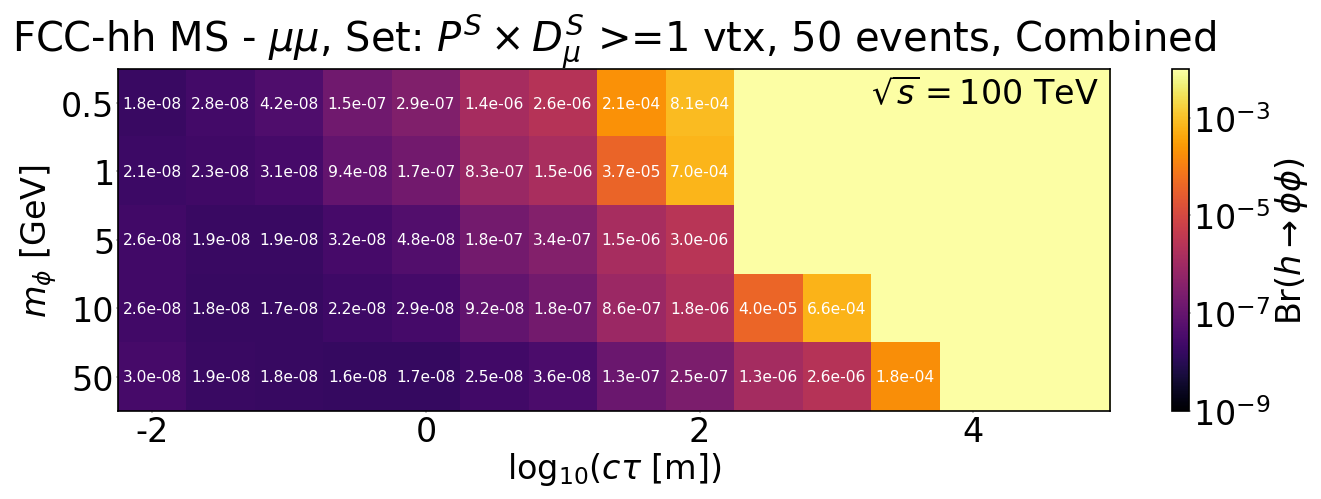}
    \caption{\small \sl Projected upper limits on the branching fraction of the Higgs boson decaying into a pair of the mediator particles, Br($h \to \phi\phi$), for 50 observed decays of the mediator particles to muons 
    The shown limits are obtained by combining the ggF, VBF and Vh productions for the Higgs boson.}
    \label{fig:mumu-100TeV-sv-combo}
\end{figure}

Fig.\,\ref{fig:mumu-100TeV-combo} shows the projected upper limits on the branching fraction, Br$(h \to \phi \phi)$, for all the four combinations of cuts described earlier ($D^H_\mu \geq$ 1\,vtx, $D^H_\mu =$ 2\,vtx, $P^H \times D^S_\mu \geq$ 1\,vtx, and $P^S \times D^S_\mu \geq$ 1\,vtx), for a range of masses and decay lengths of the mediator particle at the 100\,TeV collider, by the combination of barrel-endcap and forward MSs of the FCC-hh detector at 30\,ab$^{-1}$. The shown limits are obtained by combining the ggF, VBF and Vh modes for the Higgs boson production. With the displaced di-muons analysis of the CMS MS extended to FCC-hh, and utilising both the barrel-endcap and forward MS detectors, the best sensitivity that can be achieved using the 
$D_\mu^H \geq 1$\,vtx set of cuts is $7.6 \times 10^{-9}$ for 0.5\,GeV at $c \tau_\phi =$ 1\,cm and $5.0 \times 10^{-9}$ for 50\,GeV at $c \tau_\phi =$ 0.5\,m.
Comparing to CMS results, the decay length corresponding to the most sensitive limit remain the same at FCC-hh.
In Fig.\,\ref{fig:mumu-100TeV-sv-combo}, we show the projected limits with the isolation cut applied on $\Delta\phi_{\mu\mu,{\rm mom}}$ instead of the physical separation at the detector edge and we find similar result as for the 14\,TeV analysis.

\subsubsection{Mediator particle decaying into $b$ and $c$ quarks}

As for the CMS MS analysis, here we need to consider only those decays which occur inside the FCC-hh MS, i.e. the region defined by $d_T >$ 6\,m or $|d_z| >$ 9\,m, and $d_T <$ 9\,m and $|d_z| <$ 12\,m in the barrel and endcap MS, as well as 12\,m $< |d_z| <$ 23\,m in the forward MS. We then apply the same set of cuts as we used in the CMS MS analysis for displaced jets ($D_{jets}$), along with the same cuts for the prompt associated objects. We change the pseudo-rapidity coverage of the FCC-hh MS for both the barrel-endcap and forward regions as described in the beginning of this section.

\begin{figure}[t]
    \centering
    \includegraphics[width=0.45\textwidth]{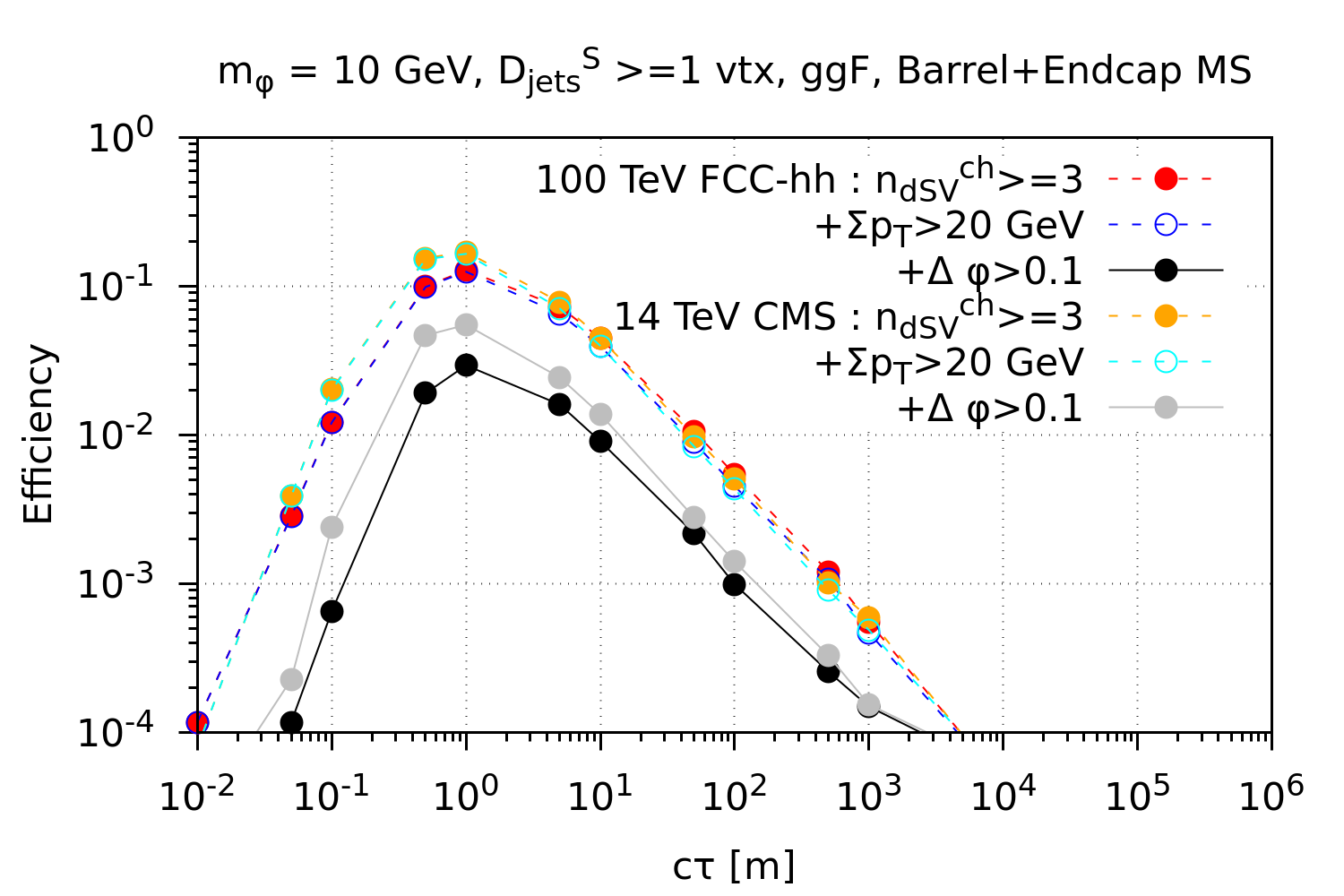} \qquad
    \includegraphics[width=0.47\textwidth]{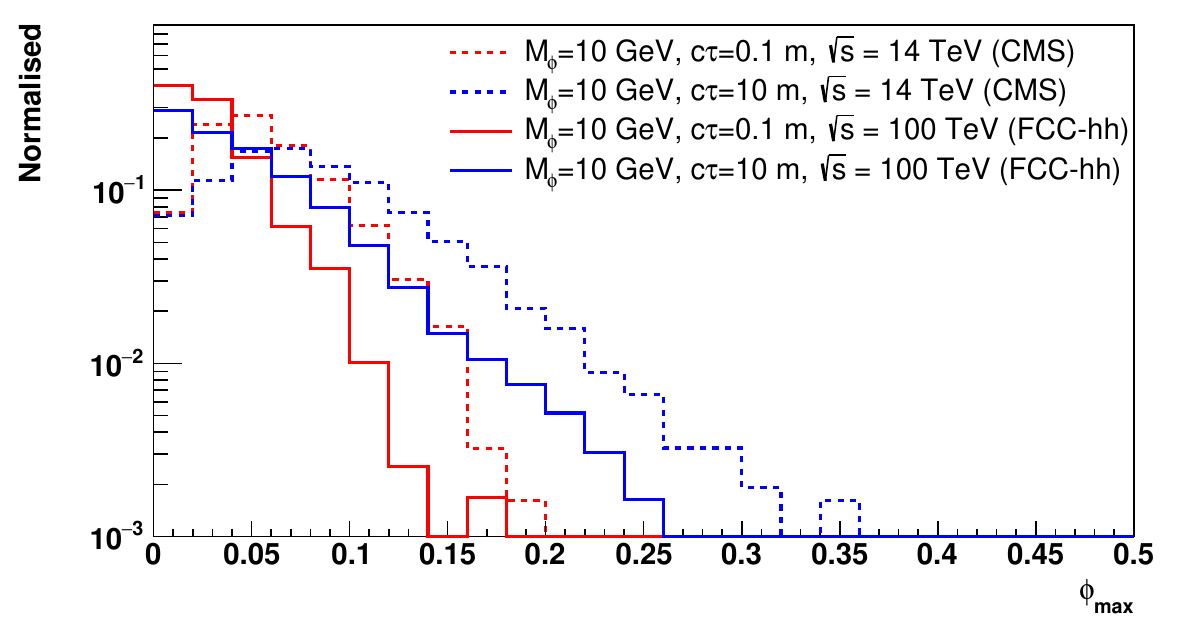}
    \caption{\small \sl {\bf Left:} Efficiency of various cuts with varying decay lengths for 10\,GeV mediator particle in the FCC-hh barrel and endcap MS along with the corresponding efficiencies obtained in the CMS MS. {\bf Right:} Comparison of the normalised distributions of $\Delta\phi_{\rm max}$ for two decay lengths for 10\,GeV mediator particle in the 14\,TeV and 100\,TeV colliders.}
    \label{fig:delphi-bb-100}
\end{figure}

To understand how the same set of cuts affect the signal at the 100\,TeV experiment, we revisit the efficiency plot for the 10\,GeV mediator particle in the ggF production mode. The left panel of Fig.\,\ref{fig:delphi-bb-100} shows the efficiencies of the $n_{\rm dSV}^{\rm ch}$, $\sum p_{T,\,dSV}$ and $\Delta\phi_{\rm max}$ cuts on the signal. We find on comparing these with the corresponding efficiencies at the 14\,TeV with the one we get at the 100\,TeV colliders, the latter has reduced by a factor of about half. The decomposition of the efficiency in individual components of the cuts show that this reduction in the efficiency is mostly due to the $\Delta\phi_{\rm max}$ cut. The left panel of the figure shows the $\Delta\phi_{\rm max}$ distribution for two benchmark decay lengths of 0.1\,m and 10\,m for the mediator particle of 10\,GeV mass at the 14 TeV\,and 100\,TeV collider experiments. 
We find that the 100\,TeV distributions fall off faster than the ones at the 14\,TeV, which explains the reduced efficiency.
For the 100\,TeV FCC-hh analysis, we use $\Delta\phi_{\rm max}>0.1$ even for the hard set of cuts.

\begin{figure}[t]
    \centering
    \includegraphics[width=0.45\textwidth]{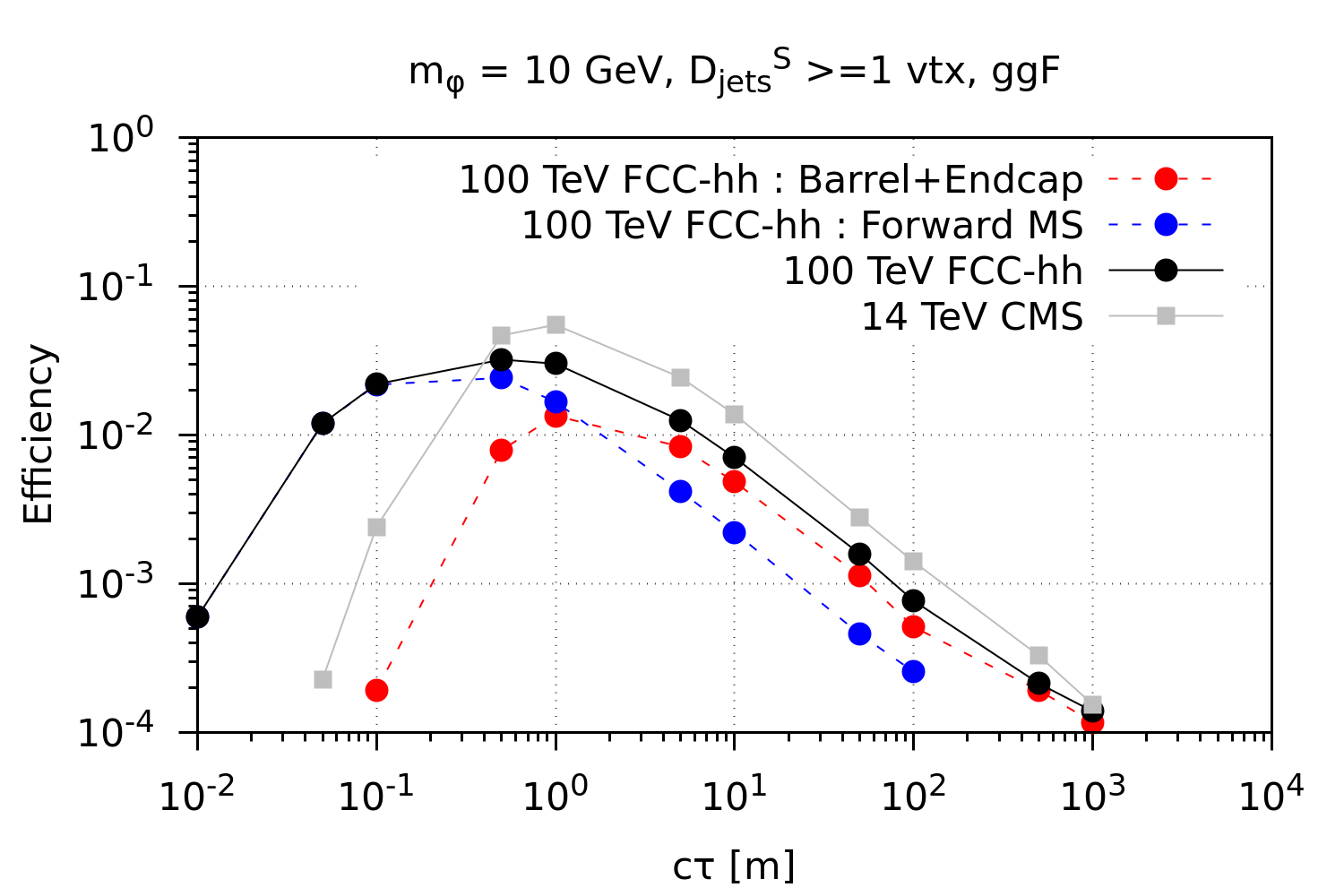} \qquad
    \includegraphics[width=0.45\textwidth]{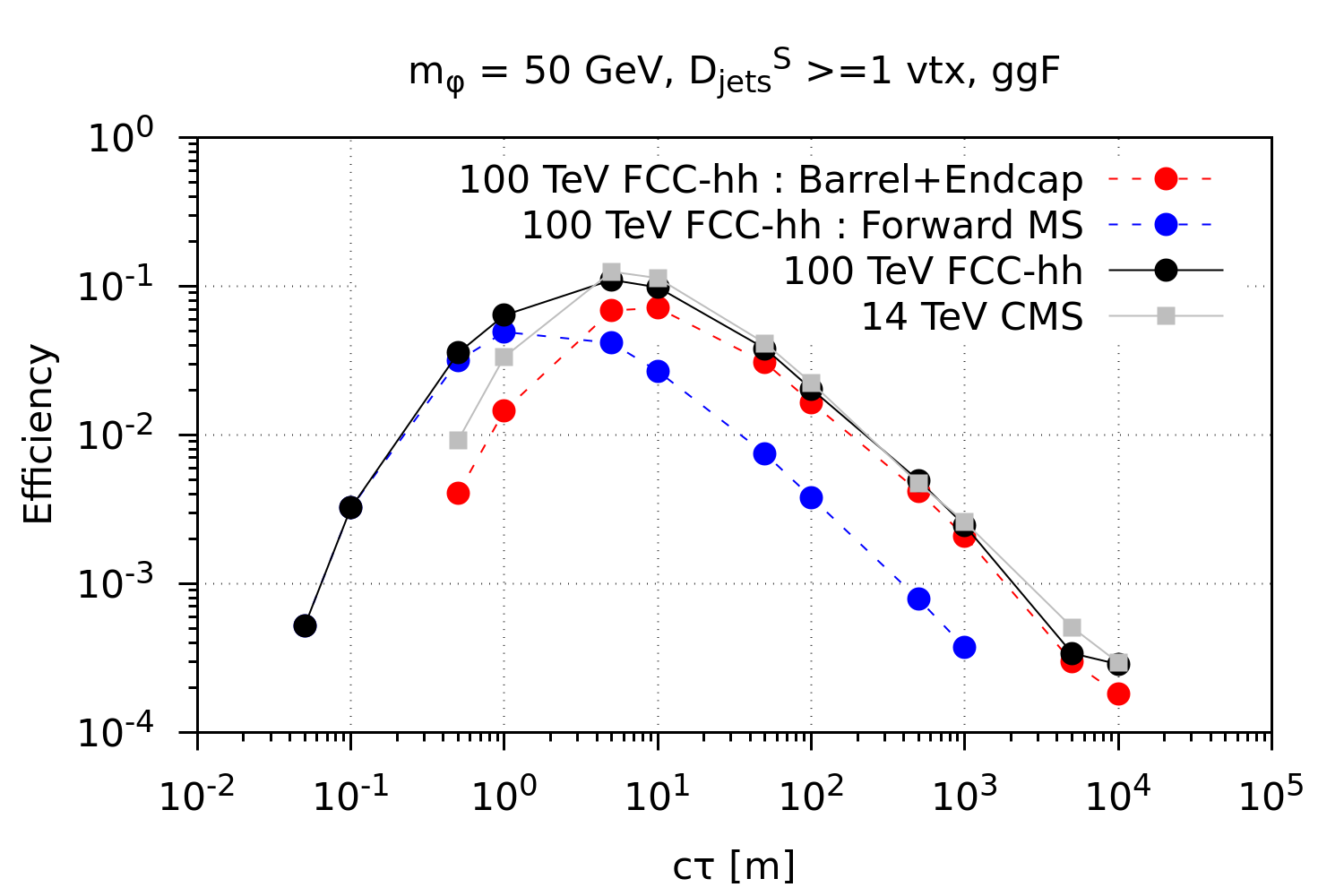} \\
    \includegraphics[width=0.45\textwidth]{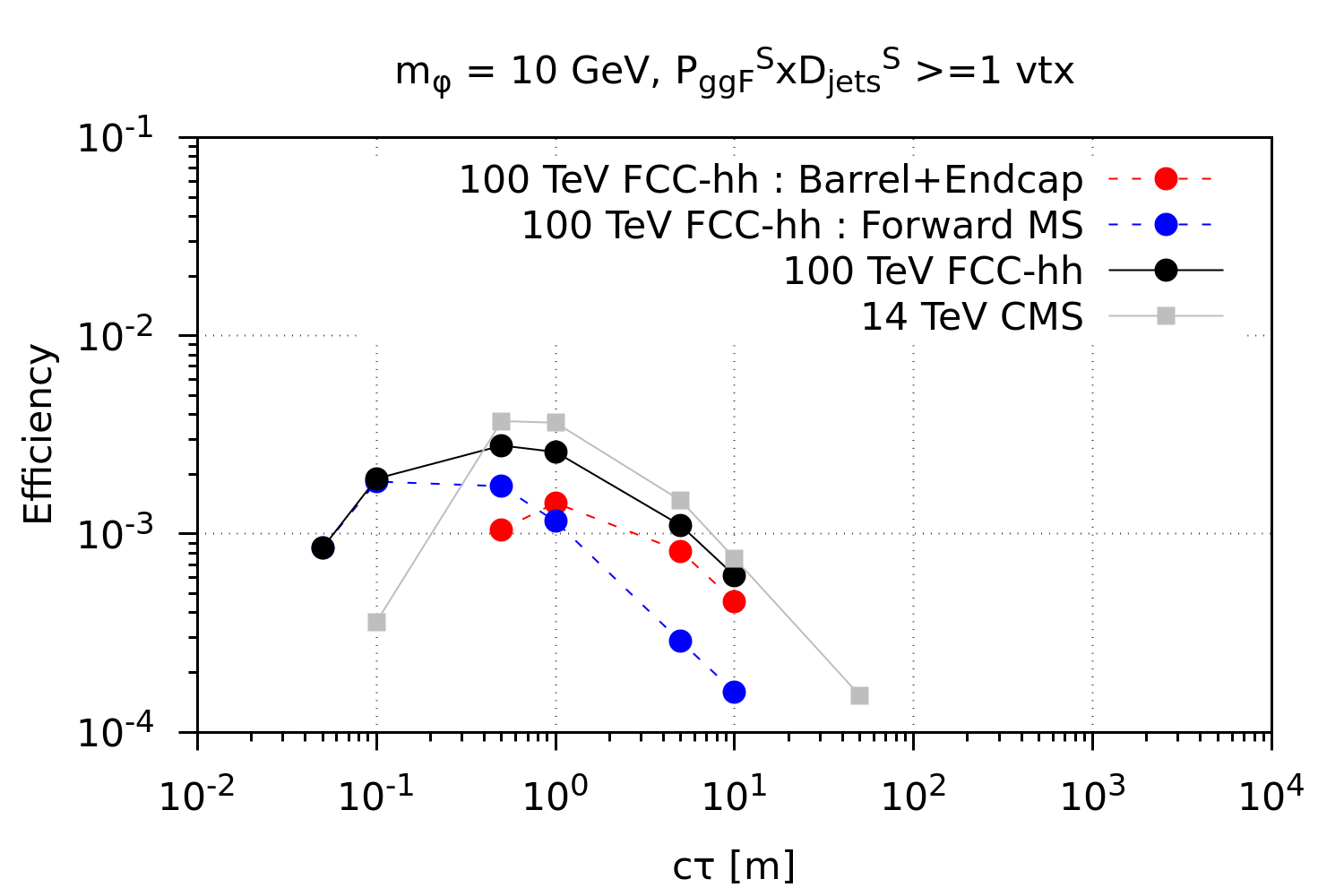} \qquad
    \includegraphics[width=0.45\textwidth]{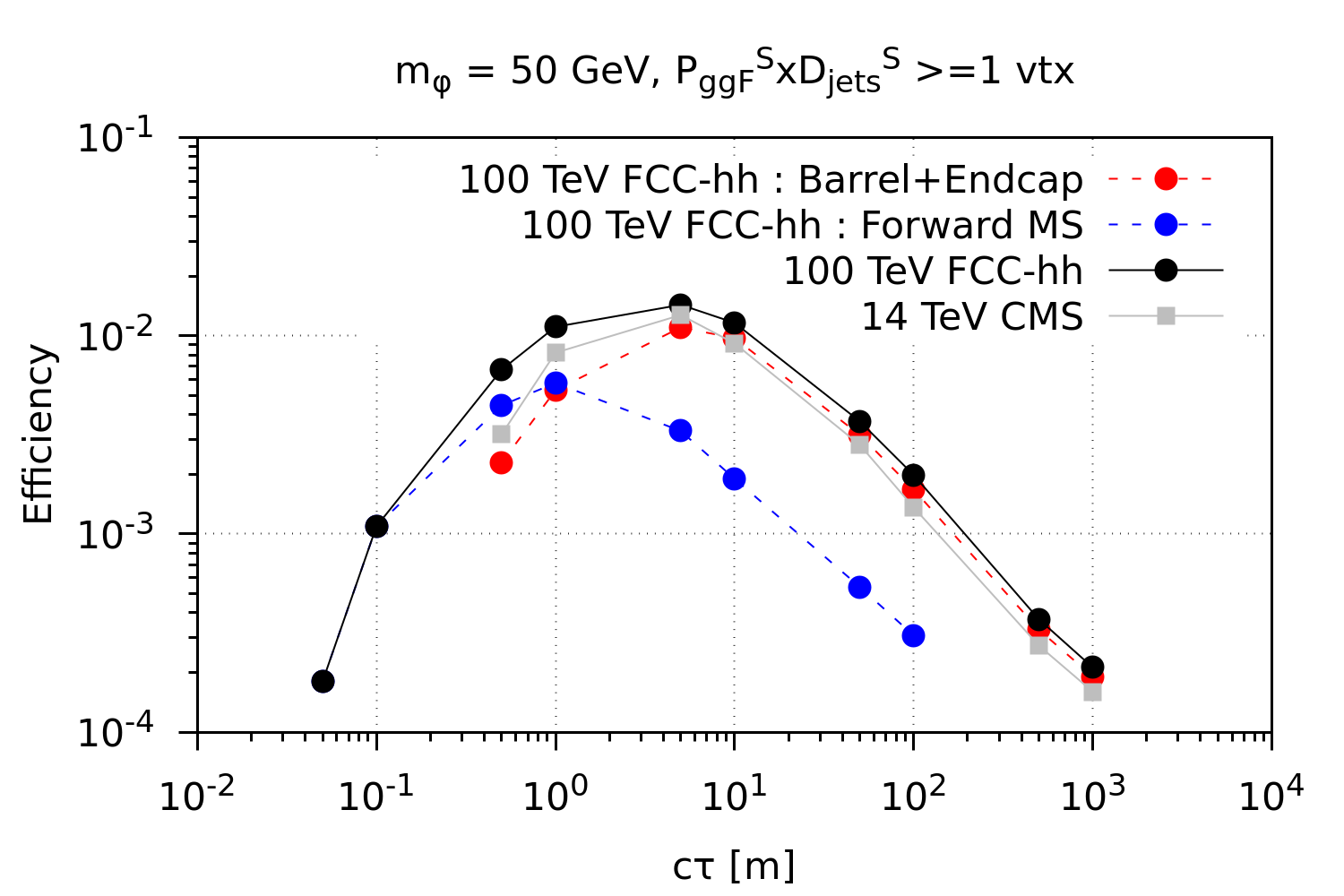}
    \caption{\small \sl Comparison of the efficiencies as a function of decay lengths, as obtained from the $D_{jets}^S \geq 1$\,vtx (top) and $P_{\rm ggF}^S \times D_{jets}^S \geq 1$\,vtx (bottom) sets of cuts between the CMS MS at 14 TeV and FCC-hh MS at 100 TeV, when a mediator of mass 10\,GeV (left) and 50\,GeV (right) decays to a pair of $b$ quarks. For the FCC-hh detector, we also show the efficiencies for the barrel and endcap MS, and the forward MS separately.}
    \label{fig:eff-bb-14vs100}
\end{figure}

Fig.\,\ref{fig:eff-bb-14vs100} shows the effect of including the forward MS in our analysis observing displaced MS clusters for two sets of cuts, i.e. $D_{jets}^S \geq 1$\,vtx and $P_{\rm ggF}^S \times D_{jets}^S \geq 1$\,vtx, for the Higgs boson produced in ggF mode. We find that for both the sets of cuts, the forward MS increases sensitivity to lower decay lengths, which is explained by revisiting Fig.\,\ref{fig:pt-eta-LLP}, where we found that the dSV cut selects more forward mediator particles with lower decay lengths, which can now pass the $\eta$ condition with the forward MS. The combined efficiencies of barrel-endcap and forward MS is almost similar to the 14\,TeV efficiencies for the 50\,GeV mediator, however, it is reduced for the 10\,GeV mediator, predominantly due to the $\Delta\phi_{\rm max}$ cut, as discussed earlier.

\begin{figure}[t]
    \centering
    \includegraphics[width=0.65\textwidth]{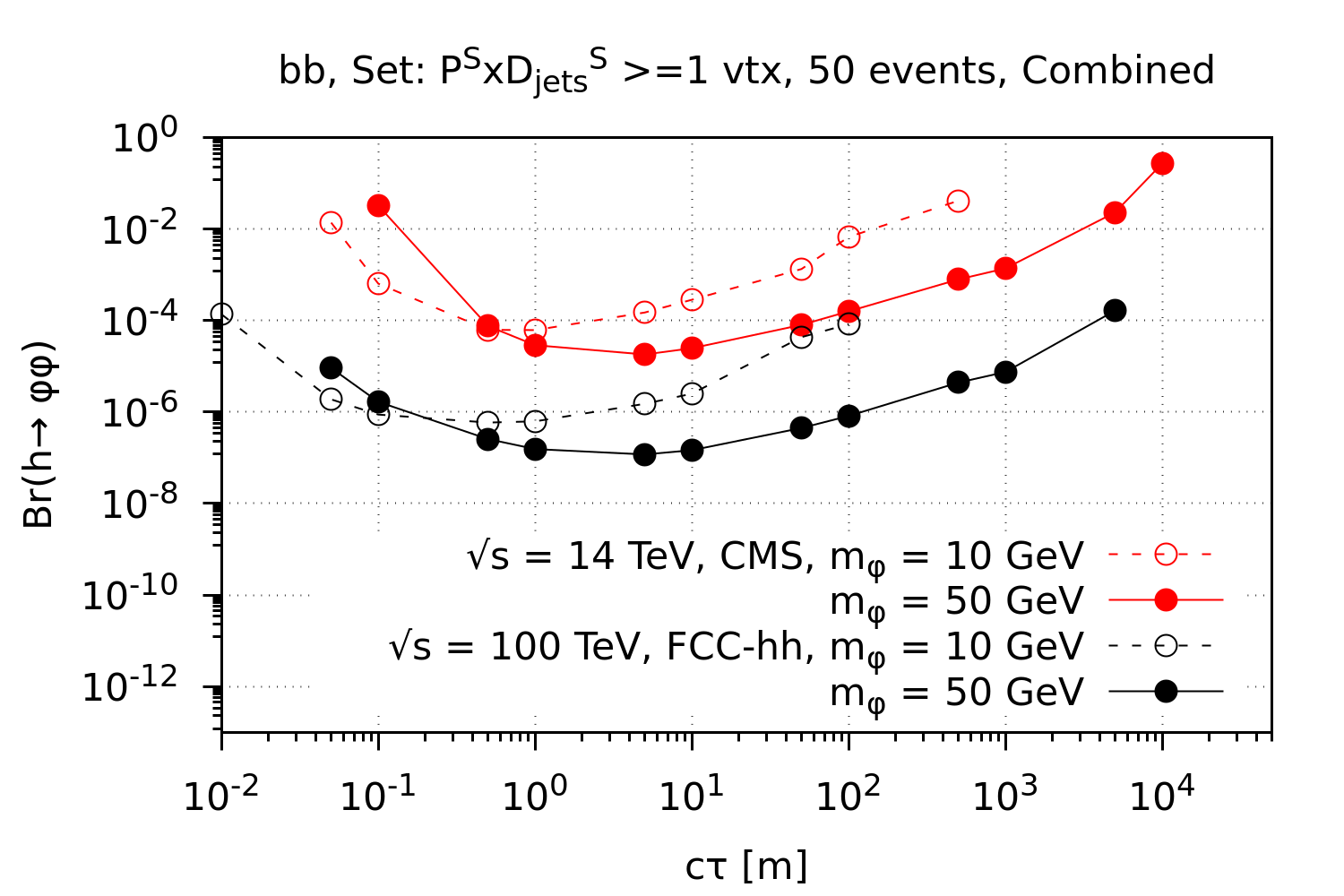}
    \caption{\small \sl Comparison of the projected upper limits on the branching fraction, Br$(h \to \phi \phi)$, for 50 observed decays of long-lived mediator particles into a pair of $b$ quarks within the CMS and the FCC-hh MS for the $P^S \times D^S_{jets} \geq$ 1\,vtx set of cuts. The shown limits are obtained by combining the ggF, VBF and Vh modes for the Higgs boson production.}
    \label{fig:bb-14vs100-combo}
\end{figure}

Nevertheless, the 100\,TeV collider experiment gains from increased cross section as well as luminosity. The cross section for the ggF production is increased from around 50\,pb to 740\,pb (enhancement of a factor of $\sim 15$), and it is expected to collect data with an integrated luminosity of 30\,ab$^{-1}$ (a factor of 10 increase compared to the HL-LHC collider experiment). We, therefore, expect the projected sensitivity to increase by a factor of around 150, however, the reduction in efficiency costs us a factor of half and $0.71$ for the $D^S \geq 1$\,vtx and $P^S \times D^S \geq 1$\,vtx sets of cuts respectively for the 10\,GeV mediator. We are therefore able to improve the 14\,TeV CMS results by a factor of $\sim 106$ with the $P^S\times D^S\geq1$\,vtx cuts for mediator mass of 10\,GeV. For the 50\,GeV mediator, there is not much reduction in the efficiencies compared to the 14\,TeV results, as can also be seen from Fig.\,\ref{fig:eff-bb-14vs100}, since it has lesser boost and effect of $\Delta\phi_{\rm max}$ cut is small, i.e. for the $D^S \geq 1$\,vtx cuts, we find a reduction by a factor of 0.88 and for the $P^S\times D^S \geq 1$\,vtx cuts, we have a slight improvement by a factor of 1.12. Fig.\,\ref{fig:bb-14vs100-combo} compares the 14\,TeV HL-LHC CMS limits with the 100\,TeV FCC-hh limits on the branching fraction, Br$(h \to \phi \phi)$, for mediator particle masses of 10\,GeV and 50\,GeV. This shows the results when all the production modes are combined, and in each of them, the $P^S \times D^S_{jets} \geq$ 1\,vtx set of cuts are used. We find the results to improve by a similar factor which we have just estimated above.

\begin{figure}[t]
    \centering
    \includegraphics[width=0.46\textwidth]{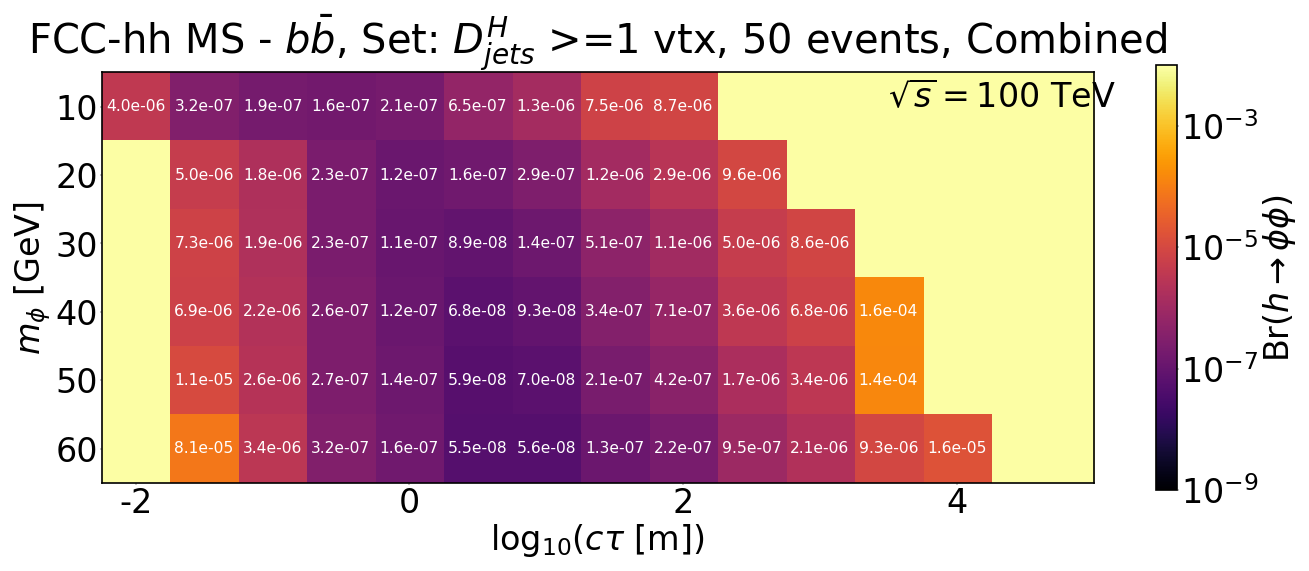} \qquad
    \includegraphics[width=0.46\textwidth]{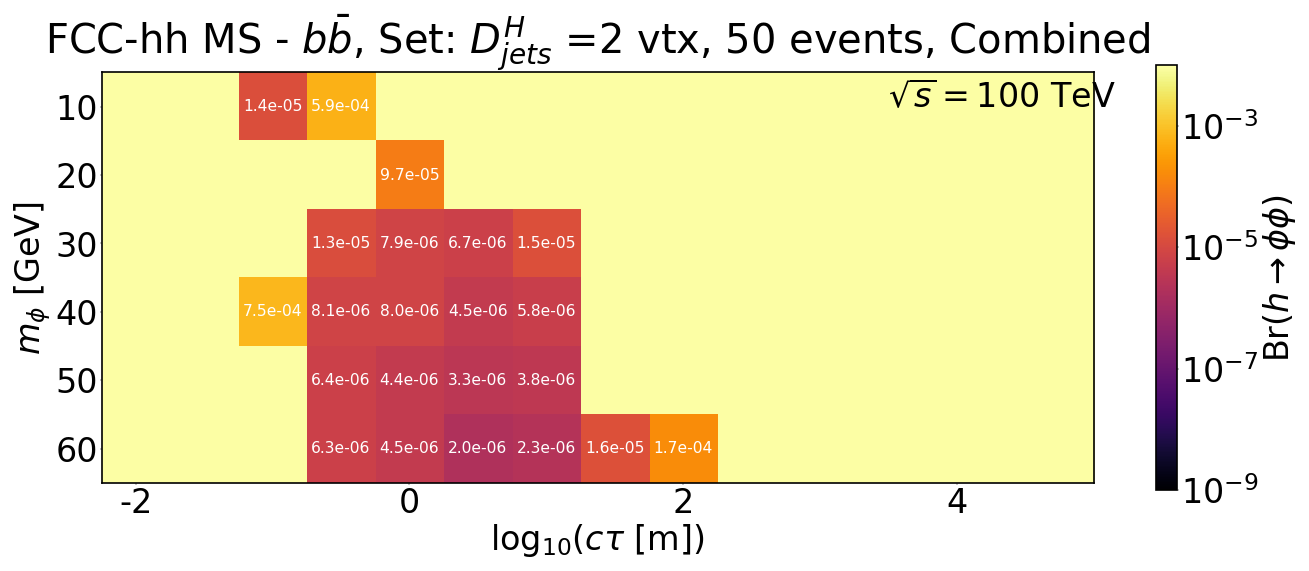} \\
    \includegraphics[width=0.46\textwidth]{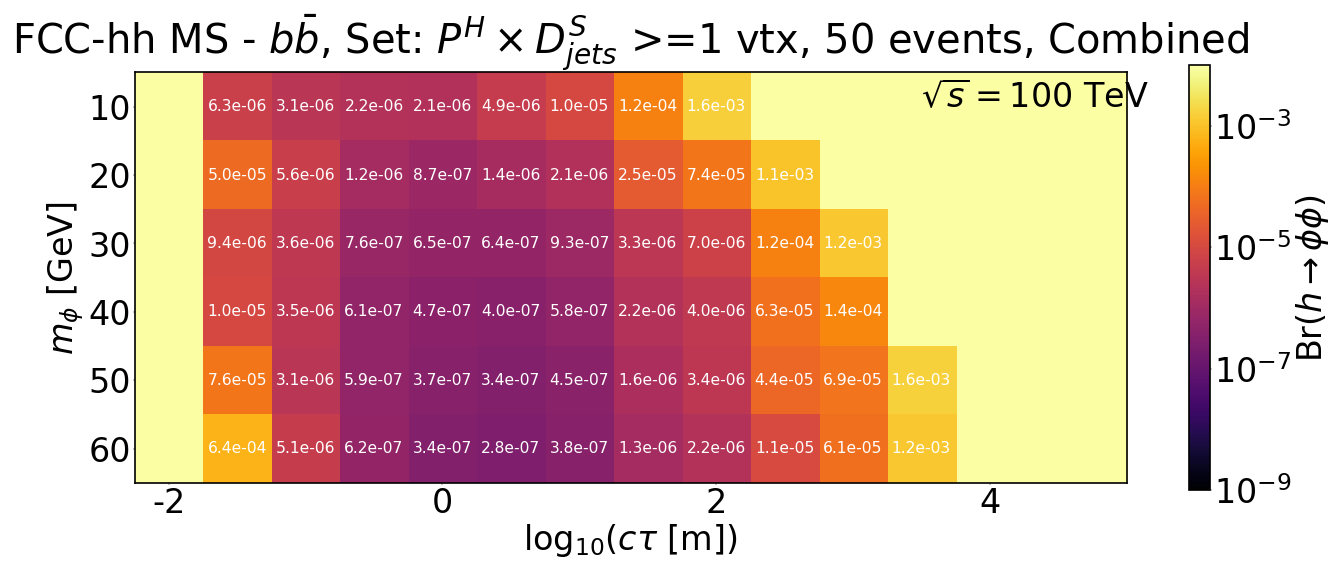} \qquad
    \includegraphics[width=0.46\textwidth]{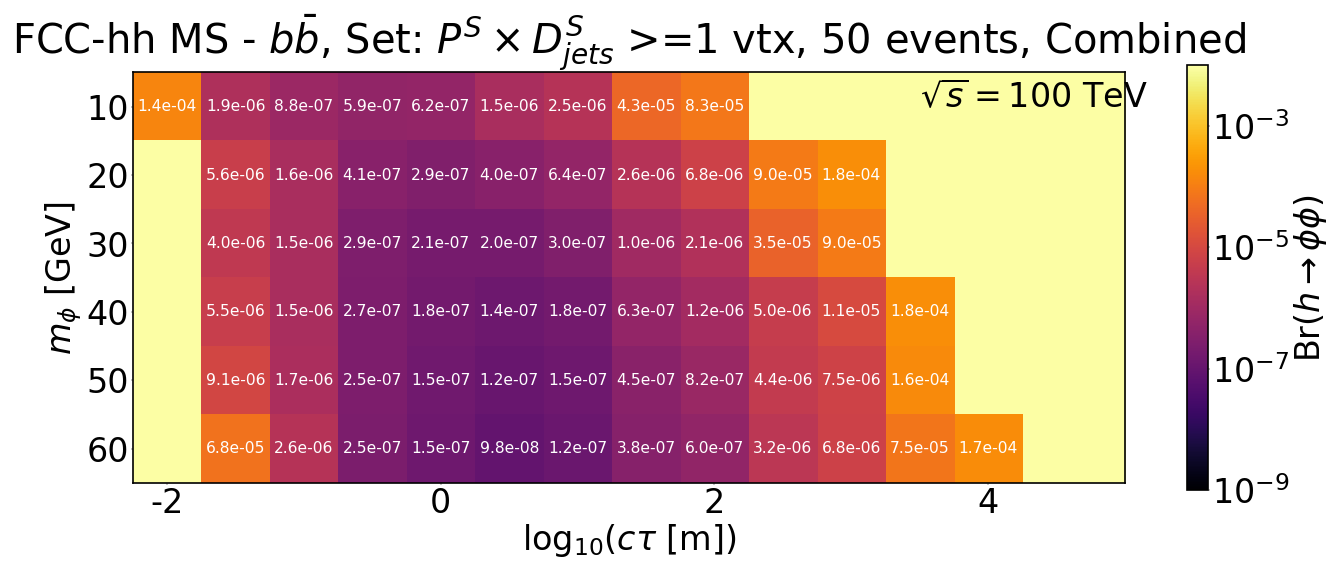}
    \caption{\small \sl Projected upper limits on the branching fraction, Br$(h \to \phi \phi)$, for 50 observed decays of long-lived mediator particles into $b$ quarks within the FCC-hh MS at the 100\,TeV collider experiment for the four sets of cuts explained in the text. The shown limits are obtained by combining the ggF, VBF and Vh production modes of the Higgs boson.}
    \label{fig:bb-100TeV-combo}
\end{figure}

\begin{figure}[t]
    \centering
    \includegraphics[width=0.46\textwidth]{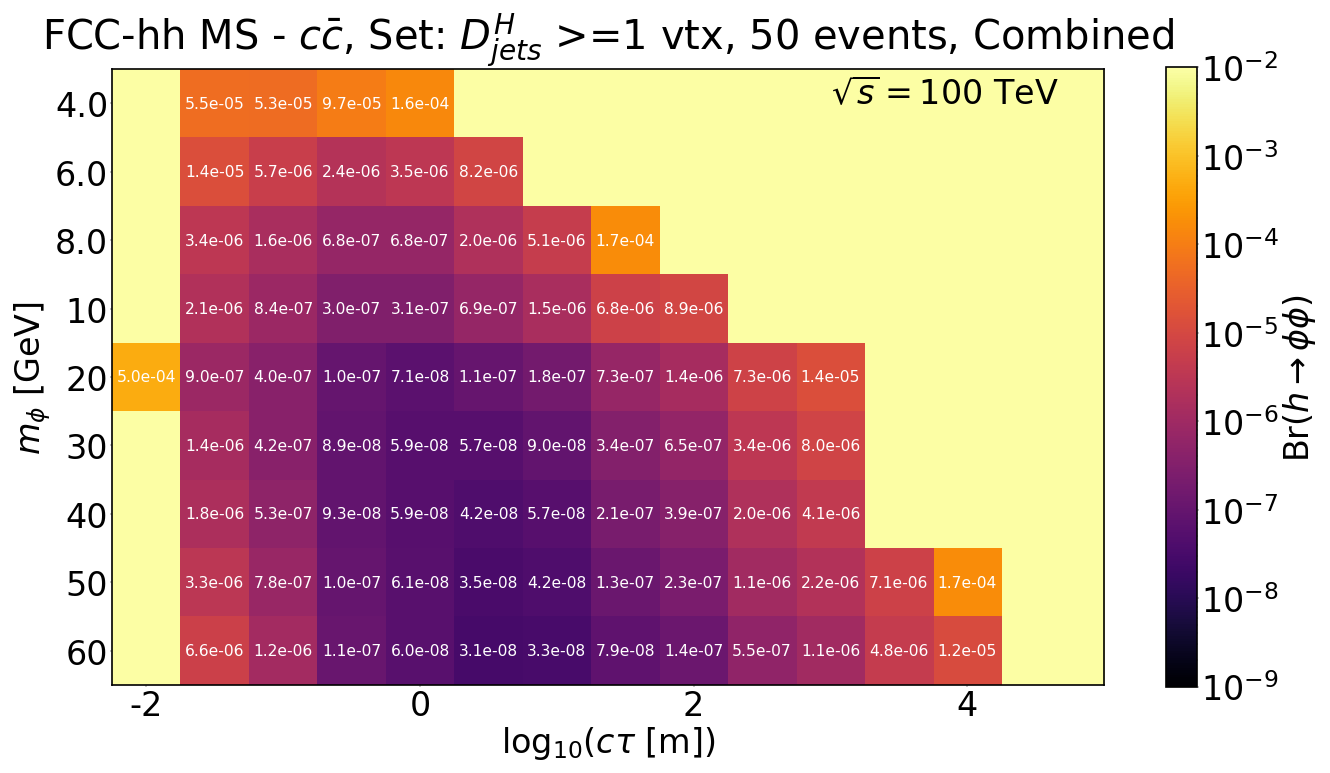} \qquad
    \includegraphics[width=0.46\textwidth]{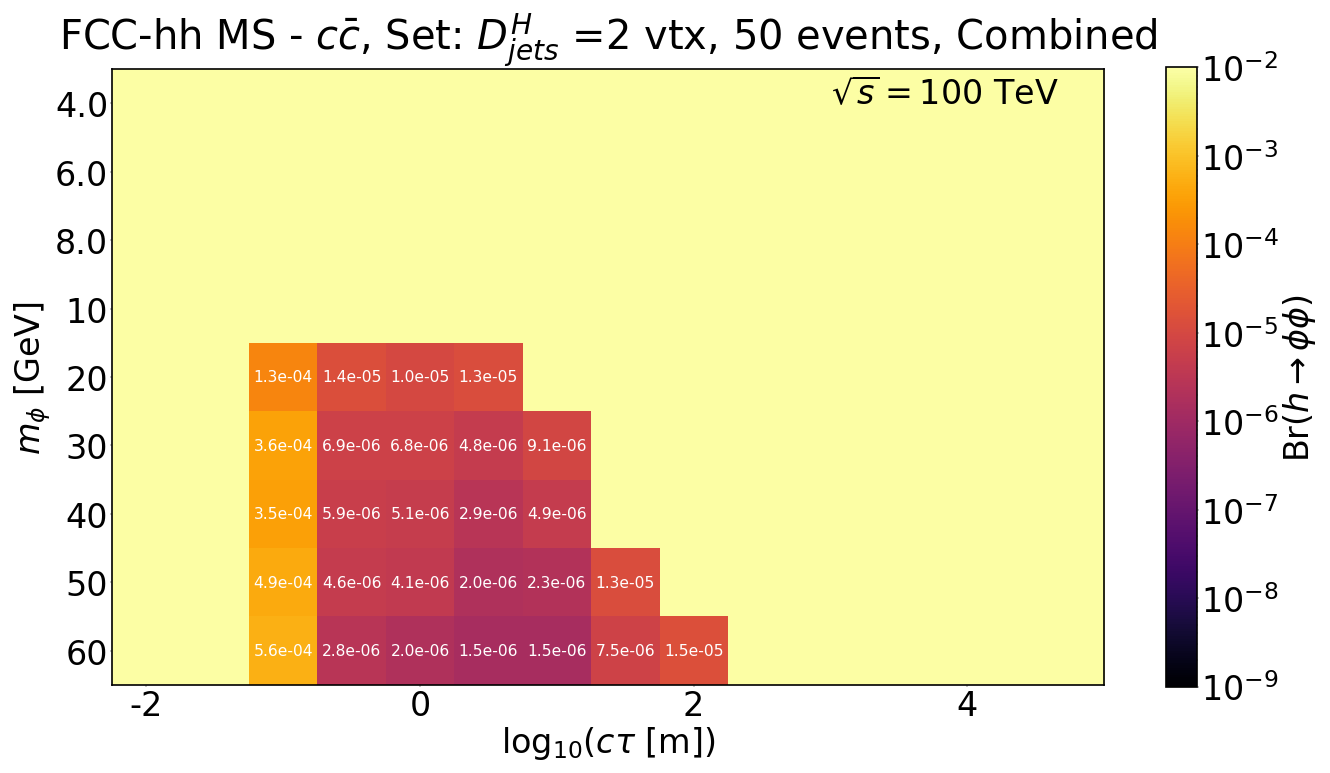} \\
    \includegraphics[width=0.46\textwidth]{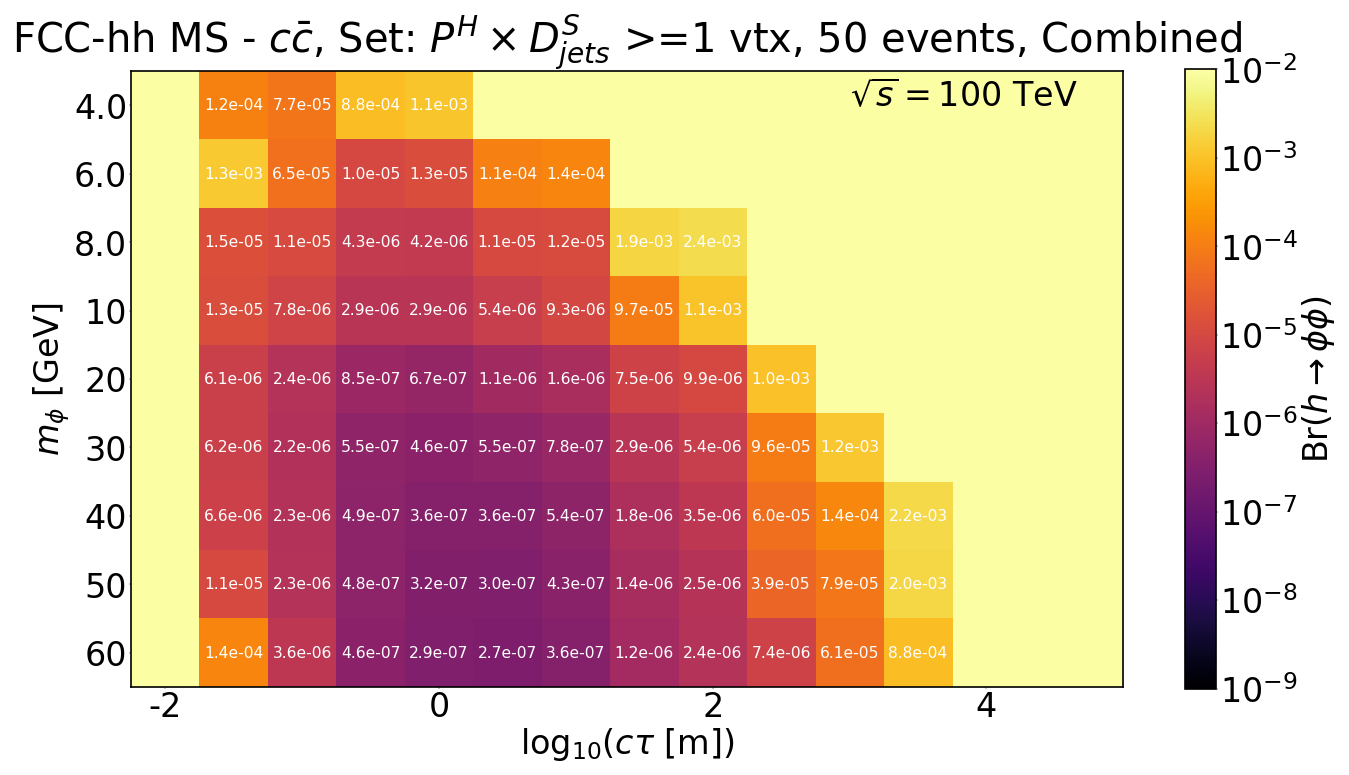} \qquad
    \includegraphics[width=0.46\textwidth]{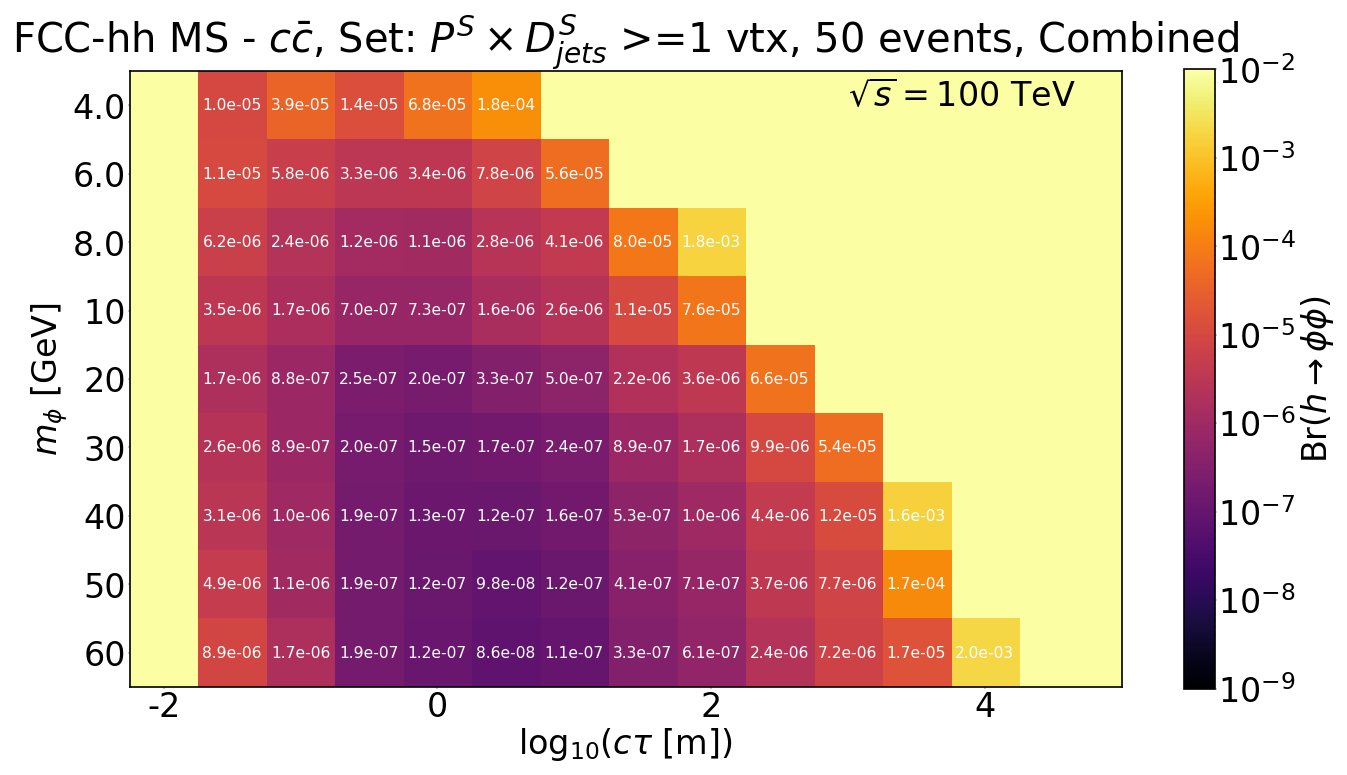}
    \caption{\small \sl Projected upper limits on the branching fraction, Br$(h \to \phi \phi)$, for 50 observed decays of long-lived mediator particles into $c$ quarks within the FCC-hh MS at the 100\,TeV collider experiment for the four sets of cuts explained in the text. The shown limits are obtained by combining the ggF, VBF and Vh production modes of the Higgs boson.}
    \label{fig:cc-100TeV-combo}
\end{figure}

Figs.\,\ref{fig:bb-100TeV-combo} and \ref{fig:cc-100TeV-combo} show the projected upper limits on the branching fraction of the Higgs boson into a pair of the long-lived mediator particles, Br$(h \to \phi \phi)$, which further decays into a pair of $b$ and $c$ quarks respectively, for all the four combinations of cuts as described earlier, for a range of mass and decay length values, using the combination of barrel-endcap and forward MS regions. 
For the decay into $b$-quarks ($c$-quarks), we achieve a limit of Br$(h \to \phi \phi) < 1.6\times 10^{-7}$ ($3.0 \times 10^{-7}$) for 10\,GeV mediator particle at $c\tau_\phi = 0.5$\,m and Br$(h \to \phi \phi) < 5.9\times 10^{-8}$ ($3.5\times10^{-8}$) for 50\,GeV mediator particle at $c \tau_\phi = 5$\,m, when we use the $D^H \geq 1$\,vtx set of cuts, followed by limits from $P^S \times D^S \geq 1$\,vtx.

\begin{figure}[t]
    \centering
    \includegraphics[width=0.6\textwidth]{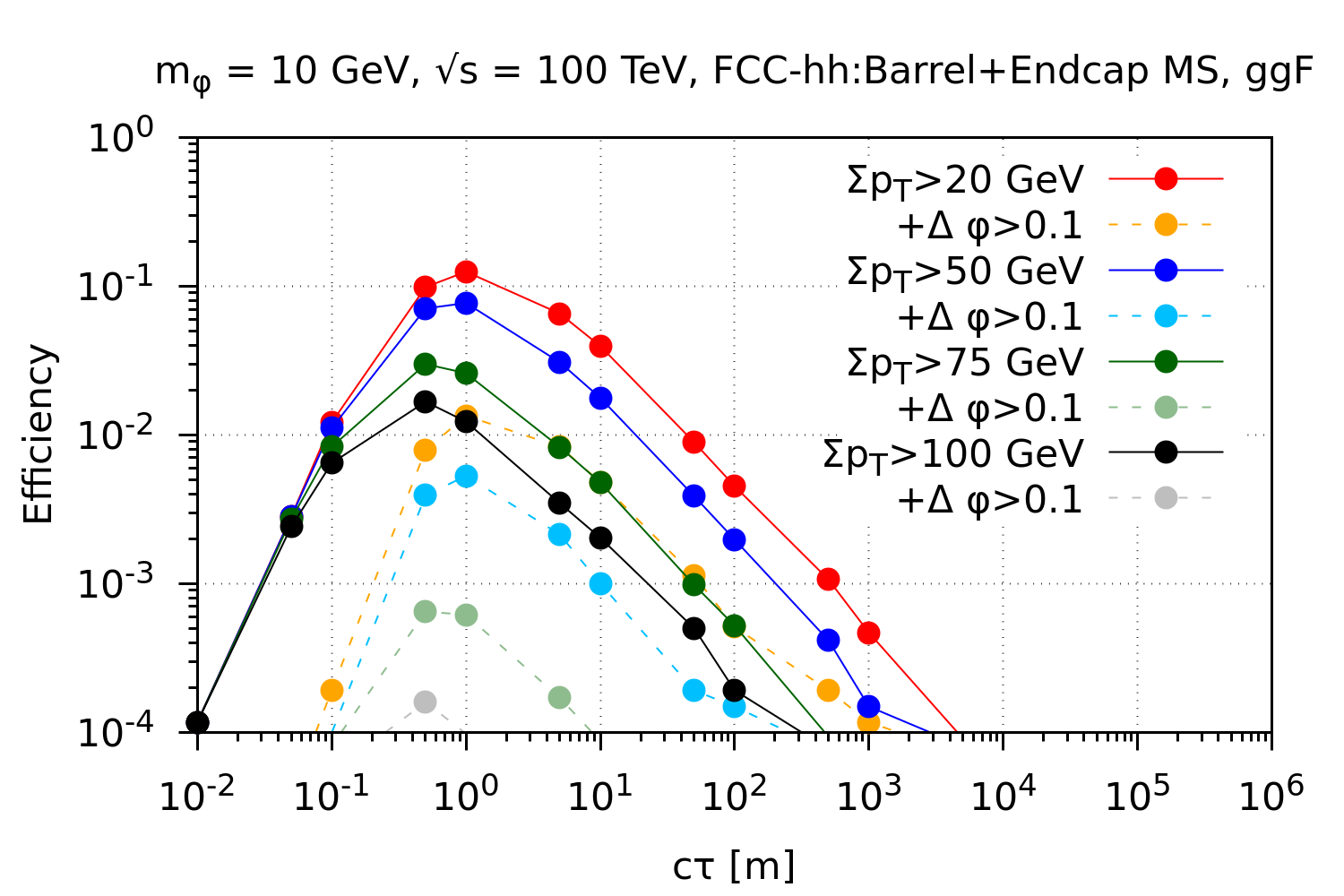}
    \caption{\small \sl Efficiencies of different $\sum p_T$ cuts, with and without $\Delta\phi_{\rm max} > 0.2$ cut in the FCC-hh barrel and endcap MS with varying decay lengths for 10\,GeV mediator particle.}
    \label{fig:eff-100TeV-bb-m10}
\end{figure}

In the 100\,TeV collider experiment, a displaced activity in the MS with energy deposition of 20\,GeV might not be enough to select such events maintaining reasonable background rates, and we might need to increase this energy threshold. In Fig.\,\ref{fig:eff-100TeV-bb-m10} shows the efficiencies of different $\sum p_T$ cuts, with and without the $\Delta\phi_{\rm max} > 0.1$ cut in the FCC-hh barrel and endcap MS with varying decay lengths for the 10\,GeV mediator particle, produced from the decay of Higgs boson in the ggF production mode. The $n_{\rm dSV}^{\rm ch} \geq 3$ cut is imposed for each of the curves shown in Fig.\,\ref{fig:eff-100TeV-bb-m10}. 
For the $\sum p_T>20$\,GeV cut used along with the $\Delta\phi_{\rm max} > 0.1$ cut, we have already seen from the left panel of Fig.\,\ref{fig:delphi-bb-100}, a reduction by a factor of 0.55 compared to the 14\,TeV efficiency at $c \tau_\phi = 1$\,m. Increasing the $\sum p_T$ cut to 50\,GeV, 75\,GeV and 100\,GeV reduces the efficiencies by factors of around 0.22, 0.02 and 0.004, respectively, when the $\Delta\phi_{\rm max}>0.1$ cut is also imposed. We expected an improvement by a factor of 150 in the projected limits in going from the 14\,TeV to 100\,TeV collider provided the efficiencies remained the same. With the computed efficiency of the $\sum p_T>75$\,GeV cut, we can improve the results only by a factor of $\sim$ 4 due to the loss in efficiency. The future dedicated LLP detectors can be expected to have a higher granularity that can help in relaxing the $\Delta\phi_{\rm max}$ cut and regaining the improvement in sensitivity. Then, removing the $\Delta\phi_{\rm max}$ cut, we can gain the full improvement by a factor of $\sim$ 150 with the $\sum p_T>75$\,GeV cut.
These can be kept in mind while finalising the design as well as the analysis strategies for the FCC-hh detector.

\subsection{DELIGHT: Proposal of dedicated LLP detectors at 100\,TeV collider}

Similar to the multitude of proposals of dedicated LLP detectors in the previous section for the HL-LHC experiment, we can also place such dedicated decay volumes near main general-purpose detectors for the 100\,TeV collider experiment to increase the sensitivity to larger decay length of LLPs. The advantage of designing such a dedicated detector in the periphery of the future 100\,TeV collider is that the collider, as well as the detectors, are not yet constructed. Therefore, we can optimise the position as well as the size of the detector to maximise its sensitivity, rather than finding empty spaces near the various IPs to place and fit the LLP detectors for the HL-LHC experiment. This gives us the scope to vary over the various parameters specifying the position and size of the decay volume of the detector to find the one which gives the best sensitivity over a large range of LLP masses and decay lengths, also being the most ergonomic among them, as we are not limited by already occupied spaces. We here propose few designs of a dedicated LLP detector, DELIGHT (\textbf{De}tector for \textbf{l}ong-l\textbf{i}ved particles at hi\textbf{gh} energy of 100 \textbf{T}eV), which is a box-type detector in the periphery of the 100\,TeV collider and present their sensitivities. 

\begin{figure}[t]
    \centering
    \includegraphics[width=0.6\textwidth]{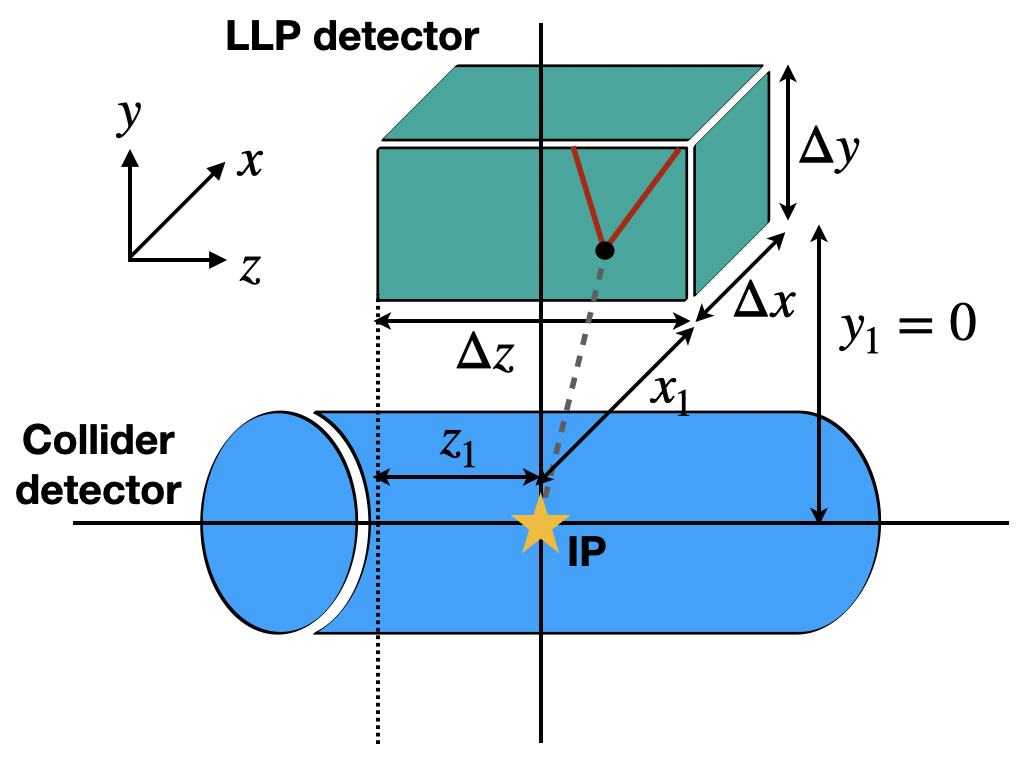}
    \caption{\small \sl Illustration of the parameters denoting position and size of the decay volumes for DELIGHT detector, a box-type LLP detector near the IP of the 100\,TeV collider.}
    \label{fig:LLP-box}
\end{figure}

Before specifying the dimensions and position of proposed LLP detector for the 100\,TeV collider experiment, we define the coordinate system that we use as follows:
\begin{align}
    \textbf{IP:} \quad & \text{Origin} \nonumber \\
    \textbf{x:} \quad & \text{Towards the centre of the collider ring} \nonumber \\
    \textbf{y:} \quad & \text{Vertically upwards} \nonumber \\
    \textbf{z:} \quad & \text{Along the beam pipe} \nonumber
\end{align}
We can choose from a box-type geometry such as the MATHUSLA and CODEX-b detectors, or a cylindrical geometry such as the FASER and ANUBIS detectors. In this paper, we focus on the former. Then, there are six parameters specifying the box, which we denote as $(x_1,y_1,z_1)$ and $(\Delta x,\Delta y,\Delta z)$ denoting the coordinates of the starting position of the box and the sizes in each direction, respectively. Fig.\,\ref{fig:LLP-box} shows the illustration of such a box-type detector with the parameters denoting the position and decay volume of the detector. Since the distribution of the LLP is uniform in the azimuthal ($\phi$) direction, we can fix it without affecting the sensitivity, which leaves us with five free parameters. We fix $y_1 = 0$, implying the detector is horizontal to the collider detector with $\phi = 0$. As we have earlier seen from the pseudo-rapidity distributions of the mediator particle, they are mostly central for lighter masses. Therefore, in order to increase the sensitivity, we propose to place the detector such that it is centered at the IP, i.e. $z_1 = -\Delta z/2$. The remaining free parameter that defines the distance of the detector from the IP is $x_1$. The closer the detector is placed near the IP, the sensitivity will be, of course, more. However, we need to keep it at a distance to provide sufficient shielding. We therefore choose $x_1 =$ 25\,m motivated from the CODEX-b detector. Revisiting the proposed design of the FCC-hh detector, we find that the collider detector extends till $R=9$\,m and will mostly have concrete and shafts. A position starting at around 25\,m in the $x$-direction around $\eta=0$ region can, therefore, be kept empty for placing a dedicated LLP detector. We understand that the 100\,TeV collider will have very high luminosity, and a distance of 25\,m might not seem enough to provide a background free environment. However, since it is still in the designing phase, we might be able to use materials that can provide significant shielding even with a smaller distance from the IP. Moreover, we will be able to use some active veto components in between the IP and the detector, such as the one proposed for the CODEX-b detector\,\cite{Gligorov:2017nwh, Aielli:2019ivi} in order to further reduce backgrounds. Nevertheless, it should be better to treat our limits as optimistic ones.

Fixing the position of the detector as above, we study three benchmark sizes:
\begin{align}
    \textbf{DELIGHT\,(A):} \quad & \text{The same as the dimensions of the MATHUSLA detector,} \nonumber \\
    & {\rm i.e.}\,\Delta x \times \Delta y \times \Delta z = 25 \times 100 \times 100\,{\rm m^3}. \nonumber \\
    \textbf{DELIGHT\,(B):}\quad & \text{Four times bigger than the MATHUSLA detector,} \nonumber \\
    & {\rm i.e.}\,\Delta x \times \Delta y \times \Delta z = 100 \times 100 \times100\,{\rm m^3}. \nonumber \\
    \textbf{DELIGHT\,(C):} \quad & \text{Twice the same decay volume as the MATHUSLA detector with} \nonumber \\
    & \text{different dimensions, i.e.}\,\Delta x \times \Delta y \times \Delta z = 200 \times 50 \times 50\,{\rm m^3}.
    \nonumber
\end{align}

The detector volume requires elements for detecting charged decay products of the LLP with good timing and spatial resolutions. The current technology of RPCs can achieve a timing resolution of 350\,ps per single gas-gap of 1\,mm, which can be further improved by combining several such layers \cite{Aielli:2019ivi}. We can also achieve a spatial resolution of at least 0.5\,cm \cite{Bauer:2019vqk}. Since the long-lived particle can decay to a wide variety of final states, the presence of some calorimeter can aid in the identification of the particles, which can enhance the sensitivity of these detectors, extending the search to LLPs decaying to photons. calorimeters will also enable mass reconstruction, which can help in reducing the SM background coming from long-lived particles like $K_L$. Placing our detector in the central region also has the added advantage of reduced backgrounds as compared to the forward direction. The possibility of integrating the detector with the FCC-hh trigger system can also be explored since we are proposing it to be placed just 25\,m away from the IP.

\begin{figure}[t]
    \centering
    \includegraphics[width=0.46\textwidth]{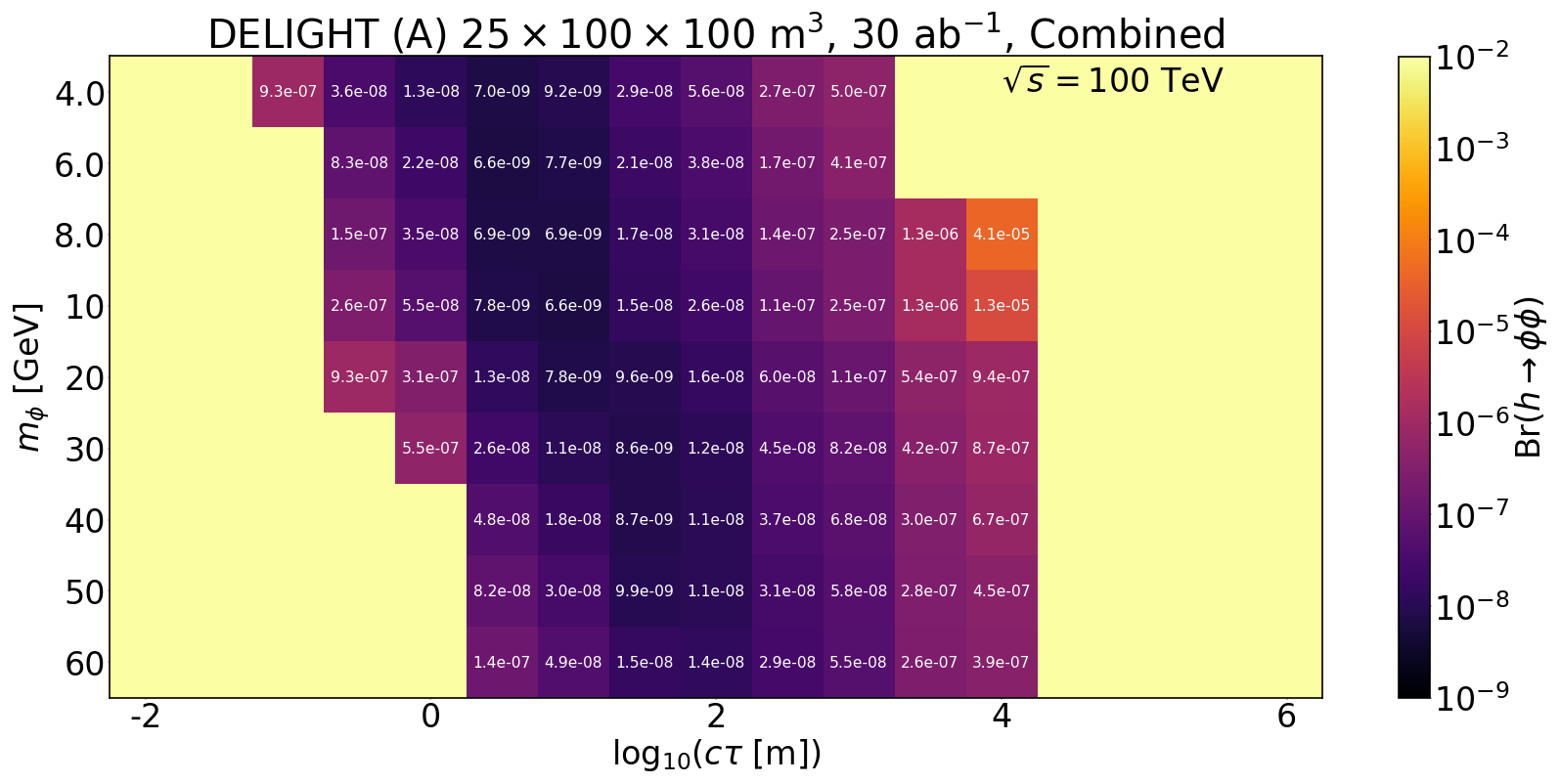} \qquad
    \includegraphics[width=0.46\textwidth]{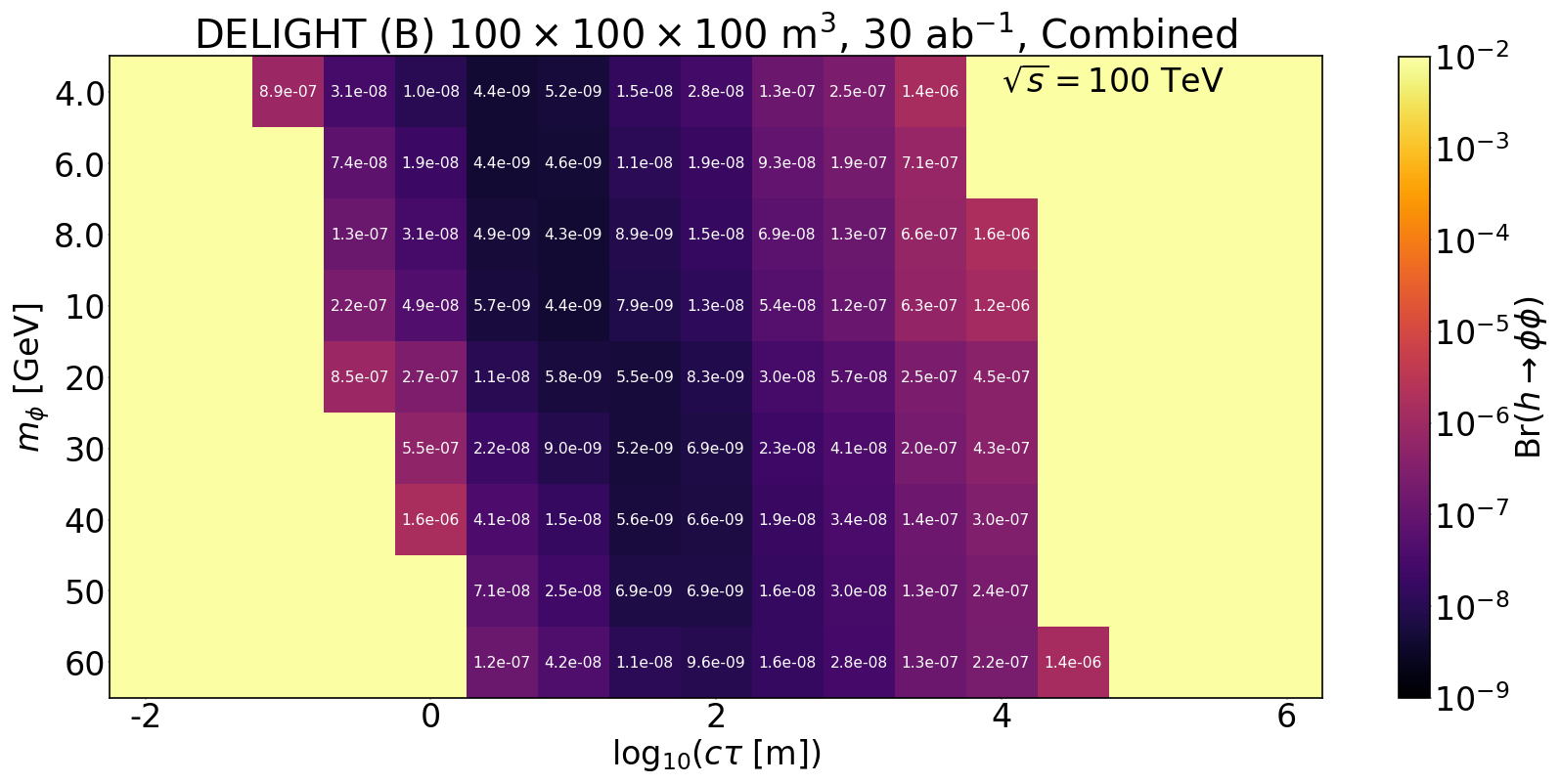} \\
    \includegraphics[width=0.46\textwidth]{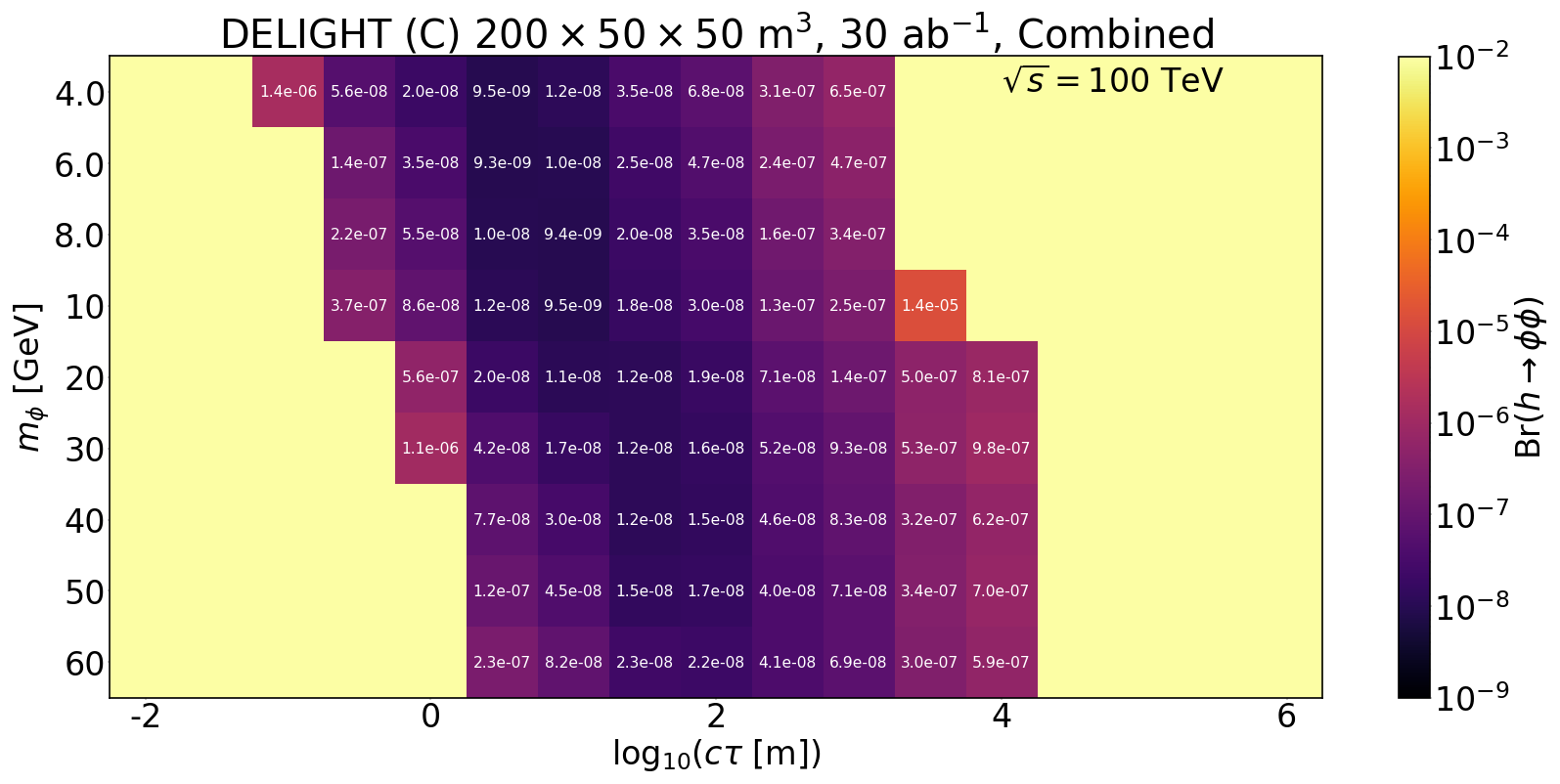}
    \caption{\small \sl Projected upper limits on the branching fraction, Br$(h \to \phi \phi)$, for 4 observed decays of long-lived mediator particles within DELIGHT\,(A), (B) and (C) at the 100\,TeV collider experiment which are described in the text. The shown limits are obtained by combining the ggF, VBF and Vh production modes of the Higgs boson.}
    \label{fig:limit-mathu100}
\end{figure}

Fig.\,\ref{fig:limit-mathu100} shows the projected upper limits on the branching fraction of the Higgs boson decaying into a pair of mediator particles, Br$(h \to \phi \phi)$, for 4 observed decays of long-lived mediator particles within the three decay volumes of the DELIGHT detectors as described above. Comparing the results from DELIGHT\,(A) with that of MATHUSLA, we observe an improvement by a factor of $\sim 540$, of which around $\sim 150$ is accounted by the increased cross-section and integrated luminosity. Another factor of $\sim$ 3--4 is gained in the efficiency by moving the detector close to the IP. The best limits come from DELIGHT\,(B), which has the highest decay volume among the three (about four times bigger than the decay volume of MATHUSLA), and the performance is better by a factor of 2 compared to DELIGHT\,(A).

DELIGHT\,(C) has twice the same decay volume as DELIGHT\,(A).
Here, the $\Delta\eta \times \Delta\phi$ coverage of the detector is determined by its $\Delta y\times\Delta z$, which is smaller for DELIGHT\,(C) compared to DELIGHT\,(A). Reducing $\Delta y$ makes sure that it has some shielding from cosmic rays, and therefore, this background is easier to be handled. Its detector volume is compensated by an increased $\Delta x$, and it is like a long tunnel along the line towards the centre of the collider ring. The limits obtained from DELIGHT\,(C) are smaller than that from DELIGHT\,(A) by a factor of around $0.8-0.9$. Therefore, if the dimension of DELIGHT\,(C) is more ergonomic and feasible than DELIGHT\,(A), since the former has smaller $\Delta y \times \Delta z$, we do not lose on the sensitivity much.


\section{Summary and conclusion}
\label{sec:summary}

Motivated by the importance of a light scalar mediator particle connecting between the dark sector and the SM and also driven by the fact that the mixing of such a particle with the SM Higgs boson has been constrained to be small by the LHC experiment, we take a closer look into the long-lived parameter space of such a mediator particle. The summary of our work with the key findings are listed below: 

\begin{itemize}
\item We have focused on the production of the mediator from the on-shell decay of the SM Higgs boson and have considered the dominant production modes of the latter in hadron colliders, {\it viz}., ggF, VBF and Vh. We also consider a variety of decay modes of the mediator particle $-$ muons, pions, Kaons, gluons, $s$ quarks, $c$ quarks, $\tau$ leptons and $b$ quarks. \textit{We have examined in this work how the interplay between a wide variety of production and decay modes of the long-lived scalar mediator affects the search sensitivity of such a particle.}

\item As reasoned in the text earlier, we focus on displaced activities in the CMS MS detector at the HL-LHC experiment, and use a combination of cuts on the prompt particles coming from production modes of the Higgs boson and the displaced SM particles coming from the mediator particle decay for the analyses. We use trigger cuts proposed for the Phase-II HL-LHC runs, and relax them when an additional displaced or prompt activity is demanded. Prospects of using such hybrid cuts could prove very beneficial for designing LLP analyses. 

\item We begin by presenting the analyses for each of the decay modes, assuming 100\% branching ratio to that particular mode. \textit{Such a model independent depiction of the result presented in this work can be translated and used by various models as well.} We present our results in the form of grids in the mass and $c \tau_\phi$ plane, with the exact value of the projected upper limits on the branching fraction, Br$(h \to \phi \phi)$, available as well, which is easier to recast than reading from the color plot. Although we do not quantify the background in this paper, we present a discussion on some of the possible backgrounds that can arise in the CMS MS search along with ways of reducing them, and elucidate the minimum number of observed events we should use for setting the limits. 

\item Next, we combine the decay modes according to the branching ratio predicted by the minimal model of the scalar mediator particle, and present the results in the $(m_\phi,\,c\tau_\phi$)- and $(m_\phi,\,\sin \theta)$-planes. The limits obtained from the displaced di-muons analysis are at least one order of magnitude better than the ones obtained for displaced activities in the muon spectrometer from mesons like pions and Kaons, gluons, and various flavors of quarks. We can achieve a sensitivity of Br$(h \to \phi \phi) < 1.7\times 10^{-5}$ for 60\,GeV mediator at decay length of 5\,m with the $P^S\times D^S \geq 1$\,vtx set of cuts (described in Section \ref{sssec:analysis}) when the mediator decays to a pair of $b$ quarks.
 
\item In addition, we study the contribution of dedicated LLP detectors such as the MATHUSLA and CODEX-b detectors in the search for long-lived scalar mediator particles coming from the decay of the Higgs boson produced in either of ggF, VBF or Vh production modes, and their complementarity with the CMS MS search. We observe that CMS and MATHUSLA together can probe decay lengths till 10$^5$\,m for a mediator of 60\,GeV mass, without any gap if the branching of Higgs boson to the mediator particles is at least $\sim$0.1\%.

\item We study the sensitivity of the muon spectrometer of the proposed 100\,TeV hadronic collider experiment, FCC-hh, and investigate the effect of including the forward MS of the FCC-hh detector to the barrel and endcap MS stations of the detector. Apart from the increased cross-section and integrated luminosity, we discuss how the signal efficiency is affected for the FCC-hh detector design. 
The forward MS increases the efficiencies and opens up sensitivity to lower decay lengths for the analyses with displaced activities in the muon spectrometer.
The factor responsible for this is the more forward distribution of long-lived mediators with higher mass and lower decay lengths once the dSV is restricted within the MS, which are not missed by the forward MS. This is discussed in detail in the text.

\item \textit{We propose the detector designs for dedicated LLP searches in the periphery of the 100\,TeV FCC-hh collider, named DELIGHT.} Since the designs for the 100\,TeV collider are under development, we have the freedom to survey the nearby regions for identifying the positions and sizes of detectors that will be more effective for LLP signals. Motivated from the distributions of direction of the scalar mediator particle produced from the decay of the Higgs boson in the colliders, we investigate the prospects of three benchmarks of detector designs. Our study suggests that these can provide an improvement by a factor of 500 or more compared to the MATHUSLA detector at the HL-LHC experiment.
\end{itemize}

We believe that our detailed analysis and findings will be useful for a range of LLP models. Our analyses for the future 100\,TeV collider can be a starting point for further detailed studies exploring their prospects for LLPs. The proposal put forth by us in this work for the DELIGHT detector near the FCC-hh collider shows promising results, motivating further studies in this direction for other LLP models.

\section*{Acknowledgement}

BB acknowledges the hospitality of Kavli IPMU, Japan, where the project was initiated.
The work of SM is supported by Grant-in-Aid for Scientific Research from the Ministry of Education, Culture, Sports, Science, and Technology (MEXT), Japan (17H02878, 20H01895, 19H05810, 20H00153), by World Premier International Research Center Initiative (WPI), MEXT, Japan (Kavli IPMU), and also by JSPS Core-to-Core Program (JPJSCCA20200002).
BB and RS thank Swagata Mukherjee, Prabhat Solanki and Rahool Kumar Barman for useful discussions.

\appendix



\section{The dedicated LLP detectors: validation \& efficiency map}
\label{app:LLP-detectors}

\begin{figure}[t]
    \centering
    \includegraphics[width=0.46\textwidth]{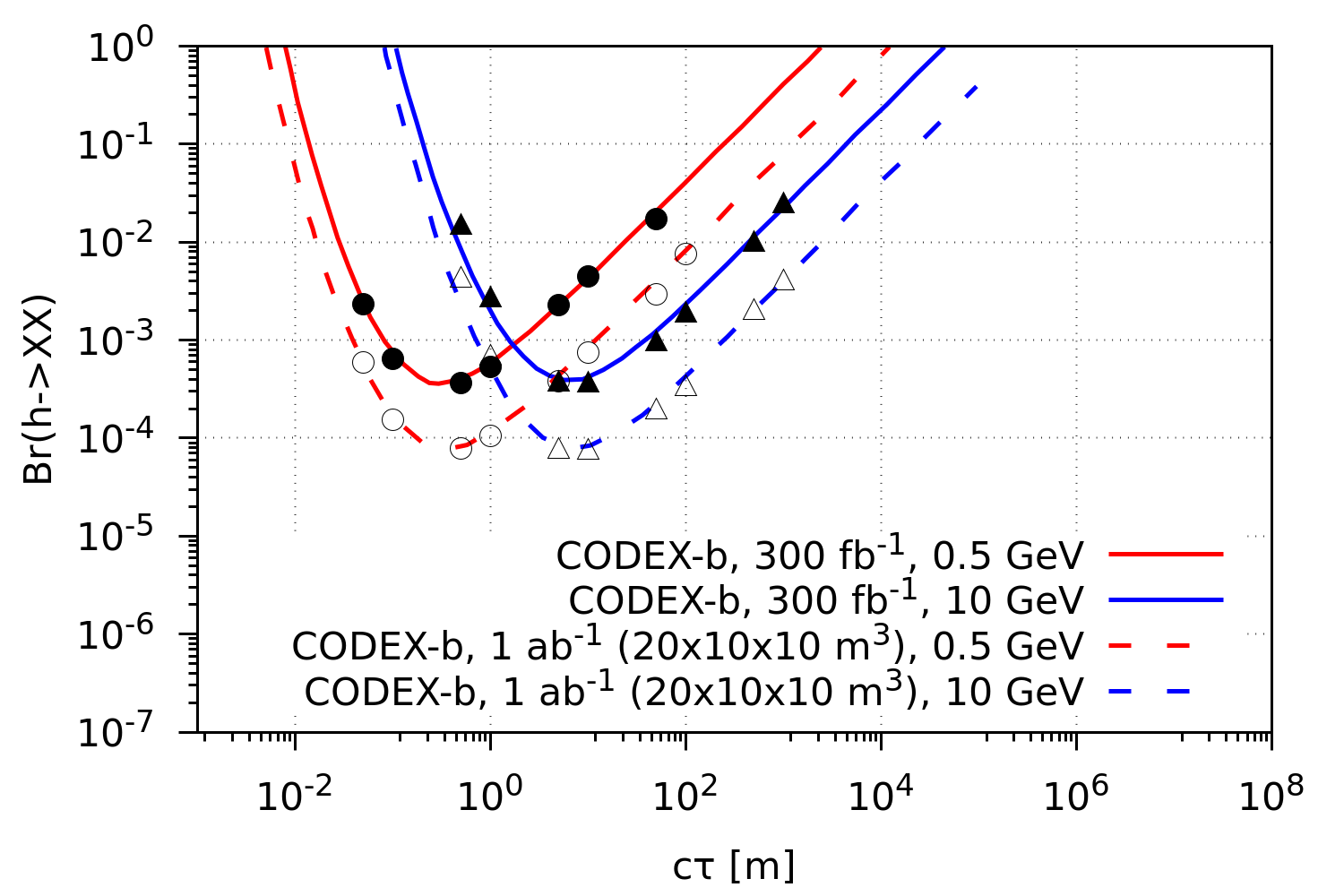}
    \qquad
    \includegraphics[width=0.46\textwidth]{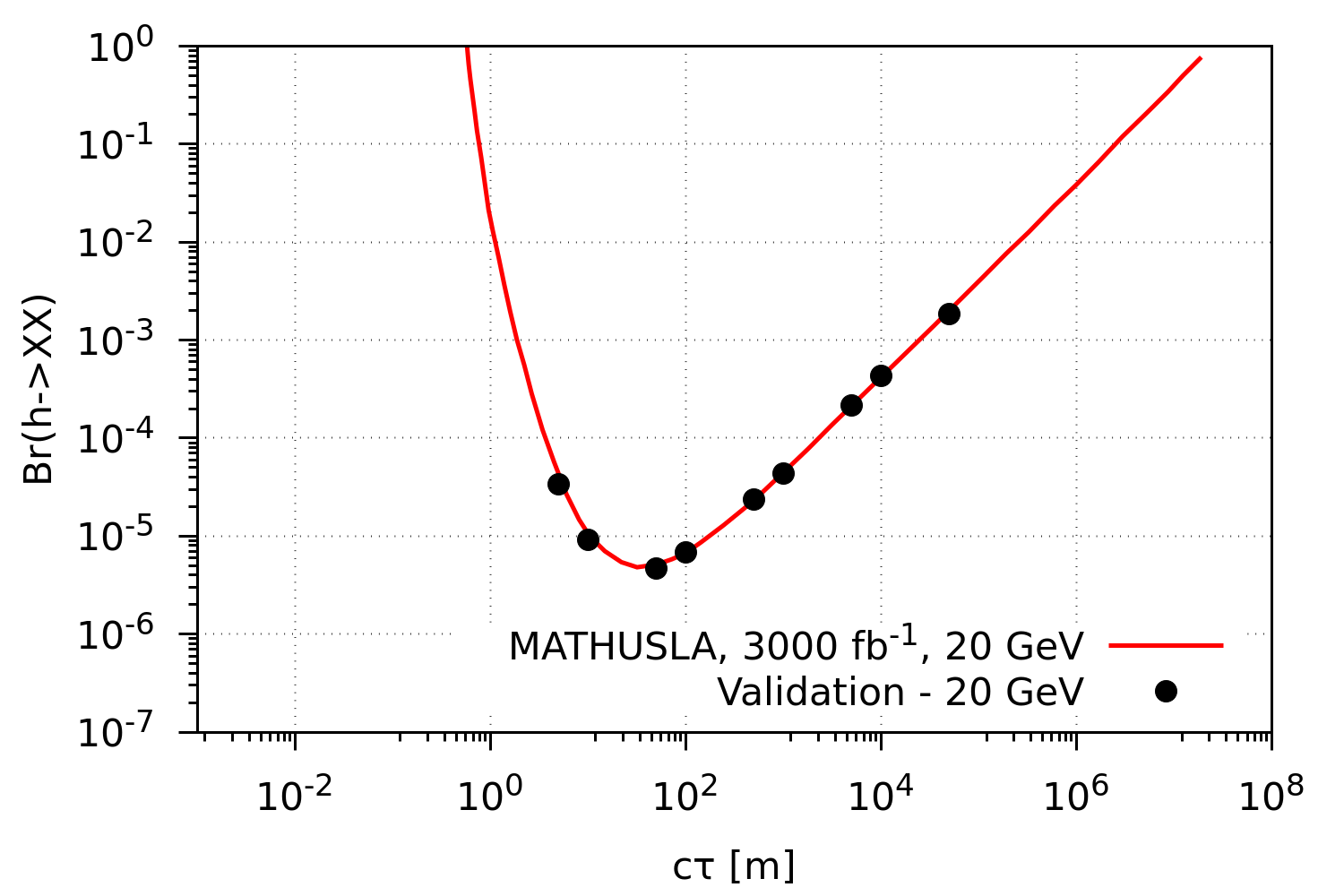}\\
    \caption{\small \sl Validation of the CODEX-b ({\it left}) and MATHUSLA ({\it right}) detectors \cite{Gligorov:2017nwh, Alpigiani:2020iam}.}
    \label{fig:val}
\end{figure}

\begin{figure}[t]
    \centering
    \includegraphics[width=0.46\textwidth]{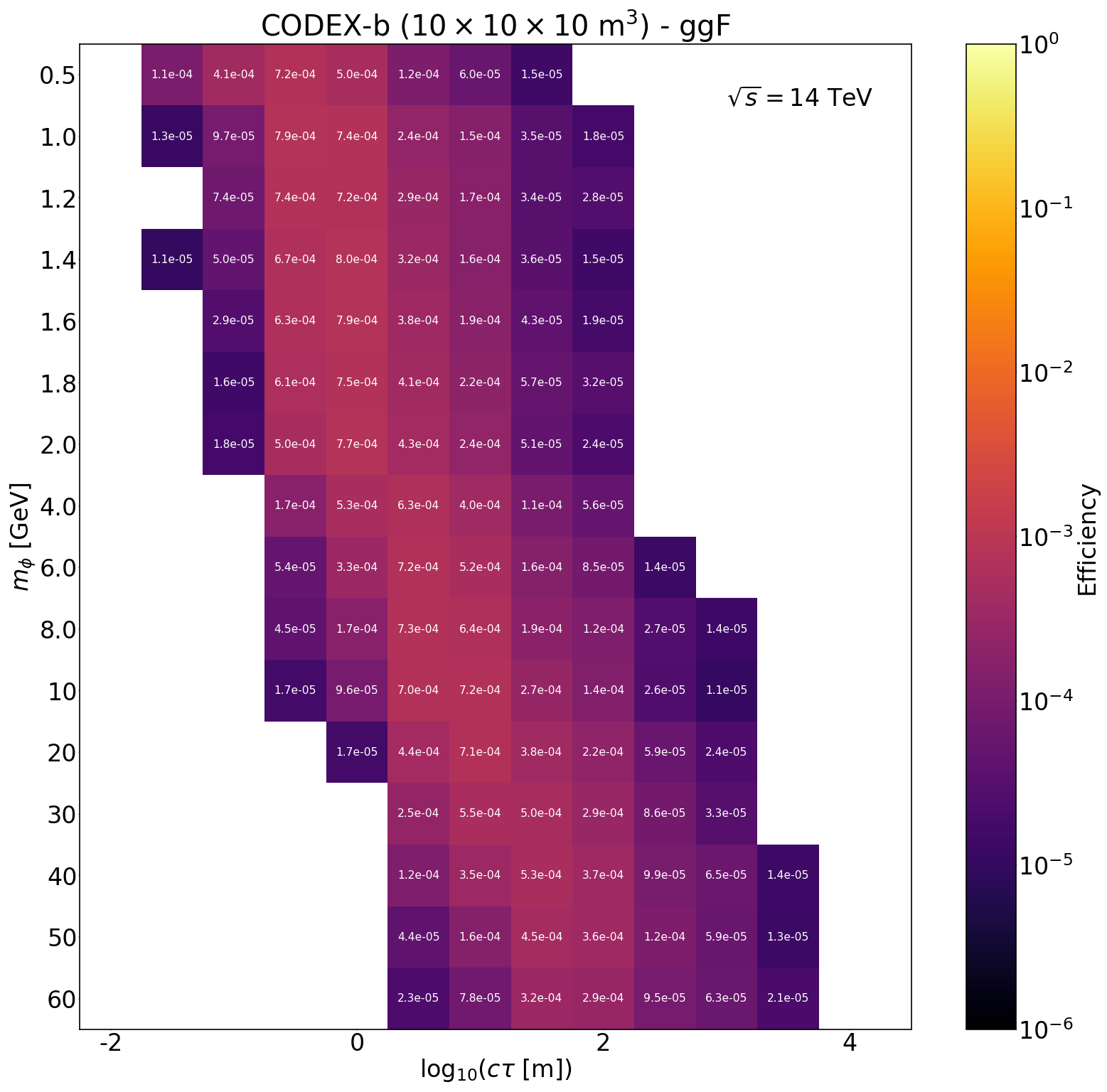} \qquad
    \includegraphics[width=0.46\textwidth]{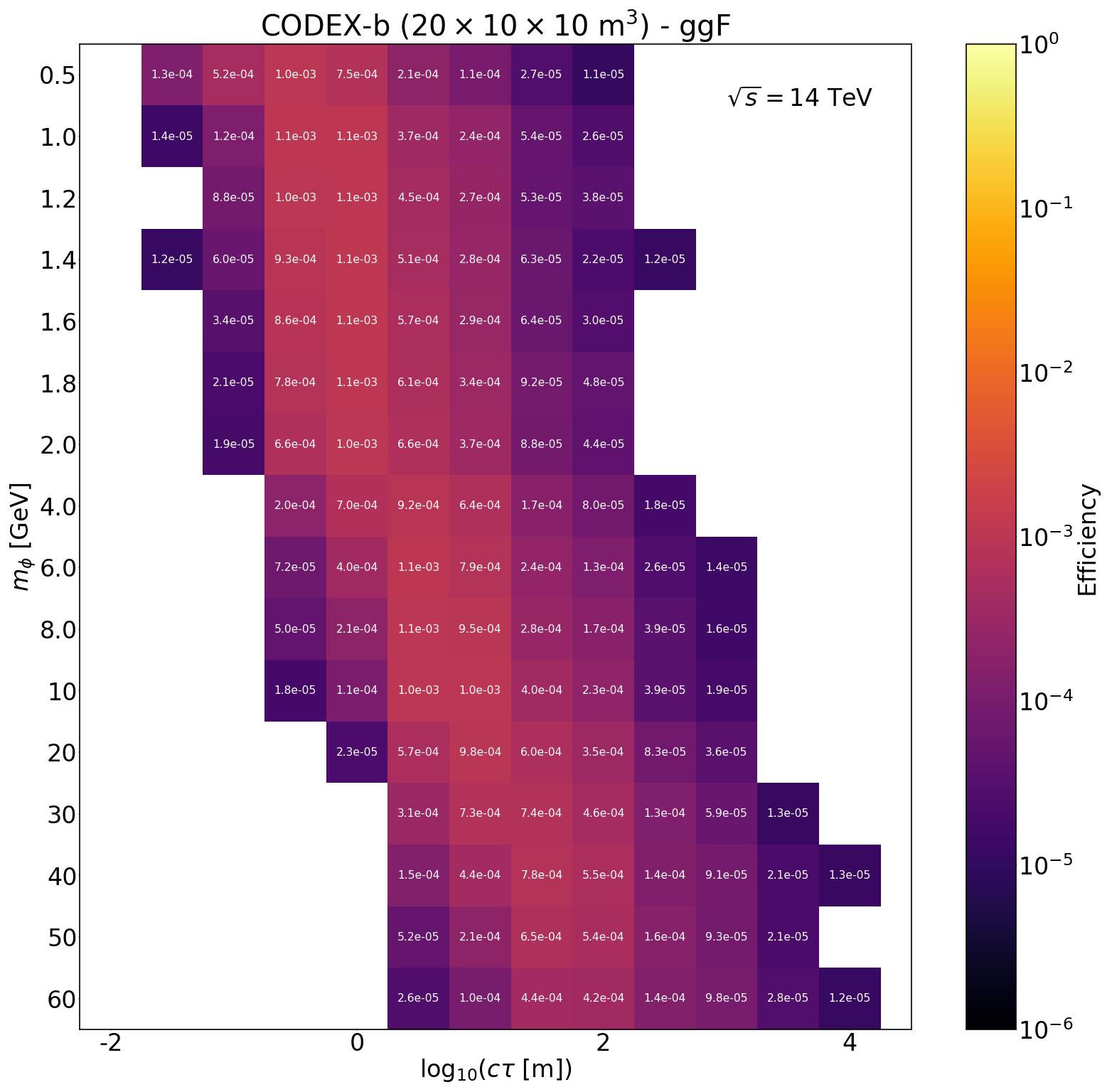} \\
    \includegraphics[width=0.46\textwidth]{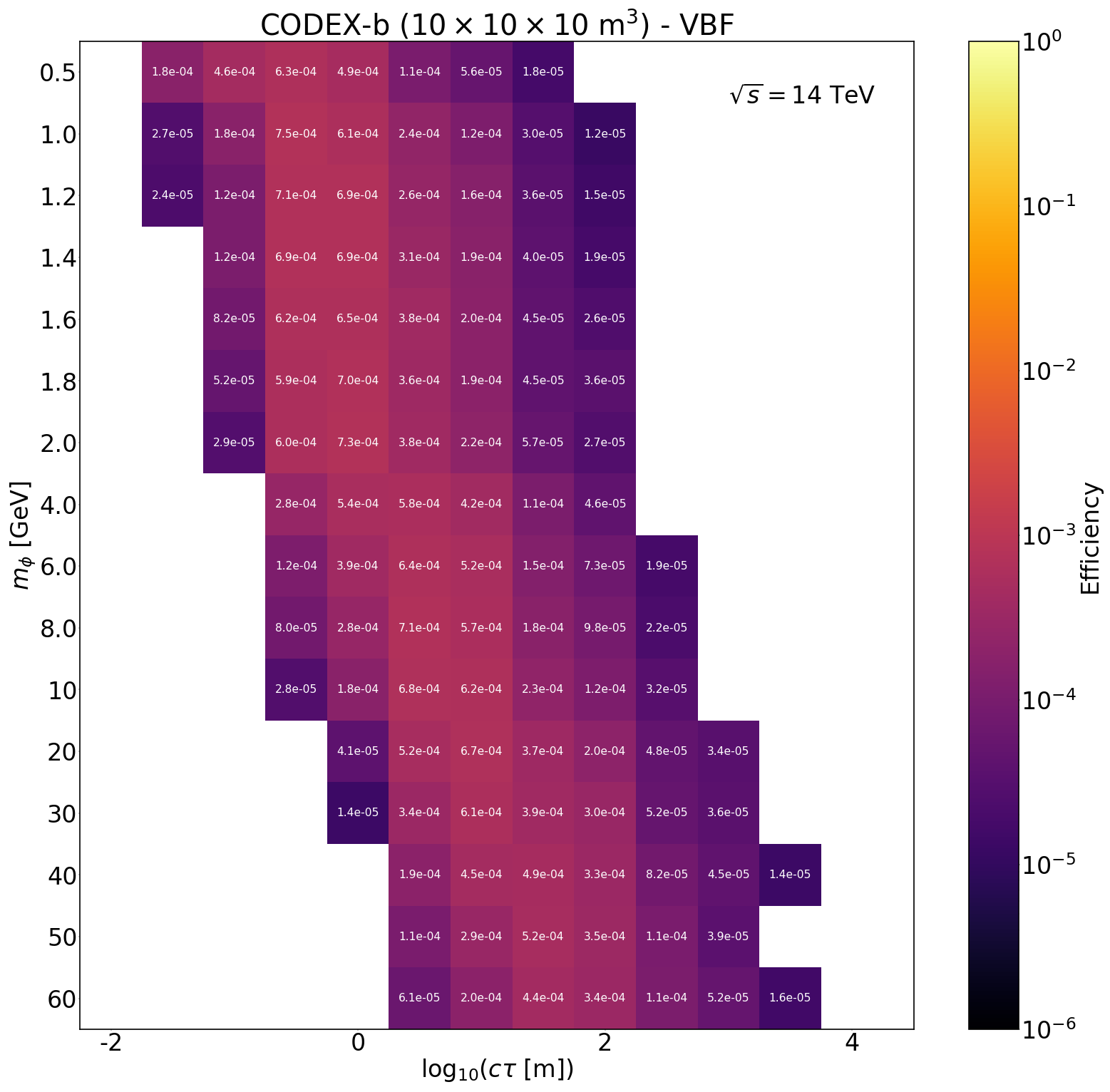} \qquad
    \includegraphics[width=0.46\textwidth]{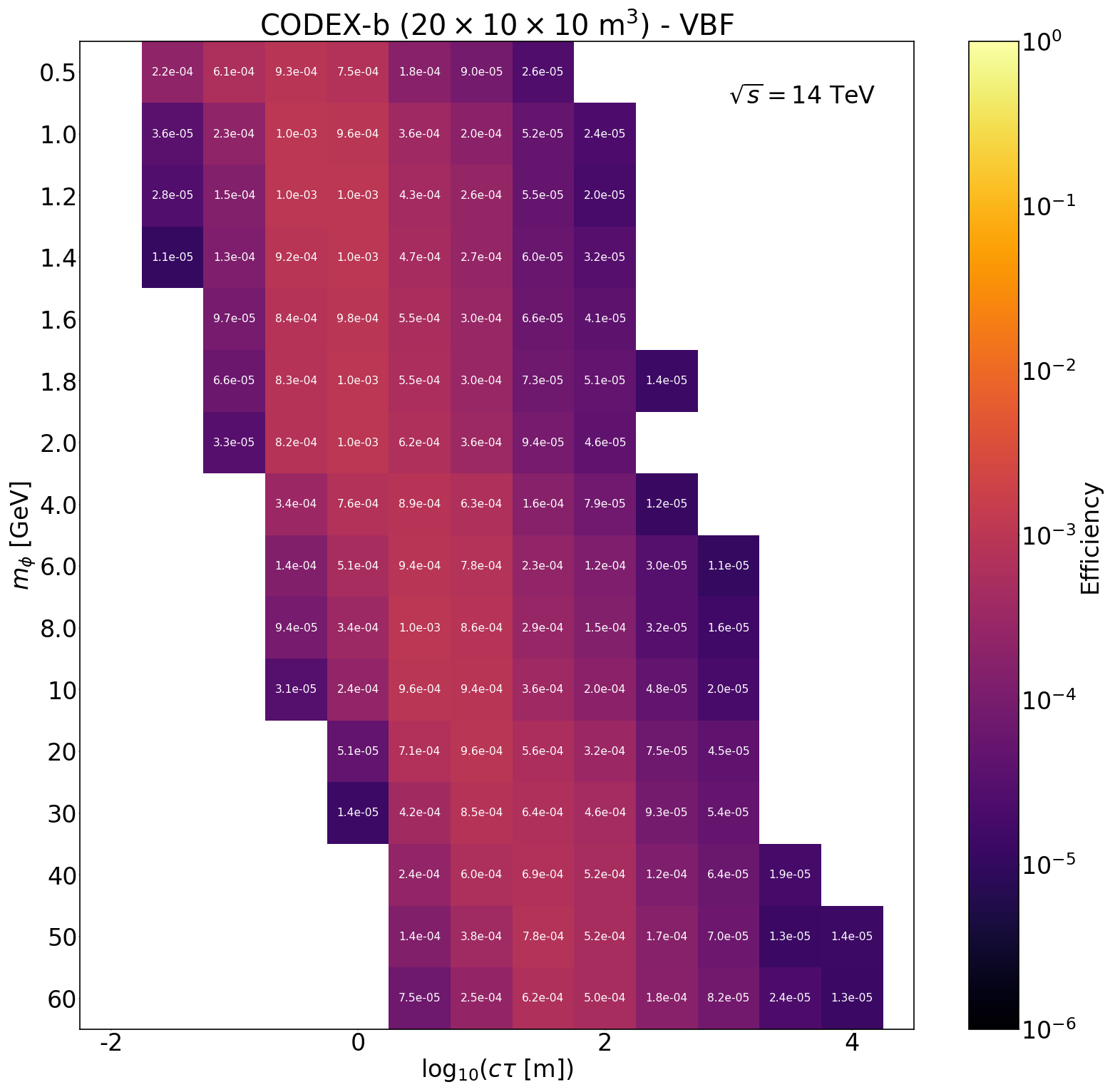} \\
    \includegraphics[width=0.46\textwidth]{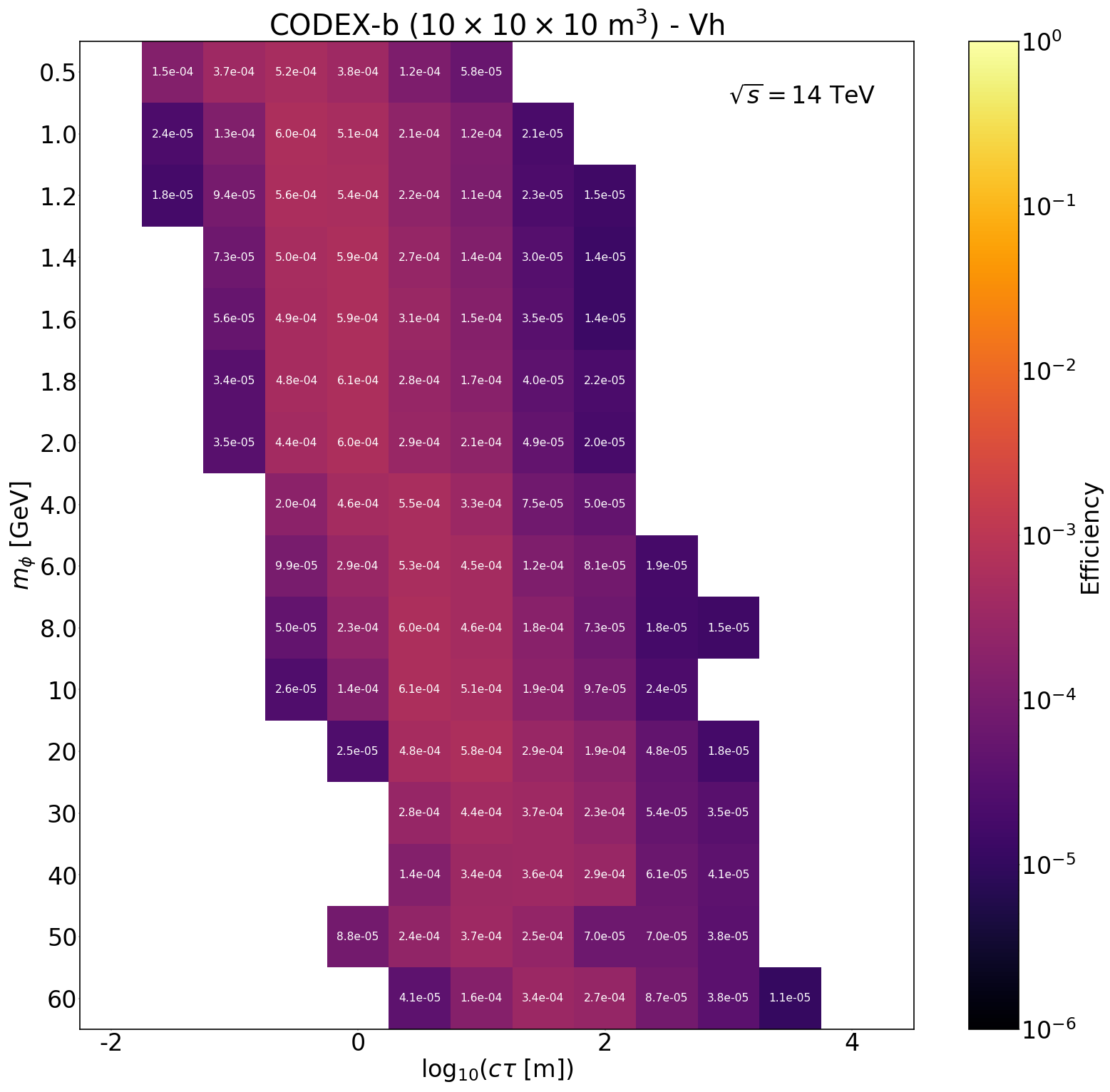} \qquad
    \includegraphics[width=0.46\textwidth]{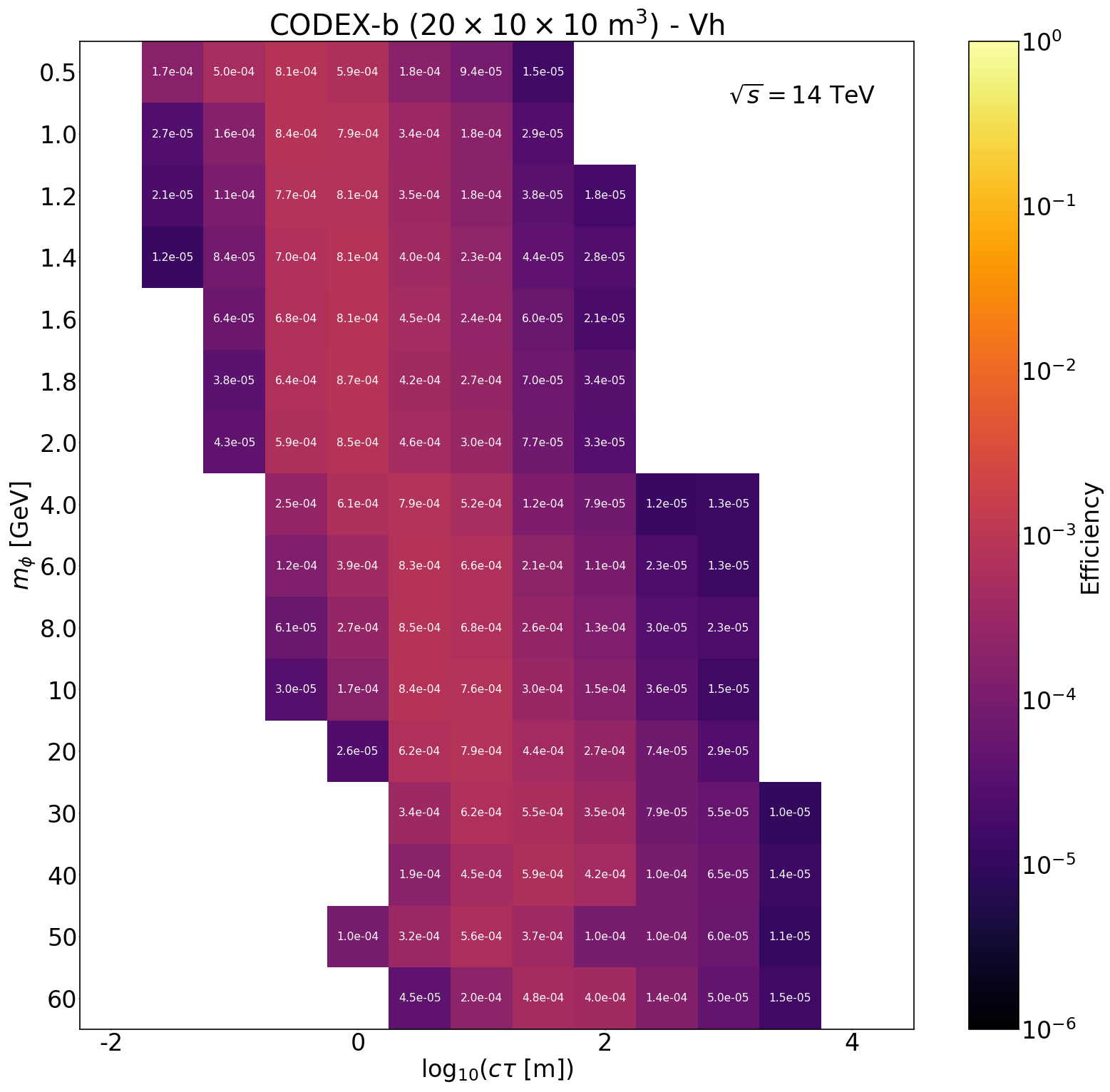}
    \caption{\small \sl Efficiency maps for decays of long-lived mediator particles within the CODEX-b detector for the ggF (top), VBF (center) and Vh (bottom) production of the Higgs boson.}
    \label{fig:eff-maps-codex}
\end{figure}

\begin{figure}[t]
    \centering
    \includegraphics[width=0.46\textwidth]{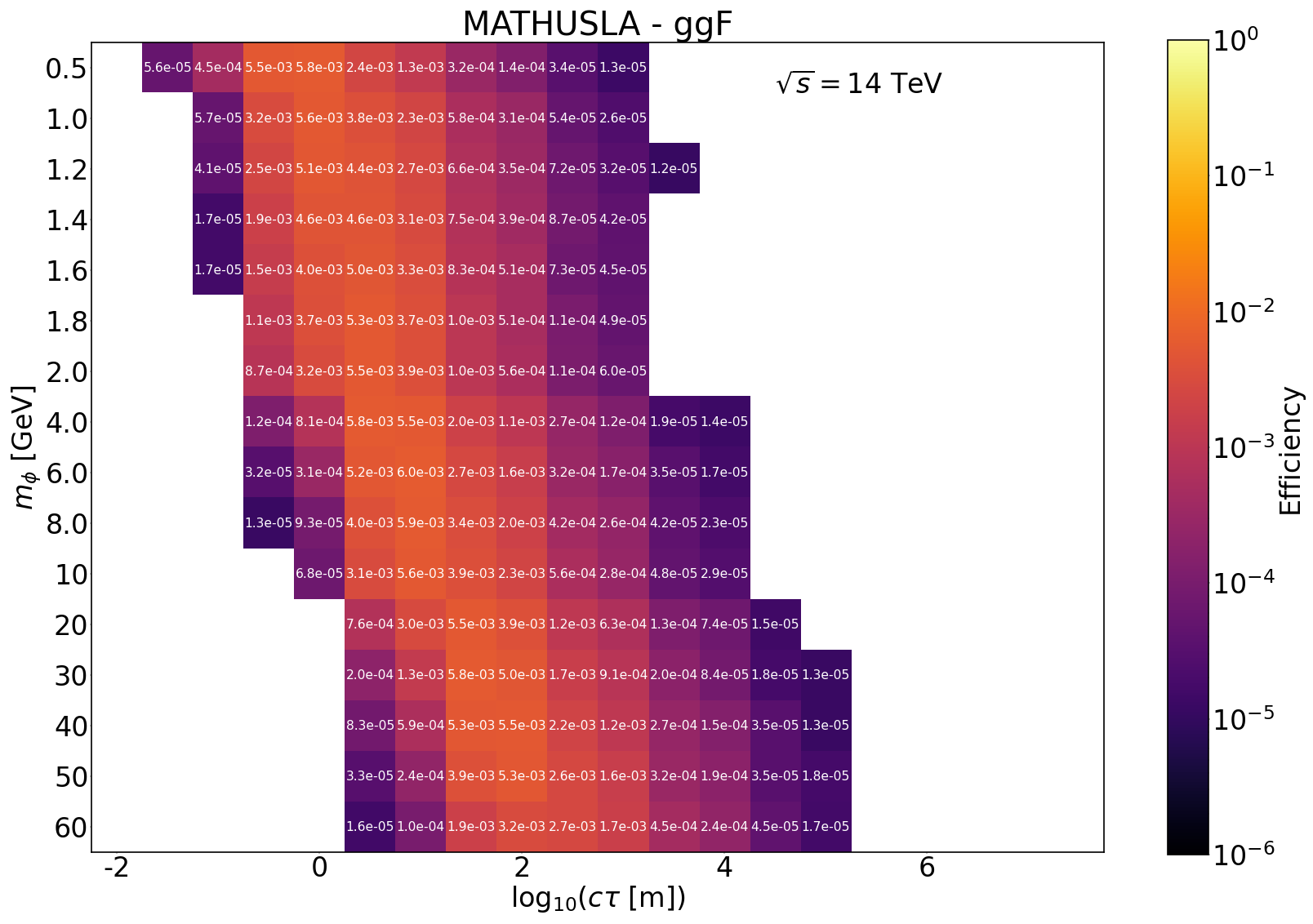} \qquad
    \includegraphics[width=0.46\textwidth]{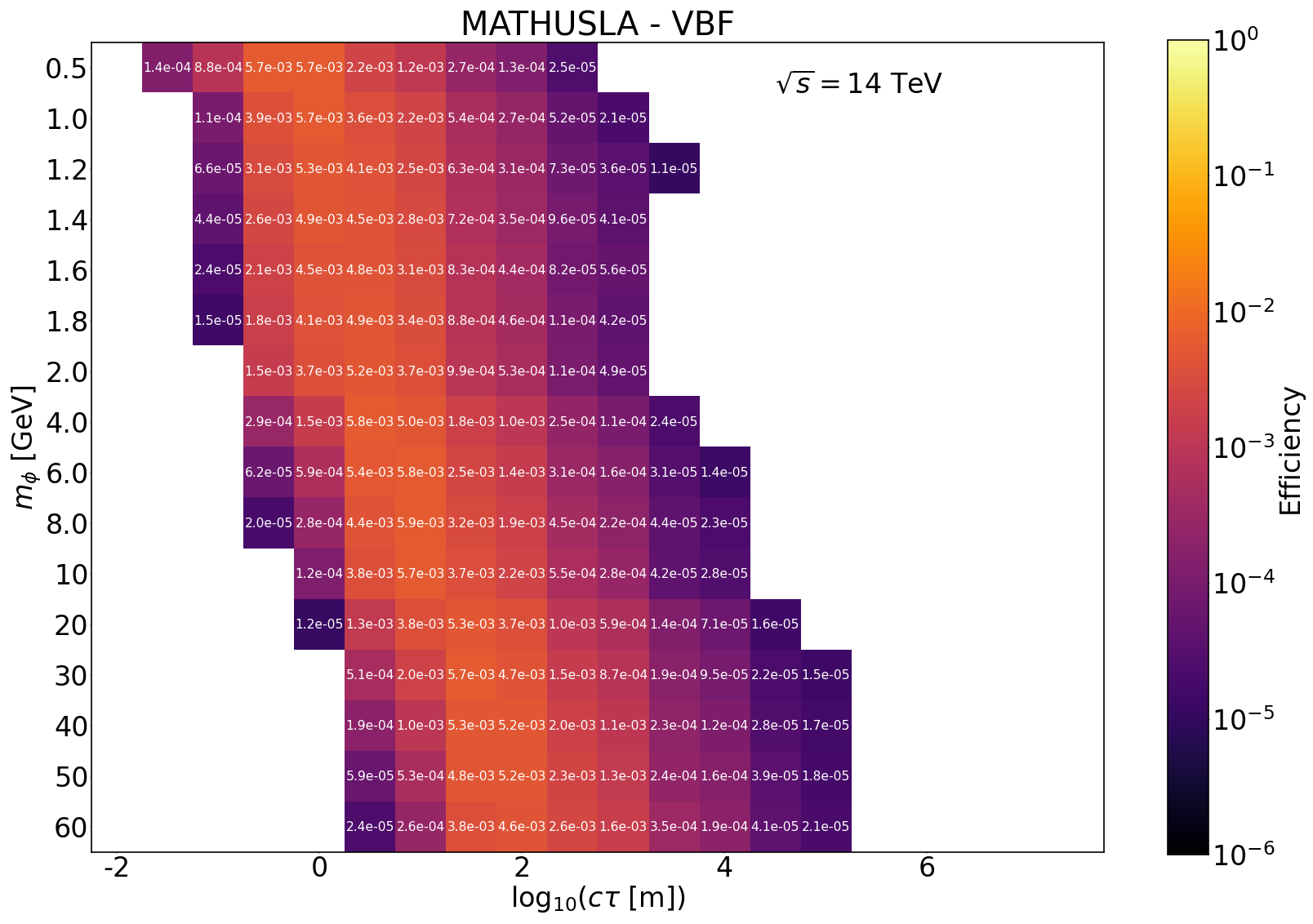}\\
    \includegraphics[width=0.46\textwidth]{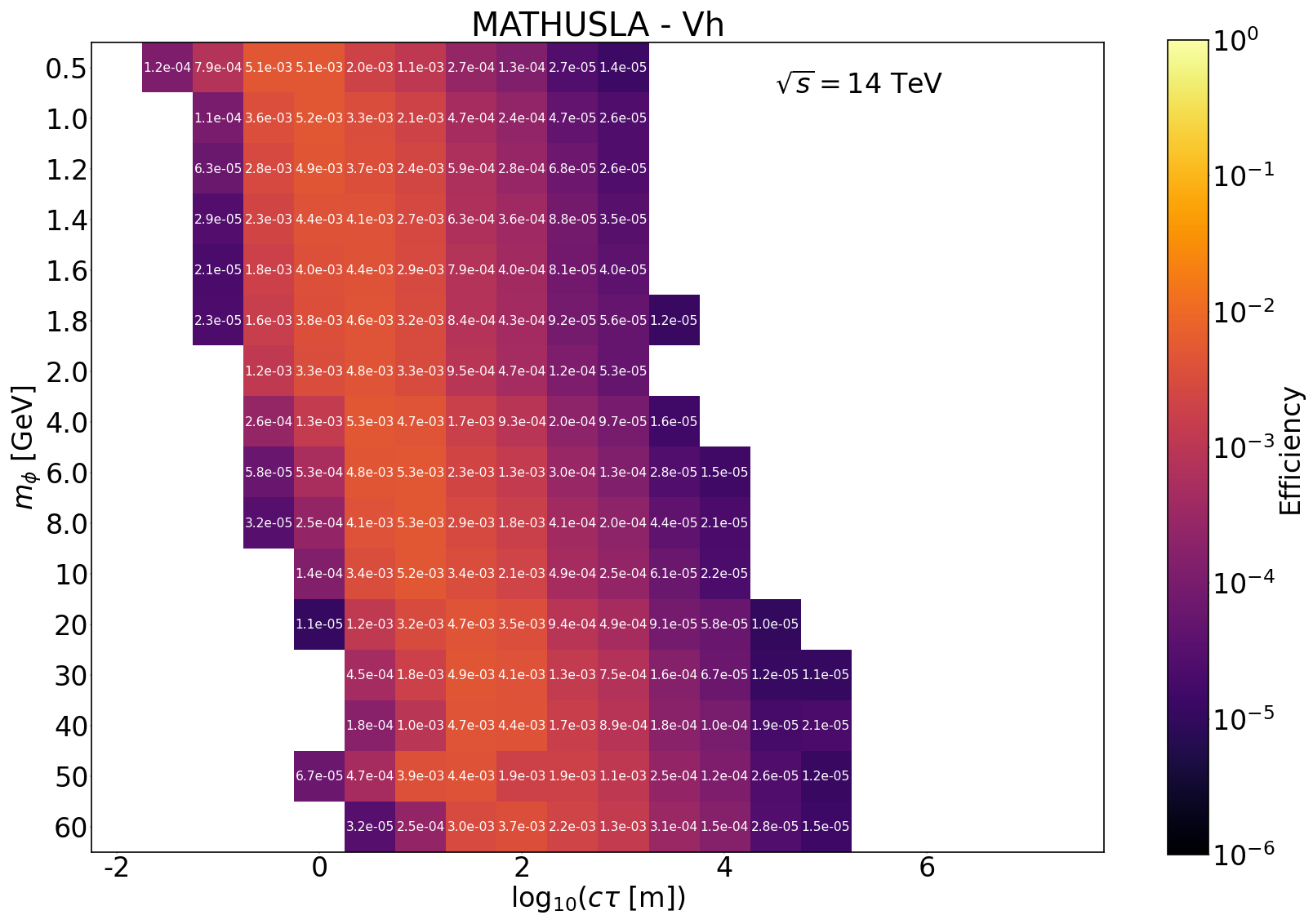}
    \caption{\small \sl Efficiency maps for decays of long-lived mediator particles within MATHUSLA for the ggF (top), VBF (center) and Vh (bottom) production of the Higgs boson.}
    \label{fig:eff-maps-mathu}
\end{figure}

The validation as well as the efficiency maps for the dedicated LLP detectors at the HL-LHC experiment are given in Fig.\,\ref{fig:val} and Figs \ref{fig:eff-maps-codex} \& \ref{fig:eff-maps-mathu}, respectively.

\section{Efficiency maps of the DELIGHT detector}

\begin{figure}[h]
    \centering
    \includegraphics[width=0.46\textwidth]{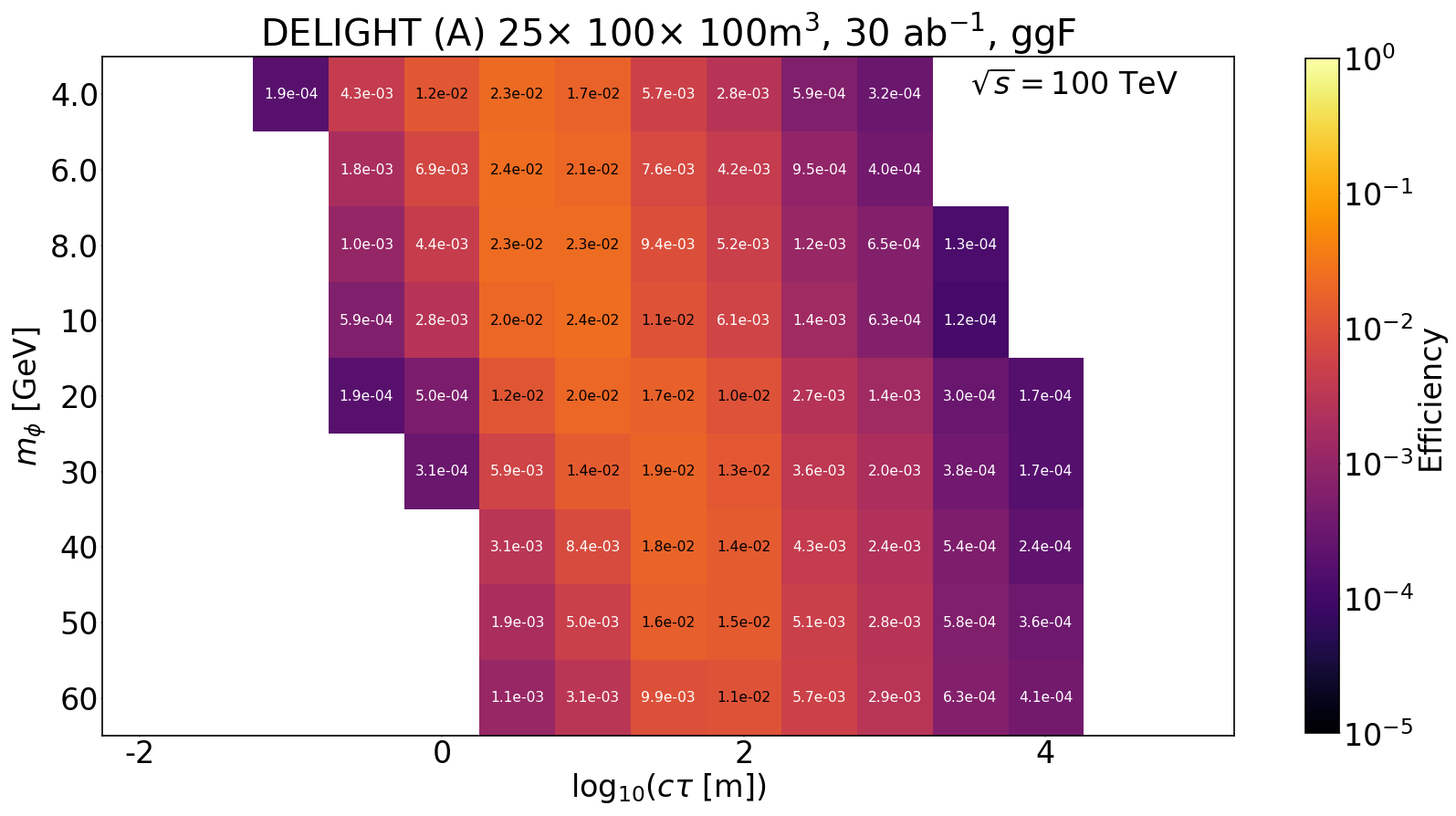} \qquad
    \includegraphics[width=0.46\textwidth]{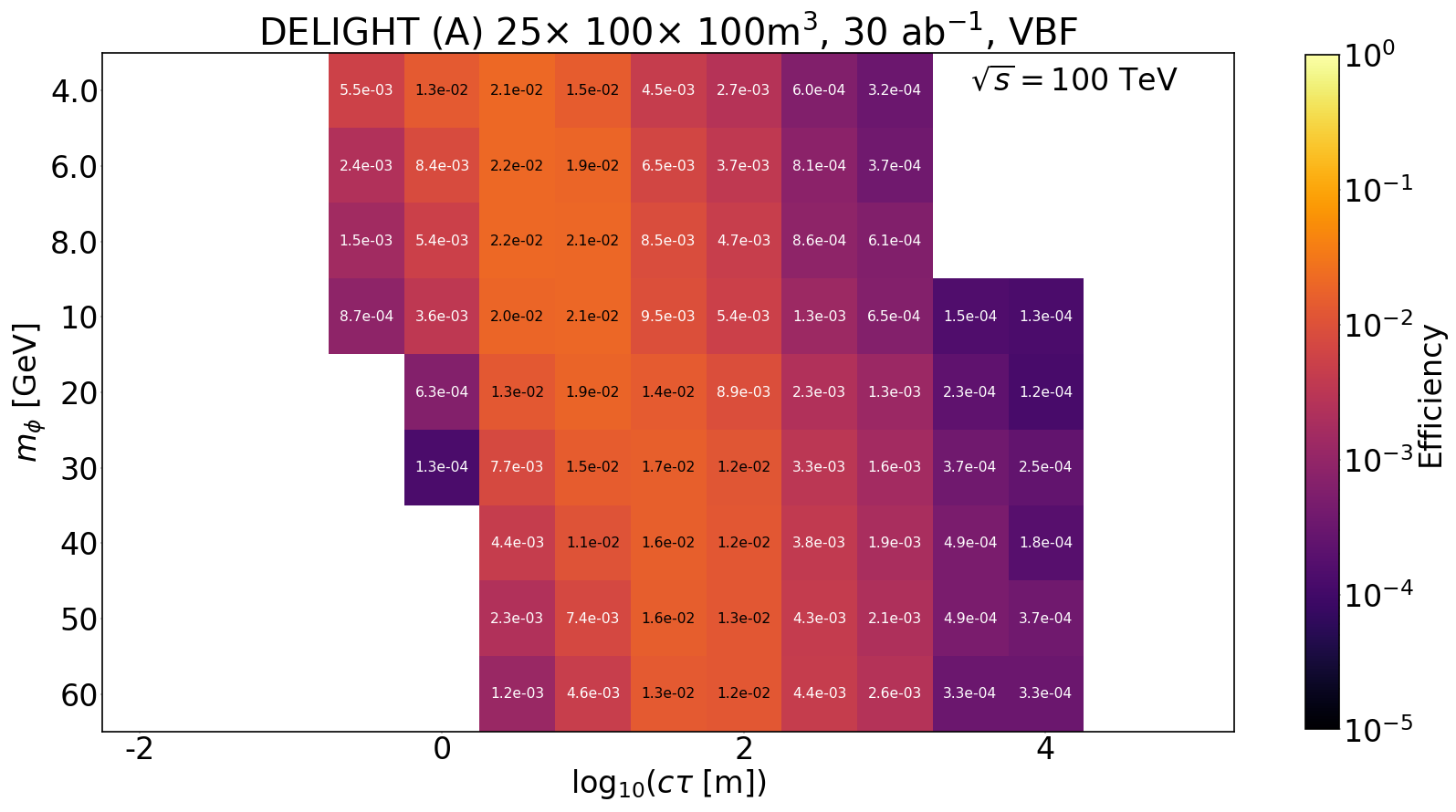} \\
    \includegraphics[width=0.46\textwidth]{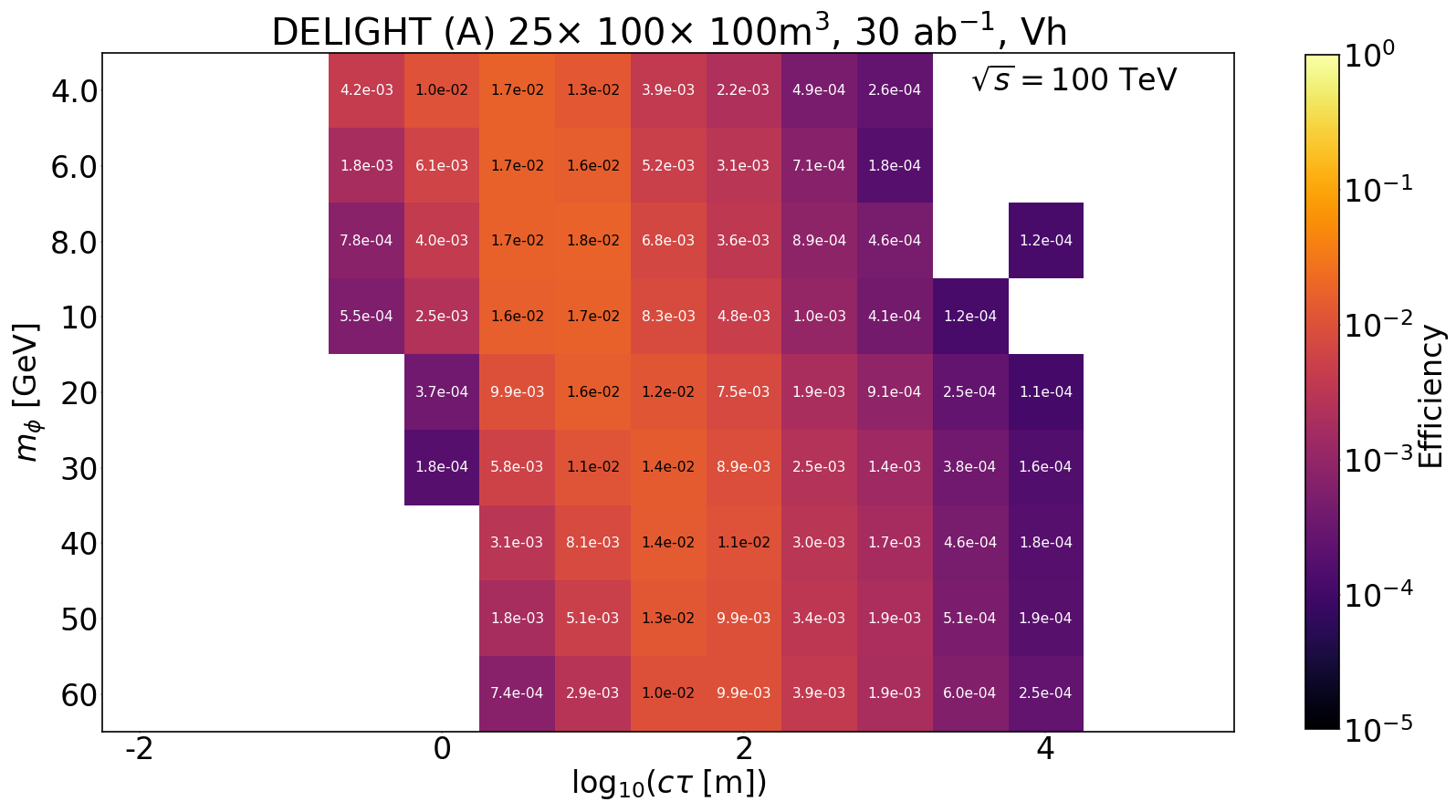}
    \caption{\small \sl Efficiency maps for decays of long-lived mediator particles within the DELIGHT\,(A) detector near the 100 TeV collider for the ggF (top panels), VBF (center panels) and Vh (bottom panels) production of the Higgs boson at 100\,TeV collider experiment.}
    \label{fig:eff-map-100TeV-LLP_A}
\end{figure}

\begin{figure}[h]
    \centering
    \includegraphics[width=0.46\textwidth]{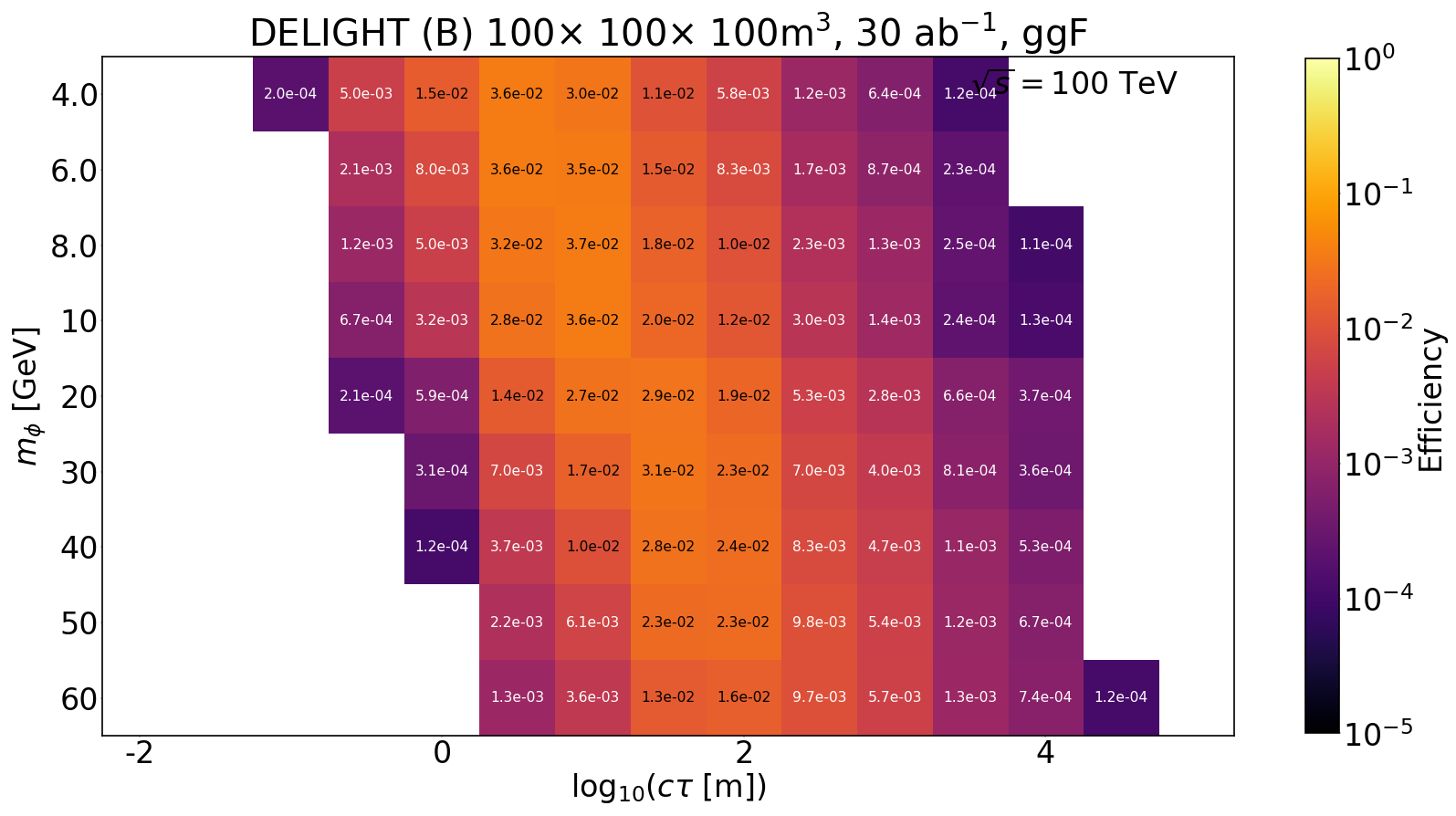} \qquad
    \includegraphics[width=0.46\textwidth]{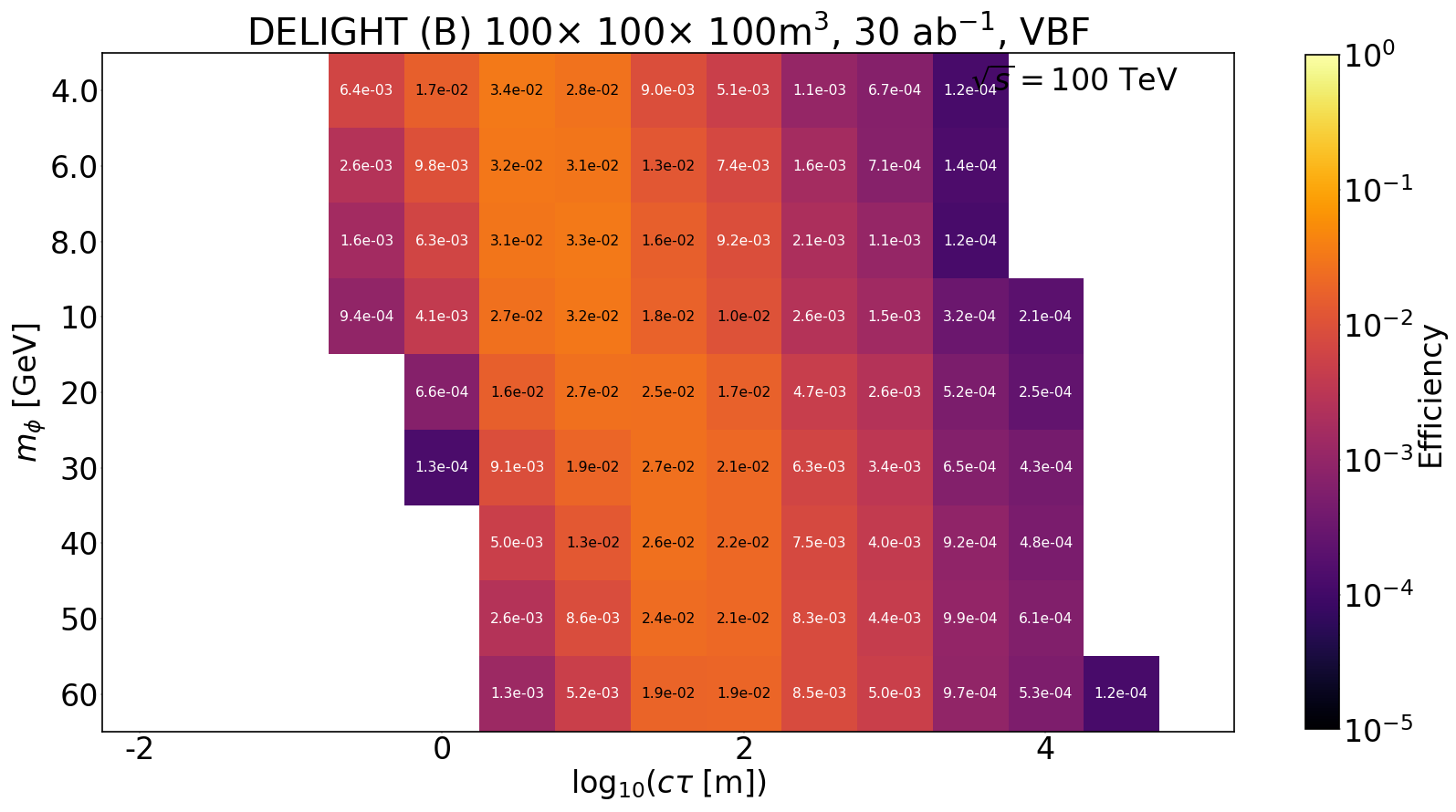} \\
    \includegraphics[width=0.46\textwidth]{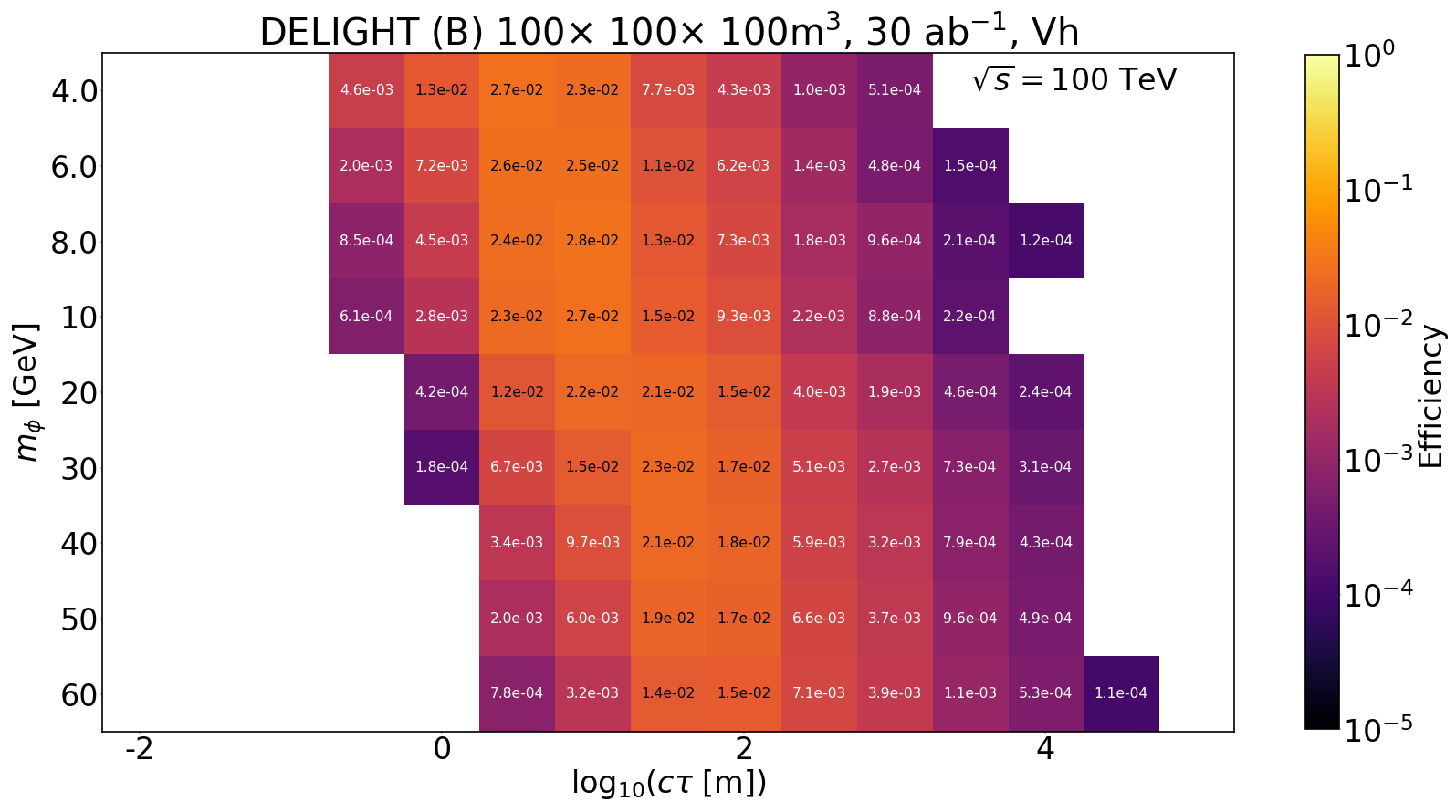}
    \caption{\small \sl Efficiency maps for decays of long-lived mediator particles within the DELIGHT\,(B) detector near the 100 TeV collider for the ggF (top panels), VBF (center panels) and Vh (bottom panels) production of the Higgs boson at 100\,TeV collider experiment.}
    \label{fig:eff-map-100TeV-LLP_B}
\end{figure}

\begin{figure}[h]
    \centering
    \includegraphics[width=0.46\textwidth]{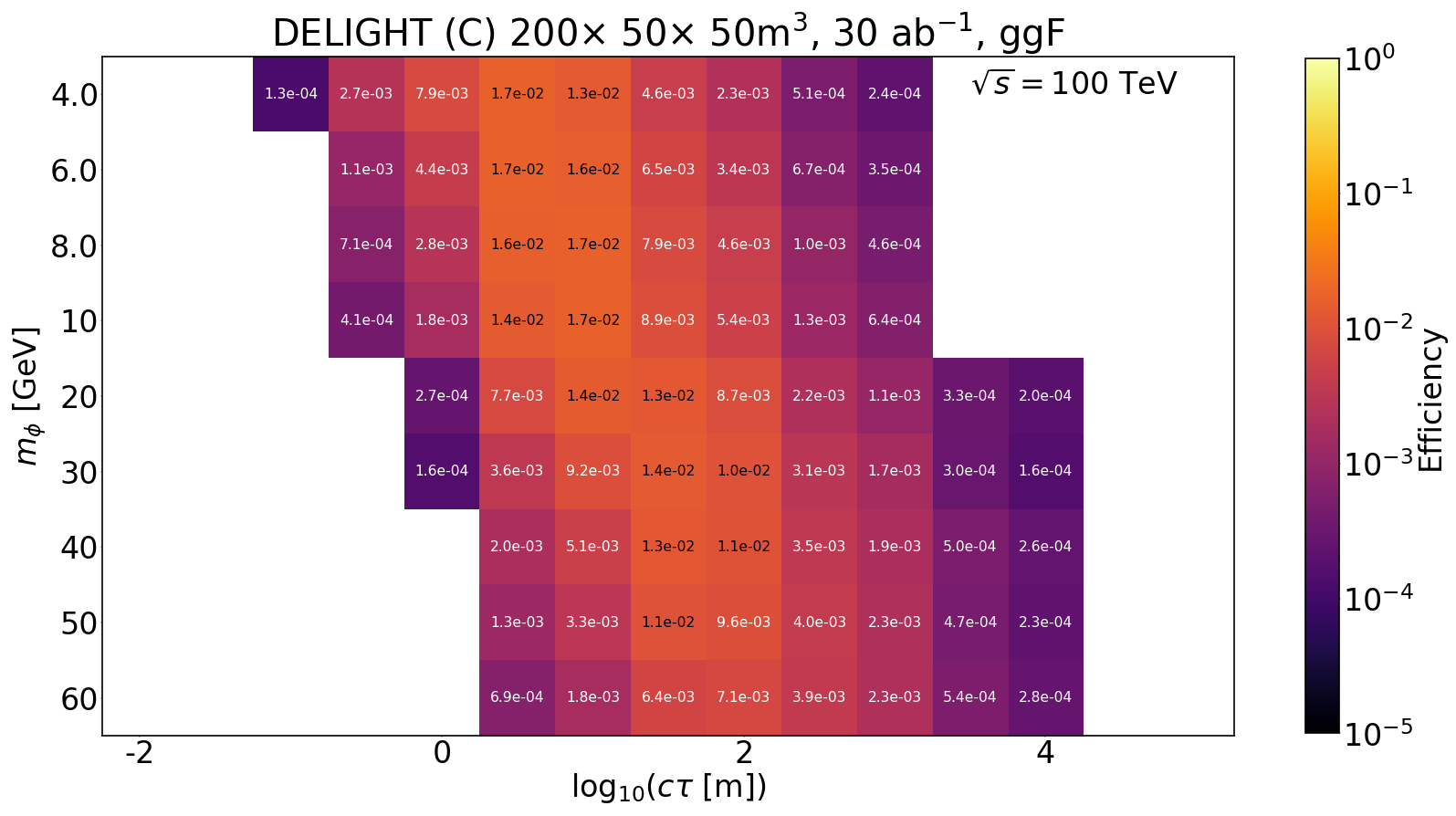} \qquad
    \includegraphics[width=0.46\textwidth]{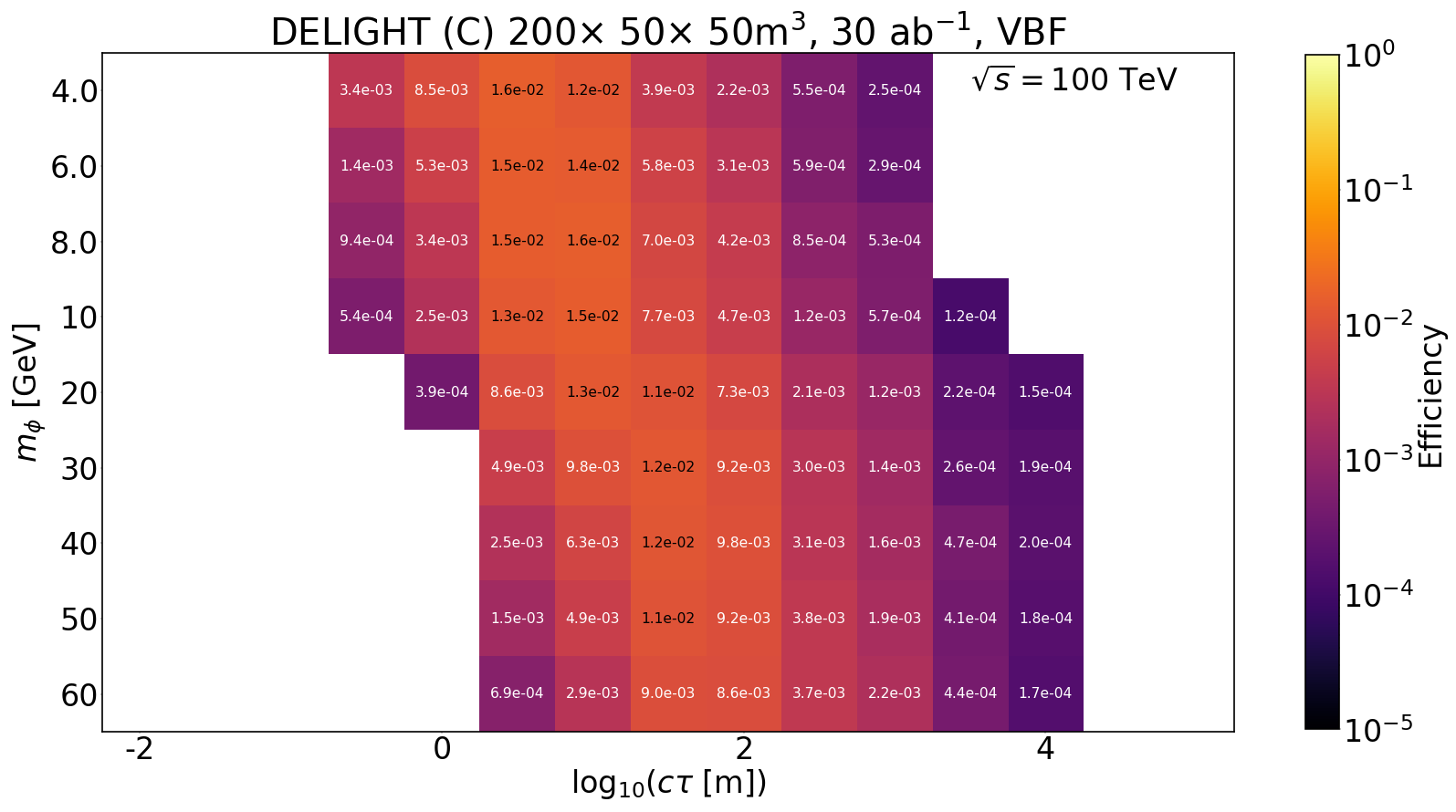} \\
    \includegraphics[width=0.46\textwidth]{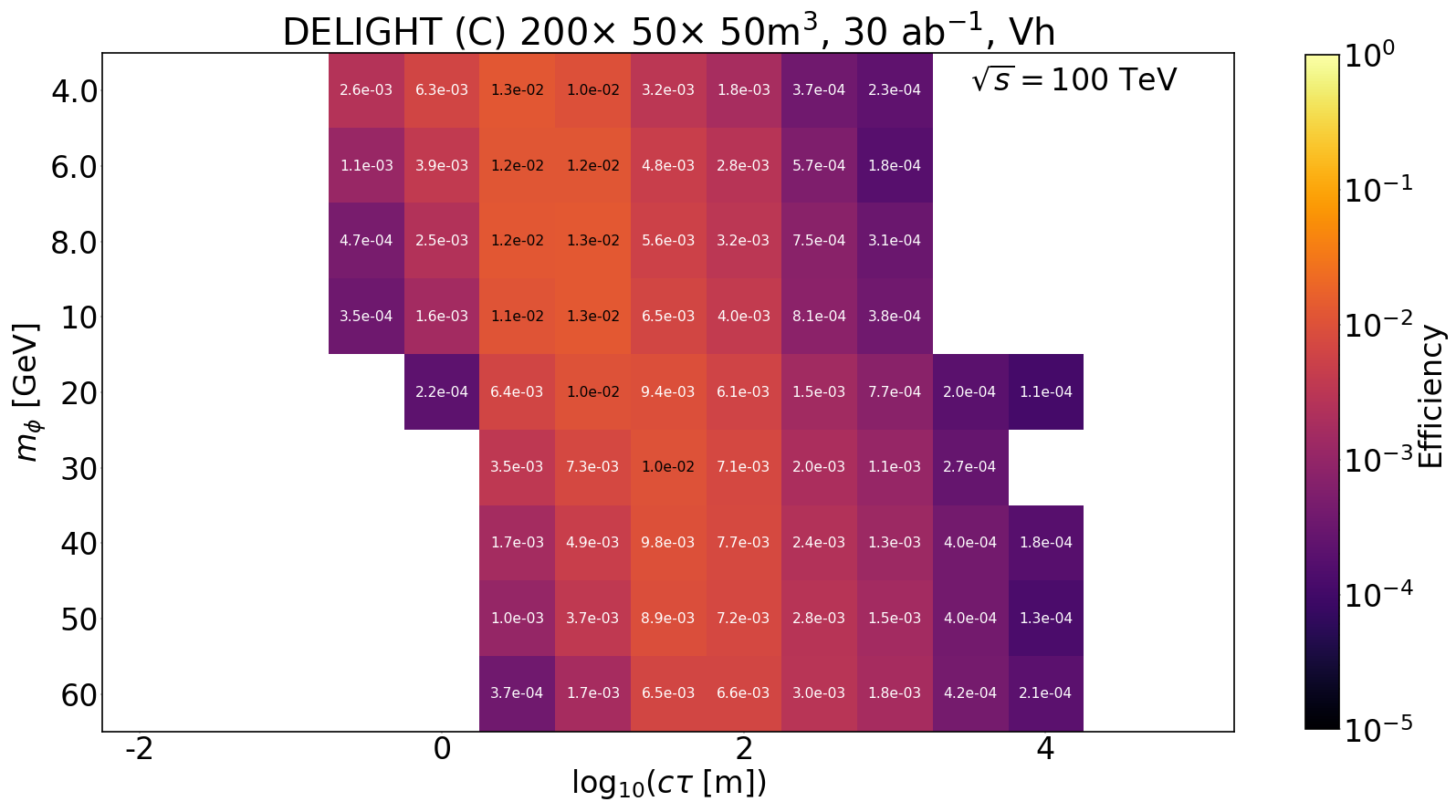}
    \caption{\small \sl Efficiency maps for decays of long-lived mediator particles within the DELIGHT\,(C) detector near the 100 TeV collider for the ggF (top panels), VBF (center panels) and Vh (bottom panels) production of the Higgs boson at 100\,TeV collider experiment.}
    \label{fig:eff-map-100TeV-LLP_C}
\end{figure}

Efficiency maps for the DELIGHT detector (A, B, C) are given in Figs.\,\ref{fig:eff-map-100TeV-LLP_A}, \ref{fig:eff-map-100TeV-LLP_B}, \ref{fig:eff-map-100TeV-LLP_C}.

\clearpage


\providecommand{\href}[2]{#2}\begingroup\raggedright\endgroup

\end{document}